\documentclass[11pt,a4paper]{article}
\pdfoutput=1

\usepackage{cite}
\usepackage{physics}
\usepackage[english]{babel}
\usepackage[utf8]{inputenc}
\usepackage{amsmath}
\usepackage{graphicx}

\usepackage[nottoc]{tocbibind}
\usepackage{amssymb}

\usepackage[colorlinks=true, allcolors=blue]{hyperref}
\usepackage{float}
\usepackage{multirow}
\usepackage{makeidx}

\newcommand{\be}{\begin{equation}}
\newcommand{\ee}{\end{equation}}
\newcommand{\bea}{\begin{eqnarray}}
\newcommand{\eea}{\end{eqnarray}}
\def\beal#1\eeal{\begin{align}#1\end{align}}   %one & only for aligning
\def\besp#1\eesp{\begin{multline}#1\end{multline}} %split an equation with first line left aligned & later right aligned

\usepackage[normalem]{ulem}

\newcommand\ie{\textit{i.e.}\ }
\newcommand\eg{\textit{e.g.}\ }
\newcommand\cf{\textit{cf.}\ }

\newcommand{\aka}{{a.k.a.}\ }

\newcommand{\viz}{{\it viz.}\ }

\title{The functional $f(R)$ approximation}

\providecommand{\keywords}[1]
{
  \small	
  \textbf{\textit{Keywords---}} #1
}
\author{Tim R. Morris and Dalius Stulga\\
\small \texttt{T.R.Morris@soton.ac.uk, D.Stulga@soton.ac.uk}\\
 \small \textit{STAG Research Centre, Department of Physics and Astronomy,}\\
  \small \textit{University of Southampton, Highfield, Southampton, SO17 1BJ, U.K.}
  }

\date{\today}
\makeindex
\begin{document}

\maketitle
\begin{abstract}
This article is a review of functional $f(R)$ approximations in the asymptotic safety approach to quantum gravity. It mostly focusses on a formulation that uses a non-adaptive cutoff, resulting in a second order differential equation. This formulation is used as an example to give a detailed explanation for how asymptotic analysis and Sturm-Liouville analysis can be used to uncover some of its most important properties. In particular, if defined appropriately for all values $-\infty<R<\infty$, one can use these methods to establish that there are at most a discrete number of fixed points, that these support a finite number of relevant operators, and that the scaling dimension of high dimension operators is universal up to parametric dependence inherited from the single-metric approximation. Formulations using adaptive cutoffs, are also reviewed, and the main differences are highlighted.   
\end{abstract}
\newpage
\tableofcontents
\newpage

\section{Introduction}

One attempted route to a quantum theory of gravity is through the asymptotic safety programme \cite{Weinberg:1980,Reuter:1996,Percacci:2017fkn,Reuter:2019byg}. Although quantum gravity based on the Einstein-Hilbert action is plagued by ultraviolet infinities that are perturbatively non-renormalizable (implying the need for an infinite number of coupling constants), a sensible theory of quantum gravity might be recovered if there exists a suitable ultraviolet fixed point \cite{Weinberg:1980}. 

The task is not just that of searching for an ultraviolet fixed point. They must also have the correct properties. Perturbatively renormalizable ones exist for example
``Conformal gravity'', based on the square of the Weyl tensor, which thus corresponds to a Gaussian ultraviolet fixed point \cite{Stelle:1976gc}. It is apparently not suitable however, because the theory is not unitary. Suitable unitary fixed points, if they exist, have to be non-perturbative. \index{unitarity} They must also satisfy phenomenological constraints, \index{phenomenology} for example they have to allow a renormalized trajectory with classical-like behaviour in the infrared, since General Relativity is confirmed by observation across many phenomena and to impressive precision. Of particular relevance for this chapter is that there should be a fixed point with a finite number of relevant directions (otherwise it would be no more predictive than the perturbatively defined theory). Preferably the theory should have only one fixed point, or at least only a finite number (otherwise again we lose predictivity). 

Functional RG (renormalization group) equation \cite{Wilson:1973,Wegner:1972ih,Polchinski:1983gv,Nicoll1977,Wetterich:1992,Morris:1993} studies,  first introduced by Wilson and Wegner many years ago \cite{Wilson:1973,Wegner:1972ih} (and called by them the ``exact RG''),  have flourished into a powerful approach for investigating this possibility. \index{renormalization group: functional} \index{renormalization group: exact} \index{renormalization group: Wilsonian}
These equations describe the flow of the Wilsonian effective action for some quantum field theory, under changes in an effective cutoff scale $k$.
The asymptotic safety literature uses almost exclusively the flow equation for $\Gamma_k$ which is, modulo minor details, the Legendre effective action (the generator of one-particle irreducible diagrams) cut off in the infrared by $k$. It was also formulated long ago \cite{Nicoll1977} (in the sharp cutoff limit) and then rediscovered for smooth cutoffs much later in refs. \cite{Wetterich:1992,Morris:1993}. Following ref. \cite{Wetterich:1992}, $\Gamma_k$ is sometimes called the ``effective average action'', however in this chapter it will simply be called an effective action.\index{effective average action}

It is not practical to solve the full functional RG equations exactly. In a situation such as this, where there are no useful small parameters, one can only proceed by considering model approximations. These always proceed from the following observation: Wilsonian effective actions can be written as a sum over operators,  where the coefficients are the couplings for these operators and they evolve with the scale $k$. 

In fact this sum should be restricted to local operators. This is the requirement of quasi-locality, which comes from the short range nature of the Kadanoff blocking step in
Wilsonian RG \cite{Wilson:1973}, when implemented in the continuum \cite{Morris:1999px,Morris:2000jj}. A related point is that the Wilsonian RG is performed in euclidean signature, so that ``{short range}'' has a sensible meaning. \index{quasi-locality} %\index{locality}  \index{Kadanoff blocking}

The problem is that for any general solution, this sum is infinite, over all possible local operators allowed by the symmetries (the space of all such couplings being known as ``theory space''). %\index{theory space} 
However, this motivates the  simplest model approximation which is to truncate drastically the infinite dimensional theory space to a handful of operators. An example is the original truncation studied by Reuter \cite{Reuter:1996}:
\be 
  \Gamma_k[g_{\mu\nu}]=\int d^4x \sqrt{g}\big( u_0(k)+u_1(k)R \big)\,, \label{EHtrunc}
\ee 
which retains only the cosmological constant term and the scalar curvature $R$ term. 
For obvious reasons this is called the ``Einstein-Hilbert truncation''. \index{truncation} Classically $u_0= -\lambda_{cc}/(8\pi G)$ and $u_1 =-1/(16\pi G)$, where $\lambda_{cc}$ is the cosmological constant and $G$ is Newton's constant, but after quantum corrections these couplings run with $k$ in the functional RG. The minus sign in $u_1$ comes from working in euclidean signature. 
%As we will see, this leads to its own problems \cite{Mitchell:2021qjr}.

Apart from RG symmetry, these truncations destroy pretty well all the properties that ought to hold. For example  scheme independence (\ie independence on choice of cutoff, or more generally universality), and modified BRST invariance \cite{Ellwanger:1994iz,Reuter:1996} (which encodes diffeomorphism invariance for the quantum field under influence of the cutoff) cannot then be recovered. Furthermore, only by keeping an infinite number of local operators can the non-local long-range nature of the (one-particle irreducible) Green's functions be recovered (see \eg ref. \cite{Morris:2018zgy}).
One has to trust that by considering ever less restrictive truncations the description gets closer to the truth. 
There are some examples that go well beyond the Einstein-Hilbert truncation by keeping a large number of operators \cite{Falls_2014,Falls:2018ylp,Falls:2017lst,Kluth:2020bdv}. These are based around polynomial truncations, \ie where everything is discarded except powers of some suitable local operators, typically the scalar curvature $R$ again, up to some maximum degree. They appear to show convergence, in particular the number of relevant operators is found to be three.

Another approximation in the asymptotic safety literature that is necessary in order to formulate diffeomorphism invariant truncations, such as eqn. \eqref{EHtrunc}, conflates the true (quantum) metric with the background metric. It is called the ``single metric'' or ``background field'' approximation, \index{single metric approximation} \index{background field approximation} and will be described in the next section. It is harder to relax this approximation in any substantive way, although see refs. \cite{Becker:2014qya,Becker:2021pwo,Morris:2016spn,Percacci:2016arh, Ohta:2017dsq,Morris:2016nda,Falls:2020tmj,Mandric:2022dte,Pawlowski:2020qer}  for some approaches. 

Whilst very encouraging results are found from multiple studies of such finite order truncations (see  \eg the review \cite{Bonanno:2020bil}), successful implementations of more powerful approximations would build confidence in the scenario. The next step is to keep an infinite number of operators.
Arguably the simplest such truncation is to keep a full function $f(R)$, making the ansatz \cite{Codello:2007bd,Machado:2007,Codello:2008,Benedetti:2012,Demmel:2012ub,Demmel:2013myx,Demmel:2014hla,Demmel:2014fk,Demmel2015b,Ohta:2015efa,Ohta2016,Percacci:2016arh,Morris:2016spn,Falls:2016msz,Ohta:2017dsq,Benedetti:2013jk,Mitchell:2021qjr}
\be
\label{fR}
\Gamma_k[g]=\int \!d^4x\sqrt{g}\, f_k(R)\,.
\ee
%(plus ghost and auxiliary field terms). 
This is the \textit{functional $f(R)$ approximation} which is the subject of this chapter. \index{functional $f(R)$ approximation} It is achieved by specialising to a maximally symmetric background manifold, either a four-sphere or four-hyperboloid.

Closely related approximations have been studied in scalar-tensor 
\cite{Percacci:2015wwa,Labus:2015ska} and unimodular \cite{Eichhorn:2015bna} gravity, and in three space-time dimensions \cite{Demmel:2014fk}.
In fact, the high order finite dimensional truncations \cite{Falls_2014,Falls:2018ylp,Falls:2017lst,Kluth:2020bdv} were developed by taking examples of these $f(R)$ equations and then further approximating to polynomial truncations. 

Note that the functional $f(R)$ approximation actually goes beyond  keeping a countably infinite number of couplings, the Taylor expansion coefficients $g_n=f^{(n)}(0)$, because \textit{a priori} the large field parts of $f(R)$ contain degrees of freedom that are unrelated to all these $g_n$. 
For example suppose that at large $R$ one finds that $f(R)\approx\exp( -a/R^2)$, where $a>0$ is some parameter. Such an $f(R)$ is in the form of a standard counter-example in mathematical analysis. It has the property that $g_n=0$ for all $n$.

As ref.  \cite{Benedetti:2012} emphasised, the truncation \eqref{fR} is as close as one can get to the Local Potential Approximation (LPA) \index{local potential approximation} \cite{Hasenfratz:1985dm,Morris:1994ki}, a successful approximation for scalar field theory in which only a general potential $V(\varphi)$ is kept for a scalar field $\varphi$ (see \eg \cite{Hasenfratz:1985dm,Morris:1994ki,Morris:1994ie,Morris:1994jc,Morris:1996xq,Morris:1998}). 
The  LPA can be viewed as the start of a systematic derivative expansion \cite{Morris:1994ie}, in which case this lowest order corresponds to regarding the field $\varphi$ as constant. In rough analogy, an approximation of form \eqref{fR} may be derived by working on a euclidean signature space of maximal symmetry, where the scalar curvature $R$ is constant. (Typically a four-sphere is chosen.) In particular, techniques that have proved successful in scalar field theory \cite{Morris:1994ki,Morris:1994ie,Morris:1994jc,Morris:1996xq,Morris:1996nx,Morris:1998} have been adapted to this very different context, and used to gain substantial insight \cite{Dietz:2012ic,Dietz:2013sba,Bridle:2013sra,Gonzalez_Martin_2017,Mitchell:2021qjr}.

The functional truncation \eqref{fR} still has the problems that were highlighted earlier for its finite dimensional counterparts. However, again one can hope that it is closer to the truth. One hint that this is in fact the case is covered at the end of this chapter. Assuming that the most recent version \cite{Mitchell:2021qjr} does have a fixed point solution, then it turns out that operators with high scaling dimension do begin to display universality --  unfortunately up to an annoying parameter that remains which is clearly caused by the single-metric approximation.\index{single metric approximation} \index{fixed singularity} \index{poor asymptotic behaviour} \index{second order formulation} \index{third order formulation}

In this chapter, it will be explained how to construct functional $f(R)$ approximations and how to interpret them. Important properties of formulations that use an adaptive cutoff \cite{Codello:2007bd,Machado:2007,Codello:2008,Benedetti:2012,Demmel:2012ub,Demmel:2013myx,Demmel:2014hla,Demmel:2014fk,Demmel2015b,Ohta:2015efa,Ohta2016,Percacci:2016arh,Morris:2016spn,Falls:2016msz,Ohta:2017dsq} will be reviewed. These result in third order differential equations, with fixed singularities and problematic asymptotic behaviour. Mostly the chapter will focus on a non-adaptive cutoff formulation \cite{Benedetti:2013jk,Mitchell:2021qjr} that results in a second order differential equation, using it as an example to give a detailed exposition of the techniques, especially asymptotic analysis and Sturm-Liouville analysis, that can be used to prove properties of functional $f(R)$ approximations. In particular, if the second order formulation is taken to apply to only one of the two spaces (sphere or hyperboloid), the fixed point solutions form a continuous set and the eigenoperator spectrum is not quantised. However, if these spaces are joined together smoothly (through flat space at their boundary), these methods establish that there are at most a discrete number of fixed points, that the fixed points support a finite number of relevant operators, and yield the result above for operators of high scaling dimension. They do not establish that such fixed points actually exist however. Such a demonstration requires more powerful numerical analysis and/or simpler fixed point formulations \cite{Mitchell:2021qjr}.

\section{Flow equations}
\label{sec:flowequations}
The starting point is the functional RG flow equation \cite{Wetterich:1992,Morris:1993}: \index{renormalization group: functional}
\be 
  \partial_t\Gamma_k=\frac{1}{2}\text{STr}\Big[(\Gamma^{(2)}_k+\mathcal{R}_k)^{-1}\partial_t\mathcal{R}_k\Big], \label{LegFlow}
\ee 
where Str is a functional trace over spacetime coordinates and indices that takes into account statistics of the fields. In momentum space this is an integral over one loop momentum. The right hand side is a one-loop integral if $\Gamma_k$ is taken to be classical. It includes all higher loops because $\Gamma_k$ is actually given by the full (all orders) effective action. $\Gamma_k^{(2)}$ (Hessian) is the second variation of 
the effective action with respect to the fields and $\mathcal{R}_k$ is an IR (infrared) cutoff for these fields. The cutoff scale $k$ is related to RG ``time'' $t$ via $t=\ln(k/\mu)$, where $\mu$ is the standard arbitrary physical energy scale that appears in RG treatments (including in perturbative quantum field theory). %\index{renormalization group time}

An essential step in Wilsonian RG is to introduce dimensionless variables by multiplication of appropriate powers of the cut-off scale $k$. In the $f(R)$ approximation the appropriate powers are just the canonical (\aka engineering) ones:
\be 
\label{dimensionless}
  \tilde{f}_k(\tilde{R}) \equiv \tilde{f}(\tilde{R},t) = k^{-4}f_k(k^2\tilde{R}), \qquad \tilde{R}={R}/{k^2} \,.
\ee 
In Wilsonian RG, one integrates out modes, starting with the high momentum modes first, by a  coarse-graining procedure. Traditionally, after integrating out the modes, one has to rescale the action back to the original UV cut-off of the theory to see how the couplings change. By working with dimensionless quantities this is taken care of automatically. %\index{dimensionless quantities by rescaling}

(From this point onwards it is convenient to drop the tilde denoting dimensionless quantities, unless otherwise specified, but the reader should assume that all the quantities are dimensionless.)

In this way solutions to the flow equation will reveal all the \index{fixed point} fixed points of the theory, \ie $t$ independent solutions ${f}({R},t) = {f}({R})$. 
%These could be Gaussian (non-interacting) or non-Gaussian (interacting) fixed points. 
Fixed points are characterized by the number of eigenoperators ${v}({R})$ (operators of definite scaling dimension) that flow into the fixed point when we increase the cutoff scale $k$. \index{eigenoperator} These are called relevant eigenoperators. Conversely irrelevant operators are the ones that flow away from the fixed point. \index{relevant} \index{irrelevant} The terms relevant (irrelevant) are common in the Wilsonian RG literature. Following Weinberg \cite{Weinberg:1980}, in asymptotic safety literature often they are referred to equivalently as essential (inessential). Eigenoperators whose couplings do not flow (in some approximation or exactly) are called marginal. We do not need to discuss them in this chapter. Exceptionally eigenoperators can appear that are ``redundant'', corresponding to a change of variables in the theory \cite{WR,Dietz:2013sba,Latorre:2000jp}. \index{marginal} \index{redundant}

Eigenoperators are found by linearising the flow equations around the fixed point and separating variables:
\be 
  f_k(R)=f(R)+\epsilon\, v(R)\,\mathrm{e}^{-\theta t} \label{eq:pert}
\ee 
where $\epsilon$ is a small parameter.  \index{eigenoperator} This turns the flow equation into an eigenvalue problem where the RG eigenvalue $\theta$ is often called a ``critical exponent'' in the asymptotic safety literature. \index{critical exponent} From its associated $v(R)$ it can be similarly classified as relevant, irrelevant, marginal or redundant. Thus if $\Re\theta>0$ then it is relevant, whilst if $\Re\theta<0$ it is irrelevant. In statistical physics, non-redundant $\theta$ can be straightforwardly related to experimentally defined and measurable critical exponents, see \eg \cite{ZinnJustin:2002ru}. If computed correctly an important property of a non-redundant $\theta$ is that it is universal, which means in particular that its value is independent of the regularisation scheme and the choice of flow equation \cite{WR}. \index{universality}

As already intimated, one is generally interested in those fixed points that have finitely many relevant operators, because their couplings become the  free parameters in the theory, and will have to be fixed by experiments. Thus, theories based around these points are predictive and are safe from UV divergences when $k\rightarrow\infty$. The goal of the asymptotic safety program is to verify if such points exist for gravity, analyse their properties and deduce their consequences, both qualitatively and quantitatively. \index{phenomenology}

%\section{$f(R)$ approximation}

Actually, the flow equation \eqref{LegFlow} requires a significant amount of adaptation to deal with the fact that quantum gravity is a gauge theory. In standard fashion, it therefore requires gauge fixing. This is commonly done by employing the \index{background field method} background field method where the full (\aka total) metric $\hat{g}_{\mu\nu}$ is split into a background $g_{\mu\nu}$ plus fluctuations (the quantum field): \index{gauge fixing}
\be 
\label{split}
  \hat{g}_{\mu\nu}={g}_{\mu\nu}+h_{\mu\nu}\,.
\ee 
In common with most of the literature, this chapter will only use a linear split,
although other, non-perturbatively better motivated, splits are possible \cite{Ohta:2015efa,Bonanno:2020bil}. Then the gauge fixing is imposed on the quantum field $h_{\mu\nu}$ in such a way that diffeomorphism invariance of the background metric ${g}_{\mu\nu}$ is retained:
\be 
  F_\mu={\nabla}^\nu h_{\mu\nu}-\frac{1}{4}{\nabla}_\mu h^\nu_\nu\,,
\ee 
where the covariant derivative and raised indices, are defined using the background metric. The process of fixing a gauge, adds the gauge fixing term
\begin{equation}
\label{gf}
    \Gamma_{gf}=\frac{1}{2\alpha}\int d^4x \sqrt{{g}} {g}^{\mu\nu}F_\mu F_\nu
\end{equation}
to the effective action, and leads also to a ghost action. In practice the Landau gauge is chosen: $\alpha\to0$. Finally, it proves useful to make a change of variables, this is explained in \eqref{TT}, and this leads to further, auxiliary, fields. 

The true solution involves arbitrarily complicated interactions to arbitrarily high order between all these fields, molified only by the symmetries (in particular  background field diffeomorphism invariance and modified BRST invariance \cite{Ellwanger:1994iz,Igarashi:2019gkm}).  The next steps in the approximation drastically truncates all of this \cite{Reuter:1996}. It can be summarised as follows. Only the one-loop contributions from the bilinear ghost and auxiliary field and fluctuation field actions are retained, \ie on the right hand side of the flow equation \eqref{LegFlow} only the Hessian from the classical action for these fields is used.  The flow of the bit of the effective action that only depends on the background metric, is therefore reproduced correctly at one loop. For the part beyond one loop, the correct Hessian in \eqref{LegFlow} for the metric,
\be 
\frac{\delta^2\Gamma_k}{\delta h_{\mu\nu}(x)\, \delta h_{\alpha\beta}(y)} \,,
\ee
is replaced by one in which the functional derivatives are with respect to the background field instead:
\be \frac{\delta^2\Gamma_k}{\delta {g}_{\mu\nu}(x)\, \delta {g}_{\alpha\beta}(y)}\,. \ee
This is the single metric, or background field, approximation. \index{single metric approximation} \index{background field approximation} It is almost always applied in asymptotic safety investigations. The review \cite{Pawlowski:2020qer} covers exceptions. It should be emphasised that already at one loop the single metric approximation is not correct, because  the dependence of the effective action on $h_{\mu\nu}$ has no direct relation to its dependence on $g_{\mu\nu}$. The replacement above would be correct only if the effective action were a functional of the full metric \eqref{split} alone, but that relation is broken at the classical level by the gauge fixing term \eqref{gf} (and corresponding ghost action). Nevertheless the replacement is attractive as a model, because it leaves us with a flow equation for $\Gamma_k[g]$  that depends only on the background metric and in a  diffeomorphism invariant way.

Now by choosing the background manifold to be one of maximal symmetry, all diffeomorphism invariants can be related either to the volume or the scalar curvature $R$, which is a constant: $\partial_\mu R=0$. In this way the effective action has been reduced to \eqref{fR}: the functional $f(R)$ approximation. \index{functional $f(R)$ approximation}

%\section{Flow equations: details}

Plugging %the functional $f(R)$ approximation  \eqref{fR} 
this with appropriately scaled fields \eqref{dimensionless} (and coordinates $\tilde{x}^\mu=k x^\mu$), into the flow equation \eqref{LegFlow}, one readily derives the form of the left hand side:
\be 
  \partial_t\Gamma_k=\int\!\! d^4{x} \sqrt{g} \big[\, \partial_t {f}_k ({R})+4{f}_k({R})-2{R}{f}'_k({R})\,\big].
\ee 
The right hand side of the flow equation depends on the detailed way the quantum corrections are handled, which differs between authors \cite{Codello:2007bd,Machado:2007,Codello:2008,Benedetti:2012,Demmel:2012ub,Demmel:2013myx,Benedetti:2013jk,Demmel:2014hla,Demmel:2014fk,Demmel2015b,Ohta:2015efa,Ohta2016,Percacci:2016arh,Morris:2016spn,Falls:2016msz,Ohta:2017dsq,Mitchell:2021qjr}. 
%However core ideas of the functional $f(R)$ approximation are common. 
%This chapter will mostly focus on a formulation \cite{Benedetti:2013jk,Mitchell:2021qjr} that results in a second order differential equation, using it as an example to give a detailed exposition of the relevant techniques required and the properties of functional $f(R)$ approximations that these uncover. Along the way we point out important differences in the rest of the literature \cite{Codello:2007bd,Machado:2007,Codello:2008,Benedetti:2012,Demmel:2012ub,Demmel:2013myx,Demmel:2014hla,Demmel:2014fk,Demmel2015b,Ohta:2015efa,Ohta2016,Percacci:2016arh,Morris:2016spn,Falls:2016msz,Ohta:2017dsq}, which use a formulation that results in a third order differential equation.
For this we need to compute the second variation of $\Gamma_k$ with respect to the fields. First, the gauge fixing term \eqref{gf} is chosen and the ghost action is derived. Then the transverse \index{transverse traceless decomposition} traceless (\aka York)  decomposition of the metric \cite{York:1973ia} is used:
\be 
\label{TT}
  h_{\mu\nu}=h^T_{\mu\nu}+{\nabla}_\mu\xi_{\nu}+{\nabla}_\nu\xi_{\mu}+{\nabla}_\mu {\nabla}_\nu\sigma+\frac{1}{d}{g}_{\mu\nu}\bar{h}\,,
\ee 
which separates physical degrees of freedom, \viz $h^T_{\mu\nu}$ and $\bar{h}$, from the unphysical ones associated with gauge degrees of freedom, namely $\xi_\mu$ and $\sigma$. These fields satisfy
\be 
  {h^T}_\mu^\mu=0, \quad {\nabla}^\mu h^T_{\mu\nu}=0, \quad {\nabla}^\mu \xi_\mu=0, \quad \bar{h}=h-{\nabla}^2\sigma.
\ee 
Expressing $\sqrt{\hat{g}}$ and $\hat{R}$, where the latter is the curvature of the full metric \eqref{split},
to quadratic order in these fields, the elements of the Hessian can be determined for these components. For example for the physical components one finds \index{Hessian}
\be 
\label{hesshT}
  {\Gamma}^{(2)}_{h^T_{\mu\nu}h^T_{\alpha\beta}}=-\frac{1}{2}\bigg[f'_k(R) \bigg(-{\nabla}^2+\frac{1}{6}{R}\bigg)+\bigg(f_k-\frac{1}{2}{R}f'_k\bigg)\bigg]\delta^{\mu\nu,\alpha\beta}\,,
\ee 
\be
\label{hesshbar}
  {\Gamma}^{(2)}_{\bar{h}\bar{h}}=\frac{1}{16}\bigg[ 9f''_k\bigg(-{\nabla}^2-\frac{{R}}{3}\bigg)^2+3f'_k\bigg(-{\nabla}^2-\frac{{R}}{3}\bigg)-\bigg({R}f'_k-2f_k\bigg)\bigg]\,,
\ee
where the right hand side is evaluated at $\hat{g}_{\mu\nu} = g_{\mu\nu}$,  in preparation for the single metric approximation. \index{single metric approximation} We can write these more compactly if we introduce \index{equation of motion}
\be 
\label{EoM}
E_k(R)=2f_k(R)-Rf'_k(R)\,,
\ee
which is the equation of motion that follows from the action \eqref{fR}, and express them instead using the natural Laplacian $\Delta_s$ for a spin $s$ component field (on a maximally symmetric background) \cite{Benedetti:2012}: \index{natural Laplacian $\Delta_s$}
\be 
\label{modifiedLaplacians}
  \Delta_0=-\nabla^2-\frac{R}{3}, \quad \Delta_1=-\nabla^2-\frac{R}{4}, \quad \Delta_2=-\nabla^2+\frac{R}{6} \,.
\ee

A similar decomposition is applied to the ghost action. In the formulation of ref. \cite{Benedetti:2012} the contribution to the Hessian coming from the gauge degrees of freedom from the metric and the ghosts cancel each other exactly. Finally, including the contributions of the auxiliary fields that encode the Jacobians due to the transverse traceless decomposition of the metric and the ghost fields, gives the full flow equation \eqref{LegFlow}  in the single metric and functional $f(R)$ approximation: \index{functional $f(R)$ approximation} %\index{flow equation}
\be 
    V\big(\partial_tf_k(R) +2E_k(R)\big)=\mathcal{T}_2+\mathcal{T}_0^{\Bar{h}}+\mathcal{T}_1^{Jac}+\mathcal{T}_0^{Jac}\,,  \label{eq:4}
\ee 
where $V=\int d^4x \sqrt{g}$ is the volume of the manifold, and the $\mathcal{T}$ objects are the following spacetime traces: \index{traces}
\be 
\label{trtensor}
\mathcal{T}_2=\text{Tr}\Big[\frac{d_t \mathcal{R}^T_k}{-f'_k(R)\Delta_2-E_k(R)/2 +2\mathcal{R}^T_k}\Big]\,,
\ee 
\begin{equation}
   \mathcal{T}_0^{\Bar{h}}=\text{Tr}\Big[\frac{8\,d_t \mathcal{R}^{\Bar{h}}_k}{9f''_k(R)\Delta_0^2+3f'_k(R)\Delta_0+E_k(R) +16\mathcal{R}^{\Bar{h}}_k}\Big]\,, \label{trhbar}
\end{equation}
\be 
\label{trJac1}
    \mathcal{T}_1^{Jac}=-\frac{1}{2}\text{Tr}\Big[\frac{d_t\mathcal{R}_k^V}{\Delta_1+\mathcal{R}^V_k}\Big] \,,
\ee 
\be\label{trJac0} 
    \mathcal{T}_0^{Jac}=\frac{1}{2}\text{Tr}\Big[\frac{d_t\mathcal{R}_{S_1}^V}{\Delta_0+R/3+\mathcal{R}^{S_1}_k}\Big]
    -\text{Tr}\Big[\frac{2\,d_t\mathcal{R}_{S_2}^V}{(3\Delta_0+R)\Delta_0+4\mathcal{R}^{S_2}_k}\Big]\,. 
\ee
The right hand side of \eqref{eq:4} has been subdivided into contributions coming from fields of different spins. The first two come from the physical spin-2 traceless part of the metric and the spin-0 trace of the metric, as the reader can see by using \eqref{hesshT} and \eqref{hesshbar} in \eqref{LegFlow}. The last two are spin-1 and spin-0 parts coming from field redefinitions.
%For convenience modified Laplacians are introduced with terms appearing naturally in the flow equations \cite{Benedetti:2013jk}:
%\be 
%\label{modifiedLaplacians}
%  \Delta_0=-\nabla^2-\frac{R}{3}, \quad \Delta_1=-\nabla^2-\frac{R}{4}, \quad \Delta_2=-\nabla^2+\frac{R}{6} \,.
%\ee 
%(The first and the last appear in the examples (\ref{hesshT},\ref{hesshbar}).) Furthermore
%The cutoffs have been chosen to depend on the $\Delta_i$, rather than simply the $-\nabla^2$ part \cite{Benedetti:2012}.
%and furthermore include a curvature correction with coefficients $\alpha_s$ \cite{Benedetti:2013jk}. 
%This last is to ensure that by adjusting these coefficients all the modes are integrated out as $k \rightarrow 0$. This in turn is guaranteed if a condition $\Delta_s+\alpha_sR > 0$ is satisfied. 

\section{Cutoff functions}
\label{sec:cutoffs}

 \index{cutoff}
One place where crucial differences occur between the different implementations is in the choice of cutoff $\mathcal{R}_k$.
There is quite a lot of freedom as these functions only need to satisfy a few key properties which ensure that they behave like momentum dependent mass terms suppressing low momentum modes:
\be 
  \lim_{p^2 \rightarrow0} \mathcal{R}_k(p^2)>0, \quad \lim_{p^2\rightarrow \infty}\mathcal{R}_k(p^2)=0, \quad \lim_{k\rightarrow0}\mathcal{R}_k(p^2)=0\,.
\ee 
The first two conditions ensure that we integrate out the UV modes first and ignore the IR modes. The last condition ensures that we are left with the standard definition of the effective action once the cutoff scale is sent to zero. 

An apparently attractive strategy is to choose cutoffs that simplify the flow equations as much as possible. \index{cutoff} ``Adaptive cutoffs'' are introduced partly with that aim \cite{Codello:2007bd,Machado:2007,Codello:2008,Benedetti:2012,Demmel:2012ub,Demmel:2013myx,Demmel:2014hla,Demmel:2014fk,Demmel2015b,Ohta:2015efa,Ohta2016,Percacci:2016arh,Morris:2016spn,Falls:2016msz,Ohta:2017dsq}. They implement the following rule for all appearances of the Laplacian operator $-\nabla^2$:
\be 
  -\nabla^2 \mapsto -\nabla^2 +k^2 r(-\nabla^2/k^2)\,, \label{replaceRule}
\ee 
where $r(z)$ is a cutoff profile function. \index{cutoff} 

Such a choice also seemingly solves an  awkward feature of euclidean quantum gravity, which is that the euclidean signature Einstein-Hilbert action \eqref{EHtrunc} has a wrong-sign \index{wrong sign} kinetic term and propagator for $\bar{h}$, the so-called conformal instability\cite{Gibbons:1978ac}. \index{conformal instability} This can be seen in the negative coefficient for $\Delta_0$ in \eqref{trhbar} in this case. By implementing \eqref{replaceRule}, the cutoff automatically adapts to this wrong sign, so that it continues to modify the propagator in the intended way: by adding a momentum dependent mass term. Indeed if this were not done, the cutoff and kinetic term would have opposite signs, resulting in a singular propagator. However, this trick does not entirely cure the problem since it results in poor asymptotic (large $R$) behaviour. \index{poor asymptotic behaviour} This issue will be briefly touched on below and in sec. \ref{sec:flowadaptivecutoff}. For further discussion, see refs.\cite{Reuter:1996,Dietz:2012ic,Dietz:2016gzg,Morris:2018mhd,Morris:2018upm,Mitchell:2021qjr}.

Technically the above replacement rule is implemented by setting 
\be 
\label{adaptive}
\mathcal{R}^\phi_k=\Gamma^{(2)}_k\big[-\nabla^2+k^2r(-\nabla^2/k^2)\big]-\Gamma^{(2)}_k[-\nabla^2]\,,
\ee
for each mode $\phi$,
so that the desired effect is created for $\Gamma^{(2)}_k[-\nabla^2]+\mathcal{R}_k$ in the flow equation \eqref{LegFlow}. Notice that the cutoff function is then of the same form as the Hessian elements themselves and thus now also depends on $f(R)$. This has a particular consequence for the scalar $\bar{h}$ mode, since $\Gamma^{(2)}_{\bar{h}\bar{h}}$ contains $f''_k(R)$, \cf eqn. \eqref{hesshbar}. It means that plugging this type of cutoff into the flow equation will result in the appearance of $Rf'''_k(R)$,  due to the presence of $d_t\mathcal{R}^{\bar{h}}_k$ in the numerator in \eqref{trhbar} and the definition \eqref{dimensionless} of $f_k(R)$. This makes the flow equation a third order differential equation, which unfortunately lacks the powerful properties found in a second order formulation (as covered in sec. \ref{sec:flowadaptivecutoff}). Furthermore, the factor of $R$ leads to a so-called ``fixed singularity'' at $R=0$. \index{fixed singularity} Third order formulations suffer from further fixed singularities and, as already mentioned, poor asymptotic behaviour, this latter leading to continuous eigenoperator spectra \cite{Dietz:2012ic}. \index{poor asymptotic behaviour} These problems will be further covered in sec. \ref{sec:flowadaptivecutoff}.

When using an  adaptive cutoff, the cutoff profile function $r(-\nabla^2/k^2)$  is almost always chosen to be the ``optimised'' profile \cite{Litim:2001} \index{cutoff}
\be 
  r(z)=(1-z)\,\theta(1-z). \label{optimised}
\ee 
The advantage of using this setup is that $d_t\mathcal{R}_k\propto \theta(1+\nabla^2/k^2)$, and thus the eigenvalues of $-\nabla^2$ are restricted to be less than $k^2$. This means that in denominators one can simply ignore the $\theta$ and thus $k^2r(-\nabla^2/k^2) \equiv k^2+\nabla^2$. Therefore the net effect in denominators is just to replace $\Gamma^{(2)}_k[-\nabla^2]$ with $\Gamma^{(2)}_k[k^2]$, massively simplifying the computation of spacetime traces.

The second order formulation \cite{Benedetti:2013jk,Mitchell:2021qjr} chooses a non-adaptive cutoff function of the form \index{cutoff}
\begin{equation}
\label{nonadaptive}
    \mathcal{R}^{\phi}_k=k^{m_\phi}c_\phi r({\Delta}_s+\alpha_s {R})
\end{equation}
where $s$ is the spin of the mode $\phi$,
$m_\phi$ is set such that the cutoff has the same dimension as $\Gamma^{(2)}$ for this mode, and $c_\phi$ is a number. In this chapter the $c_\phi$ will be taken to be positive for all fields. This is a problem for developing solutions $f_k(R)$ that approximate the perturbative quantisation of the Einstein-Hilbert action \eqref{EHtrunc} because the $\bar{h}$ Hessian has the wrong sign \index{wrong sign} there (as noted above). But again the alternative choice $c_{\bar{h}}<0$ leads to poor asymptotic behaviour at large $R$, resulting in a continuous spectrum of eigenoperators \cite{Mitchell:2021qjr}.

Notice that the cutoffs \eqref{nonadaptive} have been chosen to depend on  $\Delta_s$, rather than simply the $-\nabla^2$ part \cite{Benedetti:2012}, and furthermore include an ``endomorphism'', \index{endomorphism} a curvature correction with endomorphism coefficient $\alpha_s$ \cite{Benedetti:2013jk}. In refs. \cite{Benedetti:2013jk,Mitchell:2021qjr}  the traces are computed directly as a sum over modes. The $\alpha_s$ are there to ensure that 
\be 
\label{positiveeigenvalues}
\Delta_s+\alpha_sR > 0\,, 
\ee
for all modes, which in turn ensures that they are all integrated out as $k \rightarrow 0$, and that the flow equation does not suffer from fixed singularities. \index{fixed singularity} For these non-adaptive cutoffs, the optimised cutoff profile \eqref{optimised} brings no particular advantage. In fact on a sphere the trace is a discrete sum and sharp cutoff profiles \index{sharp cutoff} would lead to a staircase behaviour \cite{Benedetti:2012}, with an ill-defined limit as $R\rightarrow0$. Hence, a smooth (infinitely differentiable) cutoff profile \index{smooth cutoff} is used, such as \cite{Wetterich:1992}
\be 
  r(z)=\frac{z}{\exp(az^b)-1}, \qquad a>0,b\geq1\,. \label{WetCutoff}
\ee 

\section{Flow equations with adaptive cutoff}
\label{sec:flowadaptivecutoff}

\index{cutoff}
In those formulations that use an adaptive cutoff,  spacetime traces are evaluated using a heat-kernel asymptotic expansion, \index{heat kernel} apart from ref. \cite{Benedetti:2012} which uses a direct spectral sum together with a smoothing procedure (to get over the aforementioned staircase problem). As an illustration, the result of the earliest four such formulations \cite{Machado:2007,Codello:2007bd,Codello:2008} for the flow of $f\equiv f(R,t)$ on a four-sphere,
can be summarised as:
\bea\label{flow1} 
&& 384 \pi^2   \left( \partial_t f + 4 f - 2 R f^{\prime} \right) = 
\\ \nonumber
&& \quad \Big[ 5 R^2 \theta\left(1-\tfrac{R}{3}\right) 
-  \left( 12 + 4 \, R - \tfrac{61}{90} \, R^2 \right)\Big]
\Big[1 - \tfrac{R}{3} \Big]^{-1}  
+ \Sigma \\ \nonumber
&& + \Big[ 10 \, R^2 \, \theta(1-\tfrac{R}{4}) - R^2 \, \theta(1+\tfrac{R}{4}) 
-   \left( 36 + 6 \, R - \tfrac{67}{60} \, R^2 \right) \Big] 
\Big[ 1 - \tfrac{R}{4}\Big]^{-1} \\ \nonumber
&& +  \Big[ (\partial_t f' +2f'-2Rf'') \, \left( 10 - 5  R - \tfrac{271}{36}  R^2 + \tfrac{7249}{4536}  R^3 \right) 
+ f'\left( 60 - 20  R - \tfrac{271}{18}  R^2 \right)
\Big] \left[ f+f'(1  - \tfrac{R}{3}) \right]^{-1} \\ \nonumber
&& + \tfrac{5R^2}{2} \, \Big[ 
 (\partial_t f' +2f'-2Rf'') \left\{r(-\tfrac{R}{3})+2r(-\tfrac{R}{6}) \right\}
+ 2 f'\theta(1+\tfrac{R}{3}) + 4 f'\theta(1+\tfrac{R}{6}) 
\Big]  
 \left[ f+f'(1  - \tfrac{R}{3}) \right]^{-1} \\ \nonumber
&& + 
\Big[
(\partial_t f' +2f'-2Rf'') f^{\prime}
\left(6 + 3 R + \tfrac{29}{60} R^2 + \tfrac{37}{1512} \, R^3 \right)\\ \nonumber
&& \qquad+ \left( \partial_t f^{\prime \prime}  - 2 R f^{\prime \prime \prime} \right) 
\left( 27 - \tfrac{91}{20} R^2  - \tfrac{29}{30} \, R^3 - \tfrac{181}{3360} \, R^4 \right) 
 \\ \nonumber
&&  \qquad + f^{\prime \prime} \left( 216 - \tfrac{91}{5}  R^2 - \tfrac{29}{15}  R^3 \right)   
+  f^\prime \left( 36 + 12  R + \tfrac{29}{30}  R^2 \right)
\Big] \Big[ 2 f + 3 f^\prime (1-\tfrac{2}{3}R) + 9 f^{\prime \prime} (1-\tfrac{R}{3})^2 \Big]^{-1} \, .
\eea
Here the function $r$ is the optimised cutoff profile \eqref{optimised}, \index{optimised cutoff profile} which also leads to the appearance of the step functions (\aka Heaviside $\theta$ functions). In ref. \cite{Codello:2008} the equation is adapted to polynomial truncations only, which means that the step functions are all set to one. The first two lines of the right hand side are independent of $f(R,t)$ and encapsulate the contributions from the ghosts, auxiliaries, $\xi_\mu$ and $\sigma$. Here we have introduced the term $\Sigma$.
The third and fourth line arises from $h^T_{\mu\nu}$, whilst the final ratio is the contribution from ${ h}$. Unphysical modes are isolated differently in these implementations, but the changes can be summarised in the different expressions
\be
\label{Sigma2}
\Sigma = 0\,,\quad 10 \, R^2 \, \theta\left(1 -\tfrac{R}{3}\right)\,, \quad - \frac{10 R^2 (R^2-20R+54)}{(R-3)(R-4)}\,,\quad   \frac{10(11R-36)}{(R-3)(R-4)}\,.
\ee
The first, third and fourth options are derived in refs. \cite{Codello:2007bd,Codello:2008}, whilst the second option comes from  ref. \cite{Machado:2007}. We have suppressed some other details, for more discussion see ref. \cite{Dietz:2012ic}. 

Setting $\partial_tf=0$ in the above turns this flow equation into the differential equation that must be satisfied by a fixed point $f(R)$. It is a highly non-linear third-order ODE (ordinary differential equation). In the formulation \cite{Machado:2007}, the appearance of the $\theta$ functions, explicitly and in $r$, will result in jumps in $f'''(R)$ across the point where they switch on or off, but this can be accommodated.  \index{third order formulation}

A more important and generic feature is the existence of  fixed and moveable singularities. These concepts come from the mathematics of analysis of ODEs. To discuss them it is helpful to cast the fixed point ODE in ``normal'' form: \index{fixed singularity} \index{moveable singularity}
\be
\label{normal}
f'''(R)=rhs\,,
\ee
where $rhs$ (right hand side) contains no $f'''$ terms. A Taylor expansion about some generic point $R_p$ takes the form:
\be
\label{Taylor}
f(R)=f(R_p)+(R-R_p)f'(R_p)+\frac{1}{2}(R-R_p)^2f''(R_p)+\frac{1}{6}(R-R_p)^3 f'''(R_p)+\cdots\,.
\ee
Since \eqref{normal} determines the fourth coefficient in terms of the first three, we see that typically \eqref{Taylor} provides a series solution depending on three continuous real parameters, here 
\be
\label{generic}
f(R_p)\,,\quad f'(R_p)\quad{\rm and}\quad f''(R_p)\,,
\ee 
with some finite radius of convergence $\rho$ whose value also depends on these parameters. Therefore the standard mathematical result is recovered that around a generic point $R_p$ there is some domain ${\cal D}=(R_p-\rho,R_p+\rho)$ in which there is a three-parameter set of well-defined solutions. From here one can try to extend the solution to a larger domain, \eg by matching to a Taylor expansion about another point within ${\cal D}$. A typical problem, seen also in the LPA and the derivative expansion \cite{Morris:1994ki,Morris:1994ie,Morris:1996nx,Morris:1994jc,Morris:1998} and in the second order formulation \cite{Benedetti:2013jk,Mitchell:2021qjr}, is that eventually, at some point $R=R_c$, dependent on the parameters, the denominator of $rhs$ develops a zero, so that as $R\to R_c$, \eqref{normal} implies 
\be
\label{movsing}
f'''(R)= 2c/(R-R_c)+\cdots\,,
\ee
where $c$ is some constant and the ellipses contains the non-singular part. Integrating this we see that the solution typically ends in a \textit{moveable} singularity, of form \index{moveable singularity}
\be
\label{sing}
f(R)\sim c\,(R-R_c)^2\ln|R-R_c|\,,
\ee
where ``$\sim$'' means that less singular parts are neglected.

\index{fixed singularity}
As already mentioned, fixed point equations derived with adaptive cutoff present another challenge in that they also have {fixed singular points} $R_c$. These correspond in $rhs$ to explicit algebraic poles in $R$, where  the domain of interest is $R\ge0$ since the equations apply to the four-sphere. Whatever the formulation there is always one fixed singularity $R_c=0$, which is unavoidable when using an adaptive cutoff as we have seen \cite{Benedetti:2012,Dietz:2012ic}. 
Different formulations 
have different numbers and positions for the other fixed singularities (see \eg the discussion in refs. \cite{Demmel2015b,Gonzalez_Martin_2017})
 but there is always at least one more. Inspecting the  example \eqref{flow1}, we see that
$f'''$ appears once in the penultimate line in eqn. \eqref{flow1}, where it is multiplied by the polynomial 
\be 
R \left( 27 - \frac{91}{20} R^2  - \frac{29}{30} \, R^3 - \frac{181}{3360} \, R^4 \right)\,.
\ee
Thus, rearranging the fixed point equation into normal form \eqref{normal}, results in poles from the zeroes of this polynomial. Two of these are in the required domain, namely at $R_c=0$ and  $R_c=2.0065$. There are also two further single poles, at $R_c=3$ and $R_c=4$,  from the first two lines of the right hand side of  \eqref{flow1}. 

As $R$ approaches one of these $R_c$, $f$ will end at a singularity of form \eqref{sing} unless the $f$--dependent parts in $rhs$ are tuned so as to conspire to cancel the pole. Substituting the Taylor expansion \eqref{Taylor}, with $R_p=R_c$, one sees that this requirement forces some generally non-linear combination of $f(R_c)$, $f'(R_c)$ and $f''(R_c)$ to vanish. Thus, a fixed singularity imposes a constraint on the solution, reducing the number of free parameters by one. 

The inevitable fixed singularity at $R_c=0$ can thus be seen as restoring consistency since it reduces the three parameter set of solutions to a two parameter set, in agreement in this respect with what is obtained from the non-adaptive-cutoff second order formulation. 

Unfortunately, since there are a further three fixed singularities, these equations are overconstrained, and thus there are no fixed point solutions $f(R)$ that are valid over the whole range $R\ge0$.

However, these fixed singularities are artefacts of the regularisation procedure: it is possible to move them and eliminate most of them. Benedetti and Caravelli were the first to realise this, and we will refer to their version \cite{Benedetti:2012} as the ``BC'' formulation.
Before regularisation, the Jacobian trace \eqref{trJac1} has a denominator that vanishes if $\Delta_1$ vanishes. Likewise the Jacobian trace \eqref{trJac0} has a denominator that vanishes when $\Delta_0$ vanishes. Recalling the form \eqref{modifiedLaplacians} of the $\Delta_s$, and that the net effect of the adaptive optimised cutoff is to replace $-\nabla^2$ with $k^2$ in the denominator, we see that these contributions give poles $1/(1-R/4)$ and $1/(1-R/3)$ (after using \eqref{dimensionless} to scale to dimensionless quantities). These are the poles that are visible in the first two lines of the right hand side of  \eqref{flow1}. 

BC eliminate them by using an endomorphism, \index{endomorphism} namely by using $r(\Delta_s)$ instead of $r(-\nabla^2)$ \cite{Benedetti:2012} (a so-called cutoff of type II \cite{Codello:2008}). Then one is left with the $R_c=0$ singularity, and a fixed singularity at some positive $R_c$ which is due to the fact that the $\bar{h}$ trace vanishes there  \cite{Benedetti:2012,Dietz:2012ic}. These fixed singularities thus reduce $f(R)$ solutions to a one-parameter set.

Now there is still the danger of encountering a moveable singularity \eqref{sing}, and this imposes further restrictions on the remaining parameter.  Such a singularity can appear at any value of $R$, and in particular at large $R$ where the equations can then be solved analytically by developing the solution as an asymptotic expansion. In scalar field theory \cite{Morris:1994ki,Morris:1994ie,Morris:1996nx,Morris:1996xq,Morris:1994jc,Morris:1998} and in the second order formulation \cite{Mitchell:2021qjr}, what is found is that this asymptotic expansion has less than the full number of parameters expected. One can also show that the missing parameters are associated with fast growing perturbations that are incompatible with an asymptotic solution. In this way it is possible to deduce analytically the number of constraints that moveable singularities are responsible for imposing. 

The result for scalar field theory is that the parameters are fixed, typically to a handful of values \cite{Morris:1994ki,Morris:1994ie,Morris:1996nx}, corresponding to a finite set of fixed points, or in special cases a discrete infinity of fixed points \cite{Morris:1994jc}. However, there is at this stage also the possibility that there are no fixed point solutions.  The actual number of solutions then needs to be determined numerically.\footnote{Although some may be found analytically, \eg the Gaussian fixed point, or  special cases  \cite{Ohta2016}.} We will see this at work in the second order formulation in sec. \ref{sec:fixed} where we describe in detail how to find asymptotic solutions $f_{asy}(R)$.

Unfortunately for third order formulations,  asymptotic analysis typically does not find sufficient constraints  \cite{Gonzalez_Martin_2017}. For example for the BC formulation, the asymptotic solution turns out to have the maximum three parameters \cite{Dietz:2012ic}: \index{asymptotic analysis}
\be 
\label{Benasy}
f_{asy}(R) = A \,R^2 + R\left\{\frac{3}{2}A+B\cos\ln R^2 + C\sin\ln R^2\right\}+\cdots\,,
\ee
where the ellipses stand for asymptotic corrections with lower powers of $R$, and the three parameters are restricted only by the inequality:
\be
\label{safedisc}
\frac{121}{20}A^2 > B^2 + C^2\,.
\ee
Thus, one still expects to find one-parameter sets (\ie lines) of global solutions $f(R)$ in this case, and that is exactly what is found by careful numerical analysis \cite{Dietz:2012ic}. \index{numerical analysis}
Asymptotic analysis also shows that the BC formulation has continuous eigenoperator spectra. Initially it was suggested that these effects can be attributed to the fact that all eigenoperators are redundant \index{redundant} if the equation of motion \eqref{EoM} for the fixed point $f(R)$, has no solution for $R$ in the required range $R\ge0$ \cite{Dietz:2013sba}. But it is now clear that the poor behaviour is again associated to the scalar mode $\bar{h}$ \cite{Dietz:2012ic,Demmel2015b,Dietz:2016gzg}, and is one more malign effect of the conformal instability \cite{Gibbons:1978ac,Dietz:2012ic,Dietz:2016gzg}. \index{wrong sign} \index{conformal instability} In fact precisely these problems reappear in the second order formulation if one chooses $c_{\bar{h}}<0$, as already mentioned in sec. \ref{sec:cutoffs}.

\index{asymptotic analysis} \index{numerical analysis}
As emphasised in ref. \cite{Gonzalez_Martin_2017}, asymptotic analysis plays three powerful r\^oles. Firstly, as just sketched and discussed in detail in sec. \ref{sec:asymptoticsols}, it allows one to deduce the dimension of the solution space.  Secondly the asymptotic solution provides a way to validate numerical solutions since if one can integrate out far enough, the numerical solution should match the asymptotic solution, allowing a reliable determination of the asymptotic parameters. 

Finally, the asymptotic solution actually
contains only the physical part of the fixed point effective action. \index{physical part} \index{equation of state} To see this, we need to return temporarily to labelling scaled quantities with a tilde, and recall that the effective infrared cutoff $k$ is added by hand such that the physical Legendre effective action  is recovered only in the limit that this cutoff $k\to0$. This must be done while holding the physical quantities such as $R$ fixed, rather than scaled quantities $\tilde{R}$. In normal field theory, \eg scalar field theory, the analogous object is the universal scaling equation of state, which for a constant field precisely at the fixed point takes the simple form 
\be 
\label{Vscale}
V(\varphi) = A \,\varphi^{d/d_\varphi}\,,
\ee 
where $d$ is the space-time dimension and $d_{\varphi}$ is the full scaling dimension of the field (\ie incorporating also the anomalous dimension). In the current case we keep fixed the constant background scalar curvature $R$. Thus by \eqref{fR} and \eqref{dimensionless}, the only physical part of the fixed point action in this approximation is:
\be 
\label{phys}
f({R}) |_{\text{phys}} = \lim_{k\to 0} k^4\, \tilde{f}(R/k^2) = \lim_{k\to 0} k^4\, \tilde{f}_{asy}(R/k^2)\,.
\ee
For example from \eqref{Benasy}, for the BC formulation one finds:
\be 
f({R}) |_{\text{phys}} = AR^2\,.
\ee
This is invariant under changes of scale as it must be, and is a sensible answer for the scaling equation of state precisely at the fixed point. We will find the same answer from the second order formulation. 

We still have the problem that since there are one-parameter sets of fixed-point solutions, $A$ is not fixed. In third order formulations one can use the ability to add endomorphisms to try to patch this up \cite{Demmel2015b} but asymptotic analysis then shows there is actually a whole zoo of possibilities for the scaling equation of state and dimension of the solution space, depending on parameter choices in the endomorphisms \cite{Gonzalez_Martin_2017}. One can also try to extend the solution to negative $R$. This does reduce the solution space of the BC formulation to a discrete set but that set appears to be empty since no numerical solutions were then found \cite{Dietz:2012ic}. A more careful version of this strategy is also used in the second order formulation. \index{third order formulation} \index{second order formulation}

Actually one can question whether the large $\tilde{R}=R/k^2$ regime makes physical sense 
\cite{Demmel:2014fk,Demmel2015b,Ohta2016}. The problem arises when the cutoff depends on modified Laplacians, \eg as in \eqref{positiveeigenvalues}, 
where the endomorphism is added to ensure that the minimum eigenvalue is positive. It is most easily seen if we take a sharp (step function) cutoff profile, and write the minimum eigenvalue as $R\,\lambda_{min}$. Then once $k^2<R\,\lambda_{min}$, \ie $\tilde{R}>1/\lambda_{min}$, there are no more modes to be integrated out. This means that the functional behaviour in this large $\tilde{R}$ regime is meaningless since it is not describing any actual changes. However the physical Legendre effective action is only reached by taking $k\to0$, and this argument would appear to imply that such a limit is inherently ill-defined. \index{ensemble of manifolds} \index{large $R$ regime}

\index{single metric approximation}
In fact this conundrum is another artefact of the single-metric approximation \cite{Morris:2016spn}. In reality one should be integrating out over an ensemble of manifolds described by the fluctuating full metric $\hat{g}_{\mu\nu}$. The Wilsonian RG only makes sense when applied to such an ensemble. \index{renormalization group: Wilsonian} Then no matter how small $k$ is, there are always manifolds with sufficiently small curvature that their eigenvalues remain to be integrated out. It is possible to repair the single-metric approximation sufficiently in this case by retaining the scale degree of freedom $h_{\mu\nu}\propto g_{\mu\nu}$ in the fluctuation field dependence, and thus regaining an ensemble of manifolds. However the net result of such a repair is the same type of functional RG equations again, but now with a clear explanation for why the large $\tilde{R}$ regime should be trusted  \cite{Morris:2016spn,Percacci:2016arh, Ohta:2017dsq}. \index{renormalization group: functional}

We now abandon third order formulations and concentrate on a second order formulation \cite{Benedetti:2013jk,Mitchell:2021qjr}, which in almost all respects has more promising behaviour.

\section{Evaluating traces}
\label{sec: Evaluating traces}

In the formulation \cite{Benedetti:2013jk,Mitchell:2021qjr} the traces \index{traces} are evaluated by a direct spectral sum.
%, although other methods such as the heat kernel method are possible \cite{Kluth:2019vkg}. The traces are evaluated on the background metric and we have to specify what that is in order to perform the computation. We will choose to work 
In common with the rest of the literature one chooses a (globally) maximally symmetric background manifold. 
There are three to choose from: the four sphere $\mathbb{S}^4$, which has a finite volume and positive curvature, so the spectrum of the allowed modes form a discrete set that have to be summed over; the hyperboloid $\mathbb{H}^4$ which has negative curvature and infinite volume so the spectrum is continuous; and finally flat space $\mathbb{R}^4$, which is a limiting case for both of the two previous manifolds when $R\rightarrow0$. As we will see they all need to be considered. Actually they become smoothly joined together in an ensemble which thus allows the same flow equation to be defined over the entire domain $-\infty<R<\infty$.

\subsection{Sphere}
\label{sec:sphere}
On the sphere the traces are evaluated using \index{sphere}
\begin{equation}
    \text{Tr}\,W(\Delta_s)=\sum_nD_{n,s}W(\lambda_{n,s})
\end{equation}
where $\lambda_{n,s}$ are eigenvalues of the $\Delta_s$ defined in \eqref{modifiedLaplacians},
and $D_{n,s}$ are their multiplicities. Explicit values are shown in table \ref{tab1} \cite{Benedetti:2012}.
\begin{table}[t]
    \centering
    \begin{tabular}{c|c|c}
         Spin s& Eigenvalue $\lambda_{s,n}$ & Multiplicity $D_{n,s}$  \\
         \hline
         \hline
        0 & $\frac{n(n+3)-4}{12}R$&$\frac{(n+2)(n+1)(2n+3)}{6}$ \\
        \hline
        1 & $\frac{n(n+3)-4}{12}R$ & $\frac{n(n+3)(2n+3)}{2}$ \\
        \hline
        2 & $\frac{n(n+3)}{12}R$&$\frac{5(n+4)(n-1)(2n+3)}{6}$ \\
    \end{tabular}
    \caption{Values of the multiplicities and eigenvalues for evaluating the traces. }
    \label{tab1}
\end{table}
There are a few caveats. %that we need to address. 
Not all the modes contribute in the sum, for example vectors satisfying $\nabla_\mu \xi_\nu+\nabla_\nu \xi_\mu=0$ and the scalar modes $\sigma=\text{constant}$. Because of this, the tensor mode and the vector mode sums start at $n=2$, the scalar mode of the Jacobian starts at $n=1$ and the $\bar{h}$ mode starts at $n=0$.
%Earlier we have introduced additional shifts $\alpha_s$ in the action to ensure that all modes are integrated out as we lower our cutoff to $k\rightarrow 0$. This 
Now the requirement \eqref{positiveeigenvalues} means that $\lambda_{n,s}+\alpha_s R>0$ must be satisfied. For the tensor and vector modes it is sufficient to set $\alpha_2=\alpha_1=0$, however from table \ref{tab1} we see that we must have $\alpha_0>1/3$.
\subsection{Hyperboloid}
\label{sec:hyperboloid}
As already mentioned, the hyperboloid has a negative curvature, an infinite volume, and a continuous spectrum of eigenvalues. The traces on this manifold are evaluated using \cite{Camporesi:1994ga} \index{hyperboloid}
\begin{align}
  \text{Tr}\,W(\Delta_s)=
  \frac{2s+1}{8\pi^2}\int \!d^4x \sqrt{g}\ \bigg( -\frac{R^2}{12}\bigg)^2 \int _0^\infty d\lambda \bigg(\lambda^2+\big(s+\frac{1}{2}\big)^2\bigg)\lambda \tanh(\pi\lambda)W(\Delta_{\lambda,s}).
\end{align}
Even though there is now an infinite volume factor in the flow equation \eqref{eq:4}, this precise factor also appears above, % in the traces, 
so the equations still make sense once we cancel this factor from both sides. The eigenvalues of the spectrum are
\be 
  \Delta_{\lambda,s}=-\frac{R}{12}\lambda^2-\beta_sR\,, \qquad\text{where}\qquad
%\ee 
%\be 
  \beta_0=\frac{25}{48}, \quad \beta_1=\frac{25}{48},\quad \beta_2=\frac{9}{48}\,.
\ee 
Using the same flow equation, and thus the same endomorphism parameters $\alpha_s$, the requirement \eqref{positiveeigenvalues} must again be satisfied.
We can still take $\alpha_2=\alpha_1=0$, but now $\alpha_0$ also has an upper bound $\alpha_0<25/48$.
\subsection{Flat space} \label{sec:flat}
Finally, evaluating traces on flat space can be achieved by taking the limit as $R\rightarrow0$ from positive or negative side. \index{flat space}
%depending on which manifold we start. 
If we start from the positive side we first make a substitution $p=n\sqrt{R/12}$ then take $R\rightarrow0$ while keeping $p$ fixed. All Laplacians then become $\Delta_{n,s}\rightarrow p^2$ and $p^2$ can be identified as the flat space momentum. Plugging in our choice of the cutoff \eqref{nonadaptive}, and performing these substitutions, yields  
\begin{align}
    \partial_tf_k(0)+4f_k(0)&=\frac{1}{8\pi^2}\int_0^\infty dp p^3 \Bigg[ 16c_{\Bar{h}}\frac{2r(p^2)-p^2r'(p^2)}{9f_k''(0)p^4+3f_k'(0)p^2+2f_k(0)+16c_{\Bar{h}} r(p^2)} \nonumber \\
    &+10c_T \frac{r(p^2)-p^2r'(p^2)}{-f_k'(0)p^2-f_k(0)+2c_T r(p^2)}-3c_V \frac{r(p^2)-p^2r'(p^2)}{p^2+c_V r(p^2)} \nonumber \\
    &-4c_{S_2}\frac{2r(p^2)-p^2r'(p^2)}{3p^4+4c_{S_2}r(p^2)} + c_{S_1} \frac{r(p^2)-p^2r'(p^2)}{p^2+c_{S_1}r(p^2)} \Bigg]  \label{eq:flat}
\end{align}
This same equation is arrived at if we take $R\rightarrow0$ from the negative side by first setting $p=\lambda\sqrt{-R/12}$ on the hyperboloid and holding $p$ fixed. The form of these equations already give some information about the possible solutions,
%For this equation to be well defined and non-singular we need non-vanishing denominators for all $p^2\geq0$. For example in the case of diverging cutoff profile ($b>1$ in \eqref{optimised}) $r(p^2)\rightarrow+\infty$ for very low momentum, but rapidly falls to zero for $p^2>0$. To avoid singularity in the first term we see that this implies that $f_k''(0)>0$ (if $c_{\bar{h}}>0$). In the second term $f'_k(0)$ appears with a negative sign, so to avoid a vanishing denominator, the first derivative must be negative. Considerations like these can guide us through numerical searches, which in most cases is the only way to solve these type of equations.
and can help guide numerical searches \cite{Mitchell:2021qjr}. In particular, by inspection, it is clear that there are no fixed singularities, and choices for $f_k(0)$, $f'_k(0)$ and $f''_k(0)$ can be made that give well defined non-singular integrals. 

\section{Fixed point solutions} \label{sec:fixed}
The fixed point solution to the flow equation $f_k(R)=f(R)$ occurs when $\partial_tf_k(R)=0$. An advantage of the non-adaptive cutoff is that $\partial_tf_k(R)$ only appears once on the left hand side of \eqref{eq:4}, so the fixed point equation \index{fixed point} is 
\be 
  2\,VE(R)=\mathcal{T}_2+\mathcal{T}_0^{\Bar{h}}+\mathcal{T}_1^{Jac}+\mathcal{T}_0^{Jac}\,. \label{eq:9}
\ee 
Another crucial advantage is, like \eqref{eq:flat}, inspection of the trace equations \eqref{trtensor} -- \eqref{trJac0} makes clear that there are no fixed singularities any more. The flow equation is non-linear and very hard to work with, so solving the equations exactly is unfeasible. 
The strategy is to solve analytically for $f(R)$ around $R=0$ as a Taylor expansion and around $R=\pm\infty$ by an asymptotic expansion. Then numerical methods can be used to try to patch in a solution that goes smoothly from the Taylor expansion at $R=0$ to the asymptotic solutions at $R=\pm\infty$. 

\subsection{Asymptotic analysis} \label{sec:asymptoticsols}

\index{asymptotic analysis} \index{fixed point} \index{sphere} \index{hyperboloid}
We now explain in detail how to develop asymptotic solutions, using these equations as an example. In these large $R$ limits,  the equations simplify due to rapidly decaying cutoff profiles $r(z)$. At first sight, it looks like all the traces on the right hand side of the flow equation vanish and one is only left with \eqref{EoM}, the equation of motion $E(R)=0$. This is actually true on the hyperboloid and the fixed point solution is therefore the solution of $E(R)=0$ namely
\be 
\label{solE}
  f(R)=AR^2,
\ee 
where $A$ is an arbitrary constant. At any finite $R$ this is then accompanied by rapidly decaying corrections as discussed later, \cf  eqn. \eqref{asympFPH}. 

The story is different on the sphere since upon closer inspection not all of the terms in the sums vanish. \index{sphere}
There are three such terms, the $n = 0$ and $n = 1$ components from $\mathcal{T}_0^{\bar{h}}$ and the $n = 1$ of $\mathcal{T}_0^{Jac}$. 
To see this for the $n=0$ case, note that from table \ref{tab1}, $\Delta_0=-R/3$. Thus, using \eqref{nonadaptive}, the denominator of this term in the sum \eqref{trhbar} is given by 
  \be
    9f''(R)\Delta_0^2+3f'(R)\Delta_0+E(R) +16\mathcal{R}^{\Bar{h}}_k =
  R^2f''(R)-Rf'(R)+E(R)+16k^4c_\phi r([\alpha_0-\tfrac13]R) 
  \ee
  Now, assuming that the leading asymptotic behaviour is $f(R)=AR^2$, we see that the first two terms cancel each other, and likewise $E(R)$ vanishes, so we are left only with the cutoff term in the denominator. Therefore this term takes the form of 
  \be 
    \frac{1}{k^4\,r(z)}\frac{d}{dt}\left[k^4r(z)\right]=4-2z\frac{d \ln r(z)}{dz}
  \ee 
 with $z$ set equal to $z=[\alpha_0-\tfrac13]R$.

Turning to the $n=1$ components, note that from table \ref{tab1}, both $\Delta_0$ and $\Delta_1$ vanish for $n=1$. In \eqref{trhbar}, apart from the cutoff term the whole denominator therefore vanishes (because  $E(R)$ vanishes). In \eqref{trJac0} it is the second component that has a vanishing denominator apart from the cutoff term. The $S_1$ (first) component does not suffer from the same problem because there is also the $+R/3$ part in the denominator. However, the cutoff dependence is the same for the $n=1$ contributions namely $r(\alpha_0 R)$ and the numerical factors are such that these two $n=1$ contributions exactly cancel each other. 
  
Altogether then, effectively the only term on the RHS (right hand side) of the flow equation that does not vanish asymptotically is the $n=0$ component of the $\mathcal{T}^{\bar{h}}_0$ trace. This is a problem however, since  the $n=0$ component of $\mathcal{T}^{\bar{h}}_0$ contributes a term that grows at least as fast as $R^2$. This is inconsistent with the fact that the LHS (left hand side) of flow equation has been set to vanish asymptotically. Actually this analysis shows that $f(R)$ grows faster than $R^2$. For example in the best-case scenario the RHS $\sim R^2$ but that implies $f(R)\sim R^2 \ln R$ so that the LHS is left with an $E(R)\sim R^2$ to balance the contribution from the $n=0$ component of $\mathcal{T}^{\bar{h}}_0$.
  
Therefore we now assume that $f(R)$ actually grows faster than $R^2$ at large $R$. But this means we need to check again which terms in the traces have denominators that would vanish without a cutoff. By inspection none of the traces that depend on $f(R)$ can now have this issue. In particular the $n=1$ component of the $\mathcal{T}^{\bar{h}}_0$ trace no longer has a denominator that could vanish, because $E(R)$ no longer vanishes at large $R$, while for the $n=0$ component the $f''(R)$ part in the denominator now dominates at large $R$. So the only contribution that survives on the RHS at large $R$, is now the $n=1$ $S_2$ component of $\mathcal{T}^{Jac}_0$. 
  
Keeping just this term it turns out one can solve the fixed point equation in closed form, thus obtaining the correct asymptotic behaviour for general cutoff function $r(z)$. Using the values from table \ref{tab1} we have that the multiplicity of the $n=1$ component is $D_{1,0}=5$, note that $m_{S_2}=4$ and that $1/V=R^2/384\pi^2$ for the four-sphere. Thus, keeping only this leading term on the RHS of the flow equation, we have 
  \be 
    2 f(R) - Rf'(R) = \frac{R^2}{768\pi^2}\left[ -10 + 5\alpha_0 R \frac{r'(\alpha_0 R)}{r(\alpha_0 R)}\right]\,.
  \ee  
 This is exactly soluble. Indeed dividing through by $R^3$ it can be rewritten as 
  \be 
    -\frac{d}{dR}\left(\frac{f(R)}{R^2}\right) =\frac1{768\pi^2}\left[-\frac{10}{R}+5 \frac{d}{dR} \ln r(\alpha_0 R)\right]\,,
  \ee 
which can be immediately integrated to give
  \be 
    f(R) = \frac{5R^2}{768\pi^2} \ln\frac{R^2}{r(\alpha_0R)}+AR^2+o(R^2)\quad\text{as}\quad R\to+\infty\,,
  \ee 
where we included the integration constant $A$ and finally we noted that terms that grow slower than $R^2$ will be generated by iterating this asymptotic solution to higher orders, hence the $o(R^2)$ part. 
The $\ln r$ term actually dominates, \ie the large $R$ behaviour is dominated by cutoff-dependent effects.
For example using the cutoff \eqref{WetCutoff}, gives the first three terms in this series:
\be 
\label{asympFPS}
f(R) = \frac{5a\alpha_0^b}{768\pi^2}R^{2+b}+\frac{5}{768\pi^2}R^2\ln R+ AR^2+\frac{16c_{\bar{h}}}{5ab(1+b)\alpha_0^b}\left(\alpha_0-\frac13\right)\,\mathrm{e}^{-a\left(\alpha_0-\frac13\right)^b R^b}+\cdots\,.
\ee
To get the next term in the series, the solution is substituted back into the fixed point equation and the next leading correction is isolated. This leads to the last displayed correction above. It is exponentially decaying and comes from the $n=0$ term in the $\mathcal{T}^{\bar{h}}_0$ trace. One finds that other corrections decay faster  provided that $\alpha_0 < \frac56+\alpha_1$. This is satisfied thanks to the restrictions on the $\alpha_i$ parameters discussed in secs. \ref{sec:sphere} and \ref{sec:hyperboloid}. Substituting \eqref{asympFPS} back into the fixed point equation and proceeding similarly one can in principle develop the whole asymptotic series. It is an infinite series of ever faster decaying terms and is indicated by the ellipses. In particular these terms will include a power series in $A$.

At this point we have succeeded in finding consistent asymptotics. $f(R)$ does grow faster than $R^2$ on the sphere, as assumed, and using such a form in the RHS of the fixed point equation one can see that the $n=1$ $S_2$ component of $\mathcal{T}^{Jac}_0$ dominates at large $R$, which leads back to the above equation.

Recall that the fixed point equation is actually second order. But the asymptotic solutions only have one free parameter $A$, even though there should be two. To find out where the second parameter has gone we linearise about the fixed point $f(R)+\delta f(R)$ and plug it into the flow equation \eqref{eq:9} to get
\be
\label{eq:10a}
  -a_2(R)\,\delta f''(R)+a_1(R)\,\delta f'(R)+a_0(R)\,\delta f(R)=4\,\delta f(R) \,,
\ee
with 
\be
  a_2=\frac{144c_h}{V}\text{Tr}\Bigg[\frac{\Delta_0^2(2r(\Delta_0+\alpha_0R)-(\Delta_0+\alpha_0R)r'(\Delta_0+\alpha_0R))}{(9f''(R)\Delta_0^2+3f'(R)\Delta_0+E(R)+16c_hr(\Delta_0+\alpha_0R))^2}\Bigg]\,, \label{a2}
\ee
\begin{align}
    a_1&=2R-\frac{16c_h}{V}\text{Tr}\Bigg[\frac{(3\Delta_0-R)(2r(\Delta_0+\alpha_0R)-(\Delta_0+\alpha_0R)r'(\Delta_0+\alpha_0R))}{(9f''(R)\Delta_0^2+3f'(R)\Delta_0+E(R)+16c_hr(\Delta_0+\alpha_0R))^2}\Bigg] \nonumber \\
    &+\frac{2c_T}{V}\text{Tr}\Bigg[\frac{(R/2-\Delta_2)(2r(\Delta_2+\alpha_2R)-(\Delta_2+\alpha_2R)r'(\Delta_2+\alpha_2R))}{(-f'(R)\Delta_2-E(R)/2+2c_Tr(\Delta_2+\alpha_2R))^2}\Bigg]\,,
\end{align}
\begin{align}
    a_0&=\frac{32c_h}{V}\text{Tr}\Bigg[\frac{(2r(\Delta_0+\alpha_0R)-(\Delta_0+\alpha_0R)r'(\Delta_0+\alpha_0R))}{(9f''(R)\Delta_0^2+3f'(R)\Delta_0+E(R)+16c_hr(\Delta_0+\alpha_0R))^2}\Bigg] \nonumber \\
    &+\frac{2c_T}{V}\text{Tr}\Bigg[\frac{(2r(\Delta_2+\alpha_2R)-(\Delta_2+\alpha_2R)r'(\Delta_2+\alpha_2R))}{(-f'(R)\Delta_2-E(R)/2+2c_Tr(\Delta_2+\alpha_2R))^2}\Bigg]\,. \label{a0}
\end{align}
In the large $R$ limit $a_1(R) \sim 2R$ and $a_0$ and $a_2$ vanish asymptotically. Then it is tempting to simply set $a_0$ and $a_2$ to zero to find the asymptotic solution to \eqref{eq:10a}. But if this is done there is only one solution $\delta f(R)=\delta A\, R^2$. In fact this is just the leading term in an asymptotic series which is nothing but what one would derive from \eqref{asympFPS} by differentiating with respect to $A$. (Recall  that the ellipses actually contain a power series in $A$.) This asymptotic solution is an exact series solution to \eqref{eq:10a} where $a_0$ and $a_2$ are only involved in constructing the subleading corrections. To find more than the one parameter $\delta A$ in the solution to \eqref{eq:10a}, 
 $\delta f''(R)$ cannot be neglected, implying that higher derivatives must dominate over lower ones in the large $R$ limit. Hence, the other solution is one where $\delta f(R)$ can at first be neglected. Then writing \eqref{eq:10a} as
\be 
\label{Basymp}
  \frac{d}{dR}\ln \delta f'(R)=\frac{a_1(R)}{a_2(R)} \quad \implies \delta f(R)=B\int^R dR'\exp\int^{R'}dR'' \frac{a_1(R'')}{a_2(R'')},
\ee 
where $B$ is the second parameter.  
For the explicit form, $a_2$ is needed. It gets its leading contribution from the same source as the last displayed term in \eqref{asympFPS}. Using the same cutoff choice, \eqref{WetCutoff}, asymptotically 
\be 
\label{asympa2S}
a_2(R) = \frac{24576\pi^2c_{\bar{h}}}{25ab(1+b)^2\alpha_0^{2b}}\left(\alpha_0-\frac13\right)^{1+b}\!\!\! R^{1-b} \,\mathrm{e}^{-a\left(\alpha_0-\frac13\right)^b R^b} +\cdots\,.
\ee
Recalling that $a_1 =2R$ to leading order, the integrals can be evaluated by successive integration by parts, as an asymptotic series and where each term is given in closed form. 

Since this strategy is used many times in this kind of asymptotic analysis let us sketch it on the indefinite integral:
\be 
\label{trick}
\int\!\!dR\, G(R) \,\mathrm{e}^{F(R)} = \frac{G(R)}{F'(R)}\,\mathrm{e}^{F(R)} - \int\!\! dR\, \left(\frac{G(R)}{F'(R)}\right)'\,\mathrm{e}^{F(R)}\,.
\ee
The above equality follows by integration by parts, however if $F(R)$ grows at least as fast as $R$ for large $R$, where $F$ is either sign, and $G(R)$ grows or decays slower than an exponential of $R$, then the integral on the right is subleading compared to the integral on the left. Iterating this identity then evaluates the integral in the large $R$ limit as $\mathrm{e}^{F(R)}$ times an asymptotic series, the first term on the RHS being the leading term. 

In this way, using the cutoff \eqref{WetCutoff}, the solution \eqref{Basymp} on the sphere turns out to be
\be 
  \delta f(R) \sim B \exp \bigg\{ \frac{12(1+b)^2\alpha_0^{2b}}{12288\pi^2c_{\bar{h}}}\bigg(\alpha_0-\frac{1}{3}\bigg)^{-1-2b}R\, \mathrm{e}^{a(\alpha_0-1/3)^bR^b}\bigg\}\,. \label{eq:sph}
\ee 
The analysis proceeds similarly on the hyperboloid \cite{Mitchell:2021qjr}. As $R\to-\infty$, one finds: \index{hyperboloid}
\be 
\label{asympFPH}
f(R) = AR^2 +\frac{c_{S1}}{96\sqrt{3\pi a^3b^3}}\left(\frac{25}{48}-\alpha_0\right)^{\frac{5-3b}{2}}\!\!\!\left(-R\right)^{2-\frac{3b}{2}}\left\{1+ O\left(|R|^{-\frac12}\right)\right\}\,\mathrm{e}^{-a\left[\left(\alpha_0-\frac{25}{48}\right)R\right]^b}+\cdots\,.
\ee
The correction is again a decaying exponential because $\alpha_0$ is restricted to $\alpha_0<25/48$. All scalar traces (thus also $A$) contribute to the $O\left(|R|^{-\frac12}\right)$ term, and the ellipses stand for terms with faster decaying exponentials. 
%Again we ask where the other parameter has gone. The analysis proceeds in a similar fashion to that on the sphere. We have again the asymptotic perturbed fixed point equation \eqref{perturbedFP} except now:
The asymptotic behaviour of $a_2$ turns out now to be:
\be 
\label{asympa2H}
a_2(R)= \frac{4c_{\bar{h}}}{81A^2\sqrt{3\pi ab}}\left(\frac{25}{48}-\alpha_0\right)^{\frac{5-b}{2}}\!\!\!\left(-R\right)^{1-\frac{b}{2}}\,\mathrm{e}^{-a\left[\left(\alpha_0-\frac{25}{48}\right)R\right]^b}+\cdots
%a_2(R) = \frac{4c_{\bar{h}}}{81A^2}\sqrt{-\frac{R}{3\pi}}\left(\frac{25}{48}-\alpha_0\right)^2\,\mathrm{e}^{\left(\frac{25}{48}-\alpha_0\right)R}+\cdots\,.
\ee
%From it we find qualitatively similar conclusions. 
(the ellipses being faster decaying terms). And thus one finds on the hyperboloid
\be 
  \delta f(R) \sim B \exp \bigg\{ \frac{81A^2}{2c_{\bar{h}}}\sqrt{\frac{3\pi}{ab}}\bigg(\frac{25}{48}-\alpha_0\bigg)^{-\frac{b+5}{2}}(-R)^{1-b/2}\,\mathrm{e}^{a[(\alpha_0-25/48)R]^b}   \bigg\}. \label{eq:hyp}
\ee 
However, there is a problem here. Both these solutions \eqref{eq:sph} and \eqref{eq:hyp} for $\delta f(R)$,  are rapidly growing exponentials of an exponential. In the asymptotic regime, these perturbations are no longer small, thus invalidating the initial linearization assumption used to derive them. Therefore, these solutions must be discarded and thus we conclude that the fixed points have only one free parameter on both the sphere and hyperboloid.

These results allow us to draw important conclusions. Each of the asymptotic fixed point solutions, \eqref{asympFPS} and \eqref{asympFPH}, contribute one constraint on the flow equation.\footnote{For example at some initial very large $R$ we can set $2f(R)=Rf'(R)$ since this boundary condition imposes the leading behaviour \eqref{solE}. Using the subleading corrections we can furnish a more accurate Robin boundary condition at more reasonable values of $R$.} There are no boundary conditions coming from $R=0$, so we can expect one-parameter sets of fixed point solutions on both $\mathbb{S}^4$ and $\mathbb{H}^4$. At first sight this is a disappointing result for the asymptotic safety program. However, if we now use $\mathbb{R}^4$, eqn. \eqref{eq:flat}, to match smoothly between these solutions then we have two boundary conditions on a second order differential equation, one coming from the sphere and one from the hyperboloid. Thus, there can now only be at most a discrete set of solutions. In the next section more evidence will be presented for why these topologies should be considered smoothly joined together in this way.  \index{topologies smoothly joined together}

\subsection{Numerical solution}

\index{numerical analysis}
Note that this does not answer the question of whether there is more than one fixed point, or no fixed point at all, or the phenomenologically preferred answer: a single fixed point. As already mentioned, the only way to find out which of these latter possibilities is realised, is to perform an extensive numerical search for such global $f(R)$ solutions. 
As we have just seen, the asymptotic fixed point solutions \eqref{asympFPS} and \eqref{asympFPH}, on the sphere and hyperboloid respectively, depend on one single parameter, which we called $A$ on both sides. Even if there is a global solution that connects them, the value of $A$ will almost surely be different in the two different topologies.
On one of the topologies, one can scan through $A$ at some large value of $R$ and solve the equations backwards towards $R=0$. With the exponential cutoff profile \eqref{WetCutoff}, solutions have been found on the sphere this way, in a narrow region around $A=-0.01$, matching to the asymptotic series at $R\sim10$. 
%Global solutions were found on the sphere . 
%but even finding that solution required more input. As the equations are very sensitive, only by choosing to match $f'(R)$ and $f''(R)$ and then computing $f(R)$ (instead of setting $f(R)$ and $f'(R)$ from the asymptotic solution) were we able to locate a solution. 
On the hyperboloid no solution was found however, although a more comprehensive numerical analysis might find one \cite{Mitchell:2021qjr}.

\section{Eigenoperators}
So far we have analyzed the flow equation in the $f(R)$ approximation at the fixed point (where $\partial_tf(R)=0$). \index{eigenoperator}
%The question on whether a full solution exists, not just for the limiting cases where $R\rightarrow0$ or $R\rightarrow \infty$, is left to numerical analysis. Assuming that it does, 
Assuming that there is a global solution, we now turn to the question of whether the theory is predictive. This is answered by solving the eigenvalue equation and figuring out how many relevant operators \index{relevant} the fixed point solution has. Relevant operators are the ones that fall into the fixed point when increasing the cutoff scale $k$. The number of these operators corresponds to the number of parameters that will have to be fixed experimentally. 
%Hence, if the fixed point has an infinite number of relevant operators then the theory is no more predictive than a perturbatively non-renormalizable theory. 
We now prove that in this second order formulation, if we take the equations to apply simultaneously across the three spaces $\mathbb{S}^4$, $\mathbb{R}^4$ and $\mathbb{H}^4$, there are a finite number of relevant operators \cite{Benedetti:2013jk}.\index{relevant} Plugging \eqref{eq:pert} into the flow equation \eqref{eq:4} we get a second order ordinary differential eigenvalue equation: \index{eigenoperator}
\be 
  -a_2(R) \,v''(R)+a_1(R)\,v'(R)+a_0(R)\,v(R)=\lambda v(R)  \label{eq:eig}
\ee 
where the eigenvalues $\lambda=4-\theta$, $v(R)$ is the eigenoperator, and the $a_i$'s are given by eqns. (\ref{a2} -- \ref{a0}).

\subsection{Asymptotic analysis}
\label{sec:asympeigen}

\index{asymptotic analysis}
The first step is to apply asymptotic analysis to the eigenoperator equation. The procedure closely follows that for the fixed point in sec. \ref{sec:asymptoticsols}.
As already noted there, $a_0$ and $a_2$ decay exponentially fast and in the large $R$ limit $a_1\sim 2R$. Then the asymptotic form of the eigenvalue equation is: 
\be 
  \lambda\, v(R)-2R\,v'(R)=-a_2(R)\,v''(R)\,.
\ee 
Starting with the left hand side the solution is
\be 
\label{eigenpow}
  v(R)\propto |R|^{\frac{\lambda}{2}}+\cdots\,,
\ee 
where the ellipses stand for subleading corrections from the $a_i$'s, in particular from the RHS. The solutions have one parameter, the constant of proportionality. The missing parameter must come from a solution for which $v''(R)$ cannot be neglected. But this implies diverging derivatives and thus $v(R)$ can be neglected. The equation is then analogous to what we had before where the second solution is now $v(R)\sim \delta\! f(R)$ in \eqref{eq:sph} on the sphere and \eqref{eq:hyp} on the hyperboloid. 

%But for the same reason as before, these type of solutions are not physical. Since $|v(R)/f(R)|\rightarrow\infty$ as $R\rightarrow\infty$, for any fixed $\epsilon$ in \eqref{eq:pert}, we would have that the linearised solution is no longer describing the renormalization group flow close to the fixed point, invalidating our initial assumptions. The only eigenoperator solutions that  are left, are those that scale as powers of the curvature $|R|^{\frac{\lambda}{2}}$. (If this power is too large, the solution would again appear to be invalidated. Actually in this case an argument 

Now we ask whether these solutions are actually valid. 
The linearised solution  \eqref{eq:pert} is meant to describe the RG flow `close' to the fixed point. For any fixed $\epsilon$, if $|v(R)/f(R)|\to\infty$ as $R\to\pm\infty$ that is not necessarily true since linearisation is no longer valid. In this case one can set
\be 
\label{pertgen}
f_k(R)=f(R)+\epsilon\,v_k(R)\,,
\ee 
and, without linearising, ask for the correct evolution for $v_k(R)$ at large $R$. For large negative $R$ the RHS of the flow equation \eqref{eq:4} can be neglected. For large positive $R$, the RHS of the flow equation can be neglected except  for the $n=1$ $S_2$ component of $\mathcal{T}^{Jac}_0$, which however just cancels the contributions from the LHS that grow faster than $R^2$ resulting from $f(R)$, \cf \eqref{asympFPS}. Since in fact the $O(R^2)$ part of $f(R)$ also vanishes from the LHS  (on both sphere and hyperboloid), in the large $R$ regime one has
\be 
\partial_t v_k(R)-2R\,v'_k(R)+4\,v_k(R) = o( R^2)\,.
\ee
Any part of $v_k(R)$ growing at least as fast as $R^2$ is then easily solved for, and gives mean-field evolution involving some arbitrary function $v$: \index{mean field}
\be 
\label{tgen}
v_k(R) = {\rm e}^{-4t}\, v(R\,{\rm e}^{\,2t})+ o( R^2)\,.
\ee
It will be the same function $v$ that was introduced in the linearised solution \eqref{eq:pert} if one requires as boundary condition, $v_k(R)=v(R)$ at $k=\mu$. The question that remains is whether the RG evolution \eqref{tgen} is consistent with what we found by linearising.

For the power-law solution \eqref{eigenpow}, linearisation is valid at large $|R|$ if and only if $\lambda\le4$. This follows from the hyperboloid fixed point asymptotics \eqref{asympFPH},  the sphere side \eqref{asympFPS} requiring only the weaker constraint, $\lambda\le4+2b$.  On the other hand if $\lambda>4$, one can use the general perturbation \eqref{pertgen}, finding the solution \eqref{tgen}. Substituting the explicit form \eqref{eigenpow} of the boundary condition, gives:
\be 
\label{backtolinearised}
v_k(R) = v(R)\,\text{e}^{-\theta t} + o( R^2)\,,
\ee
where $\theta=4-\lambda$, \ie the linearised solution \eqref{eq:pert} is reproduced. We conclude that asymptotically, power-law eigenoperators \eqref{eigenpow} are valid solutions for any $\lambda$. Their $t$ evolution is multiplicative and given by the flow of a conjugate coupling $g(t)=\epsilon\,\text{e}^{-\theta t}$, \cf \eqref{eq:pert}. \index{multiplicative evolution}

On the other hand, the solutions that behave asymptotically as $v(R)\sim \delta\! f(R)$, are growing exponentials of exponentials. Linearisation is not valid at large $|R|$, where the $t$ dependence is given instead by \eqref{tgen}. Now we cannot separate out the $t$ dependence. Therefore, such perturbations cannot be regarded as eigenoperators evolving multiplicatively.

Excluding them leads to quantisation of the spectrum. \index{quantisation of spectrum}
This is because the large $R$ dependence \eqref{eigenpow} provides a boundary condition on both the sphere and the hyperboloid side, and  linearity provides a further boundary condition since one can choose a normalisation \eg $v(0)=1$. These three conditions over-constrain the eigenoperator equation \eqref{eq:eig} leading to quantisation of $\lambda$, \ie to a discrete eigenoperator spectrum. 

\subsection{Sturm-Liouville theory}
Sturm-Liouville type equations take the form \index{Sturm-Liouville analysis}
\be 
  Lv(R)=\lambda w(R)v(R),
\ee 
where $L$ is the self adjoint operator 
\be 
  L= -\frac{d}{dR}\Big(p(R) \frac{d}{dR} \cdot \Big)+q(R),
\ee 
with $p(R)$ and $q(R)$ being real functions and $w(R)$ also being positive. For the second order formulation, the eigenvalue equation can be put in this form. The properties of these equations will then allow us to draw conclusions about the spectrum of the eigenvalues. 

The weight function is defined as
\be 
  w(R)=\frac{1}{a_2(R)}\exp-\int^R \!\! dR'\frac{a_1(R')}{a_2(R')}\,,
\ee 
since, multiplying with the eigenvalue equation \eqref{eq:eig} and rearranging, casts it in Sturm-Liouville form:
\be 
  -\big(a_2(R)w(R)v'(R)\big)'+w(R)a_0(R)v(R)=\lambda w(R) v(R)\,.
\ee 
Notice that the trace in $a_2$ is positive. This is because the cutoff is monotonically decreasing, hence $r'(z)<0$ and $r(z)>0$, so the sign of $a_2$ depends on $c_{\bar{h}}$, which is positive. This implies that the weight function $w(R)>0$ as required.

Next we check if the operator is self-adjoint. Taking $v=v_j(R)$, multiplying by $v_i(R)$, and integrating over $R$, gives:
\be 
\label{selfadjcheck}
  -\int v_iLv_j=-\int v_i\big(a_2wv'_j\big)'+\int v_ia_0wv_j\,.
\ee 
If $L$ is self-adjoint then this should be the same for $j \leftrightarrow i$. 
The first term on the RHS can be written as
\be 
  -\int \big[v_i\big(a_2wv_j'\big)\big]'+\int \big[v_j\big(a_2wv_i'\big)\big]' -\int \big(a_2wv_i'\big)'v_j\,.
\ee 
Thus what is required is that the first two terms above cancel each other. This is automatically satisfied if $R$ is taken to have the full range since $w(R)\rightarrow0$ exponentially fast as $R\rightarrow\pm \infty$. If the differential equation is restricted to either the four-sphere or four-hyperboloid, there would be a boundary at $R=0$. The weight function does not vanish there and thus the operator $L$ would then not be self-adjoint. This is another powerful hint that the correct treatment is to smoothly join the three topologies together. Note also that none of these equations would make sense if the exponentially growing set of solutions $v(R)\sim \delta\! f(R)$ are included, where $\delta\! f(R)$ is given by \eqref{eq:sph} or \eqref{eq:hyp}.  From \eqref{Basymp} and \eqref{trick} one can see that actually these $\delta\! f(R)\sim 1/w(R)$ and thus such $v(R)$ are not square integrable under the weight function $w(R)$ since $w(R)v^2(R)\sim 1/w(R)$, which diverges at large $R$. Hence, this condition only picks out the correct solutions from the eigenvalue equation and justifies the use of Sturm-Liouville techniques. \index{topologies smoothly joined together}

Thus, when restricted to perturbations that grow only as a power at large $|R|$, the eigenvalue equation \eqref{eq:eig} is of Sturm-Liouville type. The consequences for the spectrum of the eigenvalues can be seen by a standard transformation to Liouville normal form. Define a coordinate $x$ as
\be 
  x=\int_{0}^R\frac{1}{\sqrt{a_2(R')}}dR'. \label{eq:x}
\ee 
Then $x\rightarrow\pm \infty$ as $R\rightarrow\pm \infty$ because $a_2(R)$ vanishes at large $|R|$. Defining the `wave-function'
\be 
  \psi(x)=a_2^{\frac{1}{4}}(R)w^{\frac{1}{2}}(R)v(R),
\ee 
\eqref{eq:eig} can be transformed into
\be 
  -\frac{d^2\psi(x)}{dx^2}+U(x)\psi(x)=\lambda \psi(x), \label{eq:28}
\ee 
which is just the one-dimensional Schr\"odinger equation with energy $\lambda$. The potential turns out to be \cite{Benedetti:2013jk}
\be 
  U(x)=a_0+\frac{a_1^2}{4a_2}-\frac{a'_1}{2}+a'_2\Big(\frac{a_1}{2a_2}+\frac{3a'_2}{16a_2}\Big)-\frac{a''_2}{4}\,.
\ee 
This potential has no singularities at finite $x$. Asymptotically the term proportional to $a_1^2$ will dominate for $x\rightarrow \pm \infty$ and thus the potential $U(x)\rightarrow+\infty$. This then implies the following important properties:
\begin{itemize}
    \item[1] The eigenvalues $\lambda_n$ are discrete, real and non-degenerate.
    \item[2] There exists a lowest eigenvalue $\lambda_0$ (i.e. bounded from below).
    \item[3] The only accumulation point is at infinity.
\end{itemize}
Asymptotic analysis already showed that the eigenvalues are discrete, but this Sturm-Liouville analysis allows to conclude much more.
Now it is straightforward to see that there is a finite number of relevant operators such that $\theta_n=4-\lambda_n\geq0$.\index{relevant} Indeed this is so because $\lambda_n \rightarrow \infty$ as $n\rightarrow \infty$ and because there exists a lowest eigenvalue $\lambda_0$.

But these results should be accepted with caution. Recall that to obtain them some severe approximations were used, such as the single metric approximation and the truncation to the function $f(R)$. One way to judge the validity of the results is to check the extent to which they are scheme independent (universal), in particular independent of the choice of cutoff. It turns out that the critical exponents $\theta_n$ can be solved for analytically, again by using asymptotic analysis, and this gives a precise way to answer the question of scheme dependence in this regime. \index{universality} \index{scheme independence}

From \eqref{asympa2S} and \eqref{asympa2H}, the reader can see that the leading contribution to $a_2$ takes the following form on both sphere and hyperboloid: 
\be 
\label{a2gen}
  a_2(R)=\frac{1}{G^2(R)}\,\mathrm{e}^{-2F(R)}
\ee 
where $F(R)$ is positive and proportional to $|R|^b$ and $G(R)$ goes like a power of $R$. They therefore satisfy the conditions required to use the trick \eqref{trick} on the equation \eqref{eq:x} defining $x$. Then asymptotically
\be 
  x=\frac{G(R)}{F'(R)}\,\mathrm{e}^{F(R)}+...\label{eq:10}
\ee 
where the ellipsis stand for multiplicative subleading terms. Alternatively this can be seen by differentiating \eqref{eq:x} and \eqref{eq:10} with respect to $R$. The potential can then be approximated to leading order as
\be 
  U(x)=\frac{a_1^2}{4a_2}=\frac{R^2}{a_2(R)}=\big[RF'(R)\big]^2x^2.
\ee 
Evidently $RF'(R)=bF(R)$ and thus, taking logs of \eqref{eq:10},
\be 
  U(x)=(bx\ln|x|)^2\bigg\{1+O\bigg(\frac{\ln\ln|x|}{\ln|x|}\bigg)\bigg\}\quad\quad x\rightarrow \pm \infty\,,
\ee 
where in the equation above the order of the subleading correction is also indicated. (The latter requires taking into account iterations of \eqref{trick} and the subleading corrections to $a_2$.)
Using the WKB approximation one can then find the critical exponents \index{critical exponent} for large $n$ \cite{Mitchell:2021qjr}: \index{critical exponents $\theta_n$ at large $n$}
\be 
\label{thetan}
  \theta_n=-b\,(n\ln n)\bigg\{1+O\bigg(\frac{\ln\ln n}{\ln n}\bigg)\bigg\}\quad \text{as}\quad n\to \infty\,.
\ee 
The result shows almost a linear dependence on $n$. This much is similar to extensive numerical work done on large polynomial truncations of a third order formulation up to $n\leq70$ \cite{Falls:2018ylp}. These authors find near-Gaussian scaling dimension. They use an adaptive cutoff so there is no direct comparison, and they use the optimised profile \eqref{optimised} with no free parameters in the cutoff, so universality is not tested in this way. Indeed the scaling dimension should be universal. The leading behaviour of this expression is independent of all parameters in the chosen general family of cutoffs, except one, namely the parameter $b$ in \eqref{WetCutoff}. Explicitly, it is independent of $a$ in \eqref{WetCutoff}, and of all the $c_\phi$ and $\alpha_i$. Unfortunately the dependence on $b$ still amounts to strong dependence. 

\index{single metric approximation}
Actually this remaining dependence is an artefact of the single-metric approximation \cite{Reuter:1996}.\footnote{More generally, single-field approximations are a known source of artefacts \cite{Bridle:2013sra}.} We have seen that it comes from the $R^b$ dependence of $F(R)$ in \eqref{a2gen}, equivalently \eqref{asympa2S} and \eqref{asympa2H}. This in turn arises from the cutoff dependence in eqn. \eqref{a2} and in particular the cutoff profile's dependence on $R$ (through in fact the lowest eigenvalue). To see that the dependence in \eqref{thetan} is an artefact of the single-metric approximation, imagine for the moment that the single-metric approximation was not made and yet somehow the initial ansatz \eqref{fR} still made sense. (In reality such a simple ansatz would no longer be possible because diffeomorphism invariance is replaced by BRST invariance for the quantum fields and furthermore it is badly broken, but let us overlook that for the moment.) Now the curvature  in it is the full quantum curvature $\hat{R}$,  due to the full \textit{quantum} metric $\hat{g}_{\mu\nu}$ in \eqref{split}. The trace and the cutoff in \eqref{a2} come from summing over modes on the background manifold in \eqref{LegFlow} so they depend on the \textit{background} curvature $R$. 
The Hessian in \eqref{LegFlow} will result in differentiating $f(\hat{R})$ with respect to the fluctuation field $h_{\mu\nu}$ or equivalently differentiating with respect to $\hat{g}_{\mu\nu}$. Thus ultimately the eigenoperator perturbation equation \eqref{eq:10a} would take the form:
\be
\label{eq:10ap}
  -a_2(R,\hat{R})\,\delta f''(\hat{R})+a_1(R,\hat{R})\,\delta f'(\hat{R})+a_0(R,\hat{R})\,\delta f(\hat{R})=4\,\delta f(\hat{R}) 
\ee
with in particular:
\be
  a_2=\frac{144c_h}{V}\text{Tr}\Bigg[\frac{\Delta_0^2(2r(\Delta_0+\alpha_0R)-(\Delta_0+\alpha_0R)r'(\Delta_0+\alpha_0R))}{(9f''(\hat{R})\Delta_0^2+3f'(\hat{R})\Delta_0+E(\hat{R})+16c_hr(\Delta_0+\alpha_0R))^2}\Bigg]\,. \label{a2p}
\ee
In deriving \eqref{thetan} one is interested in the large $\hat{R}$ dependence of \eqref{eq:10ap}. This depends on the large $\hat{R}$ dependence of the fixed point functional $f(\hat{R})$, and this feeds in to the coefficients $a_i(R,\hat{R})$. But there is no $\exp(-a\hat{R}^b)$ dependence because the cutoff profile $r$ depends only on the background curvature $R$, either directly or through the Laplacians whose eigenvalues only depend on the background manifold.

\newpage

\bibliographystyle{hunsrt}
\bibliography{references}

\begin{thebibliography}{10}

\bibitem{Weinberg:1980}
S.~Weinberg.
\newblock {Ultraviolet Divergences In Quantum Theories Of Gravitation}.
\newblock {\em In Hawking, S.W., Israel, W.: General Relativity; Cambridge
  University Press}, pages 790--831, 1980.

\bibitem{Reuter:1996}
M.~Reuter.
\newblock {Nonperturbative evolution equation for quantum gravity}.
\newblock {\em Phys.Rev.}, D57:971--985, 1998, hep-th/9605030.

\bibitem{Percacci:2017fkn}
Robert Percacci.
\newblock {\em {An Introduction to Covariant Quantum Gravity and Asymptotic
  Safety}}, volume~3 of {\em 100 Years of General Relativity}.
\newblock World Scientific, 2017.

\bibitem{Reuter:2019byg}
Martin Reuter and Frank Saueressig.
\newblock {\em {Quantum Gravity and the Functional Renormalization Group}}.
\newblock Cambridge University Press, 2019.

\bibitem{Stelle:1976gc}
K.~S. Stelle.
\newblock {Renormalization of Higher Derivative Quantum Gravity}.
\newblock {\em Phys. Rev.}, D16:953--969, 1977.

\bibitem{Wilson:1973}
K.G. Wilson and John~B. Kogut.
\newblock {The Renormalization group and the epsilon expansion}.
\newblock {\em Phys.Rept.}, 12:75--200, 1974.

\bibitem{Wegner:1972ih}
Franz~J. Wegner and Anthony Houghton.
\newblock {Renormalization group equation for critical phenomena}.
\newblock {\em Phys. Rev.}, A8:401--412, 1973.

\bibitem{Polchinski:1983gv}
Joseph Polchinski.
\newblock {Renormalization and Effective Lagrangians}.
\newblock {\em Nucl.Phys.}, B231:269--295, 1984.

\bibitem{Nicoll1977}
J.~F. Nicoll and T.~S. Chang.
\newblock {An Exact One Particle Irreducible Renormalization Group Generator
  for Critical Phenomena}.
\newblock {\em Phys. Lett.}, A62:287--289, 1977.

\bibitem{Wetterich:1992}
Christof Wetterich.
\newblock {Exact evolution equation for the effective potential}.
\newblock {\em Phys.Lett.}, B301:90--94, 1993.

\bibitem{Morris:1993}
Tim~R. Morris.
\newblock {The Exact renormalization group and approximate solutions}.
\newblock {\em Int.J.Mod.Phys.}, A 09:2411--2450, 1994, hep-ph/9308265.

\bibitem{Morris:1999px}
Tim~R. Morris.
\newblock {A Gauge invariant exact renormalization group. 1.}
\newblock {\em Nucl. Phys.}, B573:97--126, 2000, hep-th/9910058.

\bibitem{Morris:2000jj}
Tim~R. Morris.
\newblock {An Exact RG formulation of quantum gauge theory}.
\newblock {\em Int. J. Mod. Phys.}, A16:1899--1912, 2001, hep-th/0102120.

\bibitem{Ellwanger:1994iz}
Ulrich Ellwanger.
\newblock {Flow equations and BRS invariance for Yang-Mills theories}.
\newblock {\em Phys. Lett.}, B335:364--370, 1994, hep-th/9402077.

\bibitem{Morris:2018zgy}
Tim~R. Morris and Roberto Percacci.
\newblock {Trace anomaly and infrared cutoffs}.
\newblock {\em Phys. Rev.}, D99(10):105007, 2019, 1810.09824.

\bibitem{Falls_2014}
Kevin Falls and Daniel~F. Litim.
\newblock Black hole thermodynamics under the microscope.
\newblock {\em Physical Review D}, 89(8), Apr 2014.

\bibitem{Falls:2018ylp}
Kevin~G. Falls, Daniel~F. Litim, and Jan Schr\"oder.
\newblock {Aspects of asymptotic safety for quantum gravity}.
\newblock {\em Phys. Rev. D}, 99(12):126015, 2019, 1810.08550.

\bibitem{Falls:2017lst}
Kevin Falls, Callum~R. King, Daniel~F. Litim, Kostas Nikolakopoulos, and
  Christoph Rahmede.
\newblock {Asymptotic safety of quantum gravity beyond Ricci scalars}.
\newblock {\em Phys. Rev. D}, 97(8):086006, 2018, 1801.00162.

\bibitem{Kluth:2020bdv}
Yannick Kluth and Daniel~F. Litim.
\newblock {Fixed Points of Quantum Gravity and the Dimensionality of the UV
  Critical Surface}.
\newblock 8 2020, 2008.09181.

\bibitem{Becker:2014qya}
Daniel Becker and Martin Reuter.
\newblock {En route to Background Independence: Broken split-symmetry, and how
  to restore it with bi-metric average actions}.
\newblock {\em Annals Phys.}, 350:225--301, 2014, 1404.4537.

\bibitem{Becker:2021pwo}
Maximilian Becker and Martin Reuter.
\newblock {Background independent field quantization with sequences of
  gravity-coupled approximants. II. Metric fluctuations}.
\newblock {\em Phys. Rev. D}, 104(12):125008, 2021, 2109.09496.

\bibitem{Morris:2016spn}
Tim~R. Morris.
\newblock {Large curvature and background scale independence in single-metric
  approximations to asymptotic safety}.
\newblock {\em JHEP}, 11:160, 2016, 1610.03081.

\bibitem{Percacci:2016arh}
Roberto Percacci and Gian~Paolo Vacca.
\newblock {The background scale Ward identity in quantum gravity}.
\newblock {\em Eur. Phys. J.}, C77(1):52, 2017, 1611.07005.

\bibitem{Ohta:2017dsq}
Nobuyoshi Ohta.
\newblock {Background Scale Independence in Quantum Gravity}.
\newblock {\em PTEP}, 2017(3):033E02, 2017, 1701.01506.

\bibitem{Morris:2016nda}
Tim~R. Morris and Anthony W.~H. Preston.
\newblock {Manifestly diffeomorphism invariant classical Exact Renormalization
  Group}.
\newblock {\em JHEP}, 06:012, 2016, 1602.08993.

\bibitem{Falls:2020tmj}
Kevin Falls.
\newblock {Background independent exact renormalisation}.
\newblock 2020, 2004.11409.

\bibitem{Mandric:2022dte}
Vlad-Mihai Mandric and Tim~R. Morris.
\newblock {Properties of a proposed background independent exact
  renormalization group}.
\newblock 10 2022, 2210.00492.

\bibitem{Pawlowski:2020qer}
Jan~M. Pawlowski and Manuel Reichert.
\newblock {Quantum Gravity: A Fluctuating Point of View}.
\newblock {\em Front. in Phys.}, 8:551848, 2021, 2007.10353.

\bibitem{Bonanno:2020bil}
Alfio Bonanno, Astrid Eichhorn, Holger Gies, Jan~M. Pawlowski, Roberto
  Percacci, Martin Reuter, Frank Saueressig, and Gian~Paolo Vacca.
\newblock {Critical reflections on asymptotically safe gravity}.
\newblock {\em Front. in Phys.}, 8:269, 2020, 2004.06810.

\bibitem{Codello:2007bd}
Alessandro Codello, Roberto Percacci, and Christoph Rahmede.
\newblock {Ultraviolet properties of f(R)-gravity}.
\newblock {\em Int. J. Mod. Phys.}, A23:143--150, 2008, 0705.1769.

\bibitem{Machado:2007}
Pedro~F. Machado and Frank Saueressig.
\newblock {On the renormalization group flow of f(R)-gravity}.
\newblock {\em Phys.Rev.}, D77:124045, 2008, 0712.0445.

\bibitem{Codello:2008}
Alessandro Codello, Roberto Percacci, and Christoph Rahmede.
\newblock {Investigating the Ultraviolet Properties of Gravity with a Wilsonian
  Renormalization Group Equation}.
\newblock {\em Annals Phys.}, 324:414--469, 2009, 0805.2909.

\bibitem{Benedetti:2012}
Dario Benedetti and Francesco Caravelli.
\newblock {The Local potential approximation in quantum gravity}.
\newblock {\em JHEP}, 1206:017, 2012, 1204.3541.

\bibitem{Demmel:2012ub}
Maximilian Demmel, Frank Saueressig, and Omar Zanusso.
\newblock {Fixed-Functionals of three-dimensional Quantum Einstein Gravity}.
\newblock {\em JHEP}, 11:131, 2012, 1208.2038.

\bibitem{Demmel:2013myx}
Maximilian Demmel, Frank Saueressig, and Omar Zanusso.
\newblock {Fixed Functionals in Asymptotically Safe Gravity}.
\newblock In {\em {Proceedings, 13th Marcel Grossmann Meeting on Recent
  Developments in Theoretical and Experimental General Relativity,
  Astrophysics, and Relativistic Field Theories (MG13): Stockholm, Sweden, July
  1-7, 2012}}, pages 2227--2229, 2015, 1302.1312.

\bibitem{Demmel:2014hla}
Maximilian Demmel, Frank Saueressig, and Omar Zanusso.
\newblock {RG flows of Quantum Einstein Gravity in the linear-geometric
  approximation}.
\newblock {\em Annals Phys.}, 359:141--165, 2015, 1412.7207.

\bibitem{Demmel:2014fk}
Maximilian Demmel, Frank Saueressig, and Omar Zanusso.
\newblock {RG flows of Quantum Einstein Gravity on maximally symmetric spaces}.
\newblock {\em JHEP}, 06:026, 2014, 1401.5495.

\bibitem{Demmel2015b}
Maximilian Demmel, Frank Saueressig, and Omar Zanusso.
\newblock {A proper fixed functional for four-dimensional Quantum Einstein
  Gravity}.
\newblock {\em JHEP}, 08:113, 2015, 1504.07656.

\bibitem{Ohta:2015efa}
Nobuyoshi Ohta, Roberto Percacci, and Gian~Paolo Vacca.
\newblock {Flow equation for $f(R)$ gravity and some of its exact solutions}.
\newblock {\em Phys. Rev.}, D92(6):061501, 2015, 1507.00968.

\bibitem{Ohta2016}
Nobuyoshi Ohta, Roberto Percacci, and Gian~Paolo Vacca.
\newblock {Renormalization Group Equation and scaling solutions for f(R)
  gravity in exponential parametrization}.
\newblock {\em Eur. Phys. J.}, C76(2):46, 2016, 1511.09393.

\bibitem{Falls:2016msz}
Kevin Falls and Nobuyoshi Ohta.
\newblock {Renormalization Group Equation for $f(R)$ gravity on hyperbolic
  spaces}.
\newblock {\em Phys. Rev.}, D94(8):084005, 2016, 1607.08460.

\bibitem{Benedetti:2013jk}
Dario Benedetti.
\newblock {On the number of relevant operators in asymptotically safe gravity}.
\newblock {\em Europhys. Lett.}, 102:20007, 2013, 1301.4422.

\bibitem{Mitchell:2021qjr}
Alex Mitchell, Tim~R. Morris, and Dalius Stulga.
\newblock {Provable properties of asymptotic safety in f(R) approximation}.
\newblock {\em JHEP}, 01:041, 2022, 2111.05067.

\bibitem{Percacci:2015wwa}
Roberto Percacci and Gian~Paolo Vacca.
\newblock {Search of scaling solutions in scalar-tensor gravity}.
\newblock {\em Eur. Phys. J.}, C75(5):188, 2015, 1501.00888.

\bibitem{Labus:2015ska}
Peter Labus, Roberto Percacci, and Gian~Paolo Vacca.
\newblock {Asymptotic safety in $O(N)$ scalar models coupled to gravity}.
\newblock {\em Phys. Lett.}, B753:274--281, 2016, 1505.05393.

\bibitem{Eichhorn:2015bna}
Astrid Eichhorn.
\newblock {The Renormalization Group flow of unimodular f(R) gravity}.
\newblock {\em JHEP}, 04:096, 2015, 1501.05848.

\bibitem{Hasenfratz:1985dm}
Anna Hasenfratz and Peter Hasenfratz.
\newblock {Renormalization Group Study of Scalar Field Theories}.
\newblock {\em Nucl.Phys.}, B270:687--701, 1986.

\bibitem{Morris:1994ki}
Tim~R. Morris.
\newblock {On truncations of the exact renormalization group}.
\newblock {\em Phys.Lett.}, B334:355--362, 1994, hep-th/9405190.

\bibitem{Morris:1994ie}
Tim~R. Morris.
\newblock {Derivative expansion of the exact renormalization group}.
\newblock {\em Phys.Lett.}, B329:241--248, 1994, hep-ph/9403340.

\bibitem{Morris:1994jc}
Tim~R. Morris.
\newblock {The Renormalization group and two-dimensional multicritical
  effective scalar field theory}.
\newblock {\em Phys.Lett.}, B345:139--148, 1995, hep-th/9410141.

\bibitem{Morris:1996xq}
Tim~R. Morris.
\newblock {Three-dimensional massive scalar field theory and the derivative
  expansion of the renormalization group}.
\newblock {\em Nucl.Phys.}, B495:477--504, 1997, hep-th/9612117.

\bibitem{Morris:1998}
Tim~R. Morris.
\newblock {Elements of the continuous renormalization group}.
\newblock {\em Prog.Theor.Phys.Suppl.}, 131:395--414, 1998, hep-th/9802039.

\bibitem{Morris:1996nx}
Tim~R. Morris.
\newblock {On the fixed point structure of scalar fields}.
\newblock {\em Phys. Rev. Lett.}, 77:1658, 1996, hep-th/9601128.

\bibitem{Dietz:2012ic}
Juergen~A. Dietz and Tim~R. Morris.
\newblock {Asymptotic safety in the f(R) approximation}.
\newblock {\em JHEP}, 01:108, 2013, 1211.0955.

\bibitem{Dietz:2013sba}
Juergen~A. Dietz and Tim~R. Morris.
\newblock {Redundant operators in the exact renormalisation group and in the
  f(R) approximation to asymptotic safety}.
\newblock {\em JHEP}, 07:064, 2013, 1306.1223.

\bibitem{Bridle:2013sra}
I.~Hamzaan Bridle, Juergen~A. Dietz, and Tim~R. Morris.
\newblock {The local potential approximation in the background field
  formalism}.
\newblock {\em JHEP}, 03:093, 2014, 1312.2846.

\bibitem{Gonzalez_Martin_2017}
Sergio Gonzalez-Martin, Tim~R. Morris, and Zo{\"e}~H. Slade.
\newblock Asymptotic solutions in asymptotic safety.
\newblock {\em Physical Review D}, 95(10), May 2017.

\bibitem{WR}
F.~J. Wegner.
\newblock Some invariance properties of the renormalization group.
\newblock {\em J. Phys.}, C7:2098, 1974.

\bibitem{Latorre:2000jp}
Jose~I. Latorre and Tim~R. Morris.
\newblock {Scheme independence as an inherent redundancy in quantum field
  theory}.
\newblock {\em Int. J. Mod. Phys.}, A16:2071--2074, 2001, hep-th/0102037.

\bibitem{ZinnJustin:2002ru}
Jean Zinn-Justin.
\newblock {Quantum field theory and critical phenomena}.
\newblock {\em Int. Ser. Monogr. Phys.}, 113:1--1054, 2002.

\bibitem{Igarashi:2019gkm}
Yuji Igarashi, Katsumi Itoh, and Tim~R. Morris.
\newblock {BRST in the Exact RG}.
\newblock {\em PTEP}, 2019(10):103B01, 2019, 1904.08231.

\bibitem{York:1973ia}
James~W. York, Jr.
\newblock {Conformally invariant orthogonal decomposition of symmetric tensors
  on Riemannian manifolds and the initial value problem of general relativity}.
\newblock {\em J. Math. Phys.}, 14:456--464, 1973.

\bibitem{Gibbons:1978ac}
G.W. Gibbons, S.W. Hawking, and M.J. Perry.
\newblock {Path Integrals and the Indefiniteness of the Gravitational Action}.
\newblock {\em Nucl.Phys.}, B138:141, 1978.

\bibitem{Dietz:2016gzg}
Juergen~A. Dietz, Tim~R. Morris, and Zoe~H. Slade.
\newblock {Fixed point structure of the conformal factor field in quantum
  gravity}.
\newblock {\em Phys. Rev.}, D94(12):124014, 2016, 1605.07636.

\bibitem{Morris:2018mhd}
Tim~R. Morris.
\newblock {Renormalization group properties in the conformal sector: towards
  perturbatively renormalizable quantum gravity}.
\newblock {\em JHEP}, 08:024, 2018, 1802.04281.

\bibitem{Morris:2018upm}
Tim~R. Morris.
\newblock {Perturbatively renormalizable quantum gravity}.
\newblock {\em Int. J. Mod. Phys.}, D27(14):1847003, 2018, 1804.03834.

\bibitem{Litim:2001}
Daniel~F. Litim.
\newblock {Optimized renormalization group flows}.
\newblock {\em Phys.Rev.}, D64:105007, 2001, hep-th/0103195.

\bibitem{Camporesi:1994ga}
R.~Camporesi and A.~Higuchi.
\newblock {Spectral functions and zeta functions in hyperbolic spaces}.
\newblock {\em J. Math. Phys.}, 35:4217--4246, 1994.

\end{thebibliography}


%% This BibTeX bibliography file was created using BibDesk.
%% https://bibdesk.sourceforge.io/

%% Created for Tim R Morris at 2021-03-10 15:22:12 +0000 


%% Saved with string encoding Unicode (UTF-8) 


@comment{jabref-meta: databaseType:bibtex;}



@article{Manrique:2011jc,
	archiveprefix = {arXiv},
	author = {Manrique, Elisa and Rechenberger, Stefan and Saueressig, Frank},
	date-added = {2021-03-10 15:21:40 +0000},
	date-modified = {2021-03-10 15:21:40 +0000},
	doi = {10.1103/PhysRevLett.106.251302},
	eprint = {1102.5012},
	journal = {Phys. Rev. Lett.},
	pages = {251302},
	primaryclass = {hep-th},
	reportnumber = {MZ-TH-11-02},
	slaccitation = {%%CITATION = ARXIV:1102.5012;%%},
	title = {{Asymptotically Safe Lorentzian Gravity}},
	volume = {106},
	year = {2011},
	Bdsk-Url-1 = {https://doi.org/10.1103/PhysRevLett.106.251302}}

@article{Wang:2017brl,
	archiveprefix = {arXiv},
	author = {Wang, Anzhong},
	date-added = {2021-03-10 11:57:07 +0000},
	date-modified = {2021-03-10 12:50:47 +0000},
	doi = {10.1142/S0218271817300142},
	eprint = {1701.06087},
	journal = {Int. J. Mod. Phys.},
	number = {07},
	pages = {1730014},
	primaryclass = {gr-qc},
	slaccitation = {%%CITATION = ARXIV:1701.06087;%%},
	title = {{Ho\v rava gravity at a Lifshitz point: A progress report}},
	volume = {D26},
	year = {2017},
	Bdsk-Url-1 = {https://doi.org/10.1142/S0218271817300142}}

@article{second,
	archiveprefix = {arXiv},
	author = {Kellett, Matthew and Mitchell, Alex and Morris, Tim R.},
	date-added = {2021-03-10 11:41:02 +0000},
	date-modified = {2021-03-10 11:42:08 +0000},
	eprint = {2006.16682},
	keywords = {Kellett:2020mle},
	primaryclass = {hep-th},
	slaccitation = {%%CITATION = ARXIV:2006.16682;%%},
	title = {{The continuum limit of quantum gravity at second order in perturbation theory}},
	year = {2020}}

@article{Horava:2009uw,
	archiveprefix = {arXiv},
	author = {{Ho\v rava}, Petr},
	date-added = {2021-03-10 11:27:44 +0000},
	date-modified = {2021-03-10 11:38:41 +0000},
	doi = {10.1103/PhysRevD.79.084008},
	eprint = {0901.3775},
	journal = {Phys. Rev.},
	pages = {084008},
	primaryclass = {hep-th},
	slaccitation = {%%CITATION = ARXIV:0901.3775;%%},
	title = {{Quantum Gravity at a Lifshitz Point}},
	volume = {D79},
	year = {2009},
	Bdsk-Url-1 = {https://doi.org/10.1103/PhysRevD.79.084008}}

@article{Bojowald:2016itl,
	archiveprefix = {arXiv},
	author = {Bojowald, Martin and Brahma, Suddhasattwa},
	date-added = {2021-03-08 16:09:24 +0000},
	date-modified = {2021-03-08 16:09:24 +0000},
	doi = {10.1103/PhysRevD.98.026012},
	eprint = {1610.08850},
	journal = {Phys. Rev.},
	number = {2},
	pages = {026012},
	primaryclass = {gr-qc},
	slaccitation = {%%CITATION = ARXIV:1610.08850;%%},
	title = {{Signature change in two-dimensional black-hole models of loop quantum gravity}},
	volume = {D98},
	year = {2018},
	Bdsk-Url-1 = {https://doi.org/10.1103/PhysRevD.98.026012}}

@article{Bojowald:2018xxu,
	archiveprefix = {arXiv},
	author = {Bojowald, Martin and Brahma, Suddhasattwa and Yeom, Dong-han},
	date-added = {2021-03-08 16:09:01 +0000},
	date-modified = {2021-03-08 16:09:01 +0000},
	doi = {10.1103/PhysRevD.98.046015},
	eprint = {1803.01119},
	journal = {Phys. Rev.},
	number = {4},
	pages = {046015},
	primaryclass = {gr-qc},
	slaccitation = {%%CITATION = ARXIV:1803.01119;%%},
	title = {{Effective line elements and black-hole models in canonical loop quantum gravity}},
	volume = {D98},
	year = {2018},
	Bdsk-Url-1 = {https://doi.org/10.1103/PhysRevD.98.046015}}

@article{Perry:1993ry,
	archiveprefix = {arXiv},
	author = {Perry, Malcolm J. and Teo, Edward},
	date-added = {2021-03-08 16:08:37 +0000},
	date-modified = {2021-03-08 16:08:37 +0000},
	doi = {10.1103/PhysRevLett.70.2669},
	eprint = {hep-th/9302037},
	journal = {Phys. Rev. Lett.},
	pages = {2669-2672},
	primaryclass = {hep-th},
	reportnumber = {DAMTP-R-93-1},
	slaccitation = {%%CITATION = HEP-TH/9302037;%%},
	title = {{Nonsingularity of the exact two-dimensional string black hole}},
	volume = {70},
	year = {1993},
	Bdsk-Url-1 = {https://doi.org/10.1103/PhysRevLett.70.2669}}

@article{Stern:2018wud,
	archiveprefix = {arXiv},
	author = {Stern, A. and Xu, Chuang},
	date-added = {2021-03-08 16:07:17 +0000},
	date-modified = {2021-03-08 16:07:17 +0000},
	doi = {10.1103/PhysRevD.98.086015},
	eprint = {1808.07963},
	journal = {Phys. Rev.},
	number = {8},
	pages = {086015},
	primaryclass = {hep-th},
	slaccitation = {%%CITATION = ARXIV:1808.07963;%%},
	title = {{Signature change in matrix model solutions}},
	volume = {D98},
	year = {2018},
	Bdsk-Url-1 = {https://doi.org/10.1103/PhysRevD.98.086015}}

@article{Steinacker:2017vqw,
	archiveprefix = {arXiv},
	author = {Steinacker, Harold C.},
	date-added = {2021-03-08 16:06:45 +0000},
	date-modified = {2021-03-08 16:06:45 +0000},
	doi = {10.1007/JHEP02(2018)033},
	eprint = {1709.10480},
	journal = {JHEP},
	pages = {033},
	primaryclass = {hep-th},
	reportnumber = {UWTHPH-2017-31},
	slaccitation = {%%CITATION = ARXIV:1709.10480;%%},
	title = {{Cosmological space-times with resolved Big Bang in Yang-Mills matrix models}},
	volume = {02},
	year = {2018},
	Bdsk-Url-1 = {https://doi.org/10.1007/JHEP02(2018)033}}

@article{Chaney:2015mfa,
	archiveprefix = {arXiv},
	author = {Chaney, A. and Lu, Lei and Stern, A.},
	date-added = {2021-03-08 16:06:15 +0000},
	date-modified = {2021-03-08 16:06:15 +0000},
	doi = {10.1103/PhysRevD.92.064021},
	eprint = {1506.03505},
	journal = {Phys. Rev.},
	number = {6},
	pages = {064021},
	primaryclass = {hep-th},
	slaccitation = {%%CITATION = ARXIV:1506.03505;%%},
	title = {{Lorentzian Fuzzy Spheres}},
	volume = {D92},
	year = {2015},
	Bdsk-Url-1 = {https://doi.org/10.1103/PhysRevD.92.064021}}

@article{Ambjorn:2015qja,
	archiveprefix = {arXiv},
	author = {Ambjrn, Jan and Coumbe, Daniel N. and Gizbert-Studnicki, Jakub and Jurkiewicz, Jerzy},
	date-added = {2021-03-08 16:05:32 +0000},
	date-modified = {2021-03-08 16:05:32 +0000},
	doi = {10.1007/JHEP08(2015)033},
	eprint = {1503.08580},
	journal = {JHEP},
	note = {[JHEP08,033(2015)]},
	pages = {033},
	primaryclass = {hep-th},
	slaccitation = {%%CITATION = ARXIV:1503.08580;%%},
	title = {{Signature Change of the Metric in CDT Quantum Gravity?}},
	volume = {08},
	year = {2015},
	Bdsk-Url-1 = {https://doi.org/10.1007/JHEP08(2015)033}}

@book{john1991partial,
	author = {John, F.},
	date-added = {2020-10-14 11:32:10 +0100},
	date-modified = {2020-10-14 11:32:10 +0100},
	isbn = {9780387906096},
	lccn = {81016636},
	publisher = {Springer New York},
	series = {Applied Mathematical Sciences},
	title = {Partial Differential Equations},
	url = {https://books.google.co.uk/books?id=cBib\_bsGGLYC},
	year = {1991},
	Bdsk-Url-1 = {https://books.google.co.uk/books?id=cBib%5C_bsGGLYC}}

@book{evans10,
	abstract = {"This is the second edition of the now definitive text on partial differential equations (PDE). It offers a comprehensive survey of modern techniques in the theoretical study of PDE with particular emphasis on nonlinear equations. Its wide scope and clear exposition make it a great text for a graduate course in PDE. For this edition, the author has made numerous changes, including: a new chapter on nonlinear wave equations, more than 80 new exercises, several new sections, and a significantly expanded bibliography."--Publisher's description.},
	added-at = {2015-07-29T08:37:26.000+0200},
	address = {Providence, R.I.},
	author = {Evans, Lawrence C.},
	biburl = {https://www.bibsonomy.org/bibtex/2f5b120723ea78913e7e700ddd1a99301/ytyoun},
	date-added = {2020-10-12 17:34:31 +0100},
	date-modified = {2020-10-12 17:34:31 +0100},
	interhash = {59982ce44cc43813ccb14c0d647a59ee},
	intrahash = {f5b120723ea78913e7e700ddd1a99301},
	isbn = {9780821849743 0821849743},
	keywords = {partial.differential.equations pde textbook},
	publisher = {American Mathematical Society},
	refid = {465190110},
	timestamp = {2015-07-29T08:37:26.000+0200},
	title = {Partial differential equations},
	year = 2010}

@book{morse1953methods,
	author = {Morse, P.M.C. and Feshbach, H.},
	date-added = {2020-09-25 12:03:57 +0100},
	date-modified = {2020-09-25 12:03:57 +0100},
	lccn = {52011515},
	publisher = {McGraw-Hill},
	series = {International series in pure and applied physics},
	title = {Methods of Theoretical Physics},
	url = {https://books.google.co.uk/books?id=l8ENAQAAIAAJ},
	year = {1953},
	Bdsk-Url-1 = {https://books.google.co.uk/books?id=l8ENAQAAIAAJ}}

@article{Delamotte:2007pf,
	archiveprefix = {arXiv},
	author = {Delamotte, Bertrand},
	date-added = {2020-09-22 18:28:48 +0100},
	date-modified = {2020-09-22 18:28:48 +0100},
	doi = {10.1007/978-3-642-27320-9_2},
	eprint = {cond-mat/0702365},
	journal = {Lect. Notes Phys.},
	pages = {49-132},
	primaryclass = {cond-mat.stat-mech},
	slaccitation = {%%CITATION = COND-MAT/0702365;%%},
	title = {{An Introduction to the nonperturbative renormalization group}},
	volume = {852},
	year = {2012},
	Bdsk-Url-1 = {https://doi.org/10.1007/978-3-642-27320-9_2}}

@article{Dupuis:2020fhh,
	archiveprefix = {arXiv},
	author = {Dupuis, N. and Canet, L. and Eichhorn, A. and Metzner, W. and Pawlowski, J. M. and Tissier, M. and Wschebor, N.},
	date-added = {2020-09-22 18:21:36 +0100},
	date-modified = {2020-09-22 18:21:36 +0100},
	eprint = {2006.04853},
	primaryclass = {cond-mat.stat-mech},
	slaccitation = {%%CITATION = ARXIV:2006.04853;%%},
	title = {{The nonperturbative functional renormalization group and its applications}},
	year = {2020}}

@article{Lavrov:2019slr,
	archiveprefix = {arXiv},
	author = {Lavrov, Peter M.},
	date-added = {2020-09-08 18:22:50 +0100},
	date-modified = {2020-09-08 18:23:57 +0100},
	doi = {10.1016/j.physletb.2020.135314},
	eprint = {1911.00194v4},
	journal = {Phys. Lett.},
	pages = {135314},
	primaryclass = {hep-th},
	slaccitation = {%%CITATION = ARXIV:1911.00194;%%},
	title = {{RG and BV-formalism}},
	volume = {B803},
	year = {2020},
	Bdsk-Url-1 = {https://doi.org/10.1016/j.physletb.2020.135314}}

@article{Avramidi:1985ki,
	author = {Avramidi, I. G. and Barvinsky, A. O.},
	date-added = {2020-09-08 15:49:06 +0100},
	date-modified = {2020-09-08 15:49:06 +0100},
	doi = {10.1016/0370-2693(85)90248-5},
	journal = {Phys. Lett.},
	pages = {269-274},
	slaccitation = {%%CITATION = PHLTA,159B,269;%%},
	title = {{ASYMPTOTIC FREEDOM IN HIGHER DERIVATIVE QUANTUM GRAVITY}},
	volume = {159B},
	year = {1985},
	Bdsk-Url-1 = {https://doi.org/10.1016/0370-2693(85)90248-5}}

@article{Julve:1978xn,
	author = {Julve, J. and Tonin, M.},
	date-added = {2020-09-08 15:36:51 +0100},
	date-modified = {2020-09-08 15:36:51 +0100},
	doi = {10.1007/BF02748637},
	journal = {Nuovo Cim.},
	pages = {137-152},
	reportnumber = {IFPD 2/78},
	slaccitation = {%%CITATION = NUCIA,B46,137;%%},
	title = {{Quantum Gravity with Higher Derivative Terms}},
	volume = {B46},
	year = {1978},
	Bdsk-Url-1 = {https://doi.org/10.1007/BF02748637}}

@article{Fradkin:1981iu,
	author = {Fradkin, E. S. and Tseytlin, Arkady A.},
	date-added = {2020-09-08 15:34:09 +0100},
	date-modified = {2020-09-08 15:34:09 +0100},
	doi = {10.1016/0550-3213(82)90444-8},
	journal = {Nucl. Phys.},
	pages = {469-491},
	reportnumber = {LEBEDEV-81-70},
	slaccitation = {%%CITATION = NUPHA,B201,469;%%},
	title = {{Renormalizable asymptotically free quantum theory of gravity}},
	volume = {B201},
	year = {1982},
	Bdsk-Url-1 = {https://doi.org/10.1016/0550-3213(82)90444-8}}

@article{first,
	archiveprefix = {arXiv},
	author = {Mitchell, Alex and Morris, Tim R.},
	date-added = {2020-07-23 13:06:06 +0100},
	date-modified = {2020-07-23 13:06:27 +0100},
	doi = {10.1007/JHEP06(2020)138},
	eprint = {2004.06475},
	journal = {JHEP},
	pages = {138},
	primaryclass = {hep-th},
	slaccitation = {%%CITATION = ARXIV:2004.06475;%%},
	title = {{The continuum limit of quantum gravity at first order in perturbation theory}},
	volume = {06},
	year = {2020},
	Bdsk-Url-1 = {https://doi.org/10.1007/JHEP06(2020)138}}

@phdthesis{Gualtieri:2003dx,
	archiveprefix = {arXiv},
	author = {Gualtieri, Marco},
	date-added = {2020-07-23 11:50:21 +0100},
	date-modified = {2020-07-23 11:50:21 +0100},
	eprint = {math/0401221},
	primaryclass = {math-dg},
	school = {Oxford U.},
	slaccitation = {%%CITATION = MATH/0401221;%%},
	title = {{Generalized complex geometry}},
	year = {2003}}

@article{Hitchin:2004ut,
	archiveprefix = {arXiv},
	author = {Hitchin, Nigel},
	date-added = {2020-07-23 11:47:02 +0100},
	date-modified = {2020-07-23 11:47:02 +0100},
	doi = {10.1093/qjmath/54.3.281},
	eprint = {math/0209099},
	journal = {Quart. J. Math.},
	pages = {281-308},
	primaryclass = {math-dg},
	slaccitation = {%%CITATION = MATH/0209099;%%},
	title = {{Generalized Calabi-Yau manifolds}},
	volume = {54},
	year = {2003},
	Bdsk-Url-1 = {https://doi.org/10.1093/qjmath/54.3.281}}

@article{Hull:2009mi,
	archiveprefix = {arXiv},
	author = {Hull, Chris and Zwiebach, Barton},
	date-added = {2020-07-23 11:24:20 +0100},
	date-modified = {2020-07-23 11:24:20 +0100},
	doi = {10.1088/1126-6708/2009/09/099},
	eprint = {0904.4664},
	journal = {JHEP},
	pages = {099},
	primaryclass = {hep-th},
	reportnumber = {IMPERIAL-TP-2009-CH-02, MIT-CTP-4031},
	slaccitation = {%%CITATION = ARXIV:0904.4664;%%},
	title = {{Double Field Theory}},
	volume = {09},
	year = {2009},
	Bdsk-Url-1 = {https://doi.org/10.1088/1126-6708/2009/09/099}}

@article{Siegel:1993th,
	archiveprefix = {arXiv},
	author = {Siegel, W.},
	date-added = {2020-07-23 11:23:52 +0100},
	date-modified = {2020-07-23 11:23:52 +0100},
	doi = {10.1103/PhysRevD.48.2826},
	eprint = {hep-th/9305073},
	journal = {Phys. Rev.},
	pages = {2826-2837},
	primaryclass = {hep-th},
	reportnumber = {ITP-SB-93-28},
	slaccitation = {%%CITATION = HEP-TH/9305073;%%},
	title = {{Superspace duality in low-energy superstrings}},
	volume = {D48},
	year = {1993},
	Bdsk-Url-1 = {https://doi.org/10.1103/PhysRevD.48.2826}}

@article{Aldazabal:2013sca,
	archiveprefix = {arXiv},
	author = {Aldazabal, Gerardo and Marques, Diego and Nunez, Carmen},
	date-added = {2020-07-23 11:16:32 +0100},
	date-modified = {2020-07-23 11:16:32 +0100},
	doi = {10.1088/0264-9381/30/16/163001},
	eprint = {1305.1907},
	journal = {Class. Quant. Grav.},
	pages = {163001},
	primaryclass = {hep-th},
	slaccitation = {%%CITATION = ARXIV:1305.1907;%%},
	title = {{Double Field Theory: A Pedagogical Review}},
	volume = {30},
	year = {2013},
	Bdsk-Url-1 = {https://doi.org/10.1088/0264-9381/30/16/163001}}

@article{secondconf,
	archiveprefix = {arXiv},
	author = {Morris, Tim R.},
	date-added = {2020-06-11 11:06:54 +0100},
	date-modified = {2020-06-11 11:09:09 +0100},
	eprint = {2006.05185},
	primaryclass = {hep-th},
	slaccitation = {%%CITATION = ARXIV:2006.05185;%%},
	title = {{The continuum limit of the conformal sector at second order in perturbation theory}},
	year = {2020}}

@article{Igarashi:2019gkm,
	archiveprefix = {arXiv},
	author = {Igarashi, Yuji and Itoh, Katsumi and Morris, Tim R.},
	date-modified = {2020-05-28 17:33:20 +0100},
	doi = {10.1093/ptep/ptz099},
	eprint = {1904.08231},
	journal = {PTEP},
	number = {10},
	pages = {103B01},
	primaryclass = {hep-th},
	title = {{BRST in the Exact RG}},
	volume = {2019},
	year = {2019},
	Bdsk-Url-1 = {https://doi.org/10.1093/ptep/ptz099}}

@article{Ambjorn:2020rcn,
	archiveprefix = {arXiv},
	author = {Ambjorn, J. and Gizbert-Studnicki, J. and G{\"o}rlich, A. and Jurkiewicz, J. and Loll, R.},
	date-added = {2020-05-28 15:21:33 +0100},
	date-modified = {2020-05-28 15:21:33 +0100},
	eprint = {2002.01693},
	primaryclass = {hep-th},
	slaccitation = {%%CITATION = ARXIV:2002.01693;%%},
	title = {{Renormalization in quantum theories of geometry}},
	year = {2020}}

@article{Falls_2014,
	author = {Falls, Kevin and Litim, Daniel F.},
	date-added = {2020-05-15 17:03:52 +0100},
	date-modified = {2020-05-15 17:03:52 +0100},
	doi = {10.1103/physrevd.89.084002},
	issn = {1550-2368},
	journal = {Physical Review D},
	month = {Apr},
	number = {8},
	publisher = {American Physical Society (APS)},
	title = {Black hole thermodynamics under the microscope},
	url = {http://dx.doi.org/10.1103/PhysRevD.89.084002},
	volume = {89},
	year = {2014},
	Bdsk-Url-1 = {http://dx.doi.org/10.1103/PhysRevD.89.084002},
	Bdsk-Url-2 = {http://dx.doi.org/10.1103/physrevd.89.084002}}

@article{Pruttivarasin_2015,
	author = {Pruttivarasin, T. and Ramm, M. and Porsev, S. G. and Tupitsyn, I. I. and Safronova, M. S. and Hohensee, M. A. and H{\"a}ffner, H.},
	date-added = {2020-05-15 16:16:13 +0100},
	date-modified = {2020-05-15 16:16:13 +0100},
	doi = {10.1038/nature14091},
	issn = {1476-4687},
	journal = {Nature},
	month = {Jan},
	number = {7536},
	pages = {592--595},
	publisher = {Springer Science and Business Media LLC},
	title = {Michelson--Morley analogue for electrons using trapped ions to test Lorentz symmetry},
	url = {http://dx.doi.org/10.1038/nature14091},
	volume = {517},
	year = {2015},
	Bdsk-Url-1 = {http://dx.doi.org/10.1038/nature14091}}

@article{Pospelov_2004,
	author = {Pospelov, Maxim and Romalis, Michael},
	date-added = {2020-05-15 16:13:58 +0100},
	date-modified = {2020-05-15 16:13:58 +0100},
	doi = {10.1063/1.1784301},
	issn = {1945-0699},
	journal = {Physics Today},
	month = {Jul},
	number = {7},
	pages = {40--46},
	publisher = {AIP Publishing},
	title = {Lorentz Invariance on Trial},
	url = {http://dx.doi.org/10.1063/1.1784301},
	volume = {57},
	year = {2004},
	Bdsk-Url-1 = {http://dx.doi.org/10.1063/1.1784301}}

@article{Gonzalez_Martin_2017,
	author = {Gonzalez-Martin, Sergio and Morris, Tim R. and Slade, Zo{\"e} H.},
	date-added = {2020-05-15 15:45:14 +0100},
	date-modified = {2020-05-15 15:45:14 +0100},
	doi = {10.1103/physrevd.95.106010},
	issn = {2470-0029},
	journal = {Physical Review D},
	month = {May},
	number = {10},
	publisher = {American Physical Society (APS)},
	title = {Asymptotic solutions in asymptotic safety},
	url = {http://dx.doi.org/10.1103/PhysRevD.95.106010},
	volume = {95},
	year = {2017},
	Bdsk-Url-1 = {http://dx.doi.org/10.1103/PhysRevD.95.106010},
	Bdsk-Url-2 = {http://dx.doi.org/10.1103/physrevd.95.106010}}

@article{Bonanno:2013dja,
	archiveprefix = {arXiv},
	author = {Bonanno, Alfio and Reuter, Martin},
	date-added = {2020-05-15 13:06:50 +0100},
	date-modified = {2020-05-15 13:06:50 +0100},
	doi = {10.1103/PhysRevD.87.084019},
	eprint = {1302.2928},
	journal = {Phys. Rev.},
	number = {8},
	pages = {084019},
	primaryclass = {hep-th},
	slaccitation = {%%CITATION = ARXIV:1302.2928;%%},
	title = {{Modulated Ground State of Gravity Theories with Stabilized Conformal Factor}},
	volume = {D87},
	year = {2013},
	Bdsk-Url-1 = {https://doi.org/10.1103/PhysRevD.87.084019}}

@article{Falls:2020tmj,
	archiveprefix = {arXiv},
	author = {Falls, Kevin},
	date-added = {2020-05-05 17:25:01 +0100},
	date-modified = {2020-05-05 17:25:01 +0100},
	eprint = {2004.11409},
	primaryclass = {hep-th},
	slaccitation = {%%CITATION = ARXIV:2004.11409;%%},
	title = {{Background independent exact renormalisation}},
	year = {2020}}

@article{Gates:1983nr,
	archiveprefix = {arXiv},
	author = {Gates, S. J. and Grisaru, Marcus T. and Rocek, M. and Siegel, W.},
	date-added = {2020-04-23 11:16:23 +0100},
	date-modified = {2020-04-23 11:16:23 +0100},
	eprint = {hep-th/0108200},
	journal = {Front. Phys.},
	pages = {1-548},
	primaryclass = {hep-th},
	reportnumber = {YITP-SB-01-53},
	slaccitation = {%%CITATION = HEP-TH/0108200;%%},
	title = {{Superspace Or One Thousand and One Lessons in Supersymmetry}},
	volume = {58},
	year = {1983}}

@article{Freedman_1976,
	author = {Freedman, Daniel Z. and van Nieuwenhuizen, P. and Ferrara, S.},
	date-added = {2020-04-20 15:29:44 +0100},
	date-modified = {2020-04-20 15:29:44 +0100},
	doi = {10.1103/physrevd.13.3214},
	issn = {0556-2821},
	journal = {Physical Review D},
	month = {Jun},
	number = {12},
	pages = {3214--3218},
	publisher = {American Physical Society (APS)},
	title = {Progress toward a theory of supergravity},
	url = {http://dx.doi.org/10.1103/PhysRevD.13.3214},
	volume = {13},
	year = {1976},
	Bdsk-Url-1 = {http://dx.doi.org/10.1103/PhysRevD.13.3214},
	Bdsk-Url-2 = {http://dx.doi.org/10.1103/physrevd.13.3214}}

@article{Loll:2019rdj,
	archiveprefix = {arXiv},
	author = {Loll, R.},
	date-added = {2020-04-17 17:40:56 +0100},
	date-modified = {2020-04-17 17:40:56 +0100},
	doi = {10.1088/1361-6382/ab57c7},
	eprint = {1905.08669},
	journal = {Class. Quant. Grav.},
	number = {1},
	pages = {013002},
	primaryclass = {hep-th},
	slaccitation = {%%CITATION = ARXIV:1905.08669;%%},
	title = {{Quantum Gravity from Causal Dynamical Triangulations: A Review}},
	volume = {37},
	year = {2020},
	Bdsk-Url-1 = {https://doi.org/10.1088/1361-6382/ab57c7}}


@article{MR2248812,
	author = {Tuck, E. O.},
	date-added = {2020-04-02 17:27:27 +0100},
	date-modified = {2020-04-02 17:27:27 +0100},
	doi = {10.1017/S0004972700047511},
	fjournal = {Bulletin of the Australian Mathematical Society},
	issn = {0004-9727},
	journal = {Bull. Austral. Math. Soc.},
	mrclass = {42A38 (26D15 33B10 33B20)},
	mrnumber = {2248812},
	mrreviewer = {Stamatis Koumandos},
	number = {1},
	pages = {133--138},
	title = {On positivity of {F}ourier transforms},
	url = {https://mathscinet.ams.org/mathscinet-getitem?mr=2248812},
	volume = {74},
	year = {2006},
	Bdsk-Url-1 = {https://mathscinet.ams.org/mathscinet-getitem?mr=2248812}}

@article{Groote:1998ic,
	archiveprefix = {arXiv},
	author = {Groote, S. and Korner, J. G. and Pivovarov, A. A.},
	date-added = {2020-03-30 11:17:39 +0100},
	date-modified = {2020-03-30 11:17:39 +0100},
	doi = {10.1016/S0370-2693(98)01324-0},
	eprint = {hep-ph/9805224},
	journal = {Phys. Lett.},
	pages = {269-275},
	primaryclass = {hep-ph},
	reportnumber = {MZ-TH-98-12},
	slaccitation = {%%CITATION = HEP-PH/9805224;%%},
	title = {{A New technique for computing the spectral density of sunset type diagrams: Integral transformation in configuration space}},
	volume = {B443},
	year = {1998},
	Bdsk-Url-1 = {https://doi.org/10.1016/S0370-2693(98)01324-0}}

@article{Eichhorn:2013xr,
	archiveprefix = {arXiv},
	author = {Eichhorn, Astrid},
	date-added = {2020-03-28 19:21:58 +0000},
	date-modified = {2020-03-28 19:21:58 +0000},
	doi = {10.1088/0264-9381/30/11/115016},
	eprint = {1301.0879},
	journal = {Class. Quant. Grav.},
	pages = {115016},
	primaryclass = {gr-qc},
	slaccitation = {%%CITATION = ARXIV:1301.0879;%%},
	title = {{On unimodular quantum gravity}},
	volume = {30},
	year = {2013},
	Bdsk-Url-1 = {https://doi.org/10.1088/0264-9381/30/11/115016}}

@article{Kawai:1993mb,
	archiveprefix = {arXiv},
	author = {Kawai, Hikaru and Kitazawa, Yoshihisa and Ninomiya, Masao},
	date-added = {2020-03-28 19:08:23 +0000},
	date-modified = {2020-03-28 19:08:23 +0000},
	doi = {10.1016/0550-3213(93)90594-F},
	eprint = {hep-th/9303123},
	journal = {Nucl. Phys.},
	pages = {684-716},
	primaryclass = {hep-th},
	reportnumber = {UT-636, TIT-HEP-217, YITP-U-93-07},
	slaccitation = {%%CITATION = HEP-TH/9303123;%%},
	title = {{Ultraviolet stable fixed point and scaling relations in (2+epsilon)-dimensional quantum gravity}},
	volume = {B404},
	year = {1993},
	Bdsk-Url-1 = {https://doi.org/10.1016/0550-3213(93)90594-F}}

@article{Percacci:2017fsy,
	archiveprefix = {arXiv},
	author = {Percacci, R.},
	booktitle = {{Black Holes, Gravitational Waves and Spacetime Singularities Rome, Italy, May 9-12, 2017}},
	date-added = {2020-03-23 16:17:55 +0000},
	date-modified = {2020-03-23 16:17:55 +0000},
	doi = {10.1007/s10701-018-0189-5},
	eprint = {1712.09903},
	journal = {Found. Phys.},
	number = {10},
	pages = {1364-1379},
	primaryclass = {gr-qc},
	slaccitation = {%%CITATION = ARXIV:1712.09903;%%},
	title = {{Unimodular quantum gravity and the cosmological constant}},
	volume = {48},
	year = {2018},
	Bdsk-Url-1 = {https://doi.org/10.1007/s10701-018-0189-5}}

@article{Igarashi:2016gcf,
	archiveprefix = {arXiv},
	author = {Igarashi, Yuji and Itoh, Katsumi and Pawlowski, Jan M.},
	date-added = {2019-04-18 10:36:59 +0100},
	date-modified = {2019-04-18 10:36:59 +0100},
	doi = {10.1088/1751-8113/49/40/405401},
	eprint = {1604.08327},
	journal = {J. Phys.},
	number = {40},
	pages = {405401},
	primaryclass = {hep-th},
	slaccitation = {%%CITATION = ARXIV:1604.08327;%%},
	title = {{Functional flows in QED and the modified WardTakahashi identity}},
	volume = {A49},
	year = {2016},
	Bdsk-Url-1 = {https://doi.org/10.1088/1751-8113/49/40/405401}}

@article{Rosten:2011mf,
	archiveprefix = {arXiv},
	author = {Rosten, Oliver J.},
	date-added = {2019-03-26 19:46:22 +0000},
	date-modified = {2019-03-26 19:46:22 +0000},
	eprint = {1106.2544},
	primaryclass = {hep-th},
	slaccitation = {%%CITATION = ARXIV:1106.2544;%%},
	title = {{Relationships Between Exact RGs and some Comments on Asymptotic Safety}},
	year = {2011}}

@article{Bonanno:2018gck,
	archiveprefix = {arXiv},
	author = {Bonanno, Alfio and Platania, Alessia and Saueressig, Frank},
	date-added = {2019-02-17 17:06:02 +0000},
	date-modified = {2019-02-17 17:06:02 +0000},
	doi = {10.1016/j.physletb.2018.06.047},
	eprint = {1803.02355},
	journal = {Phys. Lett.},
	pages = {229-236},
	primaryclass = {gr-qc},
	slaccitation = {%%CITATION = ARXIV:1803.02355;%%},
	title = {{Cosmological bounds on the field content of asymptotically safe gravitymatter models}},
	volume = {B784},
	year = {2018},
	Bdsk-Url-1 = {https://doi.org/10.1016/j.physletb.2018.06.047}}

@article{Liu:2018hno,
	archiveprefix = {arXiv},
	author = {Liu, Lei-Hua and Prokopec, Tomislav and Starobinsky, Alexei A.},
	date-added = {2019-02-17 17:05:14 +0000},
	date-modified = {2019-02-17 17:05:14 +0000},
	doi = {10.1103/PhysRevD.98.043505},
	eprint = {1806.05407},
	journal = {Phys. Rev.},
	number = {4},
	pages = {043505},
	primaryclass = {gr-qc},
	slaccitation = {%%CITATION = ARXIV:1806.05407;%%},
	title = {{Inflation in an effective gravitational model and asymptotic safety}},
	volume = {D98},
	year = {2018},
	Bdsk-Url-1 = {https://doi.org/10.1103/PhysRevD.98.043505}}

@article{Gubitosi:2018gsl,
	archiveprefix = {arXiv},
	author = {Gubitosi, Giulia and Ooijer, Robin and Ripken, Chris and Saueressig, Frank},
	date-added = {2019-02-17 17:03:39 +0000},
	date-modified = {2019-02-17 17:03:39 +0000},
	doi = {10.1088/1475-7516/2018/12/004},
	eprint = {1806.10147},
	journal = {JCAP},
	number = {12},
	pages = {004},
	primaryclass = {hep-th},
	slaccitation = {%%CITATION = ARXIV:1806.10147;%%},
	title = {{Consistent early and late time cosmology from the RG flow of gravity}},
	volume = {1812},
	year = {2018},
	Bdsk-Url-1 = {https://doi.org/10.1088/1475-7516/2018/12/004}}

@article{Pawlowski:2018ixd,
	archiveprefix = {arXiv},
	author = {Pawlowski, Jan M. and Reichert, Manuel and Wetterich, Christof and Yamada, Masatoshi},
	date-added = {2019-02-17 17:00:16 +0000},
	date-modified = {2019-02-17 17:00:16 +0000},
	eprint = {1811.11706},
	primaryclass = {hep-th},
	slaccitation = {%%CITATION = ARXIV:1811.11706;%%},
	title = {{Higgs scalar potential in asymptotically safe quantum gravity}},
	year = {2018}}

@phdthesis{Alkofer:2018zze,
	archiveprefix = {arXiv},
	author = {Alkofer, Natalia},
	date-added = {2019-02-17 16:58:55 +0000},
	date-modified = {2019-02-17 16:58:55 +0000},
	eprint = {1810.03132},
	primaryclass = {hep-th},
	school = {Nijmegen U.},
	slaccitation = {%%CITATION = ARXIV:1810.03132;%%},
	title = {{Quantum Gravity from Fundamental Questions to Phenomenological Applications}},
	year = {2018}}

@article{Eichhorn:2017lry,
	archiveprefix = {arXiv},
	author = {Eichhorn, Astrid and Versteegen, Fleur},
	date-added = {2019-02-17 16:57:00 +0000},
	date-modified = {2019-02-17 16:57:00 +0000},
	doi = {10.1007/JHEP01(2018)030},
	eprint = {1709.07252},
	journal = {JHEP},
	pages = {030},
	primaryclass = {hep-th},
	slaccitation = {%%CITATION = ARXIV:1709.07252;%%},
	title = {{Upper bound on the Abelian gauge coupling from asymptotic safety}},
	volume = {01},
	year = {2018},
	Bdsk-Url-1 = {https://doi.org/10.1007/JHEP01(2018)030}}

@article{Eichhorn:2017als,
	archiveprefix = {arXiv},
	author = {Eichhorn, Astrid and Hamada, Yuta and Lumma, Johannes and Yamada, Masatoshi},
	date-added = {2019-02-17 16:55:35 +0000},
	date-modified = {2019-02-17 16:55:35 +0000},
	doi = {10.1103/PhysRevD.97.086004},
	eprint = {1712.00319},
	journal = {Phys. Rev.},
	number = {8},
	pages = {086004},
	primaryclass = {hep-th},
	slaccitation = {%%CITATION = ARXIV:1712.00319;%%},
	title = {{Quantum gravity fluctuations flatten the Planck-scale Higgs potential}},
	volume = {D97},
	year = {2018},
	Bdsk-Url-1 = {https://doi.org/10.1103/PhysRevD.97.086004}}

@article{Eichhorn:2018whv,
	archiveprefix = {arXiv},
	author = {Eichhorn, Astrid and Held, Aaron},
	date-added = {2019-02-17 16:54:05 +0000},
	date-modified = {2019-02-17 16:54:05 +0000},
	doi = {10.1103/PhysRevLett.121.151302},
	eprint = {1803.04027},
	journal = {Phys. Rev. Lett.},
	number = {15},
	pages = {151302},
	primaryclass = {hep-th},
	slaccitation = {%%CITATION = ARXIV:1803.04027;%%},
	title = {{Mass difference for charged quarks from asymptotically safe quantum gravity}},
	volume = {121},
	year = {2018},
	Bdsk-Url-1 = {https://doi.org/10.1103/PhysRevLett.121.151302}}

@article{Eichhorn:2018yfc,
	archiveprefix = {arXiv},
	author = {Eichhorn, Astrid},
	date-added = {2019-02-17 16:51:57 +0000},
	date-modified = {2019-02-17 16:51:57 +0000},
	eprint = {1810.07615},
	primaryclass = {hep-th},
	slaccitation = {%%CITATION = ARXIV:1810.07615;%%},
	title = {{An asymptotically safe guide to quantum gravity and matter}},
	year = {2018}}

@article{Eichhorn:2018nda,
	archiveprefix = {arXiv},
	author = {Eichhorn, Astrid and Lippoldt, Stefan and Schiffer, Marc},
	date-added = {2019-02-17 16:49:28 +0000},
	date-modified = {2019-02-17 16:49:28 +0000},
	eprint = {1812.08782},
	primaryclass = {hep-th},
	slaccitation = {%%CITATION = ARXIV:1812.08782;%%},
	title = {{Zooming in on fermions and quantum gravity}},
	year = {2018}}

@article{Christiansen:2017qca,
	archiveprefix = {arXiv},
	author = {Christiansen, Nicolai and Eichhorn, Astrid and Held, Aaron},
	date-added = {2019-02-17 16:38:16 +0000},
	date-modified = {2019-02-17 16:38:16 +0000},
	doi = {10.1103/PhysRevD.96.084021},
	eprint = {1705.01858},
	journal = {Phys. Rev.},
	number = {8},
	pages = {084021},
	primaryclass = {hep-th},
	slaccitation = {%%CITATION = ARXIV:1705.01858;%%},
	title = {{Is scale-invariance in gauge-Yukawa systems compatible with the graviton?}},
	volume = {D96},
	year = {2017},
	Bdsk-Url-1 = {https://doi.org/10.1103/PhysRevD.96.084021}}

@article{Eichhorn:2017eht,
	archiveprefix = {arXiv},
	author = {Eichhorn, Astrid and Held, Aaron},
	date-added = {2019-02-17 16:38:03 +0000},
	date-modified = {2019-02-17 16:38:03 +0000},
	doi = {10.1103/PhysRevD.96.086025},
	eprint = {1705.02342},
	journal = {Phys. Rev.},
	number = {8},
	pages = {086025},
	primaryclass = {gr-qc},
	slaccitation = {%%CITATION = ARXIV:1705.02342;%%},
	title = {{Viability of quantum-gravity induced ultraviolet completions for matter}},
	volume = {D96},
	year = {2017},
	Bdsk-Url-1 = {https://doi.org/10.1103/PhysRevD.96.086025}}

@article{Eichhorn:2017ylw,
	archiveprefix = {arXiv},
	author = {Eichhorn, Astrid and Held, Aaron},
	date-added = {2019-02-17 16:37:53 +0000},
	date-modified = {2019-02-17 16:37:53 +0000},
	doi = {10.1016/j.physletb.2017.12.040},
	eprint = {1707.01107},
	journal = {Phys. Lett.},
	pages = {217-221},
	primaryclass = {hep-th},
	slaccitation = {%%CITATION = ARXIV:1707.01107;%%},
	title = {{Top mass from asymptotic safety}},
	volume = {B777},
	year = {2018},
	Bdsk-Url-1 = {https://doi.org/10.1016/j.physletb.2017.12.040}}

@article{Nink:2014yya,
	archiveprefix = {arXiv},
	author = {Nink, Andreas},
	date-added = {2019-02-17 16:34:56 +0000},
	date-modified = {2019-02-17 16:34:56 +0000},
	doi = {10.1103/PhysRevD.91.044030},
	eprint = {1410.7816},
	journal = {Phys. Rev.},
	number = {4},
	pages = {044030},
	primaryclass = {hep-th},
	slaccitation = {%%CITATION = ARXIV:1410.7816;%%},
	title = {{Field Parametrization Dependence in Asymptotically Safe Quantum Gravity}},
	volume = {D91},
	year = {2015},
	Bdsk-Url-1 = {https://doi.org/10.1103/PhysRevD.91.044030}}

@article{Safari:2015dva,
	archiveprefix = {arXiv},
	author = {Safari, Mahmoud},
	date-added = {2019-02-17 16:34:07 +0000},
	date-modified = {2019-02-17 16:34:07 +0000},
	doi = {10.1140/epjc/s10052-016-4036-6},
	eprint = {1508.06244},
	journal = {Eur. Phys. J.},
	number = {4},
	pages = {201},
	primaryclass = {hep-th},
	slaccitation = {%%CITATION = ARXIV:1508.06244;%%},
	title = {{Splitting Ward identity}},
	volume = {C76},
	year = {2016},
	Bdsk-Url-1 = {https://doi.org/10.1140/epjc/s10052-016-4036-6}}

@article{Liu:2015bwa,
	archiveprefix = {arXiv},
	author = {Liu, Rex G. and Williams, Ruth M.},
	date-added = {2019-02-17 16:33:21 +0000},
	date-modified = {2019-02-17 16:33:21 +0000},
	doi = {10.1103/PhysRevD.93.023502},
	eprint = {1502.03000},
	journal = {Phys. Rev.},
	number = {2},
	pages = {023502},
	primaryclass = {gr-qc},
	slaccitation = {%%CITATION = ARXIV:1502.03000;%%},
	title = {{Regge calculus models of closed lattice universes}},
	volume = {D93},
	year = {2016},
	Bdsk-Url-1 = {https://doi.org/10.1103/PhysRevD.93.023502}}

@article{Hamber:2015jja,
	archiveprefix = {arXiv},
	author = {Hamber, Herbert W.},
	date-added = {2019-02-17 16:32:33 +0000},
	date-modified = {2019-02-17 16:32:33 +0000},
	doi = {10.1103/PhysRevD.92.064017},
	eprint = {1506.07795},
	journal = {Phys. Rev.},
	number = {6},
	pages = {064017},
	primaryclass = {hep-th},
	slaccitation = {%%CITATION = ARXIV:1506.07795;%%},
	title = {{Scaling Exponents for Lattice Quantum Gravity in Four Dimensions}},
	volume = {D92},
	year = {2015},
	Bdsk-Url-1 = {https://doi.org/10.1103/PhysRevD.92.064017}}

@article{Eichhorn:2015bna,
	archiveprefix = {arXiv},
	author = {Eichhorn, Astrid},
	date-added = {2019-02-17 16:29:49 +0000},
	date-modified = {2019-02-17 16:29:49 +0000},
	doi = {10.1007/JHEP04(2015)096},
	eprint = {1501.05848},
	journal = {JHEP},
	pages = {096},
	primaryclass = {gr-qc},
	slaccitation = {%%CITATION = ARXIV:1501.05848;%%},
	title = {{The Renormalization Group flow of unimodular f(R) gravity}},
	volume = {04},
	year = {2015},
	Bdsk-Url-1 = {https://doi.org/10.1007/JHEP04(2015)096}}

@article{Percacci:2015wwa,
	archiveprefix = {arXiv},
	author = {Percacci, Roberto and Vacca, Gian Paolo},
	date-added = {2019-02-17 16:28:56 +0000},
	date-modified = {2019-02-17 16:28:56 +0000},
	doi = {10.1140/epjc/s10052-015-3410-0},
	eprint = {1501.00888},
	journal = {Eur. Phys. J.},
	number = {5},
	pages = {188},
	primaryclass = {hep-th},
	slaccitation = {%%CITATION = ARXIV:1501.00888;%%},
	title = {{Search of scaling solutions in scalar-tensor gravity}},
	volume = {C75},
	year = {2015},
	Bdsk-Url-1 = {https://doi.org/10.1140/epjc/s10052-015-3410-0}}

@article{Falls:2016msz,
	archiveprefix = {arXiv},
	author = {Falls, Kevin and Ohta, Nobuyoshi},
	date-added = {2019-02-17 16:27:55 +0000},
	date-modified = {2019-02-17 16:27:55 +0000},
	doi = {10.1103/PhysRevD.94.084005},
	eprint = {1607.08460},
	journal = {Phys. Rev.},
	number = {8},
	pages = {084005},
	primaryclass = {hep-th},
	reportnumber = {KU-TP-067},
	slaccitation = {%%CITATION = ARXIV:1607.08460;%%},
	title = {{Renormalization Group Equation for $f(R)$ gravity on hyperbolic spaces}},
	volume = {D94},
	year = {2016},
	Bdsk-Url-1 = {https://doi.org/10.1103/PhysRevD.94.084005}}

@article{Falls:2016wsa,
	archiveprefix = {arXiv},
	author = {Falls, Kevin and Litim, Daniel F. and Nikolakopoulos, Kostas and Rahmede, Christoph},
	date-added = {2019-02-17 16:24:50 +0000},
	date-modified = {2019-02-17 16:24:50 +0000},
	doi = {10.1088/1361-6382/aac440},
	eprint = {1607.04962},
	journal = {Class. Quant. Grav.},
	number = {13},
	pages = {135006},
	primaryclass = {gr-qc},
	slaccitation = {%%CITATION = ARXIV:1607.04962;%%},
	title = {{On de Sitter solutions in asymptotically safe $f(R)$ theories}},
	volume = {35},
	year = {2018},
	Bdsk-Url-1 = {https://doi.org/10.1088/1361-6382/aac440}}

@article{Christiansen:2014raa,
	archiveprefix = {arXiv},
	author = {Christiansen, Nicolai and Knorr, Benjamin and Pawlowski, Jan M. and Rodigast, Andreas},
	date-added = {2019-02-17 16:22:33 +0000},
	date-modified = {2019-02-17 16:22:33 +0000},
	doi = {10.1103/PhysRevD.93.044036},
	eprint = {1403.1232},
	journal = {Phys. Rev.},
	number = {4},
	pages = {044036},
	primaryclass = {hep-th},
	slaccitation = {%%CITATION = ARXIV:1403.1232;%%},
	title = {{Global Flows in Quantum Gravity}},
	volume = {D93},
	year = {2016},
	Bdsk-Url-1 = {https://doi.org/10.1103/PhysRevD.93.044036}}

@book{Reuter:2019byg,
	author = {Reuter, Martin and Saueressig, Frank},
	date-added = {2019-02-17 16:18:53 +0000},
	date-modified = {2019-02-17 16:18:53 +0000},
	isbn = {9781107107328},
	publisher = {Cambridge University Press},
	slaccitation = {%%CITATION = INSPIRE-1716753;%%},
	title = {{Quantum Gravity and the Functional Renormalization Group}},
	url = {https://www.cambridge.org/academic/subjects/physics/theoretical-physics-and-mathematical-physics/quantum-gravity-and-functional-renormalization-group-road-towards-asymptotic-safety?format=HB&isbn=9781107107328},
	year = {2019},
	Bdsk-Url-1 = {https://www.cambridge.org/academic/subjects/physics/theoretical-physics-and-mathematical-physics/quantum-gravity-and-functional-renormalization-group-road-towards-asymptotic-safety?format=HB&isbn=9781107107328}}

@book{Percacci:2017fkn,
	author = {Percacci, Robert},
	date-added = {2019-02-17 16:10:03 +0000},
	date-modified = {2019-02-17 16:10:03 +0000},
	doi = {10.1142/10369},
	isbn = {9789813207172, 9789813207196, 9789813207172, 9789813207196},
	publisher = {World Scientific},
	series = {100 Years of General Relativity},
	slaccitation = {%%CITATION = INSPIRE-1620630;%%},
	title = {{An Introduction to Covariant Quantum Gravity and Asymptotic Safety}},
	volume = {3},
	year = {2017},
	Bdsk-Url-1 = {https://doi.org/10.1142/10369}}

@article{Nagy:2012ef,
	archiveprefix = {arXiv},
	author = {Nagy, Sandor},
	date-added = {2019-02-17 16:04:10 +0000},
	date-modified = {2019-02-17 16:04:10 +0000},
	doi = {10.1016/j.aop.2014.07.027},
	eprint = {1211.4151},
	journal = {Annals Phys.},
	pages = {310-346},
	primaryclass = {hep-th},
	slaccitation = {%%CITATION = ARXIV:1211.4151;%%},
	title = {{Lectures on renormalization and asymptotic safety}},
	volume = {350},
	year = {2014},
	Bdsk-Url-1 = {https://doi.org/10.1016/j.aop.2014.07.027}}

@inproceedings{Percacci:2011fr,
	archiveprefix = {arXiv},
	author = {Percacci, Roberto},
	booktitle = {{Time and Matter: Proceedings, 3rd International Conference, TAM2010, Budva, Montenegro, 4-8 October, 2010}},
	date-added = {2019-02-17 16:02:20 +0000},
	date-modified = {2019-02-17 16:02:20 +0000},
	eprint = {1110.6389},
	pages = {123-142},
	primaryclass = {hep-th},
	slaccitation = {%%CITATION = ARXIV:1110.6389;%%},
	title = {{A Short introduction to asymptotic safety}},
	year = {2011}}

@inproceedings{ZinnJustin:1975wb,
	author = {Zinn-Justin, Jean},
	booktitle = {{Functional and Probabilistic Methods in Quantum Field Theory. 1. Proceedings, 12th Winter School of Theoretical Physics, Karpacz, Feb 17-March 2, 1975}},
	date-added = {2018-12-10 10:07:40 +0000},
	date-modified = {2018-12-10 10:07:40 +0000},
	pages = {433-453},
	slaccitation = {%%CITATION = INSPIRE-105217;%%},
	title = {{Renormalization Problems in Gauge Theories}},
	year = {1975}}

@article{ZinnJustin:1974mc,
	author = {Zinn-Justin, Jean},
	booktitle = {{Proceedings, International Summer Institute for Theoretical Physics: Trends in Elementary Particle Theory: Bonn, Germany, July 29-August 9, 1974}},
	date-added = {2018-12-10 10:07:09 +0000},
	date-modified = {2018-12-10 10:07:09 +0000},
	doi = {10.1007/3-540-07160-1_1},
	journal = {Lect. Notes Phys.},
	pages = {1-39},
	reportnumber = {SACLAY-D.PH-T-74-88},
	slaccitation = {%%CITATION = LNPHA,37,1;%%},
	title = {{Renormalization of Gauge Theories}},
	volume = {37},
	year = {1975},
	Bdsk-Url-1 = {https://doi.org/10.1007/3-540-07160-1_1}}

@article{Morris:2018mhd,
	archiveprefix = {arXiv},
	author = {Morris, Tim R.},
	date-added = {2018-11-05 21:09:29 +0000},
	date-modified = {2018-11-05 21:09:29 +0000},
	doi = {10.1007/JHEP08(2018)024},
	eprint = {1802.04281},
	journal = {JHEP},
	pages = {024},
	primaryclass = {hep-th},
	slaccitation = {%%CITATION = ARXIV:1802.04281;%%},
	title = {{Renormalization group properties in the conformal sector: towards perturbatively renormalizable quantum gravity}},
	volume = {08},
	year = {2018},
	Bdsk-Url-1 = {https://doi.org/10.1007/JHEP08(2018)024}}

@article{Kellett:2018loq,
	archiveprefix = {arXiv},
	author = {Kellett, Matthew P. and Morris, Tim R.},
	date-added = {2018-11-05 21:08:47 +0000},
	date-modified = {2018-11-05 21:08:47 +0000},
	doi = {10.1088/1361-6382/aad06e},
	eprint = {1803.00859},
	journal = {Class. Quant. Grav.},
	number = {17},
	pages = {175002},
	primaryclass = {hep-th},
	slaccitation = {%%CITATION = ARXIV:1803.00859;%%},
	title = {{Renormalization group properties of the conformal mode of a torus}},
	volume = {35},
	year = {2018},
	Bdsk-Url-1 = {https://doi.org/10.1088/1361-6382/aad06e}}

@article{Morris:2018axr,
	archiveprefix = {arXiv},
	author = {Morris, Tim R.},
	date-added = {2018-11-05 21:08:35 +0000},
	date-modified = {2018-11-05 21:08:35 +0000},
	doi = {10.21468/SciPostPhys.5.4.040},
	eprint = {1806.02206},
	journal = {SciPost Phys.},
	pages = {040},
	primaryclass = {hep-th},
	slaccitation = {%%CITATION = ARXIV:1806.02206;%%},
	title = {{Quantum gravity, renormalizability and diffeomorphism invariance}},
	volume = {5},
	year = {2018},
	Bdsk-Url-1 = {https://doi.org/10.21468/SciPostPhys.5.4.040}}

@article{Ramond:1981pw,
	author = {Ramond, Pierre},
	date-added = {2018-08-03 11:58:35 +0100},
	date-modified = {2018-08-06 08:06:56 +0100},
	handle = {2027/umn.31951d00024038y},
	journal = {Front. Phys.},
	note = {[Front. Phys.74,1(1989)]},
	pages = {1-397},
	slaccitation = {%%CITATION = FRPHA,51,1;%%},
	title = {{Field Theory. A Modern Primer.}},
	volume = {51},
	year = {1981}}

@article{Lavrov:1986hr,
	author = {Lavrov, P. M. and Tyutin, I. V.},
	date-added = {2018-04-01 10:24:34 +0000},
	date-modified = {2018-04-01 10:26:00 +0000},
	journal = {Yad. Fiz.},
	pages = {1658-1666},
	slaccitation = {%%CITATION = YAFIA,41,1658;%%},
	title = {{Effective action in general gauge theories. (In Russian)}},
	volume = {41},
	year = {1985}}

@book{Henneaux:1992ig,
	author = {Henneaux, M. and Teitelboim, C.},
	date-added = {2018-03-30 20:10:58 +0000},
	date-modified = {2018-03-30 20:10:58 +0000},
	isbn = {0691037698, 9780691037691},
	journal = {Princeton, USA: Univ. Pr. (1992) 520 p},
	slaccitation = {%%CITATION = INSPIRE-345963;%%},
	title = {{Quantization of gauge systems}},
	year = {1992}}

@article{Feldbrugge:2017fcc,
	archiveprefix = {arXiv},
	author = {Feldbrugge, Job and Lehners, Jean-Luc and Turok, Neil},
	date-added = {2017-11-29 15:18:28 +0000},
	date-modified = {2017-11-29 15:18:28 +0000},
	doi = {10.1103/PhysRevLett.119.171301},
	eprint = {1705.00192},
	journal = {Phys. Rev. Lett.},
	number = {17},
	pages = {171301},
	primaryclass = {hep-th},
	slaccitation = {%%CITATION = ARXIV:1705.00192;%%},
	title = {{No smooth beginning for spacetime}},
	volume = {119},
	year = {2017},
	Bdsk-Url-1 = {https://dx.doi.org/10.1103/PhysRevLett.119.171301}}

@article{Feldbrugge:2017kzv,
	archiveprefix = {arXiv},
	author = {Feldbrugge, Job and Lehners, Jean-Luc and Turok, Neil},
	date-added = {2017-11-29 15:17:24 +0000},
	date-modified = {2017-11-29 15:17:24 +0000},
	doi = {10.1103/PhysRevD.95.103508},
	eprint = {1703.02076},
	journal = {Phys. Rev.},
	number = {10},
	pages = {103508},
	primaryclass = {hep-th},
	slaccitation = {%%CITATION = ARXIV:1703.02076;%%},
	title = {{Lorentzian Quantum Cosmology}},
	volume = {D95},
	year = {2017},
	Bdsk-Url-1 = {https://dx.doi.org/10.1103/PhysRevD.95.103508}}

@article{Feldbrugge:2017mbc,
	archiveprefix = {arXiv},
	author = {Feldbrugge, Job and Lehners, Jean-Luc and Turok, Neil},
	date-added = {2017-11-29 15:12:39 +0000},
	date-modified = {2017-11-29 15:12:39 +0000},
	eprint = {1708.05104},
	primaryclass = {hep-th},
	slaccitation = {%%CITATION = ARXIV:1708.05104;%%},
	title = {{No Rescue for the No Boundary Proposal}},
	year = {2017}}

@article{Keller:1990ej,
	author = {Keller, G. and Kopper, Christoph and Salmhofer, M.},
	date-added = {2017-11-11 17:19:01 +0000},
	date-modified = {2017-11-11 17:19:01 +0000},
	journal = {Helv. Phys. Acta},
	pages = {32-52},
	reportnumber = {MPI-PAE-PTH-65-90},
	slaccitation = {%%CITATION = HPACA,65,32;%%},
	title = {{Perturbative renormalization and effective Lagrangians in phi**4 in four-dimensions}},
	volume = {65},
	year = {1992}}

@article{Bonini:1992vh,
	archiveprefix = {arXiv},
	author = {Bonini, M. and D'Attanasio, M. and Marchesini, G.},
	date-added = {2017-11-11 17:14:04 +0000},
	date-modified = {2017-11-11 17:14:04 +0000},
	doi = {10.1016/0550-3213(93)90588-G},
	eprint = {hep-th/9301114},
	journal = {Nucl. Phys.},
	pages = {441-464},
	primaryclass = {hep-th},
	reportnumber = {UPRF-92-360},
	slaccitation = {%%CITATION = HEP-TH/9301114;%%},
	title = {{Perturbative renormalization and infrared finiteness in the Wilson renormalization group: The Massless scalar case}},
	volume = {B409},
	year = {1993},
	Bdsk-Url-1 = {http://dx.doi.org/10.1016/0550-3213(93)90588-G}}

@article{morriii,
	author = {Mitchell, Alex and Morris, Tim R.},
	date-added = {2017-10-04 10:43:33 +0000},
	date-modified = {2017-10-04 10:53:47 +0000},
	title = {in preparation},
	year = {2019}}

@article{Goldstone:1961eq,
	author = {Goldstone, J.},
	date-added = {2017-06-28 19:48:36 +0000},
	date-modified = {2017-06-28 19:48:36 +0000},
	doi = {10.1007/BF02812722},
	journal = {Nuovo Cim.},
	pages = {154-164},
	slaccitation = {%%CITATION = NUCIA,19,154;%%},
	title = {{Field Theories with Superconductor Solutions}},
	volume = {19},
	year = {1961},
	Bdsk-Url-1 = {http://dx.doi.org/10.1007/BF02812722}}

@article{Sakai:1985cs,
	author = {Sakai, N. and Senda, I.},
	booktitle = {{Tokyo Winter Seminar 1985:39}},
	date-added = {2017-06-19 16:46:27 +0000},
	date-modified = {2017-06-19 16:46:27 +0000},
	doi = {10.1143/PTP.75.692},
	journal = {Prog. Theor. Phys.},
	note = {[Erratum: Prog. Theor. Phys.77,773(1987)]},
	pages = {692},
	reportnumber = {TIT/HEP-88},
	slaccitation = {%%CITATION = PTPKA,75,692;%%},
	title = {{Vacuum Energies of String Compactified on Torus}},
	volume = {75},
	year = {1986},
	Bdsk-Url-1 = {http://dx.doi.org/10.1143/PTP.75.692}}

@article{Kikkawa:1984cp,
	author = {Kikkawa, Keiji and Yamasaki, Masami},
	date-added = {2017-06-19 16:45:17 +0000},
	date-modified = {2017-06-19 16:45:17 +0000},
	doi = {10.1016/0370-2693(84)90423-4},
	journal = {Phys. Lett.},
	pages = {357-360},
	reportnumber = {OU-HET 61},
	slaccitation = {%%CITATION = PHLTA,B149,357;%%},
	title = {{Casimir Effects in Superstring Theories}},
	volume = {B149},
	year = {1984},
	Bdsk-Url-1 = {http://dx.doi.org/10.1016/0370-2693(84)90423-4}}

@article{Green:1982sw,
	author = {Green, Michael B. and Schwarz, John H. and Brink, Lars},
	date-added = {2017-06-19 16:30:12 +0000},
	date-modified = {2017-06-19 16:30:12 +0000},
	doi = {10.1016/0550-3213(82)90336-4},
	journal = {Nucl. Phys.},
	pages = {474-492},
	reportnumber = {CALT-68-880},
	slaccitation = {%%CITATION = NUPHA,B198,474;%%},
	title = {{N=4 Yang-Mills and N=8 Supergravity as Limits of String Theories}},
	volume = {B198},
	year = {1982},
	Bdsk-Url-1 = {http://dx.doi.org/10.1016/0550-3213(82)90336-4}}

@article{Carroll:2010aj,
	archiveprefix = {arXiv},
	author = {Carroll, Sean M. and Tam, Heywood},
	date-added = {2017-06-15 04:16:28 +0000},
	date-modified = {2017-06-15 04:16:28 +0000},
	eprint = {1007.1417},
	primaryclass = {hep-th},
	reportnumber = {CALT-68-2797},
	slaccitation = {%%CITATION = ARXIV:1007.1417;%%},
	title = {{Unitary Evolution and Cosmological Fine-Tuning}},
	year = {2010}}

@article{Hollands:2002xi,
	archiveprefix = {arXiv},
	author = {Hollands, Stefan and Wald, Robert M.},
	date-added = {2017-06-15 04:14:26 +0000},
	date-modified = {2017-06-15 04:14:26 +0000},
	eprint = {hep-th/0210001},
	primaryclass = {hep-th},
	slaccitation = {%%CITATION = HEP-TH/0210001;%%},
	title = {{Comment on inflation and alternative cosmology}},
	year = {2002}}

@article{Kofman:2002cj,
	archiveprefix = {arXiv},
	author = {Kofman, Lev and Linde, Andrei D. and Mukhanov, Viatcheslav F.},
	date-added = {2017-06-15 04:12:42 +0000},
	date-modified = {2017-06-15 04:12:42 +0000},
	doi = {10.1088/1126-6708/2002/10/057},
	eprint = {hep-th/0206088},
	journal = {JHEP},
	pages = {057},
	primaryclass = {hep-th},
	reportnumber = {CITA-2002-15, SU-ITP-02-26},
	slaccitation = {%%CITATION = HEP-TH/0206088;%%},
	title = {{Inflationary theory and alternative cosmology}},
	volume = {10},
	year = {2002},
	Bdsk-Url-1 = {http://dx.doi.org/10.1088/1126-6708/2002/10/057}}

@article{Hollands:2002yb,
	archiveprefix = {arXiv},
	author = {Hollands, Stefan and Wald, Robert M.},
	date-added = {2017-06-15 04:11:17 +0000},
	date-modified = {2017-06-15 04:11:17 +0000},
	doi = {10.1023/A:1021175216055},
	eprint = {gr-qc/0205058},
	journal = {Gen. Rel. Grav.},
	pages = {2043-2055},
	primaryclass = {gr-qc},
	slaccitation = {%%CITATION = GR-QC/0205058;%%},
	title = {{An Alternative to inflation}},
	volume = {34},
	year = {2002},
	Bdsk-Url-1 = {http://dx.doi.org/10.1023/A:1021175216055}}

@article{Nielsen:1975fs,
	author = {Nielsen, N. K.},
	date-added = {2017-06-13 10:25:58 +0000},
	date-modified = {2017-06-13 10:25:58 +0000},
	doi = {10.1016/0550-3213(75)90301-6},
	journal = {Nucl. Phys.},
	pages = {173-188},
	reportnumber = {Print-75-0792 (AARHUS)},
	slaccitation = {%%CITATION = NUPHA,B101,173;%%},
	title = {{On the Gauge Dependence of Spontaneous Symmetry Breaking in Gauge Theories}},
	volume = {B101},
	year = {1975},
	Bdsk-Url-1 = {http://dx.doi.org/10.1016/0550-3213(75)90301-6}}

@article{Jackiw:1974cv,
	author = {Jackiw, R.},
	date-added = {2017-06-13 10:20:08 +0000},
	date-modified = {2017-06-13 10:20:08 +0000},
	doi = {10.1103/PhysRevD.9.1686},
	journal = {Phys. Rev.},
	pages = {1686},
	slaccitation = {%%CITATION = PHRVA,D9,1686;%%},
	title = {{Functional evaluation of the effective potential}},
	volume = {D9},
	year = {1974},
	Bdsk-Url-1 = {http://dx.doi.org/10.1103/PhysRevD.9.1686}}

@article{Falls:2017cze,
	archiveprefix = {arXiv},
	author = {Falls, Kevin},
	date-added = {2017-04-17 13:15:21 +0000},
	date-modified = {2017-04-17 13:15:21 +0000},
	eprint = {1702.03577},
	primaryclass = {hep-th},
	slaccitation = {%%CITATION = ARXIV:1702.03577;%%},
	title = {{Physical renormalisation schemes and asymptotic safety in quantum gravity}},
	year = {2017}}

@article{Henz:2016aoh,
	archiveprefix = {arXiv},
	author = {Henz, Tobias and Pawlowski, Jan Martin and Wetterich, Christof},
	date-added = {2017-04-17 12:07:34 +0000},
	date-modified = {2017-04-17 12:07:34 +0000},
	doi = {10.1016/j.physletb.2017.01.057},
	eprint = {1605.01858},
	primaryclass = {hep-th},
	slaccitation = {%%CITATION = ARXIV:1605.01858;%%},
	title = {{Scaling solutions for Dilaton Quantum Gravity}},
	year = {2016},
	Bdsk-Url-1 = {http://dx.doi.org/10.1016/j.physletb.2017.01.057}}

@article{Gies:2016con,
	archiveprefix = {arXiv},
	author = {Gies, Holger and Knorr, Benjamin and Lippoldt, Stefan and Saueressig, Frank},
	date-added = {2017-04-17 12:03:05 +0000},
	date-modified = {2017-04-17 12:03:05 +0000},
	doi = {10.1103/PhysRevLett.116.211302},
	eprint = {1601.01800},
	journal = {Phys. Rev. Lett.},
	number = {21},
	pages = {211302},
	primaryclass = {hep-th},
	slaccitation = {%%CITATION = ARXIV:1601.01800;%%},
	title = {{Gravitational Two-Loop Counterterm Is Asymptotically Safe}},
	volume = {116},
	year = {2016},
	Bdsk-Url-1 = {http://dx.doi.org/10.1103/PhysRevLett.116.211302}}

@article{Biemans:2016rvp,
	archiveprefix = {arXiv},
	author = {Biemans, Jorn and Platania, Alessia and Saueressig, Frank},
	date-added = {2017-04-17 11:59:41 +0000},
	date-modified = {2017-04-17 11:59:41 +0000},
	eprint = {1609.04813},
	primaryclass = {hep-th},
	slaccitation = {%%CITATION = ARXIV:1609.04813;%%},
	title = {{Quantum gravity on foliated spacetime - asymptotically safe and sound}},
	year = {2016}}

@article{Eichhorn:2016vvy,
	archiveprefix = {arXiv},
	author = {Eichhorn, Astrid and Lippoldt, Stefan},
	date-added = {2017-04-17 11:59:03 +0000},
	date-modified = {2017-04-17 11:59:03 +0000},
	doi = {10.1016/j.physletb.2017.01.064},
	eprint = {1611.05878},
	journal = {Phys. Lett.},
	pages = {142-146},
	primaryclass = {gr-qc},
	slaccitation = {%%CITATION = ARXIV:1611.05878;%%},
	title = {{Quantum gravity and Standard-Model-like fermions}},
	volume = {B767},
	year = {2017},
	Bdsk-Url-1 = {http://dx.doi.org/10.1016/j.physletb.2017.01.064}}

@article{Christiansen:2016sjn,
	archiveprefix = {arXiv},
	author = {Christiansen, Nicolai},
	date-added = {2017-04-17 11:57:57 +0000},
	date-modified = {2017-04-17 11:57:57 +0000},
	eprint = {1612.06223},
	primaryclass = {hep-th},
	slaccitation = {%%CITATION = ARXIV:1612.06223;%%},
	title = {{Four-Derivative Quantum Gravity Beyond Perturbation Theory}},
	year = {2016}}

@article{Denz:2016qks,
	archiveprefix = {arXiv},
	author = {Denz, Tobias and Pawlowski, Jan M. and Reichert, Manuel},
	date-added = {2017-04-17 11:57:03 +0000},
	date-modified = {2017-04-17 11:57:03 +0000},
	eprint = {1612.07315},
	primaryclass = {hep-th},
	slaccitation = {%%CITATION = ARXIV:1612.07315;%%},
	title = {{Towards apparent convergence in asymptotically safe quantum gravity}},
	year = {2016}}

@article{Biemans:2017zca,
	archiveprefix = {arXiv},
	author = {Biemans, Jorn and Platania, Alessia and Saueressig, Frank},
	date-added = {2017-04-17 11:56:26 +0000},
	date-modified = {2017-04-17 11:56:26 +0000},
	eprint = {1702.06539},
	primaryclass = {hep-th},
	slaccitation = {%%CITATION = ARXIV:1702.06539;%%},
	title = {{Renormalization group fixed points of foliated gravity-matter systems}},
	year = {2017}}

@article{Christiansen:2017gtg,
	archiveprefix = {arXiv},
	author = {Christiansen, Nicolai and Eichhorn, Astrid},
	date-added = {2017-04-17 11:55:39 +0000},
	date-modified = {2017-04-17 11:55:39 +0000},
	eprint = {1702.07724},
	primaryclass = {hep-th},
	slaccitation = {%%CITATION = ARXIV:1702.07724;%%},
	title = {{An asymptotically safe solution to the U(1) triviality problem}},
	year = {2017}}

@article{Hamada:2017rvn,
	archiveprefix = {arXiv},
	author = {Hamada, Yuta and Yamada, Masatoshi},
	date-added = {2017-04-17 11:54:43 +0000},
	date-modified = {2017-04-17 11:54:43 +0000},
	eprint = {1703.09033},
	primaryclass = {hep-th},
	slaccitation = {%%CITATION = ARXIV:1703.09033;%%},
	title = {{Asymptotic safety of higher derivative quantum gravity non-minimally coupled with a matter system}},
	year = {2017}}

@inproceedings{Demmel:2013myx,
	archiveprefix = {arXiv},
	author = {Demmel, Maximilian and Saueressig, Frank and Zanusso, Omar},
	booktitle = {{Proceedings, 13th Marcel Grossmann Meeting on Recent Developments in Theoretical and Experimental General Relativity, Astrophysics, and Relativistic Field Theories (MG13): Stockholm, Sweden, July 1-7, 2012}},
	date-added = {2017-04-17 11:39:52 +0000},
	date-modified = {2017-04-17 11:39:52 +0000},
	doi = {10.1142/9789814623995_0404},
	eprint = {1302.1312},
	pages = {2227-2229},
	primaryclass = {hep-th},
	reportnumber = {MITP-13-013},
	slaccitation = {%%CITATION = ARXIV:1302.1312;%%},
	title = {{Fixed Functionals in Asymptotically Safe Gravity}},
	url = {http://inspirehep.net/record/1217855/files/arXiv:1302.1312.pdf},
	year = {2015},
	Bdsk-Url-1 = {http://inspirehep.net/record/1217855/files/arXiv:1302.1312.pdf},
	Bdsk-Url-2 = {http://dx.doi.org/10.1142/9789814623995_0404}}

@article{Benedetti:2013nya,
	archiveprefix = {arXiv},
	author = {Benedetti, Dario and Guarnieri, Filippo},
	date-added = {2017-04-17 11:37:41 +0000},
	date-modified = {2017-04-17 11:37:41 +0000},
	doi = {10.1088/1367-2630/16/5/053051},
	eprint = {1311.1081},
	journal = {New J. Phys.},
	pages = {053051},
	primaryclass = {hep-th},
	reportnumber = {AEI-2013-248},
	slaccitation = {%%CITATION = ARXIV:1311.1081;%%},
	title = {{Brans-Dicke theory in the local potential approximation}},
	volume = {16},
	year = {2014},
	Bdsk-Url-1 = {http://dx.doi.org/10.1088/1367-2630/16/5/053051}}

@article{Bollini:1973wu,
	author = {Bollini, C. G. and Giambiagi, J. J.},
	date-added = {2017-04-05 20:24:19 +0000},
	date-modified = {2017-10-13 08:07:33 +0000},
	journal = {Acta Phys. Austriaca},
	pages = {211-215},
	slaccitation = {%%CITATION = APASA,38,211;%%},
	title = {{Evanescent couplings and compensation of Adler anomaly}},
	volume = {38},
	year = {1973}}

@article{Abbott:1981ke,
	author = {Abbott, L.F.},
	date-added = {2015-07-05 17:21:41 +0000},
	date-modified = {2015-07-05 17:21:41 +0000},
	journal = {Acta Phys.Polon.},
	pages = {33},
	reportnumber = {CERN-TH-3113},
	slaccitation = {%%CITATION = APPOA,B13,33;%%},
	title = {{Introduction to the Background Field Method}},
	volume = {B13},
	year = {1982}}

@article{Abbott:1980hw,
	author = {Abbott, L.F.},
	date-added = {2015-07-05 17:21:41 +0000},
	date-modified = {2015-07-05 17:21:41 +0000},
	doi = {10.1016/0550-3213(81)90371-0},
	journal = {Nucl.Phys.},
	pages = {189},
	reportnumber = {CERN-TH-2973},
	slaccitation = {%%CITATION = NUPHA,B185,189;%%},
	title = {{The Background Field Method Beyond One Loop}},
	volume = {B185},
	year = {1981},
	Bdsk-Url-1 = {http://dx.doi.org/10.1016/0550-3213(81)90371-0}}

@book{AbramowitzStegun,
	address = {New York},
	author = {Abramowitz, Milton and Stegun, Irene A.},
	date-added = {2016-05-10 18:09:59 +0000},
	date-modified = {2016-05-10 18:09:59 +0000},
	edition = {9th},
	keywords = {Handbook},
	pages = {503-515},
	publisher = {Dover},
	title = {Handbook of Mathematical Functions with Formulas, Graphs, and Mathematical Tables},
	year = {1964}}

@article{Adler:1982ri,
	author = {Adler, Stephen L.},
	date-added = {2015-11-30 08:56:33 +0000},
	date-modified = {2015-11-30 08:56:33 +0000},
	doi = {10.1103/RevModPhys.54.729},
	journal = {Rev. Mod. Phys.},
	note = {[Erratum: Rev. Mod. Phys.55,837(1983)]},
	pages = {729},
	reportnumber = {PRINT-81-0725-REV. (IAS,-PRINCETON), PRINT-81-0725 (IAS,-PRINCETON)},
	slaccitation = {%%CITATION = RMPHA,54,729;%%},
	title = {{Einstein Gravity as a Symmetry Breaking Effect in Quantum Field Theory}},
	volume = {54},
	year = {1982},
	Bdsk-Url-1 = {http://dx.doi.org/10.1103/RevModPhys.54.729}}

@article{Alkofer:2014raa,
	archiveprefix = {arXiv},
	author = {Alkofer, Natalia and Saueressig, Frank and Zanusso, Omar},
	date-added = {2015-11-30 08:33:10 +0000},
	date-modified = {2015-11-30 08:33:10 +0000},
	doi = {10.1103/PhysRevD.91.025025},
	eprint = {1410.7999},
	journal = {Phys. Rev.},
	number = {2},
	pages = {025025},
	primaryclass = {hep-th},
	slaccitation = {%%CITATION = ARXIV:1410.7999;%%},
	title = {{Spectral dimensions from the spectral action}},
	volume = {D91},
	year = {2015},
	Bdsk-Url-1 = {http://dx.doi.org/10.1103/PhysRevD.91.025025}}

@article{Altschul:2004yq,
	archiveprefix = {arXiv},
	author = {Altschul, B.},
	date-added = {2016-03-21 10:43:58 +0000},
	date-modified = {2016-03-21 10:43:58 +0000},
	doi = {10.1016/j.nuclphysb.2004.10.054},
	eprint = {hep-th/0403093},
	journal = {Nucl.Phys.},
	pages = {593-604},
	primaryclass = {hep-th},
	reportnumber = {IUHET-467},
	title = {Nonpolynomial normal modes of the renormalization group in the presence of a constant vector potential background},
	volume = {B705},
	year = {2005},
	Bdsk-Url-1 = {http://dx.doi.org/10.1016/j.nuclphysb.2004.10.054}}

@article{Altschul:2004gt,
	archiveprefix = {arXiv},
	author = {Altschul, B.},
	date-added = {2016-05-10 18:09:59 +0000},
	date-modified = {2016-05-10 18:09:59 +0000},
	eprint = {hep-th/0407173},
	primaryclass = {hep-th},
	reportnumber = {IUHET-474},
	title = {Asymptotically free Lorentz and CPT-violating scalar field theories},
	year = {2004}}

@article{Altschul:2005mu,
	archiveprefix = {arXiv},
	author = {Altschul, B. and Kostelecky, V. Alan},
	date-added = {2016-05-10 18:09:59 +0000},
	date-modified = {2016-05-10 18:09:59 +0000},
	doi = {10.1016/j.physletb.2005.09.018},
	eprint = {hep-th/0509068},
	journal = {Phys.Lett.},
	pages = {106-112},
	primaryclass = {hep-th},
	reportnumber = {IUHET-481},
	title = {Spontaneous Lorentz violation and nonpolynomial interactions},
	volume = {B628},
	year = {2005},
	Bdsk-Url-1 = {http://dx.doi.org/10.1016/j.physletb.2005.09.018}}

@article{Ambjorn:2012jv,
	archiveprefix = {arXiv},
	author = {Ambj{\o}rn, J. and Goerlich, A. and Jurkiewicz, J. and Loll, R.},
	date-added = {2015-07-05 17:21:41 +0000},
	date-modified = {2015-07-05 17:21:41 +0000},
	doi = {10.1016/j.physrep.2012.03.007},
	eprint = {1203.3591},
	journal = {Phys.Rept.},
	pages = {127-210},
	primaryclass = {hep-th},
	slaccitation = {%%CITATION = ARXIV:1203.3591;%%},
	title = {{Nonperturbative Quantum Gravity}},
	volume = {519},
	year = {2012},
	Bdsk-Url-1 = {http://dx.doi.org/10.1016/j.physrep.2012.03.007}}

@article{Ambjorn:2005db,
	archiveprefix = {arXiv},
	author = {Ambjorn, J. and Jurkiewicz, J. and Loll, R.},
	date-added = {2015-11-30 08:22:22 +0000},
	date-modified = {2015-11-30 08:22:22 +0000},
	doi = {10.1103/PhysRevLett.95.171301},
	eprint = {hep-th/0505113},
	journal = {Phys. Rev. Lett.},
	pages = {171301},
	primaryclass = {hep-th},
	reportnumber = {SPIN-05-05, ITP-UU-05-07},
	slaccitation = {%%CITATION = HEP-TH/0505113;%%},
	title = {{Spectral dimension of the universe}},
	volume = {95},
	year = {2005},
	Bdsk-Url-1 = {http://dx.doi.org/10.1103/PhysRevLett.95.171301}}

@article{Amelino-Camelia:2013cfa,
	archiveprefix = {arXiv},
	author = {Amelino-Camelia, Giovanni and Arzano, Michele and Gubitosi, Giulia and Magueijo, Jo{\~a}o},
	date-added = {2015-11-30 08:25:49 +0000},
	date-modified = {2015-11-30 08:25:49 +0000},
	doi = {10.1016/j.physletb.2014.07.030},
	eprint = {1311.3135},
	journal = {Phys. Lett.},
	pages = {317-320},
	primaryclass = {gr-qc},
	slaccitation = {%%CITATION = ARXIV:1311.3135;%%},
	title = {{Planck-scale dimensional reduction without a preferred frame}},
	volume = {B736},
	year = {2014},
	Bdsk-Url-1 = {http://dx.doi.org/10.1016/j.physletb.2014.07.030}}

@article{Anjana:2015ios,
	archiveprefix = {arXiv},
	author = {Anjana, V. and Harikumar, E.},
	date-added = {2015-11-30 08:24:29 +0000},
	date-modified = {2015-11-30 08:24:29 +0000},
	doi = {10.1103/PhysRevD.91.065026},
	eprint = {1501.00254},
	journal = {Phys. Rev.},
	number = {6},
	pages = {065026},
	primaryclass = {hep-th},
	slaccitation = {%%CITATION = ARXIV:1501.00254;%%},
	title = {{Spectral dimension of kappa-deformed spacetime}},
	volume = {D91},
	year = {2015},
	Bdsk-Url-1 = {http://dx.doi.org/10.1103/PhysRevD.91.065026}}

@article{Aoki:2000wm,
	author = {Aoki, K.},
	booktitle = {{Methods of renormalization group. Proceedings, Summer School on mathematical physics, Tokyo, Japan, September 23-26, 1999}},
	date-added = {2016-05-10 15:37:15 +0000},
	date-modified = {2016-05-10 15:37:15 +0000},
	doi = {10.1016/S0217-9792(00)00092-3},
	journal = {Int. J. Mod. Phys.},
	pages = {1249-1326},
	slaccitation = {%%CITATION = IMPAE,B14,1249;%%},
	title = {{Introduction to the nonperturbative renormalization group and its recent applications}},
	volume = {B14},
	year = {2000},
	Bdsk-Url-1 = {http://dx.doi.org/10.1016/S0217-9792(00)00092-3}}

@article{Arnone:2002cs,
	archiveprefix = {arXiv},
	author = {Arnone, Stefano and Gatti, Antonio and Morris, Tim R.},
	date-added = {2015-11-30 06:56:04 +0000},
	date-modified = {2015-11-30 06:56:04 +0000},
	doi = {10.1103/PhysRevD.67.085003},
	eprint = {hep-th/0209162},
	journal = {Phys. Rev.},
	pages = {085003},
	primaryclass = {hep-th},
	reportnumber = {SHEP-02-22},
	slaccitation = {%%CITATION = HEP-TH/0209162;%%},
	title = {{A Proposal for a manifestly gauge invariant and universal calculus in Yang-Mills theory}},
	volume = {D67},
	year = {2003},
	Bdsk-Url-1 = {http://dx.doi.org/10.1103/PhysRevD.67.085003}}

@article{Arnone:2002ai,
	archiveprefix = {arXiv},
	author = {Arnone, Stefano and Gatti, Antonio and Morris, Tim R.},
	booktitle = {{Renormalization group. Proceedings, 5th International Conference, RG 2002, Tatranska Strba, Slovakia, March 10-16, 2002}},
	date-added = {2015-11-30 06:56:10 +0000},
	date-modified = {2015-11-30 06:56:10 +0000},
	eprint = {hep-th/0205156},
	journal = {Acta Phys. Slov.},
	pages = {615-620},
	primaryclass = {hep-th},
	reportnumber = {SHEP-02-10},
	slaccitation = {%%CITATION = HEP-TH/0205156;%%},
	title = {{A Demonstration of scheme independence in scalar ERGs}},
	volume = {52},
	year = {2002}}

@inproceedings{Arnone:2002fa,
	archiveprefix = {arXiv},
	author = {Arnone, Stefano and Gatti, Antonio and Morris, Tim R.},
	booktitle = {{Renormalization group. Proceedings, 5th International Conference, RG 2002, Tatranska Strba, Slovakia, March 10-16, 2002}},
	date-added = {2015-11-30 06:56:08 +0000},
	date-modified = {2015-11-30 06:56:08 +0000},
	eprint = {hep-th/0207153},
	note = {[Submitted to: Acta Phys. Slov.(2002)]},
	primaryclass = {hep-th},
	reportnumber = {SHEP-02-11},
	slaccitation = {%%CITATION = HEP-TH/0207153;%%},
	title = {{A Manifestly gauge invariant exact renormalization group}},
	url = {http://alice.cern.ch/format/showfull?sysnb=2328480},
	year = {2002},
	Bdsk-Url-1 = {http://alice.cern.ch/format/showfull?sysnb=2328480}}

@inproceedings{Arnone:2002fb,
	archiveprefix = {arXiv},
	author = {Arnone, Stefano and Gatti, Antonio and Morris, Tim R.},
	booktitle = {{Renormalization group. Proceedings, 5th International Conference, RG 2002, Tatranska Strba, Slovakia, March 10-16, 2002}},
	date-added = {2015-11-30 06:56:07 +0000},
	date-modified = {2015-11-30 06:56:07 +0000},
	eprint = {hep-th/0207154},
	note = {[Submitted to: Acta Phys. Slov.(2002)]},
	primaryclass = {hep-th},
	reportnumber = {SHEP-02-12},
	slaccitation = {%%CITATION = HEP-TH/0207154;%%},
	title = {{Manifestly gauge invariant computations}},
	url = {http://alice.cern.ch/format/showfull?sysnb=2328491},
	year = {2002},
	Bdsk-Url-1 = {http://alice.cern.ch/format/showfull?sysnb=2328491}}

@article{Arnone:2002qi,
	archiveprefix = {arXiv},
	author = {Arnone, Stefano and Gatti, Antonio and Morris, Tim R.},
	date-added = {2015-11-30 06:56:12 +0000},
	date-modified = {2015-11-30 06:56:12 +0000},
	eprint = {hep-th/0209130},
	journal = {Acta Phys. Slov.},
	pages = {621-634},
	primaryclass = {hep-th},
	reportnumber = {SHEP-02-21},
	slaccitation = {%%CITATION = HEP-TH/0209130;%%},
	title = {{Towards a manifestly gauge invariant and universal calculus for Yang-Mills theory}},
	volume = {52},
	year = {2002}}

@article{Arnone:2002yh,
	archiveprefix = {arXiv},
	author = {Arnone, Stefano and Gatti, Antonio and Morris, Tim R.},
	date-added = {2015-11-30 06:56:13 +0000},
	date-modified = {2015-11-30 06:56:13 +0000},
	doi = {10.1088/1126-6708/2002/05/059},
	eprint = {hep-th/0201237},
	journal = {JHEP},
	pages = {059},
	primaryclass = {hep-th},
	reportnumber = {SHEP-02-02},
	slaccitation = {%%CITATION = HEP-TH/0201237;%%},
	title = {{Exact scheme independence at one loop}},
	volume = {05},
	year = {2002},
	Bdsk-Url-1 = {http://dx.doi.org/10.1088/1126-6708/2002/05/059}}

@article{Arnone:2003pa,
	archiveprefix = {arXiv},
	author = {Arnone, Stefano and Gatti, Antonio and Morris, Tim R. and Rosten, Oliver J.},
	date-added = {2015-11-30 06:55:06 +0000},
	date-modified = {2015-11-30 06:55:06 +0000},
	doi = {10.1103/PhysRevD.69.065009},
	eprint = {hep-th/0309242},
	journal = {Phys. Rev.},
	pages = {065009},
	primaryclass = {hep-th},
	reportnumber = {SHEP-03-17},
	slaccitation = {%%CITATION = HEP-TH/0309242;%%},
	title = {{Exact scheme independence at two loops}},
	volume = {D69},
	year = {2004},
	Bdsk-Url-1 = {http://dx.doi.org/10.1103/PhysRevD.69.065009}}

@article{Arnone:2000qd,
	archiveprefix = {arXiv},
	author = {Arnone, S. and Kubyshin, {\relax Yu}. A. and Morris, T. R. and Tighe, J. F.},
	booktitle = {{The exact renormalization group. Proceedings, 2nd Conference, Rome, Italy, September 18-22, 2000}},
	date-added = {2015-11-30 06:56:21 +0000},
	date-modified = {2015-11-30 06:56:21 +0000},
	doi = {10.1142/S0217751X0100461X},
	eprint = {hep-th/0102054},
	journal = {Int. J. Mod. Phys.},
	pages = {1989},
	primaryclass = {hep-th},
	reportnumber = {SHEP-01-07},
	slaccitation = {%%CITATION = HEP-TH/0102054;%%},
	title = {{A Gauge invariant regulator for the ERG}},
	volume = {A16},
	year = {2001},
	Bdsk-Url-1 = {http://dx.doi.org/10.1142/S0217751X0100461X}}

@inproceedings{Arnone:2000bv,
	archiveprefix = {arXiv},
	author = {Arnone, S. and Kubyshin, {\relax Yu}. A. and Morris, T. R. and Tighe, J. F.},
	booktitle = {{High energy physics and quantum field theory. Proceedings, 15th International Workshop, QFTHEP 2000, Tver, Russia, September 14-20, 2000}},
	date-added = {2015-11-30 06:56:34 +0000},
	date-modified = {2015-11-30 06:56:34 +0000},
	eprint = {hep-th/0102011},
	pages = {297-304},
	primaryclass = {hep-th},
	reportnumber = {SHEP-01-04},
	slaccitation = {%%CITATION = HEP-TH/0102011;%%},
	title = {{Gauge invariant regularization in the ERG approach}},
	url = {http://theory.sinp.msu.ru/~qfthep/2000/Proceedings/Arnone_QFTHEP.ps.gz},
	year = {2000},
	Bdsk-Url-1 = {http://theory.sinp.msu.ru/~qfthep/2000/Proceedings/Arnone_QFTHEP.ps.gz}}

@article{Arnone:2001iy,
	archiveprefix = {arXiv},
	author = {Arnone, Stefano and Kubyshin, Yuri A. and Morris, Tim R. and Tighe, John F.},
	date-added = {2015-11-30 06:56:17 +0000},
	date-modified = {2015-11-30 06:56:17 +0000},
	doi = {10.1142/S0217751X02009722},
	eprint = {hep-th/0106258},
	journal = {Int. J. Mod. Phys.},
	pages = {2283-2330},
	primaryclass = {hep-th},
	reportnumber = {SHEP-01-15},
	slaccitation = {%%CITATION = HEP-TH/0106258;%%},
	title = {{Gauge invariant regularization via SU($N|N$)}},
	volume = {A17},
	year = {2002},
	Bdsk-Url-1 = {http://dx.doi.org/10.1142/S0217751X02009722}}

@article{Arnone:2005fb,
	archiveprefix = {arXiv},
	author = {Arnone, Stefano and Morris, Tim R. and Rosten, Oliver J.},
	date-added = {2015-11-30 06:55:18 +0000},
	date-modified = {2015-11-30 06:55:18 +0000},
	doi = {10.1140/epjc/s10052-007-0258-y},
	eprint = {hep-th/0507154},
	journal = {Eur. Phys. J.},
	pages = {467-504},
	primaryclass = {hep-th},
	reportnumber = {SHEP-05-21},
	slaccitation = {%%CITATION = HEP-TH/0507154;%%},
	title = {{A Generalised manifestly gauge invariant exact renormalisation group for SU(N) Yang-Mills}},
	volume = {C50},
	year = {2007},
	Bdsk-Url-1 = {http://dx.doi.org/10.1140/epjc/s10052-007-0258-y}}

@article{Arnone:2006ie,
	archiveprefix = {arXiv},
	author = {Arnone, Stefano and Morris, Tim R. and Rosten, Oliver J.},
	booktitle = {{Universality and renormalization: From stochastic evolution to renormalization of quantum fields. Proceedings, Workshop on 'Percolation, SLE and related topics', Toronto, Canada, September 20-24, 2005, and Workshop on 'Renormalization and universality in mathematical physics', Toronto, Canada, October 18-22, 2005}},
	date-added = {2015-11-30 06:55:24 +0000},
	date-modified = {2020-04-20 14:49:55 +0100},
	eprint = {hep-th/0606181},
	journal = {Fields Inst. Commun.},
	pages = {1},
	primaryclass = {hep-th},
	slaccitation = {%%CITATION = HEP-TH/0606181;%%},
	title = {{Manifestly Gauge Invariant Exact Renormalization Group}},
	volume = {50},
	year = {2007}}

@article{Arnone:2005vd,
	archiveprefix = {arXiv},
	author = {Arnone, Stefano and Morris, Tim R. and Rosten, Oliver J.},
	date-added = {2015-11-30 06:55:14 +0000},
	date-modified = {2015-11-30 06:55:14 +0000},
	doi = {10.1088/1126-6708/2005/10/115},
	eprint = {hep-th/0505169},
	journal = {JHEP},
	pages = {115},
	primaryclass = {hep-th},
	reportnumber = {CERN-PH-TH-2004-123, ROMA-1399-04, SHEP-04-21},
	slaccitation = {%%CITATION = HEP-TH/0505169;%%},
	title = {{Manifestly gauge invariant QED}},
	volume = {10},
	year = {2005},
	Bdsk-Url-1 = {http://dx.doi.org/10.1088/1126-6708/2005/10/115}}

@article{Arzano:2014jfa,
	archiveprefix = {arXiv},
	author = {Arzano, Michele and Trzesniewski, Tomasz},
	date-added = {2015-11-30 08:31:29 +0000},
	date-modified = {2015-11-30 08:31:29 +0000},
	doi = {10.1103/PhysRevD.89.124024},
	eprint = {1404.4762},
	journal = {Phys. Rev.},
	number = {12},
	pages = {124024},
	primaryclass = {hep-th},
	slaccitation = {%%CITATION = ARXIV:1404.4762;%%},
	title = {{Diffusion on $\kappa$-Minkowski space}},
	volume = {D89},
	year = {2014},
	Bdsk-Url-1 = {http://dx.doi.org/10.1103/PhysRevD.89.124024}}

@article{Ashtekar:1991hf,
	author = {Ashtekar, A.},
	date-added = {2015-07-05 17:21:41 +0000},
	date-modified = {2015-07-05 17:21:41 +0000},
	journal = {Adv.Ser.Astrophys.Cosmol.},
	pages = {1-334},
	slaccitation = {%%CITATION = 00491,6,1;%%},
	title = {{Lectures on nonperturbative canonical gravity}},
	volume = {6},
	year = {1991}}

@article{Ashtekar:2004eh,
	archiveprefix = {arXiv},
	author = {Ashtekar, Abhay and Lewandowski, Jerzy},
	date-added = {2015-07-05 17:21:41 +0000},
	date-modified = {2015-07-05 17:21:41 +0000},
	doi = {10.1088/0264-9381/21/15/R01},
	eprint = {gr-qc/0404018},
	journal = {Class.Quant.Grav.},
	pages = {R53},
	primaryclass = {gr-qc},
	slaccitation = {%%CITATION = GR-QC/0404018;%%},
	title = {{Background independent quantum gravity: A Status report}},
	volume = {21},
	year = {2004},
	Bdsk-Url-1 = {http://dx.doi.org/10.1088/0264-9381/21/15/R01}}

@article{Ashtekar:2014kba,
	archiveprefix = {arXiv},
	author = {Ashtekar, Abhay and Reuter, Martin and Rovelli, Carlo},
	date-added = {2015-07-05 17:21:41 +0000},
	date-modified = {2015-07-05 17:21:41 +0000},
	eprint = {1408.4336},
	primaryclass = {gr-qc},
	slaccitation = {%%CITATION = ARXIV:1408.4336;%%},
	title = {{From General Relativity to Quantum Gravity}},
	year = {2014}}

@article{Avramidi1985,
	author = {Avramidi, I. G. and Barvinsky, A. O.},
	doi = {10.1016/0370-2693(85)90248-5},
	journal = {Phys. Lett.},
	owner = {trmorris},
	pages = {269-274},
	slaccitation = {%%CITATION = PHLTA,B159,269;%%},
	timestamp = {2016.01.26},
	title = {{Asymptotic freedom in higher derivative quantum gravity}},
	volume = {B159},
	year = {1985},
	Bdsk-Url-1 = {http://dx.doi.org/10.1016/0370-2693(85)90248-5}}

@article{Bagnuls:2000,
	archiveprefix = {arXiv},
	author = {Bagnuls, C. and Bervillier, C.},
	date-added = {2015-07-05 17:21:41 +0000},
	date-modified = {2015-07-05 17:21:41 +0000},
	doi = {10.1016/S0370-1573(00)00137-X},
	eprint = {hep-th/0002034},
	journal = {Phys.Rept.},
	pages = {91},
	primaryclass = {hep-th},
	reportnumber = {SACLAY-SPH-T-S00-008, SACLAY-SPH-T-S00-009},
	slaccitation = {%%CITATION = HEP-TH/0002034;%%},
	title = {{Exact renormalization group equations. An Introductory review}},
	volume = {348},
	year = {2001},
	Bdsk-Url-1 = {http://dx.doi.org/10.1016/S0370-1573(00)00137-X}}

@article{Ball:1994ji,
	archiveprefix = {arXiv},
	author = {Ball, Richard D. and Haagensen, Peter E. and Latorre, Jose, I. and Moreno, Enrique},
	date-added = {2016-05-16 09:32:21 +0000},
	date-modified = {2016-05-16 09:32:21 +0000},
	doi = {10.1016/0370-2693(95)00025-G},
	eprint = {hep-th/9411122},
	journal = {Phys. Lett.},
	pages = {80-88},
	primaryclass = {hep-th},
	reportnumber = {CERN-TH-7482-94, UB-ECM-PF-94-32, MCGILL-94-54},
	slaccitation = {%%CITATION = HEP-TH/9411122;%%},
	title = {{Scheme independence and the exact renormalization group}},
	volume = {B347},
	year = {1995},
	Bdsk-Url-1 = {http://dx.doi.org/10.1016/0370-2693(95)00025-G}}

@article{Bateman:2014lia,
	archiveprefix = {arXiv},
	author = {Bateman, James and McHardy, Ian and Merle, Alexander and Morris, Tim R. and Ulbricht, Hendrik},
	date-added = {2016-12-17 18:10:15 +0000},
	date-modified = {2016-12-17 18:10:15 +0000},
	doi = {10.1038/srep08058},
	eprint = {1405.5536},
	journal = {Sci. Rep.},
	pages = {8058},
	primaryclass = {hep-ph},
	slaccitation = {%%CITATION = ARXIV:1405.5536;%%},
	title = {{On the Existence of Low-Mass Dark Matter and its Direct Detection}},
	volume = {5},
	year = {2015},
	Bdsk-Url-1 = {http://dx.doi.org/10.1038/srep08058}}

@article{Becchi:1975nq,
	author = {Becchi, C. and Rouet, A. and Stora, R.},
	date-added = {2015-11-30 07:58:15 +0000},
	date-modified = {2015-11-30 07:58:15 +0000},
	doi = {10.1016/0003-4916(76)90156-1},
	journal = {Annals Phys.},
	pages = {287-321},
	reportnumber = {CPT-75-P.723-MARSEILLE},
	slaccitation = {%%CITATION = APNYA,98,287;%%},
	title = {{Renormalization of Gauge Theories}},
	volume = {98},
	year = {1976},
	Bdsk-Url-1 = {http://dx.doi.org/10.1016/0003-4916(76)90156-1}}

@article{Becchi:1974md,
	author = {Becchi, C. and Rouet, A. and Stora, R.},
	date-added = {2015-11-30 08:00:24 +0000},
	date-modified = {2015-11-30 08:00:24 +0000},
	doi = {10.1007/BF01614158},
	journal = {Commun. Math. Phys.},
	pages = {127-162},
	reportnumber = {CPT-74-P.634-MARSEILLE, CNRS-CPT-74-P634},
	slaccitation = {%%CITATION = CMPHA,42,127;%%},
	title = {{Renormalization of the Abelian Higgs-Kibble Model}},
	volume = {42},
	year = {1975},
	Bdsk-Url-1 = {http://dx.doi.org/10.1007/BF01614158}}

@article{Becchi:1974xu,
	author = {Becchi, C. and Rouet, A. and Stora, R.},
	date-added = {2015-11-30 07:58:07 +0000},
	date-modified = {2015-11-30 07:58:07 +0000},
	doi = {10.1016/0370-2693(74)90058-6},
	journal = {Phys. Lett.},
	pages = {344},
	reportnumber = {CPT-74/P.625-MARSEILLE},
	slaccitation = {%%CITATION = PHLTA,B52,344;%%},
	title = {{The Abelian Higgs-Kibble Model. Unitarity of the S Operator}},
	volume = {B52},
	year = {1974},
	Bdsk-Url-1 = {http://dx.doi.org/10.1016/0370-2693(74)90058-6}}

@article{Becker:2014pea,
	archiveprefix = {arXiv},
	author = {Becker, Daniel and Reuter, Martin},
	date-added = {2015-12-01 02:10:12 +0000},
	date-modified = {2015-12-01 02:10:12 +0000},
	doi = {10.1007/JHEP03(2015)065},
	eprint = {1412.0468},
	journal = {JHEP},
	pages = {065},
	primaryclass = {hep-th},
	reportnumber = {MITP-14-083},
	slaccitation = {%%CITATION = ARXIV:1412.0468;%%},
	title = {{Towards a $C$-function in 4D quantum gravity}},
	volume = {03},
	year = {2015},
	Bdsk-Url-1 = {http://dx.doi.org/10.1007/JHEP03(2015)065}}

@article{Becker:2014jua,
	archiveprefix = {arXiv},
	author = {Becker, Daniel and Reuter, Martin},
	date-added = {2015-07-05 17:21:41 +0000},
	date-modified = {2015-07-05 17:21:41 +0000},
	eprint = {1407.5848},
	primaryclass = {hep-th},
	slaccitation = {%%CITATION = ARXIV:1407.5848;%%},
	title = {{Propagating gravitons vs. dark matter in asymptotically safe quantum gravity}},
	year = {2014}}

@article{Belenchia:2015aia,
	archiveprefix = {arXiv},
	author = {Belenchia, Alessio and Benincasa, Dionigi M. T. and Marciano, Antonino and Modesto, Leonardo},
	date-added = {2015-11-30 08:24:06 +0000},
	date-modified = {2015-11-30 08:24:06 +0000},
	eprint = {1507.00330},
	primaryclass = {gr-qc},
	slaccitation = {%%CITATION = ARXIV:1507.00330;%%},
	title = {{Spectral Dimension from Causal Set Nonlocal Dynamics}},
	year = {2015}}

@article{Bell1975,
	author = {Bell, Thomas L. and Wilson, Kenneth G.},
	date-added = {2016-05-04 16:49:28 +0000},
	date-modified = {2016-05-04 16:55:27 +0000},
	journal = {Phys. Rev.},
	pages = {3431},
	title = {Finite-lattice approximations to renormalization groups},
	volume = {B11},
	year = {1975}}

@article{Benedetti:2014gja,
	archiveprefix = {arXiv},
	author = {Benedetti, Dario},
	date-added = {2016-10-10 09:44:10 +0000},
	date-modified = {2016-10-10 09:44:10 +0000},
	doi = {10.1088/1742-5468/2015/01/P01002},
	eprint = {1403.6712},
	journal = {J. Stat. Mech.},
	pages = {P01002},
	primaryclass = {cond-mat.stat-mech},
	reportnumber = {AEI-2014-008},
	slaccitation = {%%CITATION = ARXIV:1403.6712;%%},
	title = {{Critical behavior in spherical and hyperbolic spaces}},
	volume = {1501},
	year = {2015},
	Bdsk-Url-1 = {http://dx.doi.org/10.1088/1742-5468/2015/01/P01002}}

@article{Benedetti:2013jk,
	archiveprefix = {arXiv},
	author = {Benedetti, Dario},
	date-added = {2016-04-26 17:17:41 +0000},
	date-modified = {2016-04-26 17:17:41 +0000},
	doi = {10.1209/0295-5075/102/20007},
	eprint = {1301.4422},
	journal = {Europhys. Lett.},
	pages = {20007},
	primaryclass = {hep-th},
	reportnumber = {AEI-2013-032},
	slaccitation = {%%CITATION = ARXIV:1301.4422;%%},
	title = {{On the number of relevant operators in asymptotically safe gravity}},
	volume = {102},
	year = {2013},
	Bdsk-Url-1 = {http://dx.doi.org/10.1209/0295-5075/102/20007}}

@article{Benedetti:2008gu,
	archiveprefix = {arXiv},
	author = {Benedetti, Dario},
	date-added = {2015-11-30 08:26:05 +0000},
	date-modified = {2015-11-30 08:26:05 +0000},
	doi = {10.1103/PhysRevLett.102.111303},
	eprint = {0811.1396},
	journal = {Phys. Rev. Lett.},
	pages = {111303},
	primaryclass = {hep-th},
	reportnumber = {PI-QG-103},
	slaccitation = {%%CITATION = ARXIV:0811.1396;%%},
	title = {{Fractal properties of quantum spacetime}},
	volume = {102},
	year = {2009},
	Bdsk-Url-1 = {http://dx.doi.org/10.1103/PhysRevLett.102.111303}}

@article{Benedetti:2012,
	archiveprefix = {arXiv},
	author = {Benedetti, Dario and Caravelli, Francesco},
	date-added = {2015-07-05 17:21:41 +0000},
	date-modified = {2015-07-05 17:21:41 +0000},
	doi = {10.1007/JHEP06(2012)017, 10.1007/JHEP10(2012)157},
	eprint = {1204.3541},
	journal = {JHEP},
	pages = {017},
	primaryclass = {hep-th},
	reportnumber = {AEI-2012-017, PI-QG-260},
	slaccitation = {%%CITATION = ARXIV:1204.3541;%%},
	title = {{The Local potential approximation in quantum gravity}},
	volume = {1206},
	year = {2012},
	Bdsk-Url-1 = {http://dx.doi.org/10.1007/JHEP06(2012)017,%2010.1007/JHEP10(2012)157}}

@article{Benitez:2011xx,
	archiveprefix = {arXiv},
	author = {Benitez, F. and Blaizot, J.-P. and Chate, H. and Delamotte, B. and Mendez-Galain, R. and others},
	date-added = {2015-07-05 17:21:41 +0000},
	date-modified = {2015-07-05 17:21:41 +0000},
	doi = {10.1103/PhysRevE.85.026707},
	eprint = {1110.2665},
	journal = {Phys.Rev.},
	pages = {026707},
	primaryclass = {cond-mat.stat-mech},
	slaccitation = {%%CITATION = ARXIV:1110.2665;%%},
	title = {{Non-perturbative renormalization group preserving full-momentum dependence: implementation and quantitative evaluation}},
	volume = {E85},
	year = {2012},
	Bdsk-Url-1 = {http://dx.doi.org/10.1103/PhysRevE.85.026707}}

@article{Berges:2000,
	archiveprefix = {arXiv},
	author = {Berges, Juergen and Tetradis, Nikolaos and Wetterich, Christof},
	date-added = {2015-07-05 17:21:41 +0000},
	date-modified = {2015-07-05 17:21:41 +0000},
	doi = {10.1016/S0370-1573(01)00098-9},
	eprint = {hep-ph/0005122},
	journal = {Phys.Rept.},
	pages = {223-386},
	primaryclass = {hep-ph},
	reportnumber = {MIT-CTP-2980, HD-THEP-00-26},
	slaccitation = {%%CITATION = HEP-PH/0005122;%%},
	title = {{Nonperturbative renormalization flow in quantum field theory and statistical physics}},
	volume = {363},
	year = {2002},
	Bdsk-Url-1 = {http://dx.doi.org/10.1016/S0370-1573(01)00098-9}}

@article{Berges:2000ew,
	archiveprefix = {arXiv},
	author = {Berges, Juergen and Tetradis, Nikolaos and Wetterich, Christof},
	date-added = {2016-05-10 15:53:43 +0000},
	date-modified = {2016-05-10 15:53:43 +0000},
	doi = {10.1016/S0370-1573(01)00098-9},
	eprint = {hep-ph/0005122},
	journal = {Phys. Rept.},
	pages = {223-386},
	primaryclass = {hep-ph},
	reportnumber = {MIT-CTP-2980, HD-THEP-00-26},
	slaccitation = {%%CITATION = HEP-PH/0005122;%%},
	title = {{Nonperturbative renormalization flow in quantum field theory and statistical physics}},
	volume = {363},
	year = {2002},
	Bdsk-Url-1 = {http://dx.doi.org/10.1016/S0370-1573(01)00098-9}}

@article{deBerredoPeixoto:2004if,
	archiveprefix = {arXiv},
	author = {de Berredo-Peixoto, Guilherme and Shapiro, Ilya L.},
	date-added = {2016-01-26 20:43:01 +0000},
	date-modified = {2016-01-26 20:43:01 +0000},
	doi = {10.1103/PhysRevD.71.064005},
	eprint = {hep-th/0412249},
	journal = {Phys. Rev.},
	pages = {064005},
	primaryclass = {hep-th},
	slaccitation = {%%CITATION = HEP-TH/0412249;%%},
	title = {{Higher derivative quantum gravity with Gauss-Bonnet term}},
	volume = {D71},
	year = {2005},
	Bdsk-Url-1 = {http://dx.doi.org/10.1103/PhysRevD.71.064005}}

@article{Bervillier:2013,
	archiveprefix = {arXiv},
	author = {Bervillier, C.},
	date-added = {2015-07-05 17:21:41 +0000},
	date-modified = {2015-07-05 17:21:41 +0000},
	eprint = {1307.3679},
	primaryclass = {hep-th},
	slaccitation = {%%CITATION = ARXIV:1307.3679;%%},
	title = {{Revisiting the local potential approximation of the exact renormalization group equation}},
	year = {2013}}

@article{Bervillier:2013kda,
	archiveprefix = {arXiv},
	author = {Bervillier, C.},
	date-added = {2015-04-25 14:55:51 +0000},
	date-modified = {2015-04-25 14:55:51 +0000},
	doi = {10.5488/CMP.16.23003},
	eprint = {1304.4131},
	journal = {Condens. Matter Phys.},
	pages = {23003},
	primaryclass = {hep-th},
	slaccitation = {%%CITATION = ARXIV:1304.4131;%%},
	title = {{The Wilson exact renormalization group equation and the anomalous dimension parameter}},
	volume = {16},
	year = {2013},
	Bdsk-Url-1 = {http://dx.doi.org/10.5488/CMP.16.23003}}

@article{Bervillier2004a,
	archiveprefix = {arXiv},
	author = {Bervillier, C.},
	doi = {10.1016/j.physleta.2004.09.037},
	eprint = {hep-th/0405025},
	journal = {Phys. Lett.},
	owner = {trmorris},
	pages = {93-100},
	primaryclass = {hep-th},
	reportnumber = {SACLAY-SPH-T04-058, T04-058},
	slaccitation = {%%CITATION = HEP-TH/0405025;%%},
	timestamp = {2016.05.16},
	title = {{The Wilson-Polchinski exact renormalization group equation}},
	volume = {A332},
	year = {2004},
	Bdsk-Url-1 = {http://dx.doi.org/10.1016/j.physleta.2004.09.037}}

@article{Bervillier:2007,
	archiveprefix = {arXiv},
	author = {Bervillier, Claude and Juttner, Andreas and Litim, Daniel F.},
	date-added = {2015-07-05 17:21:41 +0000},
	date-modified = {2015-07-05 17:21:41 +0000},
	doi = {10.1016/j.nuclphysb.2007.03.036},
	eprint = {hep-th/0701172},
	journal = {Nucl.Phys.},
	pages = {213-226},
	primaryclass = {hep-th},
	reportnumber = {CERN-PH-TH-2007-006, SHEP-0702},
	slaccitation = {%%CITATION = HEP-TH/0701172;%%},
	title = {{High-accuracy scaling exponents in the local potential approximation}},
	volume = {B783},
	year = {2007},
	Bdsk-Url-1 = {http://dx.doi.org/10.1016/j.nuclphysb.2007.03.036}}

@article{Bonanno:2000sy,
	archiveprefix = {arXiv},
	author = {Bonanno, Alfio},
	date-added = {2016-05-10 18:09:59 +0000},
	date-modified = {2016-05-10 18:09:59 +0000},
	doi = {10.1103/PhysRevD.62.027701},
	eprint = {hep-th/0001060},
	journal = {Phys.Rev.},
	pages = {027701},
	primaryclass = {hep-th},
	reportnumber = {INFN-CT-01-00},
	title = {Nonperturbative scaling in the scalar theory},
	volume = {D62},
	year = {2000},
	Bdsk-Url-1 = {http://dx.doi.org/10.1103/PhysRevD.62.027701}}

@article{Bonanno:2012dg,
	archiveprefix = {arXiv},
	author = {Bonanno, Alfio and Guarnieri, Filippo},
	date-added = {2015-07-05 17:21:41 +0000},
	date-modified = {2015-07-05 17:21:41 +0000},
	doi = {10.1103/PhysRevD.86.105027},
	eprint = {1206.6531},
	journal = {Phys.Rev.},
	pages = {105027},
	primaryclass = {hep-th},
	slaccitation = {%%CITATION = ARXIV:1206.6531;%%},
	title = {{Universality and Symmetry Breaking in Conformally Reduced Quantum Gravity}},
	volume = {D86},
	year = {2012},
	Bdsk-Url-1 = {http://dx.doi.org/10.1103/PhysRevD.86.105027}}

@article{Branchina:2000jp,
	archiveprefix = {arXiv},
	author = {Branchina, Vincenzo},
	date-added = {2016-05-10 18:09:59 +0000},
	date-modified = {2016-05-10 18:09:59 +0000},
	doi = {10.1103/PhysRevD.64.043513},
	eprint = {hep-ph/0002013},
	journal = {Phys.Rev.},
	pages = {043513},
	primaryclass = {hep-ph},
	title = {Nonperturbative renormalization group potentials and quintessence},
	volume = {D64},
	year = {2001},
	Bdsk-Url-1 = {http://dx.doi.org/10.1103/PhysRevD.64.043513}}

@article{Branchina:2003ek,
	archiveprefix = {arXiv},
	author = {Branchina, Vincenzo and Meissner, Krzysztof A. and Veneziano, Gabriele},
	date-added = {2015-07-05 17:21:41 +0000},
	date-modified = {2015-07-05 17:21:41 +0000},
	doi = {10.1016/j.physletb.2003.09.020},
	eprint = {hep-th/0309234},
	journal = {Phys.Lett.},
	pages = {319-324},
	primaryclass = {hep-th},
	slaccitation = {%%CITATION = HEP-TH/0309234;%%},
	title = {{The Price of an exact, gauge invariant RG flow equation}},
	volume = {B574},
	year = {2003},
	Bdsk-Url-1 = {http://dx.doi.org/10.1016/j.physletb.2003.09.020}}

@article{Bridle:2013sra,
	archiveprefix = {arXiv},
	author = {Bridle, I. Hamzaan and Dietz, Juergen A. and Morris, Tim R.},
	date-added = {2016-12-17 18:09:39 +0000},
	date-modified = {2016-12-17 18:09:39 +0000},
	doi = {10.1007/JHEP03(2014)093},
	eprint = {1312.2846},
	journal = {JHEP},
	pages = {093},
	primaryclass = {hep-th},
	slaccitation = {%%CITATION = ARXIV:1312.2846;%%},
	title = {{The local potential approximation in the background field formalism}},
	volume = {03},
	year = {2014},
	Bdsk-Url-1 = {http://dx.doi.org/10.1007/JHEP03(2014)093}}

@article{Buchert:1999er,
	archiveprefix = {arXiv},
	author = {Buchert, Thomas},
	date-added = {2016-02-06 09:04:25 +0000},
	date-modified = {2016-02-06 09:04:25 +0000},
	doi = {10.1023/A:1001800617177},
	eprint = {gr-qc/9906015},
	journal = {Gen. Rel. Grav.},
	pages = {105-125},
	primaryclass = {gr-qc},
	reportnumber = {CERN-TH-99-165},
	slaccitation = {%%CITATION = GR-QC/9906015;%%},
	title = {{On average properties of inhomogeneous fluids in general relativity. 1. Dust cosmologies}},
	volume = {32},
	year = {2000},
	Bdsk-Url-1 = {http://dx.doi.org/10.1023/A:1001800617177}}

@article{Buchert:2015iva,
	archiveprefix = {arXiv},
	author = {Buchert, T. and others},
	date-added = {2016-02-06 09:16:05 +0000},
	date-modified = {2016-02-06 09:16:05 +0000},
	doi = {10.1088/0264-9381/32/21/215021},
	eprint = {1505.07800},
	journal = {Class. Quant. Grav.},
	pages = {215021},
	primaryclass = {gr-qc},
	reportnumber = {HIP-2015-17-TH},
	slaccitation = {%%CITATION = ARXIV:1505.07800;%%},
	title = {{Is there proof that backreaction of inhomogeneities is irrelevant in cosmology?}},
	volume = {32},
	year = {2015},
	Bdsk-Url-1 = {http://dx.doi.org/10.1088/0264-9381/32/21/215021}}

@article{Calcagni:2012hb,
	archiveprefix = {arXiv},
	author = {Calcagni, Gianluca},
	booktitle = {{Proceedings, 6th International School on Field Theory and Gravitation (ISFTG 2012)}},
	date-added = {2015-11-30 08:34:38 +0000},
	date-modified = {2015-11-30 08:34:38 +0000},
	doi = {10.1063/1.4756961},
	eprint = {1209.1110},
	journal = {AIP Conf. Proc.},
	pages = {31-53},
	primaryclass = {hep-th},
	slaccitation = {%%CITATION = ARXIV:1209.1110;%%},
	title = {{Introduction to multifractional spacetimes}},
	volume = {1483},
	year = {2012},
	Bdsk-Url-1 = {http://dx.doi.org/10.1063/1.4756961}}

@article{Calcagni:2014wba,
	archiveprefix = {arXiv},
	author = {Calcagni, Gianluca and Modesto, Leonardo and Nardelli, Giuseppe},
	date-added = {2015-11-30 08:33:35 +0000},
	date-modified = {2015-11-30 08:33:35 +0000},
	eprint = {1408.0199},
	primaryclass = {hep-th},
	slaccitation = {%%CITATION = ARXIV:1408.0199;%%},
	title = {{Quantum spectral dimension in quantum field theory}},
	year = {2014}}

@article{Calcagni:2014cza,
	archiveprefix = {arXiv},
	author = {Calcagni, Gianluca and Oriti, Daniele and Th{\"u}rigen, Johannes},
	date-added = {2015-11-30 08:32:30 +0000},
	date-modified = {2015-11-30 08:32:30 +0000},
	doi = {10.1103/PhysRevD.91.084047},
	eprint = {1412.8390},
	journal = {Phys. Rev.},
	number = {8},
	pages = {084047},
	primaryclass = {hep-th},
	reportnumber = {AEI-2014-028},
	slaccitation = {%%CITATION = ARXIV:1412.8390;%%},
	title = {{Dimensional flow in discrete quantum geometries}},
	volume = {D91},
	year = {2015},
	Bdsk-Url-1 = {http://dx.doi.org/10.1103/PhysRevD.91.084047}}

@article{Camporesi:1994ga,
	author = {Camporesi, R. and Higuchi, A.},
	date-added = {2016-10-10 09:40:14 +0000},
	date-modified = {2016-10-10 09:40:14 +0000},
	doi = {10.1063/1.530850},
	journal = {J. Math. Phys.},
	pages = {4217-4246},
	slaccitation = {%%CITATION = JMAPA,35,4217;%%},
	title = {{Spectral functions and zeta functions in hyperbolic spaces}},
	volume = {35},
	year = {1994},
	Bdsk-Url-1 = {http://dx.doi.org/10.1063/1.530850}}

@article{Capper1974b,
	author = {Capper, D. M. and Duff, M. J.},
	doi = {10.1007/BF02748300},
	journal = {Nuovo Cim.},
	owner = {trmorris},
	pages = {173-183},
	slaccitation = {%%CITATION = NUCIA,A23,173;%%},
	timestamp = {2016.04.27},
	title = {{Trace anomalies in dimensional regularization}},
	volume = {A23},
	year = {1974},
	Bdsk-Url-1 = {http://dx.doi.org/10.1007/BF02748300}}

@article{Carlip:2009kf,
	archiveprefix = {arXiv},
	author = {Carlip, Steven},
	booktitle = {{Proceedings, 25th Max Born Symposium: The Planck Scale}},
	date-added = {2015-11-30 08:32:10 +0000},
	date-modified = {2015-11-30 08:32:10 +0000},
	doi = {10.1063/1.3284402},
	eprint = {0909.3329},
	journal = {AIP Conf. Proc.},
	pages = {72},
	primaryclass = {gr-qc},
	slaccitation = {%%CITATION = ARXIV:0909.3329;%%},
	title = {{Spontaneous Dimensional Reduction in Short-Distance Quantum Gravity?}},
	volume = {1196},
	year = {2009},
	Bdsk-Url-1 = {http://dx.doi.org/10.1063/1.3284402}}

@article{Christiansen2015,
	archiveprefix = {arXiv},
	author = {Christiansen, Nicolai and Knorr, Benjamin and Meibohm, Jan and Pawlowski, Jan M. and Reichert, Manuel},
	doi = {10.1103/PhysRevD.92.121501},
	eprint = {1506.07016},
	journal = {Phys. Rev.},
	number = {12},
	owner = {trmorris},
	pages = {121501},
	primaryclass = {hep-th},
	slaccitation = {%%CITATION = ARXIV:1506.07016;%%},
	timestamp = {2016.04.18},
	title = {{Local Quantum Gravity}},
	volume = {D92},
	year = {2015},
	Bdsk-Url-1 = {http://dx.doi.org/10.1103/PhysRevD.92.121501}}

@article{Codello:2013fpa,
	archiveprefix = {arXiv},
	author = {Codello, Alessandro and D'Odorico, Giulio and Pagani, Carlo},
	date-added = {2015-07-05 17:21:41 +0000},
	date-modified = {2015-07-05 17:21:41 +0000},
	doi = {10.1103/PhysRevD.89.081701},
	eprint = {1304.4777},
	journal = {Phys.Rev.},
	pages = {081701},
	primaryclass = {gr-qc},
	slaccitation = {%%CITATION = ARXIV:1304.4777;%%},
	title = {{Consistent closure of renormalization group flow equations in quantum gravity}},
	volume = {D89},
	year = {2014},
	Bdsk-Url-1 = {http://dx.doi.org/10.1103/PhysRevD.89.081701}}

@article{Codello2013,
	archiveprefix = {arXiv},
	author = {Codello, A. and D'Odorico, G. and Pagani, C. and Percacci, R.},
	doi = {10.1088/0264-9381/30/11/115015},
	eprint = {1210.3284},
	journal = {Class. Quant. Grav.},
	owner = {trmorris},
	pages = {115015},
	primaryclass = {hep-th},
	slaccitation = {%%CITATION = ARXIV:1210.3284;%%},
	timestamp = {2016.04.27},
	title = {{The Renormalization Group and Weyl-invariance}},
	volume = {30},
	year = {2013},
	Bdsk-Url-1 = {http://dx.doi.org/10.1088/0264-9381/30/11/115015}}

@article{Codello2006,
	archiveprefix = {arXiv},
	author = {Codello, Alessandro and Percacci, Roberto},
	doi = {10.1103/PhysRevLett.97.221301},
	eprint = {hep-th/0607128},
	journal = {Phys. Rev. Lett.},
	owner = {trmorris},
	pages = {221301},
	primaryclass = {hep-th},
	slaccitation = {%%CITATION = HEP-TH/0607128;%%},
	timestamp = {2016.01.26},
	title = {{Fixed points of higher derivative gravity}},
	volume = {97},
	year = {2006},
	Bdsk-Url-1 = {http://dx.doi.org/10.1103/PhysRevLett.97.221301}}

@article{Codello:2008,
	archiveprefix = {arXiv},
	author = {Codello, Alessandro and Percacci, Roberto and Rahmede, Christoph},
	date-added = {2015-07-05 17:21:41 +0000},
	date-modified = {2015-07-05 17:21:41 +0000},
	doi = {10.1016/j.aop.2008.08.008},
	eprint = {0805.2909},
	journal = {Annals Phys.},
	pages = {414-469},
	primaryclass = {hep-th},
	slaccitation = {%%CITATION = ARXIV:0805.2909;%%},
	title = {{Investigating the Ultraviolet Properties of Gravity with a Wilsonian Renormalization Group Equation}},
	volume = {324},
	year = {2009},
	Bdsk-Url-1 = {http://dx.doi.org/10.1016/j.aop.2008.08.008}}

@article{Codello:2007bd,
	archiveprefix = {arXiv},
	author = {Codello, Alessandro and Percacci, Roberto and Rahmede, Christoph},
	date-added = {2016-09-01 19:09:54 +0000},
	date-modified = {2016-09-01 19:09:54 +0000},
	doi = {10.1142/S0217751X08038135},
	eprint = {0705.1769},
	journal = {Int. J. Mod. Phys.},
	pages = {143-150},
	primaryclass = {hep-th},
	slaccitation = {%%CITATION = ARXIV:0705.1769;%%},
	title = {{Ultraviolet properties of f(R)-gravity}},
	volume = {A23},
	year = {2008},
	Bdsk-Url-1 = {http://dx.doi.org/10.1142/S0217751X08038135}}

@article{Comellas:1997tf,
	archiveprefix = {arXiv},
	author = {Comellas, Jordi and Travesset, Alex},
	date-added = {2016-05-16 09:42:53 +0000},
	date-modified = {2016-05-16 09:42:53 +0000},
	doi = {10.1016/S0550-3213(97)00349-0},
	eprint = {hep-th/9701028},
	journal = {Nucl. Phys.},
	pages = {539-564},
	primaryclass = {hep-th},
	reportnumber = {UB-ECM-PF-96-21},
	slaccitation = {%%CITATION = HEP-TH/9701028;%%},
	title = {{O (N) models within the local potential approximation}},
	volume = {B498},
	year = {1997},
	Bdsk-Url-1 = {http://dx.doi.org/10.1016/S0550-3213(97)00349-0}}

@inproceedings{Coumbe:2015bka,
	archiveprefix = {arXiv},
	author = {Coumbe, Daniel},
	booktitle = {{14th Marcel Grossmann Meeting on Recent Developments in Theoretical and Experimental General Relativity, Astrophysics, and Relativistic Field Theories (MG14) Rome, Italy, July 12-18, 2015}},
	date-added = {2015-11-30 08:22:47 +0000},
	date-modified = {2015-11-30 08:22:47 +0000},
	eprint = {1509.07665},
	primaryclass = {hep-th},
	slaccitation = {%%CITATION = ARXIV:1509.07665;%%},
	title = {{What is dimensional reduction really telling us?}},
	url = {http://inspirehep.net/record/1394849/files/arXiv:1509.07665.pdf},
	year = {2015},
	Bdsk-Url-1 = {http://inspirehep.net/record/1394849/files/arXiv:1509.07665.pdf}}

@article{Coumbe:2014noa,
	archiveprefix = {arXiv},
	author = {Coumbe, D. N. and Jurkiewicz, J.},
	date-added = {2015-11-30 08:32:51 +0000},
	date-modified = {2015-11-30 08:32:51 +0000},
	doi = {10.1007/JHEP03(2015)151},
	eprint = {1411.7712},
	journal = {JHEP},
	pages = {151},
	primaryclass = {hep-th},
	slaccitation = {%%CITATION = ARXIV:1411.7712;%%},
	title = {{Evidence for Asymptotic Safety from Dimensional Reduction in Causal Dynamical Triangulations}},
	volume = {03},
	year = {2015},
	Bdsk-Url-1 = {http://dx.doi.org/10.1007/JHEP03(2015)151}}

@article{DAttanasio:1997he,
	archiveprefix = {arXiv},
	author = {D'Attanasio, Marco and Morris, Tim R.},
	doi = {10.1016/S0370-2693(97)00866-6},
	eprint = {hep-th/9704094},
	journal = {Phys. Lett.},
	pages = {363-370},
	primaryclass = {hep-th},
	reportnumber = {SHEP-97-03},
	slaccitation = {%%CITATION = HEP-TH/9704094;%%},
	title = {{Large N and the renormalization group}},
	volume = {B409},
	year = {1997},
	Bdsk-Url-1 = {http://dx.doi.org/10.1016/S0370-2693(97)00866-6}}

@article{DAttanasio:1996jd,
	archiveprefix = {arXiv},
	author = {D'Attanasio, Marco and Morris, Tim R.},
	doi = {10.1016/0370-2693(96)00411-X},
	eprint = {hep-th/9602156},
	journal = {Phys. Lett.},
	pages = {213-221},
	primaryclass = {hep-th},
	reportnumber = {SHEP-96-08, UPRF-96-458},
	slaccitation = {%%CITATION = HEP-TH/9602156;%%},
	title = {{Gauge invariance, the quantum action principle, and the renormalization group}},
	volume = {B378},
	year = {1996},
	Bdsk-Url-1 = {http://dx.doi.org/10.1016/0370-2693(96)00411-X}}

@article{Delamotte:2015aaa,
	archiveprefix = {arXiv},
	author = {Delamotte, Bertrand and Tissier, Matthieu and Wschebor, Nicol{\'a}s},
	date-added = {2016-05-04 17:06:53 +0000},
	date-modified = {2016-05-04 17:06:53 +0000},
	doi = {10.1103/PhysRevE.93.012144},
	eprint = {1501.01776},
	journal = {Phys. Rev.},
	number = {1},
	pages = {012144},
	primaryclass = {cond-mat.stat-mech},
	slaccitation = {%%CITATION = ARXIV:1501.01776;%%},
	title = {{Scale invariance implies conformal invariance for the three-dimensional Ising model}},
	volume = {E93},
	year = {2016},
	Bdsk-Url-1 = {http://dx.doi.org/10.1103/PhysRevE.93.012144}}

@article{Demmel:2015zfa,
	archiveprefix = {arXiv},
	author = {Demmel, Maximilian and Nink, Andreas},
	date-added = {2016-03-12 11:18:44 +0000},
	date-modified = {2016-03-12 11:18:44 +0000},
	doi = {10.1103/PhysRevD.92.104013},
	eprint = {1506.03809},
	journal = {Phys. Rev.},
	number = {10},
	pages = {104013},
	primaryclass = {gr-qc},
	reportnumber = {MITP-15-041},
	slaccitation = {%%CITATION = ARXIV:1506.03809;%%},
	title = {{Connections and geodesics in the space of metrics}},
	volume = {D92},
	year = {2015},
	Bdsk-Url-1 = {http://dx.doi.org/10.1103/PhysRevD.92.104013}}

@inproceedings{Demmel2015a,
	archiveprefix = {arXiv},
	author = {Demmel, Maximilian and Saueressig, Frank and Zanusso, Omar},
	booktitle = {{Proceedings, 13th Marcel Grossmann Meeting on Recent Developments in Theoretical and Experimental General Relativity, Astrophysics, and Relativistic Field Theories (MG13)}},
	doi = {10.1142/9789814623995_0404},
	eprint = {1302.1312},
	owner = {trmorris},
	pages = {2227-2229},
	primaryclass = {hep-th},
	reportnumber = {MITP-13-013},
	slaccitation = {%%CITATION = ARXIV:1302.1312;%%},
	timestamp = {2016.04.21},
	title = {{Fixed Functionals in Asymptotically Safe Gravity}},
	url = {http://inspirehep.net/record/1217855/files/arXiv:1302.1312.pdf},
	year = {2015},
	Bdsk-Url-1 = {http://inspirehep.net/record/1217855/files/arXiv:1302.1312.pdf},
	Bdsk-Url-2 = {http://dx.doi.org/10.1142/9789814623995_0404}}

@article{Demmel2015b,
	archiveprefix = {arXiv},
	author = {Demmel, Maximilian and Saueressig, Frank and Zanusso, Omar},
	doi = {10.1007/JHEP08(2015)113},
	eprint = {1504.07656},
	journal = {JHEP},
	owner = {trmorris},
	pages = {113},
	primaryclass = {hep-th},
	slaccitation = {%%CITATION = ARXIV:1504.07656;%%},
	timestamp = {2016.04.01},
	title = {{A proper fixed functional for four-dimensional Quantum Einstein Gravity}},
	volume = {08},
	year = {2015},
	Bdsk-Url-1 = {http://dx.doi.org/10.1007/JHEP08(2015)113}}

@article{Demmel:2014fk,
	archiveprefix = {arXiv},
	author = {Demmel, Maximilian and Saueressig, Frank and Zanusso, Omar},
	date-added = {2016-04-26 15:32:28 +0000},
	date-modified = {2016-04-26 15:34:22 +0000},
	doi = {10.1007/JHEP06(2014)026},
	eprint = {1401.5495},
	journal = {JHEP},
	pages = {026},
	primaryclass = {hep-th},
	slaccitation = {%%CITATION = ARXIV:1401.5495;%%},
	title = {{RG flows of Quantum Einstein Gravity on maximally symmetric spaces}},
	volume = {06},
	year = {2014},
	Bdsk-Url-1 = {http://dx.doi.org/10.1007/JHEP06(2014)026}}

@article{Demmel:2012ub,
	archiveprefix = {arXiv},
	author = {Demmel, Maximilian and Saueressig, Frank and Zanusso, Omar},
	date-added = {2016-04-26 15:08:58 +0000},
	date-modified = {2016-04-26 15:08:58 +0000},
	doi = {10.1007/JHEP11(2012)131},
	eprint = {1208.2038},
	journal = {JHEP},
	pages = {131},
	primaryclass = {hep-th},
	reportnumber = {MZ-TH-12-34},
	slaccitation = {%%CITATION = ARXIV:1208.2038;%%},
	title = {{Fixed-Functionals of three-dimensional Quantum Einstein Gravity}},
	volume = {11},
	year = {2012},
	Bdsk-Url-1 = {http://dx.doi.org/10.1007/JHEP11(2012)131}}

@book{DeWitt:1992cy,
	address = {Cambridge, UK},
	author = {DeWitt, Bryce S.},
	date-added = {2016-02-17 22:00:34 +0000},
	date-modified = {2016-02-17 22:00:34 +0000},
	isbn = {9781139240512, 9780521423779},
	publisher = {Cambridge Univ. Press},
	series = {Cambridge monographs on mathematical physics},
	slaccitation = {%%CITATION = INSPIRE-340701;%%},
	title = {{Supermanifolds}},
	url = {http://www.cambridge.org/mw/academic/subjects/physics/theoretical-physics-and-mathematical-physics/supermanifolds-2nd-edition?format=AR},
	year = {2012},
	Bdsk-Url-1 = {http://www.cambridge.org/mw/academic/subjects/physics/theoretical-physics-and-mathematical-physics/supermanifolds-2nd-edition?format=AR}}

@article{DeWitt:1967yk,
	author = {DeWitt, Bryce S.},
	date-added = {2016-01-05 09:37:39 +0000},
	date-modified = {2016-01-05 09:37:39 +0000},
	doi = {10.1103/PhysRev.160.1113},
	journal = {Phys. Rev.},
	pages = {1113-1148},
	slaccitation = {%%CITATION = PHRVA,160,1113;%%},
	title = {{Quantum Theory of Gravity. 1. The Canonical Theory}},
	volume = {160},
	year = {1967},
	Bdsk-Url-1 = {http://dx.doi.org/10.1103/PhysRev.160.1113}}

@article{Dietz:2015owa,
	archiveprefix = {arXiv},
	author = {Dietz, Juergen A. and Morris, Tim R.},
	date-added = {2016-12-17 18:10:09 +0000},
	date-modified = {2016-12-17 18:10:09 +0000},
	doi = {10.1007/JHEP04(2015)118},
	eprint = {1502.07396},
	journal = {JHEP},
	pages = {118},
	primaryclass = {hep-th},
	slaccitation = {%%CITATION = ARXIV:1502.07396;%%},
	title = {{Background independent exact renormalization group for conformally reduced gravity}},
	volume = {04},
	year = {2015},
	Bdsk-Url-1 = {http://dx.doi.org/10.1007/JHEP04(2015)118}}

@article{Dietz:2012ic,
	archiveprefix = {arXiv},
	author = {Dietz, Juergen A. and Morris, Tim R.},
	date-added = {2016-12-17 18:10:38 +0000},
	date-modified = {2016-12-17 18:10:38 +0000},
	doi = {10.1007/JHEP01(2013)108},
	eprint = {1211.0955},
	journal = {JHEP},
	pages = {108},
	primaryclass = {hep-th},
	reportnumber = {SHEP-12-26},
	slaccitation = {%%CITATION = ARXIV:1211.0955;%%},
	title = {{Asymptotic safety in the f(R) approximation}},
	volume = {01},
	year = {2013},
	Bdsk-Url-1 = {http://dx.doi.org/10.1007/JHEP01(2013)108}}

@article{Dietz:2013sba,
	archiveprefix = {arXiv},
	author = {Dietz, Juergen A. and Morris, Tim R.},
	date-added = {2016-12-17 18:10:24 +0000},
	date-modified = {2016-12-17 18:10:24 +0000},
	doi = {10.1007/JHEP07(2013)064},
	eprint = {1306.1223},
	journal = {JHEP},
	pages = {064},
	primaryclass = {hep-th},
	slaccitation = {%%CITATION = ARXIV:1306.1223;%%},
	title = {{Redundant operators in the exact renormalisation group and in the f(R) approximation to asymptotic safety}},
	volume = {07},
	year = {2013},
	Bdsk-Url-1 = {http://dx.doi.org/10.1007/JHEP07(2013)064}}

@article{Dietz:2016gzg,
	archiveprefix = {arXiv},
	author = {Dietz, Juergen A. and Morris, Tim R. and Slade, Zoe H.},
	date-added = {2016-12-17 18:08:14 +0000},
	date-modified = {2016-12-17 18:08:14 +0000},
	doi = {10.1103/PhysRevD.94.124014},
	eprint = {1605.07636},
	journal = {Phys. Rev.},
	number = {12},
	pages = {124014},
	primaryclass = {hep-th},
	slaccitation = {%%CITATION = ARXIV:1605.07636;%%},
	title = {{Fixed point structure of the conformal factor field in quantum gravity}},
	volume = {D94},
	year = {2016},
	Bdsk-Url-1 = {http://dx.doi.org/10.1103/PhysRevD.94.124014}}

@book{Dirac-primary,
	author = {Dirac, Paul A. M.},
	owner = {trmorris},
	publisher = {Belfer Graduate School of Science, New York, reprinted by Dover},
	series = {Belfer Graduate School of Science Monographs Series 2},
	timestamp = {2016.02.23},
	title = {Lectures on quantum mechanics},
	year = {2001}}

@article{Dirac1950,
	author = {Dirac, P. A. M.},
	doi = {10.4153/cjm-1950-012-1},
	issn = {0008-414X},
	journal = {Canad. J. Math.},
	month = {Jan},
	number = {0},
	owner = {trmorris},
	pages = {129--148},
	publisher = {Canadian Mathematical Society},
	timestamp = {2016.02.23},
	title = {Generalized Hamiltonian dynamics},
	url = {http://dx.doi.org/10.4153/CJM-1950-012-1},
	volume = {2},
	year = {1950},
	Bdsk-Url-1 = {http://dx.doi.org/10.4153/CJM-1950-012-1},
	Bdsk-Url-2 = {http://dx.doi.org/10.4153/cjm-1950-012-1}}

@article{Distler:1988jt,
	author = {Distler, Jacques and Kawai, Hikaru},
	date-added = {2015-07-05 17:21:41 +0000},
	date-modified = {2015-07-05 17:21:41 +0000},
	doi = {10.1016/0550-3213(89)90354-4},
	journal = {Nucl.Phys.},
	pages = {509},
	reportnumber = {CLNS-88-853},
	slaccitation = {%%CITATION = NUPHA,B321,509;%%},
	title = {{Conformal Field Theory and 2D Quantum Gravity Or Who's Afraid of Joseph Liouville?}},
	volume = {B321},
	year = {1989},
	Bdsk-Url-1 = {http://dx.doi.org/10.1016/0550-3213(89)90354-4}}

@article{Dona:2015tnf,
	archiveprefix = {arXiv},
	author = {Don{\`a}, Pietro and Eichhorn, Astrid and Labus, Peter and Percacci, Roberto},
	date-added = {2016-03-13 13:52:24 +0000},
	date-modified = {2016-03-13 13:52:24 +0000},
	doi = {10.1103/PhysRevD.93.044049},
	eprint = {1512.01589},
	journal = {Phys. Rev.},
	number = {4},
	pages = {044049},
	primaryclass = {gr-qc},
	slaccitation = {%%CITATION = ARXIV:1512.01589;%%},
	title = {{Asymptotic safety in an interacting system of gravity and scalar matter}},
	volume = {D93},
	year = {2016},
	Bdsk-Url-1 = {http://dx.doi.org/10.1103/PhysRevD.93.044049}}

@article{Dona:2014pla,
	archiveprefix = {arXiv},
	author = {Don{\`a}, P. and Eichhorn, Astrid and Percacci, Roberto},
	date-added = {2015-07-05 17:21:41 +0000},
	date-modified = {2015-07-05 17:21:41 +0000},
	eprint = {1410.4411},
	primaryclass = {gr-qc},
	slaccitation = {%%CITATION = ARXIV:1410.4411;%%},
	title = {{Consistency of matter models with asymptotically safe quantum gravity}},
	year = {2014}}

@article{Dona:2013qba,
	archiveprefix = {arXiv},
	author = {Don\`a, Pietro and Eichhorn, Astrid and Percacci, Roberto},
	doi = {10.1103/PhysRevD.89.084035},
	eprint = {1311.2898},
	journal = {Phys.Rev.},
	number = {8},
	pages = {084035},
	primaryclass = {hep-th},
	slaccitation = {%%CITATION = ARXIV:1311.2898;%%},
	title = {{Matter matters in asymptotically safe quantum gravity}},
	volume = {D89},
	year = {2014},
	Bdsk-Url-1 = {http://dx.doi.org/10.1103/PhysRevD.89.084035}}

@article{Donkin:2012ud,
	archiveprefix = {arXiv},
	author = {Donkin, Ivan and Pawlowski, Jan M.},
	date-added = {2015-07-05 17:21:41 +0000},
	date-modified = {2015-07-05 17:21:41 +0000},
	eprint = {1203.4207},
	primaryclass = {hep-th},
	slaccitation = {%%CITATION = ARXIV:1203.4207;%%},
	title = {{The phase diagram of quantum gravity from diffeomorphism-invariant RG-flows}},
	year = {2012}}

@article{Donoghue:1994dn,
	archiveprefix = {arXiv},
	author = {Donoghue, John F.},
	date-added = {2015-12-01 03:35:53 +0000},
	date-modified = {2015-12-01 03:35:53 +0000},
	doi = {10.1103/PhysRevD.50.3874},
	eprint = {gr-qc/9405057},
	journal = {Phys. Rev.},
	pages = {3874-3888},
	primaryclass = {gr-qc},
	reportnumber = {UMHEP-408},
	slaccitation = {%%CITATION = GR-QC/9405057;%%},
	title = {{General relativity as an effective field theory: The leading quantum corrections}},
	volume = {D50},
	year = {1994},
	Bdsk-Url-1 = {http://dx.doi.org/10.1103/PhysRevD.50.3874}}

@article{Dou:1997fg,
	archiveprefix = {arXiv},
	author = {Dou, Djamel and Percacci, Roberto},
	date-added = {2016-09-01 19:09:25 +0000},
	date-modified = {2016-09-01 19:09:25 +0000},
	doi = {10.1088/0264-9381/15/11/011},
	eprint = {hep-th/9707239},
	journal = {Class. Quant. Grav.},
	pages = {3449-3468},
	primaryclass = {hep-th},
	slaccitation = {%%CITATION = HEP-TH/9707239;%%},
	title = {{The running gravitational couplings}},
	volume = {15},
	year = {1998},
	Bdsk-Url-1 = {http://dx.doi.org/10.1088/0264-9381/15/11/011}}

@article{Duff1994,
	archiveprefix = {arXiv},
	author = {Duff, M. J.},
	booktitle = {{Conference on Highlights of Particle and Condensed Matter Physics (SALAMFEST) Trieste, Italy, March 8-12, 1993}},
	doi = {10.1088/0264-9381/11/6/004},
	eprint = {hep-th/9308075},
	journal = {Class. Quant. Grav.},
	owner = {trmorris},
	pages = {1387-1404},
	primaryclass = {hep-th},
	reportnumber = {CTP-TAMU-06-93, C93-03-08},
	slaccitation = {%%CITATION = HEP-TH/9308075;%%},
	timestamp = {2016.04.27},
	title = {{Twenty years of the Weyl anomaly}},
	volume = {11},
	year = {1994},
	Bdsk-Url-1 = {http://dx.doi.org/10.1088/0264-9381/11/6/004}}

@article{Duff1977,
	author = {Duff, M. J.},
	doi = {10.1016/0550-3213(77)90410-2},
	journal = {Nucl. Phys.},
	owner = {trmorris},
	pages = {334-348},
	reportnumber = {Print-77-0361 (QUEEN MARY COLL.)},
	slaccitation = {%%CITATION = NUPHA,B125,334;%%},
	timestamp = {2016.04.27},
	title = {{Observations on Conformal Anomalies}},
	volume = {B125},
	year = {1977},
	Bdsk-Url-1 = {http://dx.doi.org/10.1016/0550-3213(77)90410-2}}

@article{Eichhorn:2010tb,
	archiveprefix = {arXiv},
	author = {Eichhorn, Astrid and Gies, Holger},
	date-added = {2015-07-05 17:21:41 +0000},
	date-modified = {2015-07-05 17:21:41 +0000},
	doi = {10.1103/PhysRevD.81.104010},
	eprint = {1001.5033},
	journal = {Phys.Rev.},
	pages = {104010},
	primaryclass = {hep-th},
	slaccitation = {%%CITATION = ARXIV:1001.5033;%%},
	title = {{Ghost anomalous dimension in asymptotically safe quantum gravity}},
	volume = {D81},
	year = {2010},
	Bdsk-Url-1 = {http://dx.doi.org/10.1103/PhysRevD.81.104010}}

@article{Einstein1919,
	author = {Einstein, Albert},
	journal = {Sitzungsber. Preuss. Akad. Wiss. Berlin (Math. Phys.) English translation in {``}The principle of relativity{``}, by A. Einstein et al (Dover)},
	owner = {trmorris},
	pages = {433},
	slaccitation = {%%CITATION = SPWPA,1919,433;%%},
	timestamp = {2016.01.05},
	title = {{Do gravitational fields play an essential part in the structure of the elementary particles of matter?}},
	volume = {1919},
	year = {1919}}

@article{Ellwanger1994a,
	archiveprefix = {arXiv},
	author = {Ellwanger, Ulrich},
	booktitle = {{Proceedings, Workshop on Quantum field theoretical aspects of high energy physics}},
	doi = {10.1007/BF01555911},
	eprint = {hep-ph/9308260},
	journal = {Z. Phys.},
	note = {[,206(1993)]},
	owner = {trmorris},
	pages = {503-510},
	primaryclass = {hep-ph},
	reportnumber = {HD-THEP-93-30},
	slaccitation = {%%CITATION = HEP-PH/9308260;%%},
	timestamp = {2016.04.29},
	title = {{Flow equations for N point functions and bound states}},
	volume = {C62},
	year = {1994},
	Bdsk-Url-1 = {http://dx.doi.org/10.1007/BF01555911}}

@article{Ettle:2007qc,
	archiveprefix = {arXiv},
	author = {Ettle, James H. and Fu, Chih-Hao and Fudger, Jonathan P. and Mansfield, Paul R. W. and Morris, Tim R.},
	date-added = {2016-12-17 18:10:55 +0000},
	date-modified = {2016-12-17 18:10:55 +0000},
	doi = {10.1088/1126-6708/2007/05/011},
	eprint = {hep-th/0703286},
	journal = {JHEP},
	pages = {011},
	primaryclass = {hep-th},
	reportnumber = {SHEP-07-09},
	slaccitation = {%%CITATION = HEP-TH/0703286;%%},
	title = {{S-matrix equivalence theorem evasion and dimensional regularisation with the canonical MHV Lagrangian}},
	volume = {05},
	year = {2007},
	Bdsk-Url-1 = {http://dx.doi.org/10.1088/1126-6708/2007/05/011}}

@article{Ettle:2006bw,
	archiveprefix = {arXiv},
	author = {Ettle, James H. and Morris, Tim R.},
	date-added = {2016-12-17 18:10:47 +0000},
	date-modified = {2016-12-17 18:10:47 +0000},
	doi = {10.1088/1126-6708/2006/08/003},
	eprint = {hep-th/0605121},
	journal = {JHEP},
	pages = {003},
	primaryclass = {hep-th},
	reportnumber = {SHEP-06-17},
	slaccitation = {%%CITATION = HEP-TH/0605121;%%},
	title = {{Structure of the MHV-rules Lagrangian}},
	volume = {08},
	year = {2006},
	Bdsk-Url-1 = {http://dx.doi.org/10.1088/1126-6708/2006/08/003}}

@article{Ettle:2008ey,
	archiveprefix = {arXiv},
	author = {Ettle, James H. and Morris, Tim R. and Xiao, Zhiguang},
	date-added = {2016-12-17 18:10:58 +0000},
	date-modified = {2016-12-17 18:10:58 +0000},
	doi = {10.1088/1126-6708/2008/08/103},
	eprint = {0805.0239},
	journal = {JHEP},
	pages = {103},
	primaryclass = {hep-th},
	reportnumber = {SHEP-08-20},
	slaccitation = {%%CITATION = ARXIV:0805.0239;%%},
	title = {{The MHV QCD Lagrangian}},
	volume = {08},
	year = {2008},
	Bdsk-Url-1 = {http://dx.doi.org/10.1088/1126-6708/2008/08/103}}

@article{Evans:2006eq,
	archiveprefix = {arXiv},
	author = {Evans, Nick and Morris, Tim R. and Rosten, Oliver J.},
	date-added = {2016-12-17 18:10:51 +0000},
	date-modified = {2016-12-17 18:10:51 +0000},
	doi = {10.1016/j.physletb.2006.02.055},
	eprint = {hep-th/0601114},
	journal = {Phys. Lett.},
	pages = {148-150},
	primaryclass = {hep-th},
	reportnumber = {SHEP-06-04},
	slaccitation = {%%CITATION = HEP-TH/0601114;%%},
	title = {{Gauge invariant regularization in the AdS/CFT correspondence and ghost D-branes}},
	volume = {B635},
	year = {2006},
	Bdsk-Url-1 = {http://dx.doi.org/10.1016/j.physletb.2006.02.055}}

@article{Evans:2015zwa,
	archiveprefix = {arXiv},
	author = {Evans, Nick and Morris, Tim R. and Scott, Marc},
	date-added = {2016-02-06 09:06:00 +0000},
	date-modified = {2016-02-06 09:06:00 +0000},
	doi = {10.1103/PhysRevD.93.025019},
	eprint = {1507.02965},
	journal = {Phys. Rev.},
	number = {2},
	pages = {025019},
	primaryclass = {hep-ph},
	slaccitation = {%%CITATION = ARXIV:1507.02965;%%},
	title = {{Translational symmetry breaking in field theories and the cosmological constant}},
	volume = {D93},
	year = {2016},
	Bdsk-Url-1 = {http://dx.doi.org/10.1103/PhysRevD.93.025019}}

@article{Faddeev:1967fc,
	author = {Faddeev, L. D. and Popov, V. N.},
	date-added = {2015-11-30 07:56:21 +0000},
	date-modified = {2015-11-30 07:56:21 +0000},
	doi = {10.1016/0370-2693(67)90067-6},
	journal = {Phys. Lett.},
	pages = {29-30},
	slaccitation = {%%CITATION = PHLTA,B25,29;%%},
	title = {{Feynman Diagrams for the Yang-Mills Field}},
	volume = {B25},
	year = {1967},
	Bdsk-Url-1 = {http://dx.doi.org/10.1016/0370-2693(67)90067-6}}

@article{Falls:2015cta,
	archiveprefix = {arXiv},
	author = {Falls, Kevin},
	date-added = {2015-04-25 14:28:54 +0000},
	date-modified = {2015-04-25 14:28:54 +0000},
	eprint = {1503.06233},
	primaryclass = {hep-th},
	slaccitation = {%%CITATION = ARXIV:1503.06233;%%},
	title = {{Critical scaling in quantum gravity from the renormalisation group}},
	year = {2015}}

@article{Falls:2015qga,
	archiveprefix = {arXiv},
	author = {Falls, Kevin},
	date-added = {2015-07-05 17:21:41 +0000},
	date-modified = {2015-07-05 17:21:41 +0000},
	eprint = {1501.05331},
	primaryclass = {hep-th},
	slaccitation = {%%CITATION = ARXIV:1501.05331;%%},
	title = {{On the renormalisation of Newton's constant}},
	year = {2015}}

@article{Falls:2014tra,
	archiveprefix = {arXiv},
	author = {Falls, Kevin and Litim, Daniel F. and Nikolakopoulos, Konstantinos and Rahmede, Christoph},
	date-added = {2016-05-23 09:14:30 +0000},
	date-modified = {2016-05-23 09:14:30 +0000},
	doi = {10.1103/PhysRevD.93.104022},
	eprint = {1410.4815},
	journal = {Phys. Rev.},
	number = {10},
	pages = {104022},
	primaryclass = {hep-th},
	reportnumber = {DO-TH-14-26, KA-TP-2014-30},
	slaccitation = {%%CITATION = ARXIV:1410.4815;%%},
	title = {{Further evidence for asymptotic safety of quantum gravity}},
	volume = {D93},
	year = {2016},
	Bdsk-Url-1 = {http://dx.doi.org/10.1103/PhysRevD.93.104022}}

@article{Falls:2013bv,
	archiveprefix = {arXiv},
	author = {Falls, K. and Litim, D. F. and Nikolakopoulos, K. and Rahmede, C.},
	date-added = {2016-10-10 10:15:41 +0000},
	date-modified = {2016-10-10 10:15:41 +0000},
	eprint = {1301.4191},
	primaryclass = {hep-th},
	reportnumber = {DO-TH-13-02, KA-TP-01-2013},
	slaccitation = {%%CITATION = ARXIV:1301.4191;%%},
	title = {{A bootstrap towards asymptotic safety}},
	year = {2013}}

@article{Fu:2009nh,
	archiveprefix = {arXiv},
	author = {Fu, Chih-Hao and Fudger, Jonathan and Mansfield, Paul R. W. and Morris, Tim R. and Xiao, Zhiguang},
	date-added = {2016-12-17 18:11:02 +0000},
	date-modified = {2016-12-17 18:11:02 +0000},
	doi = {10.1088/1126-6708/2009/06/035},
	eprint = {0902.1906},
	journal = {JHEP},
	pages = {035},
	primaryclass = {hep-th},
	reportnumber = {SHEP-09-06},
	slaccitation = {%%CITATION = ARXIV:0902.1906;%%},
	title = {{S-matrix equivalence restored}},
	volume = {06},
	year = {2009},
	Bdsk-Url-1 = {http://dx.doi.org/10.1088/1126-6708/2009/06/035}}

@phdthesis{Gatti:2002kc,
	archiveprefix = {arXiv},
	author = {Gatti, Antonio},
	date-added = {2015-11-30 07:01:13 +0000},
	date-modified = {2015-11-30 07:01:13 +0000},
	eprint = {hep-th/0301201},
	primaryclass = {hep-th},
	school = {Southampton U.},
	slaccitation = {%%CITATION = HEP-TH/0301201;%%},
	title = {{A Gauge invariant flow equation}},
	url = {http://alice.cern.ch/format/showfull?sysnb=2361695},
	year = {2002},
	Bdsk-Url-1 = {http://alice.cern.ch/format/showfull?sysnb=2361695}}

@phdthesis{Giasemidis:2013axa,
	archiveprefix = {arXiv},
	author = {Giasemidis, Georgios},
	date-added = {2015-11-30 08:34:12 +0000},
	date-modified = {2015-11-30 08:34:12 +0000},
	eprint = {1310.8109},
	primaryclass = {hep-th},
	school = {Oxford U.},
	slaccitation = {%%CITATION = ARXIV:1310.8109;%%},
	title = {{Spectral dimension in graph models of causal quantum gravity}},
	url = {http://inspirehep.net/record/1262690/files/arXiv:1310.8109.pdf},
	year = {2013},
	Bdsk-Url-1 = {http://inspirehep.net/record/1262690/files/arXiv:1310.8109.pdf}}

@article{Gibbons:1978ac,
	author = {Gibbons, G.W. and Hawking, S.W. and Perry, M.J.},
	date-added = {2015-07-05 17:21:41 +0000},
	date-modified = {2015-07-05 17:21:41 +0000},
	doi = {10.1016/0550-3213(78)90161-X},
	journal = {Nucl.Phys.},
	pages = {141},
	reportnumber = {PRINT-78-0375 (CAMBRIDGE)},
	slaccitation = {%%CITATION = NUPHA,B138,141;%%},
	title = {{Path Integrals and the Indefiniteness of the Gravitational Action}},
	volume = {B138},
	year = {1978},
	Bdsk-Url-1 = {http://dx.doi.org/10.1016/0550-3213(78)90161-X}}

@article{Gies:2006wv,
	archiveprefix = {arXiv},
	author = {Gies, Holger},
	date-added = {2015-07-05 17:21:41 +0000},
	date-modified = {2015-07-05 17:21:41 +0000},
	doi = {10.1007/978-3-642-27320-9_6},
	eprint = {hep-ph/0611146},
	journal = {Lect.Notes Phys.},
	pages = {287-348},
	primaryclass = {hep-ph},
	slaccitation = {%%CITATION = HEP-PH/0611146;%%},
	title = {{Introduction to the functional RG and applications to gauge theories}},
	volume = {852},
	year = {2012},
	Bdsk-Url-1 = {http://dx.doi.org/10.1007/978-3-642-27320-9_6}}

@article{Gies:2002af,
	archiveprefix = {arXiv},
	author = {Gies, Holger},
	date-added = {2015-07-05 17:21:41 +0000},
	date-modified = {2015-07-05 17:21:41 +0000},
	doi = {10.1103/PhysRevD.66.025006},
	eprint = {hep-th/0202207},
	journal = {Phys.Rev.},
	pages = {025006},
	primaryclass = {hep-th},
	reportnumber = {CERN-TH-2002-047},
	slaccitation = {%%CITATION = HEP-TH/0202207;%%},
	title = {{Running coupling in Yang-Mills theory: A flow equation study}},
	volume = {D66},
	year = {2002},
	Bdsk-Url-1 = {http://dx.doi.org/10.1103/PhysRevD.66.025006}}

@article{Gies:2000xr,
	archiveprefix = {arXiv},
	author = {Gies, Holger},
	date-added = {2016-05-10 18:09:59 +0000},
	date-modified = {2016-05-10 18:09:59 +0000},
	doi = {10.1103/PhysRevD.63.065011},
	eprint = {hep-th/0009041},
	journal = {Phys.Rev.},
	pages = {065011},
	primaryclass = {hep-th},
	title = {Flow equation for Halpern-Huang directions of scalar O(N) models},
	volume = {D63},
	year = {2001},
	Bdsk-Url-1 = {http://dx.doi.org/10.1103/PhysRevD.63.065011}}

@article{Gies:2015tca,
	archiveprefix = {arXiv},
	author = {Gies, Holger and Knorr, Benjamin and Lippoldt, Stefan},
	date-added = {2016-03-12 12:04:45 +0000},
	date-modified = {2016-03-12 12:04:45 +0000},
	doi = {10.1103/PhysRevD.92.084020},
	eprint = {1507.08859},
	journal = {Phys. Rev.},
	number = {8},
	pages = {084020},
	primaryclass = {hep-th},
	slaccitation = {%%CITATION = ARXIV:1507.08859;%%},
	title = {{Generalized Parametrization Dependence in Quantum Gravity}},
	volume = {D92},
	year = {2015},
	Bdsk-Url-1 = {http://dx.doi.org/10.1103/PhysRevD.92.084020}}

@article{Gies:2015cka,
	archiveprefix = {arXiv},
	author = {Gies, Holger and Lippoldt, Stefan},
	date-added = {2015-07-05 17:21:41 +0000},
	date-modified = {2015-07-05 17:21:41 +0000},
	eprint = {1502.00918},
	primaryclass = {hep-th},
	slaccitation = {%%CITATION = ARXIV:1502.00918;%%},
	title = {{Global surpluses of spin-base invariant fermions}},
	year = {2015}}

@article{Gies:2013noa,
	archiveprefix = {arXiv},
	author = {Gies, Holger and Lippoldt, Stefan},
	date-added = {2015-07-05 17:21:41 +0000},
	date-modified = {2015-07-05 17:21:41 +0000},
	doi = {10.1103/PhysRevD.89.064040},
	eprint = {1310.2509},
	journal = {Phys.Rev.},
	number = {6},
	pages = {064040},
	primaryclass = {hep-th},
	slaccitation = {%%CITATION = ARXIV:1310.2509;%%},
	title = {{Fermions in gravity with local spin-base invariance}},
	volume = {D89},
	year = {2014},
	Bdsk-Url-1 = {http://dx.doi.org/10.1103/PhysRevD.89.064040}}

@article{Gies:2009hq,
	archiveprefix = {arXiv},
	author = {Gies, Holger and Scherer, Michael M.},
	date-added = {2016-05-10 18:09:59 +0000},
	date-modified = {2016-05-10 18:09:59 +0000},
	doi = {10.1140/epjc/s10052-010-1256-z},
	eprint = {0901.2459},
	journal = {Eur.Phys.J.},
	pages = {387-402},
	primaryclass = {hep-th},
	title = {Asymptotic safety of simple Yukawa systems},
	volume = {C66},
	year = {2010},
	Bdsk-Url-1 = {http://dx.doi.org/10.1140/epjc/s10052-010-1256-z}}

@article{Goroff:1985th,
	author = {Goroff, Marc H. and Sagnotti, Augusto},
	date-added = {2016-12-17 17:34:36 +0000},
	date-modified = {2016-12-17 17:34:36 +0000},
	doi = {10.1016/0550-3213(86)90193-8},
	journal = {Nucl. Phys.},
	pages = {709-736},
	reportnumber = {CALT-68-1289, LBL-19995, UCB-PTH-85-34},
	slaccitation = {%%CITATION = NUPHA,B266,709;%%},
	title = {{The Ultraviolet Behavior of Einstein Gravity}},
	volume = {B266},
	year = {1986},
	Bdsk-Url-1 = {http://dx.doi.org/10.1016/0550-3213(86)90193-8}}

@article{Goroff:1985sz,
	author = {Goroff, Marc H. and Sagnotti, Augusto},
	date-added = {2016-12-17 17:37:59 +0000},
	date-modified = {2016-12-17 17:37:59 +0000},
	doi = {10.1016/0370-2693(85)91470-4},
	journal = {Phys. Lett.},
	pages = {81-86},
	reportnumber = {CALT-68-1263, UCB-PTH-85/18, LBL-19512},
	slaccitation = {%%CITATION = PHLTA,B160,81;%%},
	title = {{Quantum Gravity at Two Loops}},
	volume = {B160},
	year = {1985},
	Bdsk-Url-1 = {http://dx.doi.org/10.1016/0370-2693(85)91470-4}}

@article{Green:2015bma,
	archiveprefix = {arXiv},
	author = {Green, Stephen R. and Wald, Robert M.},
	date-added = {2016-02-06 09:14:19 +0000},
	date-modified = {2016-02-06 09:14:19 +0000},
	eprint = {1506.06452},
	primaryclass = {gr-qc},
	slaccitation = {%%CITATION = ARXIV:1506.06452;%%},
	title = {{Comments on Backreaction}},
	year = {2015}}

@article{Groh:2010ta,
	archiveprefix = {arXiv},
	author = {Groh, Kai and Saueressig, Frank},
	date-added = {2015-07-05 17:21:41 +0000},
	date-modified = {2015-07-05 17:21:41 +0000},
	doi = {10.1088/1751-8113/43/36/365403},
	eprint = {1001.5032},
	journal = {J.Phys.},
	pages = {365403},
	primaryclass = {hep-th},
	slaccitation = {%%CITATION = ARXIV:1001.5032;%%},
	title = {{Ghost wave-function renormalization in Asymptotically Safe Quantum Gravity}},
	volume = {A43},
	year = {2010},
	Bdsk-Url-1 = {http://dx.doi.org/10.1088/1751-8113/43/36/365403}}

@article{Halpern:1997gn,
	archiveprefix = {arXiv},
	author = {Halpern, Kenneth},
	date-added = {2016-05-10 18:09:59 +0000},
	date-modified = {2016-05-10 18:09:59 +0000},
	doi = {10.1103/PhysRevD.57.6337},
	eprint = {hep-th/9708124},
	journal = {Phys.Rev.},
	pages = {6337-6341},
	primaryclass = {hep-th},
	title = {Cross-section and effective potential in asymptotically free scalar field theories},
	volume = {D57},
	year = {1998},
	Bdsk-Url-1 = {http://dx.doi.org/10.1103/PhysRevD.57.6337}}

@article{HH2ndpaper,
	author = {Halpern, Kenneth and Huang, Kerson},
	date-added = {2016-05-10 18:09:59 +0000},
	date-modified = {2016-05-10 18:09:59 +0000},
	doi = {10.1103/PhysRevD.53.3252},
	issue = {6},
	journal = {Phys. Rev. D},
	month = {Mar},
	numpages = {0},
	pages = {3252--3259},
	publisher = {American Physical Society},
	title = {Nontrivial directions for scalar fields},
	url = {http://link.aps.org/doi/10.1103/PhysRevD.53.3252},
	volume = {53},
	year = {1996},
	Bdsk-Url-1 = {http://link.aps.org/doi/10.1103/PhysRevD.53.3252},
	Bdsk-Url-2 = {http://dx.doi.org/10.1103/PhysRevD.53.3252}}

@article{HHReply,
	author = {Halpern, Kenneth and Huang, Kerson},
	date-added = {2016-05-10 18:09:59 +0000},
	date-modified = {2016-05-10 18:09:59 +0000},
	doi = {10.1103/PhysRevLett.77.1659},
	issue = {8},
	journal = {Phys. Rev. Lett.},
	month = {Aug},
	numpages = {0},
	pages = {1659--1659},
	publisher = {American Physical Society},
	title = {Halpern and Huang Reply:},
	url = {http://link.aps.org/doi/10.1103/PhysRevLett.77.1659},
	volume = {77},
	year = {1996},
	Bdsk-Url-1 = {http://link.aps.org/doi/10.1103/PhysRevLett.77.1659},
	Bdsk-Url-2 = {http://dx.doi.org/10.1103/PhysRevLett.77.1659}}

@article{HHOrig,
	author = {Halpern, Kenneth and Huang, Kerson},
	date-added = {2016-05-10 18:09:59 +0000},
	date-modified = {2016-05-10 18:09:59 +0000},
	doi = {10.1103/PhysRevLett.74.3526},
	issue = {18},
	journal = {Phys. Rev. Lett.},
	month = {May},
	numpages = {0},
	pages = {3526--3529},
	publisher = {American Physical Society},
	title = {Fixed-Point Structure of Scalar Fields},
	url = {http://link.aps.org/doi/10.1103/PhysRevLett.74.3526},
	volume = {74},
	year = {1995},
	Bdsk-Url-1 = {http://link.aps.org/doi/10.1103/PhysRevLett.74.3526},
	Bdsk-Url-2 = {http://dx.doi.org/10.1103/PhysRevLett.74.3526}}

@article{Hamber:2013rb,
	archiveprefix = {arXiv},
	author = {Hamber, Herbert W. and Toriumi, Reiko},
	date-added = {2015-07-05 17:21:41 +0000},
	date-modified = {2015-07-05 17:21:41 +0000},
	doi = {10.1142/S0218271813300231},
	eprint = {1301.6259},
	journal = {Int.J.Mod.Phys.},
	number = {13},
	pages = {1330023},
	primaryclass = {hep-th},
	slaccitation = {%%CITATION = ARXIV:1301.6259;%%},
	title = {{Inconsistencies from a Running Cosmological Constant}},
	volume = {D22},
	year = {2013},
	Bdsk-Url-1 = {http://dx.doi.org/10.1142/S0218271813300231}}

@article{Bridle:2016nsu,
	archiveprefix = {arXiv},
	author = {Hamzaan Bridle, I. and Morris, Tim R.},
	date-added = {2016-12-17 18:09:42 +0000},
	date-modified = {2016-12-17 18:09:42 +0000},
	doi = {10.1103/PhysRevD.94.065040},
	eprint = {1605.06075},
	journal = {Phys. Rev.},
	pages = {065040},
	primaryclass = {hep-th},
	slaccitation = {%%CITATION = ARXIV:1605.06075;%%},
	title = {{Fate of nonpolynomial interactions in scalar field theory}},
	volume = {D94},
	year = {2016},
	Bdsk-Url-1 = {http://dx.doi.org/10.1103/PhysRevD.94.065040}}

@article{Hasenfratz:1987,
	author = {Hasenfratz, Anna and Hasenfratz, Peter},
	date-added = {2015-07-05 17:21:41 +0000},
	date-modified = {2015-07-05 17:21:41 +0000},
	doi = {10.1016/0550-3213(88)90224-6},
	journal = {Nucl.Phys.},
	pages = {1},
	reportnumber = {BUTP-87/10},
	slaccitation = {%%CITATION = NUPHA,B295,1;%%},
	title = {{Singular Renormalization Group Transformations And First Order Phase Transformations}},
	volume = {B295},
	year = {1988},
	Bdsk-Url-1 = {http://dx.doi.org/10.1016/0550-3213(88)90224-6}}

@article{Hasenfratz:1985dm,
	author = {Hasenfratz, Anna and Hasenfratz, Peter},
	date-added = {2015-07-05 17:21:41 +0000},
	date-modified = {2015-07-05 17:21:41 +0000},
	doi = {10.1016/0550-3213(86)90573-0},
	journal = {Nucl.Phys.},
	pages = {687-701},
	reportnumber = {BUTP-85/26-BERN},
	slaccitation = {%%CITATION = NUPHA,B270,687;%%},
	title = {{Renormalization Group Study of Scalar Field Theories}},
	volume = {B270},
	year = {1986},
	Bdsk-Url-1 = {http://dx.doi.org/10.1016/0550-3213(86)90573-0}}

@article{Hasenfratz:1993sp,
	archiveprefix = {arXiv},
	author = {Hasenfratz, P. and Niedermayer, F.},
	date-added = {2015-06-01 10:59:57 +0000},
	date-modified = {2015-06-01 10:59:57 +0000},
	doi = {10.1016/0550-3213(94)90261-5},
	eprint = {hep-lat/9308004},
	journal = {Nucl.Phys.},
	pages = {785-814},
	primaryclass = {hep-lat},
	reportnumber = {BUTP-93-17},
	slaccitation = {%%CITATION = HEP-LAT/9308004;%%},
	title = {{Perfect lattice action for asymptotically free theories}},
	volume = {B414},
	year = {1994},
	Bdsk-Url-1 = {http://dx.doi.org/10.1016/0550-3213(94)90261-5}}

@inproceedings{Hooft:1993gx,
	archiveprefix = {arXiv},
	author = {'t Hooft, Gerard},
	booktitle = {{Salamfest 1993:0284-296}},
	date-added = {2015-11-30 08:21:22 +0000},
	date-modified = {2015-11-30 08:21:22 +0000},
	eprint = {gr-qc/9310026},
	pages = {0284-296},
	primaryclass = {gr-qc},
	reportnumber = {THU-93-26},
	slaccitation = {%%CITATION = GR-QC/9310026;%%},
	title = {{Dimensional reduction in quantum gravity}},
	year = {1993}}

@article{Huang:2013zaa,
	archiveprefix = {arXiv},
	author = {Huang, Kerson},
	date-added = {2016-05-10 18:09:59 +0000},
	date-modified = {2016-05-10 18:09:59 +0000},
	doi = {10.1142/S021775X13300500},
	eprint = {1310.5533},
	journal = {Int. J. Mod. Phys. A,},
	pages = {1330050},
	primaryclass = {physics.hist-ph},
	title = {A Critical History of Renormalization},
	volume = {28},
	year = {2013},
	Bdsk-Url-1 = {http://dx.doi.org/10.1142/S021775X13300500}}

@article{Huang:2011xg,
	archiveprefix = {arXiv},
	author = {Huang, Kerson and Low, Hwee-Boon and Tung, Roh-Suan},
	date-added = {2016-05-10 18:09:59 +0000},
	date-modified = {2016-05-10 18:09:59 +0000},
	doi = {10.1088/0264-9381/29/15/155014},
	eprint = {1106.5282},
	journal = {Class.Quant.Grav.},
	pages = {155014},
	primaryclass = {gr-qc},
	title = {Scalar Field Cosmology I: Asymptotic Freedom and the Initial-Value Problem},
	volume = {29},
	year = {2012},
	Bdsk-Url-1 = {http://dx.doi.org/10.1088/0264-9381/29/15/155014}}

@article{Huang:2011xha,
	archiveprefix = {arXiv},
	author = {Huang, Kerson and Low, Hwee-Boon and Tung, Roh-Suan},
	date-added = {2016-05-10 18:09:59 +0000},
	date-modified = {2016-05-10 18:09:59 +0000},
	doi = {10.1142/S0217751X12501540},
	eprint = {1106.5283},
	journal = {Int.J.Mod.Phys.},
	pages = {1250154},
	primaryclass = {gr-qc},
	related = {Huang:2011xg},
	title = {Scalar Field Cosmology II: Superfluidity and Quantum Turbulence},
	volume = {A27},
	year = {2012},
	Bdsk-Url-1 = {http://dx.doi.org/10.1142/S0217751X12501540}}

@article{Huang:2010qn,
	archiveprefix = {arXiv},
	author = {Huang, Kerson and Low, Hwee-Boon and Tung, Roh-Suan},
	date-added = {2016-05-10 18:09:59 +0000},
	date-modified = {2016-05-10 18:09:59 +0000},
	eprint = {1011.4012},
	primaryclass = {gr-qc},
	reportnumber = {IASA001},
	title = {Cosmology of an asymptotically free scalar field with spontaneous symmetry breaking},
	year = {2010}}

@book{Ince1956,
	author = {Ince, E.L.},
	owner = {trmorris},
	timestamp = {2016.03.21},
	title = {Ordinary differential equations},
	year = {1956}}

@article{NCS,
	author = {J.F. Nicoll, T.S. Chang and H.E. Stanley},
	date-added = {2015-07-05 17:21:41 +0000},
	date-modified = {2015-07-05 17:21:41 +0000},
	journal = {Phys. Lett.},
	pages = {7},
	title = {A Differential Generator for the Free Energy and the Magnetization Equation of A Differential Generator for the Free Energy and the Magnetization Equation of State},
	volume = {57A},
	year = {1976}}

@article{Kadanoff:1966wm,
	author = {Kadanoff, L. P.},
	date-added = {2015-11-30 07:03:10 +0000},
	date-modified = {2015-11-30 07:03:10 +0000},
	journal = {Physics},
	pages = {263-272},
	slaccitation = {%%CITATION = PYCSA,2,263;%%},
	title = {{Scaling laws for Ising models near T(c)}},
	volume = {2},
	year = {1966}}

@article{Kopietz:2010zz,
	author = {Kopietz, Peter and Bartosch, Lorenz and Schutz, Florian},
	date-added = {2016-05-10 15:56:53 +0000},
	date-modified = {2016-05-10 15:56:53 +0000},
	doi = {10.1007/978-3-642-05094-7},
	journal = {Lect. Notes Phys.},
	pages = {1-380},
	slaccitation = {%%CITATION = LNPHA,798,1;%%},
	title = {{Introduction to the functional renormalization group}},
	volume = {798},
	year = {2010},
	Bdsk-Url-1 = {http://dx.doi.org/10.1007/978-3-642-05094-7}}

@article{Labus:2016lkh,
	archiveprefix = {arXiv},
	author = {Labus, Peter and Morris, Tim R. and Slade, Zo{\"e} H.},
	date-added = {2016-12-17 18:09:51 +0000},
	date-modified = {2016-12-17 18:09:51 +0000},
	doi = {10.1103/PhysRevD.94.024007},
	eprint = {1603.04772},
	journal = {Phys. Rev.},
	number = {2},
	pages = {024007},
	primaryclass = {hep-th},
	slaccitation = {%%CITATION = ARXIV:1603.04772;%%},
	title = {{Background independence in a background dependent renormalization group}},
	volume = {D94},
	year = {2016},
	Bdsk-Url-1 = {http://dx.doi.org/10.1103/PhysRevD.94.024007}}

@article{Labus:2015ska,
	archiveprefix = {arXiv},
	author = {Labus, Peter and Percacci, Roberto and Vacca, Gian Paolo},
	date-added = {2016-03-12 11:53:18 +0000},
	date-modified = {2016-03-12 11:53:18 +0000},
	doi = {10.1016/j.physletb.2015.12.022},
	eprint = {1505.05393},
	journal = {Phys. Lett.},
	pages = {274-281},
	primaryclass = {hep-th},
	slaccitation = {%%CITATION = ARXIV:1505.05393;%%},
	title = {{Asymptotic safety in $O(N)$ scalar models coupled to gravity}},
	volume = {B753},
	year = {2016},
	Bdsk-Url-1 = {http://dx.doi.org/10.1016/j.physletb.2015.12.022}}

@article{Latorre:2000jp,
	archiveprefix = {arXiv},
	author = {Latorre, Jose I. and Morris, Tim R.},
	booktitle = {{The exact renormalization group. Proceedings, 2nd Conference, Rome, Italy, September 18-22, 2000}},
	date-added = {2015-11-30 06:56:28 +0000},
	date-modified = {2015-11-30 06:56:28 +0000},
	doi = {10.1142/S0217751X01004724},
	eprint = {hep-th/0102037},
	journal = {Int. J. Mod. Phys.},
	pages = {2071-2074},
	primaryclass = {hep-th},
	slaccitation = {%%CITATION = HEP-TH/0102037;%%},
	title = {{Scheme independence as an inherent redundancy in quantum field theory}},
	volume = {A16},
	year = {2001},
	Bdsk-Url-1 = {http://dx.doi.org/10.1142/S0217751X01004724}}

@article{Latorre:2000qc,
	archiveprefix = {arXiv},
	author = {Latorre, Jose I. and Morris, Tim R.},
	date-added = {2015-11-30 06:56:39 +0000},
	date-modified = {2015-11-30 06:56:39 +0000},
	doi = {10.1088/1126-6708/2000/11/004},
	eprint = {hep-th/0008123},
	journal = {JHEP},
	pages = {004},
	primaryclass = {hep-th},
	reportnumber = {SHEP-00-09},
	slaccitation = {%%CITATION = HEP-TH/0008123;%%},
	title = {{Exact scheme independence}},
	volume = {11},
	year = {2000},
	Bdsk-Url-1 = {http://dx.doi.org/10.1088/1126-6708/2000/11/004}}

@article{Lauscher:2005qz,
	archiveprefix = {arXiv},
	author = {Lauscher, O. and Reuter, M.},
	date-added = {2015-11-30 08:24:53 +0000},
	date-modified = {2015-11-30 08:24:53 +0000},
	doi = {10.1088/1126-6708/2005/10/050},
	eprint = {hep-th/0508202},
	journal = {JHEP},
	pages = {050},
	primaryclass = {hep-th},
	reportnumber = {MZ-TH-05-09},
	slaccitation = {%%CITATION = HEP-TH/0508202;%%},
	title = {{Fractal spacetime structure in asymptotically safe gravity}},
	volume = {10},
	year = {2005},
	Bdsk-Url-1 = {http://dx.doi.org/10.1088/1126-6708/2005/10/050}}

@article{Lauscher:2001ya,
	archiveprefix = {arXiv},
	author = {Lauscher, O. and Reuter, M.},
	date-added = {2015-07-05 17:21:41 +0000},
	date-modified = {2015-07-05 17:21:41 +0000},
	doi = {10.1103/PhysRevD.65.025013},
	eprint = {hep-th/0108040},
	journal = {Phys.Rev.},
	pages = {025013},
	primaryclass = {hep-th},
	reportnumber = {MZ-TH-01-21},
	slaccitation = {%%CITATION = HEP-TH/0108040;%%},
	title = {{Ultraviolet fixed point and generalized flow equation of quantum gravity}},
	volume = {D65},
	year = {2002},
	Bdsk-Url-1 = {http://dx.doi.org/10.1103/PhysRevD.65.025013}}

@article{Lauscher:2002sq,
	archiveprefix = {arXiv},
	author = {Lauscher, O. and Reuter, M.},
	date-added = {2015-07-05 17:21:41 +0000},
	date-modified = {2015-07-05 17:21:41 +0000},
	doi = {10.1103/PhysRevD.66.025026},
	eprint = {hep-th/0205062},
	journal = {Phys.Rev.},
	pages = {025026},
	primaryclass = {hep-th},
	reportnumber = {MZ-TH-02-07},
	slaccitation = {%%CITATION = HEP-TH/0205062;%%},
	title = {{Flow equation of quantum Einstein gravity in a higher derivative truncation}},
	volume = {D66},
	year = {2002},
	Bdsk-Url-1 = {http://dx.doi.org/10.1103/PhysRevD.66.025026}}

@book{LebedevSpecialFunctions,
	author = {Lebedev, N. N.},
	date-added = {2016-05-10 18:09:59 +0000},
	date-modified = {2016-05-10 18:09:59 +0000},
	editor = {Silverman, Richard A.},
	isbn = {0486606244},
	note = {Originally published by Prentice-Hall in 1965},
	pages = {260-271},
	publisher = {Dover Publications, New York},
	title = {Special Functions and Their Applications},
	year = {1972}}

@article{Lee:1972fj,
	author = {Lee, B. W. and Zinn-Justin, Jean},
	date-added = {2016-02-17 16:54:40 +0000},
	date-modified = {2016-02-17 16:54:40 +0000},
	doi = {10.1103/PhysRevD.5.3121},
	journal = {Phys. Rev.},
	pages = {3121-3137},
	reportnumber = {FERMILAB-PUB-72-031-T, NAL-THY-31},
	slaccitation = {%%CITATION = PHRVA,D5,3121;%%},
	title = {{Spontaneously Broken Gauge Symmetries. 1. Preliminaries}},
	volume = {D5},
	year = {1972},
	Bdsk-Url-1 = {http://dx.doi.org/10.1103/PhysRevD.5.3121}}

@article{Lippoldt:2015cea,
	archiveprefix = {arXiv},
	author = {Lippoldt, Stefan},
	date-added = {2015-07-05 17:21:41 +0000},
	date-modified = {2015-07-05 17:21:41 +0000},
	eprint = {1502.05607},
	primaryclass = {hep-th},
	slaccitation = {%%CITATION = ARXIV:1502.05607;%%},
	title = {{Spin-base invariance of Fermions in arbitrary dimensions}},
	year = {2015}}

@other{DanielTalk,
	author = {Litim, Daniel F.},
	date-added = {2016-05-23 09:33:49 +0000},
	date-modified = {2016-05-23 19:06:31 +0000},
	title = {\textrm{talk at Southampton and available also as {DELTA} 13 conference report online at http://www.thphys.uni-heidelberg.de/$\sim$smp/{D}elta/Delta13/talks/Delta13\_{L}itim.pdf}},
	url = {http://www.thphys.uni-heidelberg.de/$\sim$smp/Delta/Delta13/talks/Delta13\_Litim.pdf},
	Bdsk-Url-1 = {http://www.thphys.uni-heidelberg.de/$%5Csim$smp/Delta/Delta13/talks/Delta13%5C_Litim.pdf}}

@article{opt3,
	archiveprefix = {arXiv},
	author = {Litim, Daniel F.},
	date-added = {2015-07-05 17:21:41 +0000},
	date-modified = {2015-07-05 17:21:41 +0000},
	doi = {10.1142/S0217751X01004748},
	eprint = {hep-th/0104221},
	journal = {Int.J.Mod.Phys.},
	pages = {2081-2088},
	primaryclass = {hep-th},
	reportnumber = {CERN-TH-2001-013},
	slaccitation = {%%CITATION = HEP-TH/0104221;%%},
	title = {{Mind the Gap}},
	volume = {A16},
	year = {{(2001)}},
	Bdsk-Url-1 = {http://dx.doi.org/10.1142/S0217751X01004748}}

@article{opt1,
	archiveprefix = {arXiv},
	author = {Litim, Daniel F.},
	date-added = {2015-07-05 17:21:41 +0000},
	date-modified = {2015-07-05 17:21:41 +0000},
	doi = {10.1016/S0370-2693(00)00748-6},
	eprint = {hep-th/0005245},
	journal = {Phys.Lett.},
	pages = {92-99},
	primaryclass = {hep-th},
	reportnumber = {HD-THEP-00-28},
	slaccitation = {%%CITATION = HEP-TH/0005245;%%},
	title = {{Optimization of the Exact Renormalization Group}},
	volume = {B486},
	year = {{(2000)}},
	Bdsk-Url-1 = {http://dx.doi.org/10.1016/S0370-2693(00)00748-6}}

@article{Litim:2011cp,
	archiveprefix = {arXiv},
	author = {Litim, Daniel F.},
	date-added = {2015-07-05 17:21:41 +0000},
	date-modified = {2015-07-05 17:21:41 +0000},
	eprint = {1102.4624},
	journal = {Phil.Trans.Roy.Soc.Lond.},
	pages = {2759-2778},
	primaryclass = {hep-th},
	reportnumber = {INT-PUB-11-004},
	slaccitation = {%%CITATION = ARXIV:1102.4624;%%},
	title = {{Renormalisation group and the Planck scale}},
	volume = {A369},
	year = {2011}}

@article{Litim2007,
	archiveprefix = {arXiv},
	author = {Litim, Daniel F.},
	eprint = {0810.3675},
	journal = {PoS},
	owner = {trmorris},
	pages = {024},
	primaryclass = {hep-th},
	slaccitation = {%%CITATION = ARXIV:0810.3675;%%},
	timestamp = {2015.07.14},
	title = {{Fixed Points of Quantum Gravity and the Renormalisation Group}},
	volume = {QG-Ph},
	year = {2007}}

@article{Litim2005,
	archiveprefix = {arXiv},
	author = {Litim, Daniel F.},
	doi = {10.1088/1126-6708/2005/07/005},
	eprint = {hep-th/0503096},
	journal = {JHEP},
	owner = {trmorris},
	pages = {005},
	primaryclass = {hep-th},
	reportnumber = {SHEP-04-31, CERN-PH-TH-2005-042},
	slaccitation = {%%CITATION = HEP-TH/0503096;%%},
	timestamp = {2016.05.16},
	title = {{Universality and the renormalisation group}},
	volume = {07},
	year = {2005},
	Bdsk-Url-1 = {http://dx.doi.org/10.1088/1126-6708/2005/07/005}}

@article{Litim:2002cf,
	archiveprefix = {arXiv},
	author = {Litim, Daniel F.},
	date-added = {2015-07-05 17:21:41 +0000},
	date-modified = {2015-07-05 17:21:41 +0000},
	doi = {10.1016/S0550-3213(02)00186-4},
	eprint = {hep-th/0203006},
	journal = {Nucl.Phys.},
	pages = {128-158},
	primaryclass = {hep-th},
	reportnumber = {CERN-TH-2002-023},
	slaccitation = {%%CITATION = HEP-TH/0203006;%%},
	title = {{Critical exponents from optimized renormalization group flows}},
	volume = {B631},
	year = {2002},
	Bdsk-Url-1 = {http://dx.doi.org/10.1016/S0550-3213(02)00186-4}}

@article{Litim:2001,
	archiveprefix = {arXiv},
	author = {Litim, Daniel F.},
	date-added = {2015-07-05 17:21:41 +0000},
	date-modified = {2015-07-05 17:21:41 +0000},
	doi = {10.1103/PhysRevD.64.105007},
	eprint = {hep-th/0103195},
	journal = {Phys.Rev.},
	pages = {105007},
	primaryclass = {hep-th},
	reportnumber = {CERN-TH-2001-084},
	slaccitation = {%%CITATION = HEP-TH/0103195;%%},
	title = {{Optimized renormalization group flows}},
	volume = {D64},
	year = {2001},
	Bdsk-Url-1 = {http://dx.doi.org/10.1103/PhysRevD.64.105007}}

@article{Litim:2002ce,
	archiveprefix = {arXiv},
	author = {Litim, Daniel F. and Pawlowski, Jan M.},
	date-added = {2015-07-05 17:21:41 +0000},
	date-modified = {2015-07-05 17:21:41 +0000},
	eprint = {hep-th/0203005},
	journal = {JHEP},
	pages = {049},
	primaryclass = {hep-th},
	reportnumber = {CERN-TH-2002-049, FAU-TP3-02-06},
	slaccitation = {%%CITATION = HEP-TH/0203005;%%},
	title = {{Renormalization group flows for gauge theories in axial gauges}},
	volume = {0209},
	year = {2002}}

@article{Litim:2002hj,
	archiveprefix = {arXiv},
	author = {Litim, Daniel F. and Pawlowski, Jan M.},
	date-added = {2015-07-05 17:21:41 +0000},
	date-modified = {2015-07-05 17:21:41 +0000},
	doi = {10.1016/S0370-2693(02)02693-X},
	eprint = {hep-th/0208216},
	journal = {Phys.Lett.},
	pages = {279-286},
	primaryclass = {hep-th},
	reportnumber = {CERN-TH-2002-213, FAU-TP3-02-24},
	slaccitation = {%%CITATION = HEP-TH/0208216;%%},
	title = {{Wilsonian flows and background fields}},
	volume = {B546},
	year = {2002},
	Bdsk-Url-1 = {http://dx.doi.org/10.1016/S0370-2693(02)02693-X}}

@article{Litim1999,
	archiveprefix = {arXiv},
	author = {Litim, Daniel F. and Pawlowski, Jan M.},
	booktitle = {{QCD'98. Proceedings, 3rd Euroconference, 6th Conference, Montpellier, France, July 2-8, 1998}},
	doi = {10.1016/S0920-5632(99)00187-5},
	eprint = {hep-th/9809020},
	journal = {Nucl. Phys. Proc. Suppl.},
	owner = {trmorris},
	pages = {325-328},
	primaryclass = {hep-th},
	reportnumber = {DIAS-STP-98-08, ECM-UB-PF-98-20},
	slaccitation = {%%CITATION = HEP-TH/9809020;%%},
	timestamp = {2016.02.23},
	title = {{On gauge invariance and Ward identities for the Wilsonian renormalization group}},
	volume = {74},
	year = {1999},
	Bdsk-Url-1 = {http://dx.doi.org/10.1016/S0920-5632(99)00187-5}}

@article{Litim:1998nf,
	archiveprefix = {arXiv},
	author = {Litim, Daniel F. and Pawlowski, Jan M.},
	date-added = {2015-07-05 17:21:41 +0000},
	date-modified = {2015-07-05 17:21:41 +0000},
	eprint = {hep-th/9901063},
	pages = {168-185},
	primaryclass = {hep-th},
	reportnumber = {DIAS-STP-99-01, ECM-UB-PF-99-02},
	slaccitation = {%%CITATION = HEP-TH/9901063;%%},
	title = {{On gauge invariant Wilsonian flows}},
	year = {1998}}

@article{Litim1998,
	archiveprefix = {arXiv},
	author = {Litim, Daniel F. and Pawlowski, Jan M.},
	doi = {10.1016/S0370-2693(98)00761-8},
	eprint = {hep-th/9802064},
	journal = {Phys. Lett.},
	owner = {trmorris},
	pages = {181-188},
	primaryclass = {hep-th},
	reportnumber = {DIAS-STP-98-03, ECM-UB-PF-98-03, FSUJ-TPI-01-98, IMPERIAL-TP-97-98-22},
	slaccitation = {%%CITATION = HEP-TH/9802064;%%},
	timestamp = {2016.02.23},
	title = {{Flow equations for Yang-Mills theories in general axial gauges}},
	volume = {B435},
	year = {1998},
	Bdsk-Url-1 = {http://dx.doi.org/10.1016/S0370-2693(98)00761-8}}

@article{Lizana:2015hqb,
	archiveprefix = {arXiv},
	author = {Lizana, J. M. and Morris, T. R. and Perez-Victoria, M.},
	date-added = {2016-12-17 18:10:01 +0000},
	date-modified = {2016-12-17 18:10:01 +0000},
	doi = {10.1007/JHEP03(2016)198},
	eprint = {1511.04432},
	journal = {JHEP},
	pages = {198},
	primaryclass = {hep-th},
	slaccitation = {%%CITATION = ARXIV:1511.04432;%%},
	title = {{Holographic renormalisation group flows and renormalisation from a Wilsonian perspective}},
	volume = {03},
	year = {2016},
	Bdsk-Url-1 = {http://dx.doi.org/10.1007/JHEP03(2016)198}}

@article{Machado:2009ph,
	archiveprefix = {arXiv},
	author = {Machado, Pedro F. and Percacci, R.},
	date-added = {2015-07-05 17:21:41 +0000},
	date-modified = {2015-07-05 17:21:41 +0000},
	doi = {10.1103/PhysRevD.80.024020},
	eprint = {0904.2510},
	journal = {Phys.Rev.},
	pages = {024020},
	primaryclass = {hep-th},
	reportnumber = {ITP-UU-09-16, SPIN-09-16},
	slaccitation = {%%CITATION = ARXIV:0904.2510;%%},
	title = {{Conformally reduced quantum gravity revisited}},
	volume = {D80},
	year = {2009},
	Bdsk-Url-1 = {http://dx.doi.org/10.1103/PhysRevD.80.024020}}

@article{Machado:2007,
	archiveprefix = {arXiv},
	author = {Machado, Pedro F. and Saueressig, Frank},
	date-added = {2015-07-05 17:21:41 +0000},
	date-modified = {2015-07-05 17:21:41 +0000},
	doi = {10.1103/PhysRevD.77.124045},
	eprint = {0712.0445},
	journal = {Phys.Rev.},
	pages = {124045},
	primaryclass = {hep-th},
	reportnumber = {ITP-UU-07-63, SPIN-07-48, SPHT-T07-154},
	slaccitation = {%%CITATION = ARXIV:0712.0445;%%},
	title = {{On the renormalization group flow of f(R)-gravity}},
	volume = {D77},
	year = {2008},
	Bdsk-Url-1 = {http://dx.doi.org/10.1103/PhysRevD.77.124045}}

@article{Manrique:2008zw,
	archiveprefix = {arXiv},
	author = {Manrique, Elisa and Reuter, Martin},
	date-added = {2015-07-05 17:21:41 +0000},
	date-modified = {2015-07-05 17:21:41 +0000},
	doi = {10.1103/PhysRevD.79.025008},
	eprint = {0811.3888},
	journal = {Phys.Rev.},
	pages = {025008},
	primaryclass = {hep-th},
	reportnumber = {MZ-TH-08-32},
	slaccitation = {%%CITATION = ARXIV:0811.3888;%%},
	title = {{Bare Action and Regularized Functional Integral of Asymptotically Safe Quantum Gravity}},
	volume = {D79},
	year = {{(2009)}},
	Bdsk-Url-1 = {http://dx.doi.org/10.1103/PhysRevD.79.025008}}

@article{Manrique:2009tj,
	archiveprefix = {arXiv},
	author = {Manrique, Elisa and Reuter, Martin},
	date-added = {2015-07-05 17:21:41 +0000},
	date-modified = {2015-07-05 17:21:41 +0000},
	eprint = {0905.4220},
	journal = {PoS},
	pages = {001},
	primaryclass = {hep-th},
	slaccitation = {%%CITATION = ARXIV:0905.4220;%%},
	title = {{Bare versus Effective Fixed Point Action in Asymptotic Safety: The Reconstruction Problem}},
	volume = {CLAQG08},
	year = {2011}}

@article{Manrique:2009uh,
	archiveprefix = {arXiv},
	author = {Manrique, Elisa and Reuter, Martin},
	date-added = {2015-07-05 17:21:41 +0000},
	date-modified = {2015-07-05 17:21:41 +0000},
	doi = {10.1016/j.aop.2009.11.009},
	eprint = {0907.2617},
	journal = {Annals Phys.},
	pages = {785-815},
	primaryclass = {gr-qc},
	reportnumber = {MZ-TH-09-17},
	slaccitation = {%%CITATION = ARXIV:0907.2617;%%},
	title = {{Bimetric Truncations for Quantum Einstein Gravity and Asymptotic Safety}},
	volume = {325},
	year = {2010},
	Bdsk-Url-1 = {http://dx.doi.org/10.1016/j.aop.2009.11.009}}

@article{Manrique:2010am,
	archiveprefix = {arXiv},
	author = {Manrique, Elisa and Reuter, Martin and Saueressig, Frank},
	date-added = {2015-07-05 17:21:41 +0000},
	date-modified = {2015-07-05 17:21:41 +0000},
	doi = {10.1016/j.aop.2010.11.006},
	eprint = {1006.0099},
	journal = {Annals Phys.},
	pages = {463-485},
	primaryclass = {hep-th},
	reportnumber = {MZ-TH-10-19},
	slaccitation = {%%CITATION = ARXIV:1006.0099;%%},
	title = {{Bimetric Renormalization Group Flows in Quantum Einstein Gravity}},
	volume = {326},
	year = {2011},
	Bdsk-Url-1 = {http://dx.doi.org/10.1016/j.aop.2010.11.006}}

@article{Manrique:2010mq,
	archiveprefix = {arXiv},
	author = {Manrique, Elisa and Reuter, Martin and Saueressig, Frank},
	date-added = {2015-07-05 17:21:41 +0000},
	date-modified = {2015-07-05 17:21:41 +0000},
	doi = {10.1016/j.aop.2010.11.003},
	eprint = {1003.5129},
	journal = {Annals Phys.},
	pages = {440-462},
	primaryclass = {hep-th},
	reportnumber = {MZ-TH-10-08},
	slaccitation = {%%CITATION = ARXIV:1003.5129;%%},
	title = {{Matter Induced Bimetric Actions for Gravity}},
	volume = {326},
	year = {2011},
	Bdsk-Url-1 = {http://dx.doi.org/10.1016/j.aop.2010.11.003}}

@article{Modesto:2008jz,
	archiveprefix = {arXiv},
	author = {Modesto, Leonardo},
	date-added = {2015-11-30 08:25:14 +0000},
	date-modified = {2015-11-30 08:25:14 +0000},
	doi = {10.1088/0264-9381/26/24/242002},
	eprint = {0812.2214},
	journal = {Class. Quant. Grav.},
	pages = {242002},
	primaryclass = {gr-qc},
	slaccitation = {%%CITATION = ARXIV:0812.2214;%%},
	title = {{Fractal Structure of Loop Quantum Gravity}},
	volume = {26},
	year = {2009},
	Bdsk-Url-1 = {http://dx.doi.org/10.1088/0264-9381/26/24/242002}}

@article{Modesto:2009qc,
	archiveprefix = {arXiv},
	author = {Modesto, Leonardo and Nicolini, Piero},
	date-added = {2015-11-30 08:26:28 +0000},
	date-modified = {2015-11-30 08:26:28 +0000},
	doi = {10.1103/PhysRevD.81.104040},
	eprint = {0912.0220},
	journal = {Phys. Rev.},
	pages = {104040},
	primaryclass = {hep-th},
	slaccitation = {%%CITATION = ARXIV:0912.0220;%%},
	title = {{Spectral dimension of a quantum universe}},
	volume = {D81},
	year = {2010},
	Bdsk-Url-1 = {http://dx.doi.org/10.1103/PhysRevD.81.104040}}

@article{Morris:2016spn,
	archiveprefix = {arXiv},
	author = {Morris, Tim R.},
	date-added = {2016-12-17 18:07:19 +0000},
	date-modified = {2016-12-17 18:07:19 +0000},
	doi = {10.1007/JHEP11(2016)160},
	eprint = {1610.03081},
	journal = {JHEP},
	pages = {160},
	primaryclass = {hep-th},
	slaccitation = {%%CITATION = ARXIV:1610.03081;%%},
	title = {{Large curvature and background scale independence in single-metric approximations to asymptotic safety}},
	volume = {11},
	year = {2016},
	Bdsk-Url-1 = {http://dx.doi.org/10.1007/JHEP11(2016)160}}

@article{Morris:2011sn,
	archiveprefix = {arXiv},
	author = {Morris, Tim R.},
	date-added = {2016-12-17 18:11:04 +0000},
	date-modified = {2016-12-17 18:11:04 +0000},
	doi = {10.1088/0954-3899/39/4/045010},
	eprint = {1110.3266},
	journal = {J. Phys.},
	pages = {045010},
	primaryclass = {hep-ph},
	reportnumber = {arXiv:1110.2463, SHEP-11-30, SHEP-11-31},
	slaccitation = {%%CITATION = ARXIV:1110.3266;%%},
	title = {{Superluminal velocity through near-maximal neutrino oscillations or by being off shell}},
	volume = {G39},
	year = {2012},
	Bdsk-Url-1 = {http://dx.doi.org/10.1088/0954-3899/39/4/045010}}

@article{Morris:2005ck,
	archiveprefix = {arXiv},
	author = {Morris, Tim R.},
	date-added = {2015-06-01 10:42:57 +0000},
	date-modified = {2015-06-01 10:42:57 +0000},
	doi = {10.1088/1126-6708/2005/07/027},
	eprint = {hep-th/0503161},
	journal = {JHEP},
	pages = {027},
	primaryclass = {hep-th},
	reportnumber = {SHEP-0509},
	slaccitation = {%%CITATION = HEP-TH/0503161;%%},
	title = {{Equivalence of local potential approximations}},
	volume = {0507},
	year = {2005},
	Bdsk-Url-1 = {http://dx.doi.org/10.1088/1126-6708/2005/07/027}}

@article{Morris:2000jj,
	archiveprefix = {arXiv},
	author = {Morris, Tim R.},
	booktitle = {{The exact renormalization group. Proceedings, 2nd Conference, Rome, Italy, September 18-22, 2000}},
	date-added = {2015-11-30 06:56:19 +0000},
	date-modified = {2015-11-30 06:56:19 +0000},
	doi = {10.1142/S0217751X01004554},
	eprint = {hep-th/0102120},
	journal = {Int. J. Mod. Phys.},
	pages = {1899-1912},
	primaryclass = {hep-th},
	reportnumber = {SHEP-01-09},
	slaccitation = {%%CITATION = HEP-TH/0102120;%%},
	title = {{An Exact RG formulation of quantum gauge theory}},
	volume = {A16},
	year = {2001},
	Bdsk-Url-1 = {http://dx.doi.org/10.1142/S0217751X01004554}}

@article{Morris:1999px,
	archiveprefix = {arXiv},
	author = {Morris, Tim R.},
	date-added = {2015-11-30 06:56:43 +0000},
	date-modified = {2015-11-30 06:56:43 +0000},
	doi = {10.1016/S0550-3213(99)00821-4},
	eprint = {hep-th/9910058},
	journal = {Nucl. Phys.},
	pages = {97-126},
	primaryclass = {hep-th},
	reportnumber = {SHEP-99-09},
	slaccitation = {%%CITATION = HEP-TH/9910058;%%},
	title = {{A Gauge invariant exact renormalization group. 1.}},
	volume = {B573},
	year = {2000},
	Bdsk-Url-1 = {http://dx.doi.org/10.1016/S0550-3213(99)00821-4}}

@article{Morris:2000fs,
	archiveprefix = {arXiv},
	author = {Morris, Tim R.},
	date-added = {2015-11-30 06:56:42 +0000},
	date-modified = {2015-11-30 06:56:42 +0000},
	doi = {10.1088/1126-6708/2000/12/012},
	eprint = {hep-th/0006064},
	journal = {JHEP},
	pages = {012},
	primaryclass = {hep-th},
	reportnumber = {SHEP-00-04},
	slaccitation = {%%CITATION = HEP-TH/0006064;%%},
	title = {{A Gauge invariant exact renormalization group. 2.}},
	volume = {12},
	year = {2000},
	Bdsk-Url-1 = {http://dx.doi.org/10.1088/1126-6708/2000/12/012}}

@article{Morris:1996kn,
	archiveprefix = {arXiv},
	author = {Morris, T. R.},
	booktitle = {{3rd International Conference on Renormalization Group (RG 96) Dubna, Russia, August 26-31, 1996}},
	date-added = {2016-05-16 09:38:59 +0000},
	date-modified = {2016-05-16 09:38:59 +0000},
	doi = {10.1142/S0217979298000752},
	eprint = {hep-th/9610012},
	journal = {Int. J. Mod. Phys.},
	pages = {1343-1354},
	primaryclass = {hep-th},
	reportnumber = {SHEP-96-25, C96-08-26.1},
	slaccitation = {%%CITATION = HEP-TH/9610012;%%},
	title = {{Properties of derivative expansion approximations to the renormalization group}},
	volume = {B12},
	year = {1998},
	Bdsk-Url-1 = {http://dx.doi.org/10.1142/S0217979298000752}}

@article{Morris:1998,
	archiveprefix = {arXiv},
	author = {Morris, Tim R.},
	date-added = {2015-07-05 17:21:41 +0000},
	date-modified = {2015-07-05 17:21:41 +0000},
	doi = {10.1143/PTPS.131.395},
	eprint = {hep-th/9802039},
	journal = {Prog.Theor.Phys.Suppl.},
	pages = {395-414},
	primaryclass = {hep-th},
	slaccitation = {%%CITATION = HEP-TH/9802039;%%},
	title = {{Elements of the continuous renormalization group}},
	volume = {131},
	year = {1998},
	Bdsk-Url-1 = {http://dx.doi.org/10.1143/PTPS.131.395}}

@inproceedings{Morris:1998kz,
	archiveprefix = {arXiv},
	author = {Morris, Tim R.},
	booktitle = {{The exact renormalization group. Proceedings, Workshop, Faro, Portugal, September 10-12, 1998}},
	date-added = {2015-11-30 06:56:46 +0000},
	date-modified = {2015-11-30 06:56:46 +0000},
	eprint = {hep-th/9810104},
	pages = {1-40},
	primaryclass = {hep-th},
	reportnumber = {SHEP-98-12},
	slaccitation = {%%CITATION = HEP-TH/9810104;%%},
	title = {{A Manifestly gauge invariant exact renormalization group}},
	url = {http://alice.cern.ch/format/showfull?sysnb=0293277},
	year = {1998},
	Bdsk-Url-1 = {http://alice.cern.ch/format/showfull?sysnb=0293277}}

@article{Morris:1996xq,
	archiveprefix = {arXiv},
	author = {Morris, Tim R.},
	date-added = {2016-05-10 18:09:59 +0000},
	date-modified = {2016-05-10 18:09:59 +0000},
	doi = {10.1016/S0550-3213(97)00233-2},
	eprint = {hep-th/9612117},
	journal = {Nucl.Phys.},
	pages = {477-504},
	primaryclass = {hep-th},
	reportnumber = {SHEP-96-36},
	slaccitation = {%%CITATION = HEP-TH/9612117;%%},
	title = {{Three-dimensional massive scalar field theory and the derivative expansion of the renormalization group}},
	volume = {B495},
	year = {1997},
	Bdsk-Url-1 = {http://dx.doi.org/10.1016/S0550-3213(97)00233-2}}

@article{Morris:1995af,
	archiveprefix = {arXiv},
	author = {Morris, Tim R.},
	date-added = {2016-05-16 09:34:15 +0000},
	date-modified = {2016-05-16 09:34:15 +0000},
	doi = {10.1016/0550-3213(95)00541-2},
	eprint = {hep-th/9508017},
	journal = {Nucl. Phys.},
	pages = {477-503},
	primaryclass = {hep-th},
	reportnumber = {SHEP-95-21},
	slaccitation = {%%CITATION = HEP-TH/9508017;%%},
	title = {{Momentum scale expansion of sharp cutoff flow equations}},
	volume = {B458},
	year = {1996},
	Bdsk-Url-1 = {http://dx.doi.org/10.1016/0550-3213(95)00541-2}}

@article{Morris:1996nx,
	archiveprefix = {arXiv},
	author = {Morris, Tim R.},
	date-added = {2016-03-21 10:44:28 +0000},
	date-modified = {2016-03-21 10:44:28 +0000},
	doi = {10.1103/PhysRevLett.77.1658},
	eprint = {hep-th/9601128},
	journal = {Phys. Rev. Lett.},
	pages = {1658},
	primaryclass = {hep-th},
	reportnumber = {SHEP-96-02},
	slaccitation = {%%CITATION = HEP-TH/9601128;%%},
	title = {{On the fixed point structure of scalar fields}},
	volume = {77},
	year = {1996},
	Bdsk-Url-1 = {http://dx.doi.org/10.1103/PhysRevLett.77.1658}}

@article{Morris:1994jc,
	archiveprefix = {arXiv},
	author = {Morris, Tim R.},
	date-added = {2015-07-05 17:21:41 +0000},
	date-modified = {2015-07-05 17:21:41 +0000},
	doi = {10.1016/0370-2693(94)01603-A},
	eprint = {hep-th/9410141},
	journal = {Phys.Lett.},
	pages = {139-148},
	primaryclass = {hep-th},
	reportnumber = {CERN-TH-7403-94, SHEP-94-95-04},
	slaccitation = {%%CITATION = HEP-TH/9410141;%%},
	title = {{The Renormalization group and two-dimensional multicritical effective scalar field theory}},
	volume = {B345},
	year = {1995},
	Bdsk-Url-1 = {http://dx.doi.org/10.1016/0370-2693(94)01603-A}}

@article{Morris:1995he,
	archiveprefix = {arXiv},
	author = {Morris, Tim R.},
	date-added = {2015-11-30 06:58:32 +0000},
	date-modified = {2015-11-30 06:58:32 +0000},
	doi = {10.1016/0370-2693(95)00913-6},
	eprint = {hep-th/9503225},
	journal = {Phys. Lett.},
	pages = {225-231},
	primaryclass = {hep-th},
	reportnumber = {SHEP-95-07},
	slaccitation = {%%CITATION = HEP-TH/9503225;%%},
	title = {{Noncompact pure gauge QED in 3-D is free}},
	volume = {B357},
	year = {1995},
	Bdsk-Url-1 = {http://dx.doi.org/10.1016/0370-2693(95)00913-6}}

@article{Morris:1993,
	archiveprefix = {arXiv},
	author = {Morris, Tim R.},
	date-added = {2015-07-05 17:21:41 +0000},
	date-modified = {2015-07-05 17:21:41 +0000},
	doi = {10.1142/S0217751X94000972},
	eprint = {hep-ph/9308265},
	journal = {Int.J.Mod.Phys.},
	pages = {2411-2450},
	primaryclass = {hep-ph},
	reportnumber = {CERN-TH-6977-93, SHEP-92-93-27},
	slaccitation = {%%CITATION = HEP-PH/9308265;%%},
	title = {{The Exact renormalization group and approximate solutions}},
	volume = {A 09},
	year = {1994},
	Bdsk-Url-1 = {http://dx.doi.org/10.1142/S0217751X94000972}}

@article{Morris:1994ie,
	archiveprefix = {arXiv},
	author = {Morris, Tim R.},
	date-added = {2015-07-05 17:21:41 +0000},
	date-modified = {2015-07-05 17:21:41 +0000},
	doi = {10.1016/0370-2693(94)90767-6},
	eprint = {hep-ph/9403340},
	journal = {Phys.Lett.},
	pages = {241-248},
	primaryclass = {hep-ph},
	reportnumber = {CERN-TH-7203-94, SHEP-93-94-16},
	slaccitation = {%%CITATION = HEP-PH/9403340;%%},
	title = {{Derivative expansion of the exact renormalization group}},
	volume = {B329},
	year = {1994},
	Bdsk-Url-1 = {http://dx.doi.org/10.1016/0370-2693(94)90767-6}}

@article{Morris:1994ki,
	archiveprefix = {arXiv},
	author = {Morris, Tim R.},
	date-added = {2015-07-05 17:21:41 +0000},
	date-modified = {2015-07-05 17:21:41 +0000},
	doi = {10.1016/0370-2693(94)90700-5},
	eprint = {hep-th/9405190},
	journal = {Phys.Lett.},
	pages = {355-362},
	primaryclass = {hep-th},
	reportnumber = {CERN-TH-7281-94, SHEP-93-94-23},
	slaccitation = {%%CITATION = HEP-TH/9405190;%%},
	title = {{On truncations of the exact renormalization group}},
	volume = {B334},
	year = {1994},
	Bdsk-Url-1 = {http://dx.doi.org/10.1016/0370-2693(94)90700-5}}

@article{Morris:2016nda,
	archiveprefix = {arXiv},
	author = {Morris, Tim R. and Preston, Anthony W. H.},
	date-added = {2016-12-17 18:09:56 +0000},
	date-modified = {2016-12-17 18:09:56 +0000},
	doi = {10.1007/JHEP06(2016)012},
	eprint = {1602.08993},
	journal = {JHEP},
	pages = {012},
	primaryclass = {hep-th},
	slaccitation = {%%CITATION = ARXIV:1602.08993;%%},
	title = {{Manifestly diffeomorphism invariant classical Exact Renormalization Group}},
	volume = {06},
	year = {2016},
	Bdsk-Url-1 = {http://dx.doi.org/10.1007/JHEP06(2016)012}}

@article{Morris:2005tv,
	archiveprefix = {arXiv},
	author = {Morris, Tim R. and Rosten, Oliver J.},
	date-added = {2015-11-30 06:55:20 +0000},
	date-modified = {2015-11-30 06:55:20 +0000},
	doi = {10.1103/PhysRevD.73.065003},
	eprint = {hep-th/0508026},
	journal = {Phys. Rev.},
	pages = {065003},
	primaryclass = {hep-th},
	reportnumber = {SHEP-05-22},
	slaccitation = {%%CITATION = HEP-TH/0508026;%%},
	title = {{A Manifestly gauge invariant, continuum calculation of the SU(N) Yang-Mills two-loop beta function}},
	volume = {D73},
	year = {2006},
	Bdsk-Url-1 = {http://dx.doi.org/10.1103/PhysRevD.73.065003}}

@article{Morris:2006in,
	archiveprefix = {arXiv},
	author = {Morris, Tim R. and Rosten, Oliver J.},
	date-added = {2015-11-30 06:55:26 +0000},
	date-modified = {2015-11-30 06:55:26 +0000},
	doi = {10.1088/0305-4470/39/37/020},
	eprint = {hep-th/0606189},
	journal = {J. Phys.},
	pages = {11657-11681},
	primaryclass = {hep-th},
	reportnumber = {SHEP-06-20},
	slaccitation = {%%CITATION = HEP-TH/0606189;%%},
	title = {{Manifestly gauge invariant QCD}},
	volume = {A39},
	year = {2006},
	Bdsk-Url-1 = {http://dx.doi.org/10.1088/0305-4470/39/37/020}}

@article{Morris:2015oca,
	archiveprefix = {arXiv},
	author = {Morris, Tim R. and Slade, Zo{\"e} H.},
	date-added = {2016-03-21 10:20:14 +0000},
	date-modified = {2016-03-21 10:20:14 +0000},
	doi = {10.1007/JHEP11(2015)094},
	eprint = {1507.08657},
	journal = {JHEP},
	pages = {094},
	primaryclass = {hep-th},
	slaccitation = {%%CITATION = ARXIV:1507.08657;%%},
	title = {{Solutions to the reconstruction problem in asymptotic safety}},
	volume = {11},
	year = {2015},
	Bdsk-Url-1 = {http://dx.doi.org/10.1007/JHEP11(2015)094}}

@article{Morris2001,
	archiveprefix = {arXiv},
	author = {Morris, Tim R. and Tighe, John F.},
	booktitle = {{The exact renormalization group. Proceedings, 2nd Conference, Rome, Italy, September 18-22, 2000}},
	doi = {10.1142/S0217751X01004761},
	eprint = {hep-th/0102027},
	journal = {Int. J. Mod. Phys.},
	owner = {trmorris},
	pages = {2095-2100},
	primaryclass = {hep-th},
	reportnumber = {SHEP-01-06},
	slaccitation = {%%CITATION = HEP-TH/0102027;%%},
	timestamp = {2016.02.23},
	title = {{Convergence of derivative expansions in scalar field theory}},
	volume = {A16},
	year = {2001},
	Bdsk-Url-1 = {http://dx.doi.org/10.1142/S0217751X01004761}}

@article{Morris1999,
	archiveprefix = {arXiv},
	author = {Morris, Tim R. and Tighe, John F.},
	doi = {10.1088/1126-6708/1999/08/007},
	eprint = {hep-th/9906166},
	journal = {JHEP},
	owner = {trmorris},
	pages = {007},
	primaryclass = {hep-th},
	reportnumber = {SHEP-99-06},
	slaccitation = {%%CITATION = HEP-TH/9906166;%%},
	timestamp = {2016.02.23},
	title = {{Convergence of derivative expansions of the renormalization group}},
	volume = {08},
	year = {1999},
	Bdsk-Url-1 = {http://dx.doi.org/10.1088/1126-6708/1999/08/007}}

@article{Morris:1997xj,
	archiveprefix = {arXiv},
	author = {Morris, Tim R. and Turner, Michael D.},
	date-added = {2016-04-12 16:00:58 +0000},
	date-modified = {2016-04-12 16:00:58 +0000},
	doi = {10.1016/S0550-3213(97)00640-8},
	eprint = {hep-th/9704202},
	journal = {Nucl. Phys.},
	pages = {637-661},
	primaryclass = {hep-th},
	reportnumber = {SHEP-97-06},
	slaccitation = {%%CITATION = HEP-TH/9704202;%%},
	title = {{Derivative expansion of the renormalization group in O(N) scalar field theory}},
	volume = {B509},
	year = {1998},
	Bdsk-Url-1 = {http://dx.doi.org/10.1016/S0550-3213(97)00640-8}}

@article{Morris:2008uc,
	archiveprefix = {arXiv},
	author = {Morris, Tim R. and Xiao, Zhiguang},
	date-added = {2016-12-17 18:11:00 +0000},
	date-modified = {2016-12-17 18:11:00 +0000},
	doi = {10.1088/1126-6708/2008/12/028},
	eprint = {0810.3684},
	journal = {JHEP},
	pages = {028},
	primaryclass = {hep-th},
	slaccitation = {%%CITATION = ARXIV:0810.3684;%%},
	title = {{The Canonical transformation and massive CSW vertices for MHV-SQCD}},
	volume = {12},
	year = {2008},
	Bdsk-Url-1 = {http://dx.doi.org/10.1088/1126-6708/2008/12/028}}

@article{Nicoll1977,
	author = {Nicoll, J. F. and Chang, T. S.},
	doi = {10.1016/0375-9601(77)90417-0},
	journal = {Phys. Lett.},
	owner = {trmorris},
	pages = {287-289},
	slaccitation = {%%CITATION = PHLTA,A62,287;%%},
	timestamp = {2016.04.29},
	title = {{An Exact One Particle Irreducible Renormalization Group Generator for Critical Phenomena}},
	volume = {A62},
	year = {1977},
	Bdsk-Url-1 = {http://dx.doi.org/10.1016/0375-9601(77)90417-0}}

@article{Nicoll:1976ft,
	author = {Nicoll, J. F. and Chang, T. S. and Stanley, H. E.},
	date-added = {2016-05-16 09:20:59 +0000},
	date-modified = {2016-05-16 09:20:59 +0000},
	doi = {10.1103/PhysRevA.13.1251},
	journal = {Phys. Rev.},
	pages = {1251-1264},
	slaccitation = {%%CITATION = PHRVA,A13,1251;%%},
	title = {{Exact and Approximate Differential Renormalization Group Generators}},
	volume = {A13},
	year = {1976},
	Bdsk-Url-1 = {http://dx.doi.org/10.1103/PhysRevA.13.1251}}

@article{Nicoll:1974zz,
	author = {Nicoll, J. F. and Chang, T. S. and Stanley, H. E.},
	date-added = {2016-04-03 13:00:19 +0000},
	date-modified = {2016-04-03 13:00:19 +0000},
	doi = {10.1103/PhysRevLett.33.540},
	journal = {Phys. Rev. Lett.},
	pages = {540-543},
	slaccitation = {%%CITATION = PRLTA,33,540;%%},
	title = {{Approximate Renormalization Group Based on the Wegner-Houghton Differential Generator}},
	volume = {33},
	year = {1974},
	Bdsk-Url-1 = {http://dx.doi.org/10.1103/PhysRevLett.33.540}}

@article{Niedermaier:2006wt,
	author = {Niedermaier, Max and Reuter, Martin},
	date-added = {2015-07-05 17:21:41 +0000},
	date-modified = {2015-07-05 17:21:41 +0000},
	doi = {10.12942/lrr-2006-5},
	journal = {Living Rev.Rel.},
	pages = {5-173},
	slaccitation = {%%CITATION = 00222,9,5;%%},
	title = {{The Asymptotic Safety Scenario in Quantum Gravity}},
	volume = {9},
	year = {2006},
	Bdsk-Url-1 = {http://dx.doi.org/10.12942/lrr-2006-5}}

@article{Ohta2016,
	archiveprefix = {arXiv},
	author = {Ohta, Nobuyoshi and Percacci, Roberto and Vacca, Gian Paolo},
	doi = {10.1140/epjc/s10052-016-3895-1},
	eprint = {1511.09393},
	journal = {Eur. Phys. J.},
	number = {2},
	owner = {trmorris},
	pages = {46},
	primaryclass = {hep-th},
	slaccitation = {%%CITATION = ARXIV:1511.09393;%%},
	timestamp = {2016.04.21},
	title = {{Renormalization Group Equation and scaling solutions for f(R) gravity in exponential parametrization}},
	volume = {C76},
	year = {2016},
	Bdsk-Url-1 = {http://dx.doi.org/10.1140/epjc/s10052-016-3895-1}}

@article{Ohta:2015efa,
	archiveprefix = {arXiv},
	author = {Ohta, Nobuyoshi and Percacci, Roberto and Vacca, Gian Paolo},
	date-added = {2016-10-10 10:27:53 +0000},
	date-modified = {2016-10-10 10:27:53 +0000},
	doi = {10.1103/PhysRevD.92.061501},
	eprint = {1507.00968},
	journal = {Phys. Rev.},
	number = {6},
	pages = {061501},
	primaryclass = {hep-th},
	slaccitation = {%%CITATION = ARXIV:1507.00968;%%},
	title = {{Flow equation for $f(R)$ gravity and some of its exact solutions}},
	volume = {D92},
	year = {2015},
	Bdsk-Url-1 = {http://dx.doi.org/10.1103/PhysRevD.92.061501}}

@book{OlverAsymptotics,
	author = {Olver, F. W. J.},
	date-added = {2016-05-10 18:09:59 +0000},
	date-modified = {2016-05-10 18:09:59 +0000},
	isbn = {012525850X},
	note = {Computer Science and Applied Mathematics},
	pages = {156-169,254-260},
	publisher = {Academic Press, New York-London},
	title = {Asymptotics and special functions},
	year = {1974}}

@article{Padmanabhan:2015vma,
	archiveprefix = {arXiv},
	author = {Padmanabhan, T. and Chakraborty, Sumanta and Kothawala, Dawood},
	date-added = {2015-11-30 08:23:43 +0000},
	date-modified = {2015-11-30 08:23:43 +0000},
	eprint = {1507.05669},
	primaryclass = {gr-qc},
	slaccitation = {%%CITATION = ARXIV:1507.05669;%%},
	title = {{Renormalized spacetime is two-dimensional at the Planck scale}},
	year = {2015}}

@article{Pagani:2013fca,
	archiveprefix = {arXiv},
	author = {Pagani, Carlo and Percacci, Roberto},
	date-added = {2015-07-05 17:21:41 +0000},
	date-modified = {2015-07-05 17:21:41 +0000},
	eprint = {1312.7767},
	primaryclass = {hep-th},
	slaccitation = {%%CITATION = ARXIV:1312.7767;%%},
	title = {{Quantization and fixed points of non-integrable Weyl theory}},
	year = {2013}}

@article{Parisi:1979ka,
	author = {Parisi, G. and Sourlas, N.},
	date-added = {2015-11-30 08:20:52 +0000},
	date-modified = {2015-11-30 08:20:52 +0000},
	doi = {10.1103/PhysRevLett.43.744},
	journal = {Phys. Rev. Lett.},
	pages = {744},
	reportnumber = {LPTENS-79-13},
	slaccitation = {%%CITATION = PRLTA,43,744;%%},
	title = {{Random Magnetic Fields, Supersymmetry and Negative Dimensions}},
	volume = {43},
	year = {1979},
	Bdsk-Url-1 = {http://dx.doi.org/10.1103/PhysRevLett.43.744}}

@article{Pawlowski:2005xe,
	archiveprefix = {arXiv},
	author = {Pawlowski, Jan M.},
	date-added = {2015-07-05 17:21:41 +0000},
	date-modified = {2015-07-05 17:21:41 +0000},
	doi = {10.1016/j.aop.2007.01.007},
	eprint = {hep-th/0512261},
	journal = {Annals Phys.},
	pages = {2831-2915},
	primaryclass = {hep-th},
	reportnumber = {HD-THEP-05-28},
	slaccitation = {%%CITATION = HEP-TH/0512261;%%},
	title = {{Aspects of the functional renormalisation group}},
	volume = {322},
	year = {2007},
	Bdsk-Url-1 = {http://dx.doi.org/10.1016/j.aop.2007.01.007}}

@article{Pawlowski:2003sk,
	archiveprefix = {arXiv},
	author = {Pawlowski, Jan M.},
	date-added = {2015-07-05 17:21:41 +0000},
	date-modified = {2015-07-05 17:21:41 +0000},
	eprint = {hep-th/0310018},
	primaryclass = {hep-th},
	reportnumber = {FAU-TP3-03-10},
	slaccitation = {%%CITATION = HEP-TH/0310018;%%},
	title = {{Geometrical effective action and Wilsonian flows}},
	year = {2003}}

@article{Pelissetto:2000ek,
	archiveprefix = {arXiv},
	author = {Pelissetto, Andrea and Vicari, Ettore},
	date-added = {2015-07-05 17:21:41 +0000},
	date-modified = {2015-07-05 17:21:41 +0000},
	doi = {10.1016/S0370-1573(02)00219-3},
	eprint = {cond-mat/0012164},
	journal = {Phys.Rept.},
	pages = {549-727},
	primaryclass = {cond-mat},
	slaccitation = {%%CITATION = COND-MAT/0012164;%%},
	title = {{Critical phenomena and renormalization group theory}},
	volume = {368},
	year = {2002},
	Bdsk-Url-1 = {http://dx.doi.org/10.1016/S0370-1573(02)00219-3}}

@article{Percacci:2011uf,
	archiveprefix = {arXiv},
	author = {Percacci, R.},
	date-added = {2016-05-04 10:32:52 +0000},
	date-modified = {2016-05-04 10:32:52 +0000},
	doi = {10.1088/1367-2630/13/12/125013},
	eprint = {1110.6758},
	journal = {New J. Phys.},
	pages = {125013},
	primaryclass = {hep-th},
	slaccitation = {%%CITATION = ARXIV:1110.6758;%%},
	title = {{Renormalization group flow of Weyl invariant dilaton gravity}},
	volume = {13},
	year = {2011},
	Bdsk-Url-1 = {http://dx.doi.org/10.1088/1367-2630/13/12/125013}}

@article{Percacci2007,
	archiveprefix = {arXiv},
	author = {Percacci, Roberto},
	eprint = {0709.3851},
	primaryclass = {hep-th},
	slaccitation = {%%CITATION = ARXIV:0709.3851;%%},
	title = {{Asymptotic Safety}},
	year = {2007}}

@article{Percacci:2003jz,
	archiveprefix = {arXiv},
	author = {Percacci, Roberto and Perini, Daniele},
	date-added = {2016-05-10 18:09:59 +0000},
	date-modified = {2016-05-10 18:09:59 +0000},
	doi = {10.1103/PhysRevD.68.044018},
	eprint = {hep-th/0304222},
	journal = {Phys.Rev.},
	pages = {044018},
	primaryclass = {hep-th},
	title = {Asymptotic safety of gravity coupled to matter},
	volume = {D68},
	year = {2003},
	Bdsk-Url-1 = {http://dx.doi.org/10.1103/PhysRevD.68.044018}}

@article{Periwal:1995hw,
	archiveprefix = {arXiv},
	author = {Periwal, Vipul},
	date-added = {2016-05-10 18:09:59 +0000},
	date-modified = {2016-05-10 18:09:59 +0000},
	doi = {10.1142/S0217732396002885},
	eprint = {hep-th/9512108},
	journal = {Mod.Phys.Lett.},
	pages = {2915-2920},
	primaryclass = {hep-th},
	reportnumber = {PUPT-1567},
	title = {Halpern-Huang directions in effective scalar field theory},
	volume = {A11},
	year = {1996},
	Bdsk-Url-1 = {http://dx.doi.org/10.1142/S0217732396002885}}

@article{Pietrykowski:2012nc,
	archiveprefix = {arXiv},
	author = {Pietrykowski, Artur R.},
	date-added = {2016-05-10 18:09:59 +0000},
	date-modified = {2016-05-10 18:09:59 +0000},
	doi = {10.1103/PhysRevD.87.024026},
	eprint = {1210.0507},
	journal = {Phys.Rev.},
	pages = {024026},
	primaryclass = {hep-th},
	title = {Interacting Scalar Fields in the Context of Effective Quantum Gravity},
	volume = {D87},
	year = {2013},
	Bdsk-Url-1 = {http://dx.doi.org/10.1103/PhysRevD.87.024026}}

@article{Polchinski:1983gv,
	author = {Polchinski, Joseph},
	date-added = {2015-07-05 17:21:41 +0000},
	date-modified = {2015-07-05 17:21:41 +0000},
	doi = {10.1016/0550-3213(84)90287-6},
	journal = {Nucl.Phys.},
	pages = {269-295},
	reportnumber = {HUTP-83-A018},
	slaccitation = {%%CITATION = NUPHA,B231,269;%%},
	title = {{Renormalization and Effective Lagrangians}},
	volume = {B231},
	year = {1984},
	Bdsk-Url-1 = {http://dx.doi.org/10.1016/0550-3213(84)90287-6}}

@article{Polonyi:2001se,
	archiveprefix = {arXiv},
	author = {Polonyi, Janos},
	date-added = {2016-05-10 15:53:26 +0000},
	date-modified = {2016-05-10 15:53:26 +0000},
	doi = {10.2478/BF02475552},
	eprint = {hep-th/0110026},
	journal = {Central Eur. J. Phys.},
	pages = {1-71},
	primaryclass = {hep-th},
	slaccitation = {%%CITATION = HEP-TH/0110026;%%},
	title = {{Lectures on the functional renormalization group method}},
	volume = {1},
	year = {2003},
	Bdsk-Url-1 = {http://dx.doi.org/10.2478/BF02475552}}

@article{Preston:2014tua,
	archiveprefix = {arXiv},
	author = {Preston, Anthony W. H. and Morris, Tim R.},
	date-added = {2016-02-06 09:06:05 +0000},
	date-modified = {2016-02-06 09:06:05 +0000},
	doi = {10.1088/1475-7516/2014/09/017},
	eprint = {1406.5398},
	journal = {JCAP},
	pages = {017},
	primaryclass = {gr-qc},
	slaccitation = {%%CITATION = ARXIV:1406.5398;%%},
	title = {{Cosmological back-reaction in modified gravity and its implications for dark energy}},
	volume = {1409},
	year = {2014},
	Bdsk-Url-1 = {http://dx.doi.org/10.1088/1475-7516/2014/09/017}}

@article{Rasanen:2003fy,
	archiveprefix = {arXiv},
	author = {Rasanen, Syksy},
	date-added = {2016-02-06 09:05:00 +0000},
	date-modified = {2016-02-06 09:05:00 +0000},
	doi = {10.1088/1475-7516/2004/02/003},
	eprint = {astro-ph/0311257},
	journal = {JCAP},
	pages = {003},
	primaryclass = {astro-ph},
	slaccitation = {%%CITATION = ASTRO-PH/0311257;%%},
	title = {{Dark energy from backreaction}},
	volume = {0402},
	year = {2004},
	Bdsk-Url-1 = {http://dx.doi.org/10.1088/1475-7516/2004/02/003}}

@article{Reuter:1996,
	archiveprefix = {arXiv},
	author = {Reuter, M.},
	date-added = {2015-07-05 17:21:41 +0000},
	date-modified = {2015-07-05 17:21:41 +0000},
	doi = {10.1103/PhysRevD.57.971},
	eprint = {hep-th/9605030},
	journal = {Phys.Rev.},
	pages = {971-985},
	primaryclass = {hep-th},
	reportnumber = {DESY-96-080},
	slaccitation = {%%CITATION = HEP-TH/9605030;%%},
	title = {{Nonperturbative evolution equation for quantum gravity}},
	volume = {D57},
	year = {1998},
	Bdsk-Url-1 = {http://dx.doi.org/10.1103/PhysRevD.57.971}}

@article{Reuter:2012,
	archiveprefix = {arXiv},
	author = {Reuter, Martin and Saueressig, Frank},
	date-added = {2015-07-05 17:21:41 +0000},
	date-modified = {2015-07-05 17:21:41 +0000},
	doi = {10.1088/1367-2630/14/5/055022},
	eprint = {1202.2274},
	journal = {New J.Phys.},
	pages = {055022},
	primaryclass = {hep-th},
	slaccitation = {%%CITATION = ARXIV:1202.2274;%%},
	title = {{Quantum Einstein Gravity}},
	volume = {14},
	year = {2012},
	Bdsk-Url-1 = {http://dx.doi.org/10.1088/1367-2630/14/5/055022}}

@article{Reuter:1997gx,
	archiveprefix = {arXiv},
	author = {Reuter, M. and Wetterich, C.},
	date-added = {2015-07-05 17:21:41 +0000},
	date-modified = {2015-07-05 17:21:41 +0000},
	doi = {10.1103/PhysRevD.56.7893},
	eprint = {hep-th/9708051},
	journal = {Phys.Rev.},
	pages = {7893-7916},
	primaryclass = {hep-th},
	slaccitation = {%%CITATION = HEP-TH/9708051;%%},
	title = {{Gluon condensation in nonperturbative flow equations}},
	volume = {D56},
	year = {1997},
	Bdsk-Url-1 = {http://dx.doi.org/10.1103/PhysRevD.56.7893}}

@article{Reuter:1993kw,
	author = {Reuter, M. and Wetterich, C.},
	date-added = {2015-07-05 17:21:41 +0000},
	date-modified = {2015-07-05 17:21:41 +0000},
	doi = {10.1016/0550-3213(94)90543-6},
	journal = {Nucl.Phys.},
	pages = {181-214},
	reportnumber = {DESY-93-152, HD-THEP-93-40},
	slaccitation = {%%CITATION = NUPHA,B417,181;%%},
	title = {{Effective average action for gauge theories and exact evolution equations}},
	volume = {B417},
	year = {1994},
	Bdsk-Url-1 = {http://dx.doi.org/10.1016/0550-3213(94)90543-6}}

@article{Reuter:1994sg,
	author = {Reuter, M. and Wetterich, C.},
	date-added = {2015-07-05 17:21:41 +0000},
	date-modified = {2015-07-05 17:21:41 +0000},
	doi = {10.1016/0550-3213(94)90278-X},
	journal = {Nucl.Phys.},
	pages = {291-324},
	reportnumber = {DESY-94-017, HD-THEP-93-41},
	slaccitation = {%%CITATION = NUPHA,B427,291;%%},
	title = {{Exact evolution equation for scalar electrodynamics}},
	volume = {B427},
	year = {1994},
	Bdsk-Url-1 = {http://dx.doi.org/10.1016/0550-3213(94)90278-X}}

@article{Reuter:2008qx,
	archiveprefix = {arXiv},
	author = {Reuter, Martin and Weyer, Holger},
	date-added = {2015-07-05 17:21:41 +0000},
	date-modified = {2015-07-05 17:21:41 +0000},
	doi = {10.1103/PhysRevD.80.025001},
	eprint = {0804.1475},
	journal = {Phys.Rev.},
	pages = {025001},
	primaryclass = {hep-th},
	reportnumber = {MZ-TH-08-12},
	slaccitation = {%%CITATION = ARXIV:0804.1475;%%},
	title = {{Conformal sector of Quantum Einstein Gravity in the local potential approximation: Non-Gaussian fixed point and a phase of unbroken diffeomorphism invariance}},
	volume = {D80},
	year = {2009},
	Bdsk-Url-1 = {http://dx.doi.org/10.1103/PhysRevD.80.025001}}

@article{Reuter:2008wj,
	archiveprefix = {arXiv},
	author = {Reuter, Martin and Weyer, Holger},
	date-added = {2015-07-05 17:21:41 +0000},
	date-modified = {2015-07-05 17:21:41 +0000},
	doi = {10.1103/PhysRevD.79.105005},
	eprint = {0801.3287},
	journal = {Phys.Rev.},
	pages = {105005},
	primaryclass = {hep-th},
	reportnumber = {MZ-TH-08-04},
	slaccitation = {%%CITATION = ARXIV:0801.3287;%%},
	title = {{Background Independence and Asymptotic Safety in Conformally Reduced Gravity}},
	volume = {D79},
	year = {2009},
	Bdsk-Url-1 = {http://dx.doi.org/10.1103/PhysRevD.79.105005}}

@article{Reuter:2009kq,
	archiveprefix = {arXiv},
	author = {Reuter, Martin and Weyer, Holger},
	date-added = {2015-07-05 17:21:41 +0000},
	date-modified = {2015-07-05 17:21:41 +0000},
	doi = {10.1007/s10714-008-0744-z},
	eprint = {0903.2971},
	journal = {Gen.Rel.Grav.},
	pages = {983-1011},
	primaryclass = {hep-th},
	reportnumber = {MZ-TH-08-23},
	slaccitation = {%%CITATION = ARXIV:0903.2971;%%},
	title = {{The Role of Background Independence for Asymptotic Safety in Quantum Einstein Gravity}},
	volume = {41},
	year = {2009},
	Bdsk-Url-1 = {http://dx.doi.org/10.1007/s10714-008-0744-z}}

@article{Riedel:1986re,
	author = {Riedel, E. and Golner, G. R. and Newman, K. E.},
	date-added = {2016-05-04 16:58:00 +0000},
	date-modified = {2016-05-04 16:58:00 +0000},
	doi = {10.1016/0003-4916(85)90341-0},
	journal = {Annals Phys.},
	pages = {178-238},
	slaccitation = {%%CITATION = APNYA,161,178;%%},
	title = {{Scaling Field Representation of Wilson's Exact Renormalization Group Equation}},
	volume = {161},
	year = {1985},
	Bdsk-Url-1 = {http://dx.doi.org/10.1016/0003-4916(85)90341-0}}

@article{Rosten:2010,
	archiveprefix = {arXiv},
	author = {Rosten, Oliver J.},
	date-added = {2015-07-05 17:21:41 +0000},
	date-modified = {2015-07-05 17:21:41 +0000},
	doi = {10.1016/j.physrep.2011.12.003},
	eprint = {1003.1366},
	journal = {Phys.Rept.},
	pages = {177-272},
	primaryclass = {hep-th},
	slaccitation = {%%CITATION = ARXIV:1003.1366;%%},
	title = {{Fundamentals of the Exact Renormalization Group}},
	volume = {511},
	year = {2012},
	Bdsk-Url-1 = {http://dx.doi.org/10.1016/j.physrep.2011.12.003}}

@article{Rosten:2010vm,
	archiveprefix = {arXiv},
	author = {Rosten, Oliver J.},
	date-added = {2015-11-30 06:59:25 +0000},
	date-modified = {2015-11-30 06:59:25 +0000},
	doi = {10.1016/j.physrep.2011.12.003},
	eprint = {1003.1366},
	journal = {Phys. Rept.},
	pages = {177-272},
	primaryclass = {hep-th},
	slaccitation = {%%CITATION = ARXIV:1003.1366;%%},
	title = {{Fundamentals of the Exact Renormalization Group}},
	volume = {511},
	year = {2012},
	Bdsk-Url-1 = {http://dx.doi.org/10.1016/j.physrep.2011.12.003}}

@article{Rosten:2011ty,
	archiveprefix = {arXiv},
	author = {Rosten, Oliver J.},
	booktitle = {{Proceedings, Workshop on The many faces of QCD (FacesQCD2010)}},
	date-added = {2015-11-30 06:59:28 +0000},
	date-modified = {2015-11-30 06:59:28 +0000},
	eprint = {1102.3091},
	journal = {PoS},
	pages = {035},
	primaryclass = {hep-th},
	slaccitation = {%%CITATION = ARXIV:1102.3091;%%},
	title = {{Aspects of Manifest Gauge Invariance}},
	volume = {FACESQCD},
	year = {2010}}

@article{Rosten:2008zp,
	archiveprefix = {arXiv},
	author = {Rosten, Oliver J.},
	date-added = {2015-11-30 06:59:35 +0000},
	date-modified = {2015-11-30 06:59:35 +0000},
	doi = {10.1016/j.physletb.2008.03.006},
	eprint = {0801.2462},
	journal = {Phys. Lett.},
	pages = {237-243},
	primaryclass = {hep-th},
	reportnumber = {DIAS-STP-08-01},
	slaccitation = {%%CITATION = ARXIV:0801.2462;%%},
	title = {{A Resummable beta-Function for Massless QED}},
	volume = {B662},
	year = {2008},
	Bdsk-Url-1 = {http://dx.doi.org/10.1016/j.physletb.2008.03.006}}

@article{Rosten:2006pd,
	archiveprefix = {arXiv},
	author = {Rosten, Oliver J.},
	date-added = {2015-11-30 06:59:39 +0000},
	date-modified = {2015-11-30 06:59:39 +0000},
	doi = {10.1016/j.physletb.2006.12.057},
	eprint = {hep-th/0611323},
	journal = {Phys. Lett.},
	pages = {466-469},
	primaryclass = {hep-th},
	reportnumber = {DIAS-STP-06-22},
	slaccitation = {%%CITATION = HEP-TH/0611323;%%},
	title = {{Universality From Very General Nonperturbative Flow Equations in QCD}},
	volume = {B645},
	year = {2007},
	Bdsk-Url-1 = {http://dx.doi.org/10.1016/j.physletb.2006.12.057}}

@article{Rosten:2005ep,
	archiveprefix = {arXiv},
	author = {Rosten, Oliver J.},
	booktitle = {{Renormalization Group 2005 (RG 2005) Helsinki, Finland, August 30-September 3, 2005}},
	date-added = {2015-11-30 06:59:47 +0000},
	date-modified = {2015-11-30 06:59:47 +0000},
	doi = {10.1088/0305-4470/39/25/S24},
	eprint = {hep-th/0511107},
	journal = {J. Phys.},
	pages = {8141-8156},
	primaryclass = {hep-th},
	reportnumber = {SHEP-05-34},
	slaccitation = {%%CITATION = HEP-TH/0511107;%%},
	title = {{Scheme independence to all loops}},
	volume = {A39},
	year = {2006},
	Bdsk-Url-1 = {http://dx.doi.org/10.1088/0305-4470/39/25/S24}}

@article{Rosten:2005ka,
	archiveprefix = {arXiv},
	author = {Rosten, Oliver J.},
	date-added = {2015-11-30 06:59:49 +0000},
	date-modified = {2015-11-30 06:59:49 +0000},
	doi = {10.1088/0305-4470/39/27/010},
	eprint = {hep-th/0507166},
	journal = {J. Phys.},
	pages = {8699-8726},
	primaryclass = {hep-th},
	reportnumber = {SHEP-05-23},
	slaccitation = {%%CITATION = HEP-TH/0507166;%%},
	title = {{A Primer for manifestly gauge invariant computations in SU(N) Yang-Mills}},
	volume = {A39},
	year = {2006},
	Bdsk-Url-1 = {http://dx.doi.org/10.1088/0305-4470/39/27/010}}

@article{Rosten:2006qx,
	archiveprefix = {arXiv},
	author = {Rosten, Oliver J.},
	date-added = {2015-11-30 06:59:42 +0000},
	date-modified = {2015-11-30 06:59:42 +0000},
	doi = {10.1103/PhysRevD.74.125006},
	eprint = {hep-th/0604183},
	journal = {Phys. Rev.},
	pages = {125006},
	primaryclass = {hep-th},
	reportnumber = {SHEP-06-15},
	slaccitation = {%%CITATION = HEP-TH/0604183;%%},
	title = {{General Computations Without Fixing the Gauge}},
	volume = {D74},
	year = {2006},
	Bdsk-Url-1 = {http://dx.doi.org/10.1103/PhysRevD.74.125006}}

@article{Rosten:2006tk,
	archiveprefix = {arXiv},
	author = {Rosten, Oliver J.},
	date-added = {2015-11-30 06:59:43 +0000},
	date-modified = {2015-11-30 06:59:43 +0000},
	doi = {10.1142/S0217751X06033040},
	eprint = {hep-th/0602229},
	journal = {Int. J. Mod. Phys.},
	pages = {4627-4762},
	primaryclass = {hep-th},
	reportnumber = {SHEP-06-09},
	slaccitation = {%%CITATION = HEP-TH/0602229;%%},
	title = {{A Manifestly gauge invariant and universal calculus for SU(N) Yang-Mills}},
	volume = {A21},
	year = {2006},
	Bdsk-Url-1 = {http://dx.doi.org/10.1142/S0217751X06033040}}

@phdthesis{Rosten:2005qs,
	archiveprefix = {arXiv},
	author = {Rosten, Oliver Jacob},
	date-added = {2015-11-30 06:59:51 +0000},
	date-modified = {2015-11-30 06:59:51 +0000},
	eprint = {hep-th/0506162},
	primaryclass = {hep-th},
	reportnumber = {UMI-C823832, SHEP-0519},
	school = {Southampton U.},
	slaccitation = {%%CITATION = HEP-TH/0506162;%%},
	title = {{The Manifestly gauge invariant exact renormalisation group}},
	url = {http://wwwlib.umi.com/dissertations/fullcit?pC823832},
	year = {2005},
	Bdsk-Url-1 = {http://wwwlib.umi.com/dissertations/fullcit?pC823832}}

@inproceedings{Rosten:2004aw,
	archiveprefix = {arXiv},
	author = {Rosten, Oliver J. and Morris, Tim R. and Arnone, Stefano},
	booktitle = {{13th International Seminar on High-Energy Physics: Quarks 2004 Pushkinskie Gory, Russia, May 24-30, 2004}},
	date-added = {2015-11-30 06:55:09 +0000},
	date-modified = {2015-11-30 06:55:09 +0000},
	eprint = {hep-th/0409042},
	primaryclass = {hep-th},
	reportnumber = {SHEP-04-25},
	slaccitation = {%%CITATION = HEP-TH/0409042;%%},
	title = {{The Gauge invariant ERG}},
	url = {http://www.slac.stanford.edu/econf/C0405241/proceedings/FT/rosten.pdf},
	year = {2004},
	Bdsk-Url-1 = {http://www.slac.stanford.edu/econf/C0405241/proceedings/FT/rosten.pdf}}

@article{Rovelli:2008zza,
	author = {Rovelli, Carlo},
	date-added = {2015-07-05 17:21:41 +0000},
	date-modified = {2015-07-05 17:21:41 +0000},
	journal = {Living Rev.Rel.},
	pages = {5},
	slaccitation = {%%CITATION = 00222,11,5;%%},
	title = {{Loop quantum gravity}},
	volume = {11},
	year = {2008}}

@book{Rovelli:2004tv,
	author = {Rovelli, Carlo},
	date-added = {2016-11-30 12:57:00 +0000},
	date-modified = {2016-11-30 12:57:00 +0000},
	journal = {Cambridge, UK: Univ. Pr. (2004) 455 p},
	slaccitation = {%%CITATION = INSPIRE-659635;%%},
	title = {{Quantum gravity}},
	url = {http://www.cambridge.org/uk/catalogue/catalogue.asp?isbn=0521837332},
	year = {2004},
	Bdsk-Url-1 = {http://www.cambridge.org/uk/catalogue/catalogue.asp?isbn=0521837332}}

@article{Saltas:2014cta,
	archiveprefix = {arXiv},
	author = {Saltas, Ippocratis D.},
	date-added = {2015-04-25 14:25:01 +0000},
	date-modified = {2015-04-25 14:25:01 +0000},
	doi = {10.1103/PhysRevD.90.124052},
	eprint = {1410.6163},
	journal = {Phys.Rev.},
	number = {12},
	pages = {124052},
	primaryclass = {hep-th},
	slaccitation = {%%CITATION = ARXIV:1410.6163;%%},
	title = {{UV structure of quantum unimodular gravity}},
	volume = {D90},
	year = {2014},
	Bdsk-Url-1 = {http://dx.doi.org/10.1103/PhysRevD.90.124052}}

@article{Slavnov:1972qb,
	author = {Slavnov, A. A.},
	date-added = {2016-02-17 16:52:46 +0000},
	date-modified = {2016-02-17 16:52:46 +0000},
	journal = {Teor. Mat. Fiz.},
	pages = {305-328},
	slaccitation = {%%CITATION = TMFZA,10,305;%%},
	title = {{Massive gauge fields}},
	volume = {10},
	year = {1972}}

@article{Slavnov:1972sq,
	author = {Slavnov, A. A.},
	date-added = {2016-02-17 16:52:31 +0000},
	date-modified = {2016-02-17 16:52:31 +0000},
	doi = {10.1007/BF01035526},
	journal = {Teor. Mat. Fiz.},
	pages = {174-177},
	slaccitation = {%%CITATION = TMFZA,13,174;%%},
	title = {{Invariant regularization of gauge theories}},
	volume = {13},
	year = {1972},
	Bdsk-Url-1 = {http://dx.doi.org/10.1007/BF01035526}}

@article{Sokal:1994,
	archiveprefix = {arXiv},
	author = {Sokal, Alan D. and van Enter, A.C.D. and Fernandez, R.},
	date-added = {2015-07-05 17:21:41 +0000},
	date-modified = {2015-07-05 17:21:41 +0000},
	eprint = {hep-lat/9210032},
	journal = {J.Statist.Phys.},
	pages = {879-1167},
	primaryclass = {hep-lat},
	slaccitation = {%%CITATION = HEP-LAT/9210032;%%},
	title = {{Regularity properties and pathologies of position space renormalization group transformations: Scope and limitations of Gibbsian theory}},
	volume = {72},
	year = {1994}}

@article{Sotiriou:2011mu,
	archiveprefix = {arXiv},
	author = {Sotiriou, Thomas P. and Visser, Matt and Weinfurtner, Silke},
	date-added = {2015-11-30 08:31:50 +0000},
	date-modified = {2015-11-30 08:31:50 +0000},
	doi = {10.1103/PhysRevLett.107.131303},
	eprint = {1105.5646},
	journal = {Phys. Rev. Lett.},
	pages = {131303},
	primaryclass = {gr-qc},
	slaccitation = {%%CITATION = ARXIV:1105.5646;%%},
	title = {{Spectral dimension as a probe of the ultraviolet continuum regime of causal dynamical triangulations}},
	volume = {107},
	year = {2011},
	Bdsk-Url-1 = {http://dx.doi.org/10.1103/PhysRevLett.107.131303}}

@article{Stelle:1976gc,
	author = {Stelle, K. S.},
	date-added = {2015-11-30 08:55:29 +0000},
	date-modified = {2015-11-30 08:55:29 +0000},
	doi = {10.1103/PhysRevD.16.953},
	journal = {Phys. Rev.},
	pages = {953-969},
	reportnumber = {PRINT-76-1059 (BRANDEIS)},
	slaccitation = {%%CITATION = PHRVA,D16,953;%%},
	title = {{Renormalization of Higher Derivative Quantum Gravity}},
	volume = {D16},
	year = {1977},
	Bdsk-Url-1 = {http://dx.doi.org/10.1103/PhysRevD.16.953}}

@article{Stojkovic:2014lha,
	archiveprefix = {arXiv},
	author = {Stojkovic, Dejan},
	date-added = {2015-11-30 08:33:52 +0000},
	date-modified = {2015-11-30 08:33:52 +0000},
	doi = {10.1142/S0217732313300346},
	eprint = {1406.2696},
	journal = {Mod. Phys. Lett.},
	pages = {1330034},
	primaryclass = {gr-qc},
	slaccitation = {%%CITATION = ARXIV:1406.2696;%%},
	title = {{Vanishing dimensions}},
	volume = {A28},
	year = {2013},
	Bdsk-Url-1 = {http://dx.doi.org/10.1142/S0217732313300346}}

@article{Thiemann:2007zz,
	archiveprefix = {arXiv},
	author = {Thiemann, Thomas},
	date-added = {2015-07-05 17:21:41 +0000},
	date-modified = {2015-07-05 17:21:41 +0000},
	eprint = {gr-qc/0110034},
	primaryclass = {gr-qc},
	reportnumber = {AEI-2001-119},
	slaccitation = {%%CITATION = GR-QC/0110034;%%},
	title = {{Modern canonical quantum general relativity}},
	year = {2001}}

@article{Unruh1989a,
	author = {Unruh, W. G.},
	doi = {10.1103/PhysRevD.40.1048},
	journal = {Phys. Rev.},
	owner = {trmorris},
	pages = {1048},
	reportnumber = {NSF-ITP-88-169},
	slaccitation = {%%CITATION = PHRVA,D40,1048;%%},
	timestamp = {2016.01.05},
	title = {{A Unimodular Theory of Canonical Quantum Gravity}},
	volume = {D40},
	year = {1989},
	Bdsk-Url-1 = {http://dx.doi.org/10.1103/PhysRevD.40.1048}}

@article{V:2015mza,
	archiveprefix = {arXiv},
	author = {V, Anjana},
	date-added = {2015-11-30 08:23:16 +0000},
	date-modified = {2015-11-30 08:23:16 +0000},
	eprint = {1509.06892},
	primaryclass = {hep-th},
	slaccitation = {%%CITATION = ARXIV:1509.06892;%%},
	title = {{Diffusion in kappa deformed space and Spectral Dimension}},
	year = {2015}}

@article{Vacca:2011fx,
	archiveprefix = {arXiv},
	author = {Vacca, G.P. and Zambelli, L.},
	date-added = {2015-07-05 17:21:41 +0000},
	date-modified = {2015-07-05 17:21:41 +0000},
	doi = {10.1103/PhysRevD.83.125024},
	eprint = {1103.2219},
	journal = {Phys.Rev.},
	pages = {125024},
	primaryclass = {hep-th},
	slaccitation = {%%CITATION = ARXIV:1103.2219;%%},
	title = {{Functional RG flow equation: regularization and coarse-graining in phase space}},
	volume = {D83},
	year = {2011},
	Bdsk-Url-1 = {http://dx.doi.org/10.1103/PhysRevD.83.125024}}

@inproceedings{Visser:2015mur,
	archiveprefix = {arXiv},
	author = {Visser, Matt},
	date-added = {2016-02-06 09:14:50 +0000},
	date-modified = {2016-02-06 09:14:50 +0000},
	eprint = {1512.05729},
	primaryclass = {gr-qc},
	slaccitation = {%%CITATION = ARXIV:1512.05729;%%},
	title = {{Buchert coarse-graining and the classical energy conditions}},
	url = {http://inspirehep.net/record/1410047/files/arXiv:1512.05729.pdf},
	year = {2015},
	Bdsk-Url-1 = {http://inspirehep.net/record/1410047/files/arXiv:1512.05729.pdf}}

@incollection{WegnerBook1,
	author = {Wegner, Franz J.},
	booktitle = {Phase Transitions and Critical Phenomena},
	date-added = {2016-05-14 10:27:40 +0000},
	date-modified = {2016-05-14 10:31:40 +0000},
	editor = {Domb, C. and Green, M. S.},
	pages = {7},
	publisher = {Academic Press, New York},
	title = {The critical stage, General Aspects},
	volume = {VI},
	year = {1976}}

@article{WR,
	author = {F.~J.~Wegner},
	date-added = {2015-07-05 17:21:41 +0000},
	date-modified = {2015-07-05 17:21:41 +0000},
	journal = {J. Phys.},
	pages = {2098},
	title = {Some invariance properties of the renormalization group},
	volume = {C7},
	year = {1974}}

@article{Wegner:1972my,
	author = {Wegner, Franz J.},
	date-added = {2016-05-14 10:26:56 +0000},
	date-modified = {2016-05-14 10:34:03 +0000},
	doi = {10.1103/PhysRevB.5.4529},
	journal = {Phys. Rev.},
	pages = {4529-4536},
	slaccitation = {%%CITATION = PHRVA,B5,4529;%%},
	title = {{Corrections to scaling laws}},
	volume = {B5},
	year = {1972},
	Bdsk-Url-1 = {http://dx.doi.org/10.1103/PhysRevB.5.4529}}

@article{Wegner:1972ih,
	author = {Wegner, Franz J. and Houghton, Anthony},
	date-added = {2015-11-30 07:51:02 +0000},
	date-modified = {2015-11-30 07:51:02 +0000},
	doi = {10.1103/PhysRevA.8.401},
	journal = {Phys. Rev.},
	pages = {401-412},
	slaccitation = {%%CITATION = PHRVA,A8,401;%%},
	title = {{Renormalization group equation for critical phenomena}},
	volume = {A8},
	year = {1973},
	Bdsk-Url-1 = {http://dx.doi.org/10.1103/PhysRevA.8.401}}

@article{Weinberg:1980,
	author = {S. Weinberg},
	date-added = {2015-07-05 17:21:41 +0000},
	date-modified = {2015-07-05 17:21:41 +0000},
	journal = {In Hawking, S.W., Israel, W.: General Relativity; Cambridge University Press},
	pages = {790-831},
	title = {{Ultraviolet Divergences In Quantum Theories Of Gravitation}},
	year = {1980}}

@article{Wetterich:1992,
	author = {Wetterich, Christof},
	date-added = {2015-07-05 17:21:41 +0000},
	date-modified = {2015-07-05 17:21:41 +0000},
	doi = {10.1016/0370-2693(93)90726-X},
	journal = {Phys.Lett.},
	pages = {90-94},
	reportnumber = {HD-THEP-92-61},
	slaccitation = {%%CITATION = PHLTA,B301,90;%%},
	title = {{Exact evolution equation for the effective potential}},
	volume = {B301},
	year = {1993},
	Bdsk-Url-1 = {http://dx.doi.org/10.1016/0370-2693(93)90726-X}}

@article{Wilson:1973,
	author = {Wilson, K.G. and Kogut, John B.},
	date-added = {2015-07-05 17:21:41 +0000},
	date-modified = {2015-07-05 17:21:41 +0000},
	doi = {10.1016/0370-1573(74)90023-4},
	journal = {Phys.Rept.},
	pages = {75-200},
	slaccitation = {%%CITATION = PRPLC,12,75;%%},
	title = {{The Renormalization group and the epsilon expansion}},
	volume = {12},
	year = {1974},
	Bdsk-Url-1 = {http://dx.doi.org/10.1016/0370-1573(74)90023-4}}

@article{Wilson:1974mb,
	author = {Wilson, Kenneth G.},
	date-added = {2015-11-30 07:51:54 +0000},
	date-modified = {2015-11-30 07:51:54 +0000},
	doi = {10.1103/RevModPhys.47.773},
	journal = {Rev. Mod. Phys.},
	pages = {773},
	reportnumber = {CLNS-296},
	slaccitation = {%%CITATION = RMPHA,47,773;%%},
	title = {{The Renormalization Group: Critical Phenomena and the Kondo Problem}},
	volume = {47},
	year = {1975},
	Bdsk-Url-1 = {http://dx.doi.org/10.1103/RevModPhys.47.773}}

@article{Wilson:1971bg,
	author = {Wilson, Kenneth G.},
	date-added = {2015-11-30 07:45:12 +0000},
	date-modified = {2015-11-30 07:45:12 +0000},
	doi = {10.1103/PhysRevB.4.3174},
	journal = {Phys. Rev.},
	pages = {3174-3183},
	slaccitation = {%%CITATION = PHRVA,B4,3174;%%},
	title = {{Renormalization group and critical phenomena. 1. Renormalization group and the Kadanoff scaling picture}},
	volume = {B4},
	year = {1971},
	Bdsk-Url-1 = {http://dx.doi.org/10.1103/PhysRevB.4.3174}}

@article{Wilson:1971dh,
	author = {Wilson, Kenneth G.},
	date-added = {2015-11-30 07:50:08 +0000},
	date-modified = {2015-11-30 07:50:08 +0000},
	doi = {10.1103/PhysRevB.4.3184},
	journal = {Phys. Rev.},
	pages = {3184-3205},
	slaccitation = {%%CITATION = PHRVA,B4,3184;%%},
	title = {{Renormalization group and critical phenomena. 2. Phase space cell analysis of critical behavior}},
	volume = {B4},
	year = {1971},
	Bdsk-Url-1 = {http://dx.doi.org/10.1103/PhysRevB.4.3184}}

@article{Wilson:1971dc,
	author = {Wilson, Kenneth G. and Fisher, Michael E.},
	date-added = {2016-05-11 18:24:12 +0000},
	date-modified = {2016-05-11 18:24:12 +0000},
	doi = {10.1103/PhysRevLett.28.240},
	journal = {Phys. Rev. Lett.},
	pages = {240-243},
	slaccitation = {%%CITATION = PRLTA,28,240;%%},
	title = {{Critical exponents in 3.99 dimensions}},
	volume = {28},
	year = {1972},
	Bdsk-Url-1 = {http://dx.doi.org/10.1103/PhysRevLett.28.240}}

@article{York:1973ia,
	author = {York, Jr., James W.},
	date-added = {2016-09-01 19:07:02 +0000},
	date-modified = {2016-09-01 19:08:18 +0000},
	doi = {10.1063/1.1666338},
	journal = {J. Math. Phys.},
	pages = {456-464},
	slaccitation = {%%CITATION = JMAPA,14,456;%%},
	title = {{Conformally invariant orthogonal decomposition of symmetric tensors on Riemannian manifolds and the initial value problem of general relativity}},
	volume = {14},
	year = {1973},
	Bdsk-Url-1 = {http://dx.doi.org/10.1063/1.1666338}}

@book{Zach1,
	author = {E. C. Zachmanoglou and D. W. Thoe},
	date-added = {2016-04-28 10:38:42 +0000},
	date-modified = {2016-04-28 10:40:26 +0000},
	publisher = {Courier Dover Publications},
	title = {Introduction to Partial Differential Equations With Applications},
	year = {1986}}

@article{Zumbach:1994vg,
	author = {Zumbach, G.},
	date-added = {2016-05-16 09:30:44 +0000},
	date-modified = {2016-05-16 09:30:44 +0000},
	doi = {10.1016/0550-3213(94)90011-6},
	journal = {Nucl. Phys.},
	pages = {754-770},
	slaccitation = {%%CITATION = NUPHA,B413,754;%%},
	title = {{The Renormalization group in the local potential approximation and its applications to the O(n) model}},
	volume = {B413},
	year = {1994},
	Bdsk-Url-1 = {http://dx.doi.org/10.1016/0550-3213(94)90011-6}}

@article{Mohammedi2016,
	archiveprefix = {arXiv},
	author = {Mohammedi, N. and Morris, Tim. R.},
	eprint = {1606.06637},
	owner = {trmorris},
	primaryclass = {quant-ph},
	slaccitation = {%%CITATION = ARXIV:1606.06637;%%},
	timestamp = {2017.03.22},
	title = {{A solvable double well}},
	year = {2016}}

@book{Gradshteyn1980,
	author = {Gradshteyn, I. S. and Ryzhik, I. M.},
	journal = {New York},
	owner = {trmorris},
	pages = {951},
	publisher = {Academic Press, Inc. New York},
	timestamp = {2017.03.22},
	title = {Tables of integrals, series and products (4th Ed)},
	year = {1980}}

@article{tHooft:1974toh,
	author = {'t Hooft, Gerard and Veltman, M. J. G.},
	journal = {Ann. Inst. H. Poincare Phys. Theor.},
	owner = {trmorris},
	pages = {69-94},
	slaccitation = {%%CITATION = INSPIRE-95368;%%},
	timestamp = {2017.03.31},
	title = {{One loop divergencies in the theory of gravitation}},
	volume = {A20},
	year = {1974}}

@article{vandeVen:1991gw,
	author = {van de Ven, Anton E. M.},
	booktitle = {{Conference on Strings and Symmetries Stony Brook, New York, May 20-25, 1991}},
	doi = {10.1016/0550-3213(92)90011-Y},
	journal = {Nucl. Phys.},
	owner = {trmorris},
	pages = {309-366},
	reportnumber = {DESY-91-115, ITP-SB-91-52},
	slaccitation = {%%CITATION = NUPHA,B378,309;%%},
	timestamp = {2017.03.31},
	title = {{Two loop quantum gravity}},
	volume = {B378},
	year = {1992},
	Bdsk-Url-1 = {http://dx.doi.org/10.1016/0550-3213(92)90011-Y}}

@article{Mohr:2015ccw,
	archiveprefix = {arXiv},
	author = {Mohr, Peter J. and Newell, David B. and Taylor, Barry N.},
	doi = {10.1103/RevModPhys.88.035009},
	eprint = {1507.07956},
	journal = {Rev. Mod. Phys.},
	number = {3},
	owner = {trmorris},
	pages = {035009},
	primaryclass = {physics.atom-ph},
	slaccitation = {%%CITATION = ARXIV:1507.07956;%%},
	timestamp = {2017.03.31},
	title = {{CODATA Recommended Values of the Fundamental Physical Constants: 2014}},
	volume = {88},
	year = {2016},
	Bdsk-Url-1 = {http://dx.doi.org/10.1103/RevModPhys.88.035009}}

@article{Demmel:2014hla,
	archiveprefix = {arXiv},
	author = {Demmel, Maximilian and Saueressig, Frank and Zanusso, Omar},
	doi = {10.1016/j.aop.2015.04.018},
	eprint = {1412.7207},
	journal = {Annals Phys.},
	owner = {trmorris},
	pages = {141-165},
	primaryclass = {hep-th},
	slaccitation = {%%CITATION = ARXIV:1412.7207;%%},
	timestamp = {2017.04.11},
	title = {{RG flows of Quantum Einstein Gravity in the linear-geometric approximation}},
	volume = {359},
	year = {2015},
	Bdsk-Url-1 = {http://dx.doi.org/10.1016/j.aop.2015.04.018}}

@article{Vilkovisky:1984st,
	author = {Vilkovisky, G. A.},
	doi = {10.1016/0550-3213(84)90228-1},
	journal = {Nucl. Phys.},
	owner = {trmorris},
	pages = {125-137},
	slaccitation = {%%CITATION = NUPHA,B234,125;%%},
	timestamp = {2017.04.11},
	title = {{The Unique Effective Action in Quantum Field Theory}},
	volume = {B234},
	year = {1984},
	Bdsk-Url-1 = {http://dx.doi.org/10.1016/0550-3213(84)90228-1}}

@article{DeWitt:1998eq,
	archiveprefix = {arXiv},
	author = {DeWitt, Bryce S. and Molina-Paris, C.},
	doi = {10.1142/S0217732398002631},
	eprint = {hep-th/9808163},
	journal = {Mod. Phys. Lett.},
	owner = {trmorris},
	pages = {2475-2478},
	primaryclass = {hep-th},
	reportnumber = {LA-UR-98-1014},
	slaccitation = {%%CITATION = HEP-TH/9808163;%%},
	timestamp = {2017.04.11},
	title = {{Quantum gravity without ghosts}},
	volume = {A13},
	year = {1998},
	Bdsk-Url-1 = {http://dx.doi.org/10.1142/S0217732398002631}}

@article{Benedetti:2011ct,
	archiveprefix = {arXiv},
	author = {Benedetti, Dario},
	doi = {10.1088/1367-2630/14/1/015005},
	eprint = {1107.3110},
	journal = {New J. Phys.},
	owner = {trmorris},
	pages = {015005},
	primaryclass = {hep-th},
	reportnumber = {AEI-2011-043},
	slaccitation = {%%CITATION = ARXIV:1107.3110;%%},
	timestamp = {2017.04.11},
	title = {{Asymptotic safety goes on shell}},
	volume = {14},
	year = {2012},
	Bdsk-Url-1 = {http://dx.doi.org/10.1088/1367-2630/14/1/015005}}

@article{Belavin:1975fg,
	author = {Belavin, A. A. and Polyakov, Alexander M. and Schwartz, A. S. and Tyupkin, {\relax Yu}. S.},
	doi = {10.1016/0370-2693(75)90163-X},
	journal = {Phys. Lett.},
	owner = {trmorris},
	pages = {85-87},
	slaccitation = {%%CITATION = PHLTA,B59,85;%%},
	timestamp = {2017.05.11},
	title = {{Pseudoparticle Solutions of the Yang-Mills Equations}},
	volume = {B59},
	year = {1975},
	Bdsk-Url-1 = {http://dx.doi.org/10.1016/0370-2693(75)90163-X}}

@article{tHooft:1976snw,
	author = {'t Hooft, Gerard},
	doi = {10.1103/PhysRevD.18.2199.3, 10.1103/PhysRevD.14.3432},
	journal = {Phys. Rev.},
	note = {[Erratum: Phys. Rev.D18,2199(1978)]},
	owner = {trmorris},
	pages = {3432-3450},
	reportnumber = {PRINT-76-0551 (HARVARD)},
	slaccitation = {%%CITATION = PHRVA,D14,3432;%%},
	timestamp = {2017.05.11},
	title = {{Computation of the Quantum Effects Due to a Four-Dimensional Pseudoparticle}},
	volume = {D14},
	year = {1976},
	Bdsk-Url-1 = {http://dx.doi.org/10.1103/PhysRevD.18.2199.3,%2010.1103/PhysRevD.14.3432}}

@article{tHooft:1977xjm,
	author = {'t Hooft, Gerard},
	booktitle = {{15th Erice School of Subnuclear Physics: The Why's of Subnuclear Physics Erice, Italy, July 23-August 10, 1977}},
	journal = {Subnucl. Ser.},
	owner = {trmorris},
	pages = {943},
	reportnumber = {PRINT-77-0723 (UTRECHT)},
	slaccitation = {%%CITATION = SUSEE,15,943;%%},
	timestamp = {2017.05.11},
	title = {{Can We Make Sense Out of Quantum Chromodynamics?}},
	volume = {15},
	year = {1979}}

@article{Hasenfratz:1989pk,
	author = {Hasenfratz, P. and Leutwyler, H.},
	doi = {10.1016/0550-3213(90)90603-B},
	journal = {Nucl. Phys.},
	owner = {trmorris},
	pages = {241-284},
	reportnumber = {BUTP-89/28-BERN},
	slaccitation = {%%CITATION = NUPHA,B343,241;%%},
	timestamp = {2017.06.07},
	title = {{Goldstone Boson Related Finite Size Effects in Field Theory and Critical Phenomena With O($N$) Symmetry}},
	volume = {B343},
	year = {1990},
	Bdsk-Url-1 = {http://dx.doi.org/10.1016/0550-3213(90)90603-B}}

@article{Percacci:2016arh,
	archiveprefix = {arXiv},
	author = {Percacci, Roberto and Vacca, Gian Paolo},
	doi = {10.1140/epjc/s10052-017-4619-x},
	eprint = {1611.07005},
	journal = {Eur. Phys. J.},
	number = {1},
	owner = {trmorris},
	pages = {52},
	primaryclass = {hep-th},
	slaccitation = {%%CITATION = ARXIV:1611.07005;%%},
	timestamp = {2017.06.22},
	title = {{The background scale Ward identity in quantum gravity}},
	volume = {C77},
	year = {2017},
	Bdsk-Url-1 = {http://dx.doi.org/10.1140/epjc/s10052-017-4619-x}}

@article{Ohta:2017dsq,
	archiveprefix = {arXiv},
	author = {Ohta, Nobuyoshi},
	doi = {10.1093/ptep/ptx020},
	eprint = {1701.01506},
	journal = {PTEP},
	number = {3},
	owner = {trmorris},
	pages = {033E02},
	primaryclass = {hep-th},
	reportnumber = {KU-TP-069},
	slaccitation = {%%CITATION = ARXIV:1701.01506;%%},
	timestamp = {2017.06.22},
	title = {{Background Scale Independence in Quantum Gravity}},
	volume = {2017},
	year = {2017},
	Bdsk-Url-1 = {http://dx.doi.org/10.1093/ptep/ptx020}}

@article{Becker:2014qya,
	archiveprefix = {arXiv},
	author = {Becker, Daniel and Reuter, Martin},
	doi = {10.1016/j.aop.2014.07.023},
	eprint = {1404.4537},
	journal = {Annals Phys.},
	owner = {trmorris},
	pages = {225-301},
	primaryclass = {hep-th},
	slaccitation = {%%CITATION = ARXIV:1404.4537;%%},
	timestamp = {2017.06.22},
	title = {{En route to Background Independence: Broken split-symmetry, and how to restore it with bi-metric average actions}},
	volume = {350},
	year = {2014},
	Bdsk-Url-1 = {http://dx.doi.org/10.1016/j.aop.2014.07.023}}

@article{Appelquist:1974tg,
	author = {Appelquist, Thomas and Carazzone, J.},
	doi = {10.1103/PhysRevD.11.2856},
	journal = {Phys. Rev.},
	owner = {trmorris},
	pages = {2856},
	reportnumber = {Print-74-1486 (HARVARD)},
	slaccitation = {%%CITATION = PHRVA,D11,2856;%%},
	timestamp = {2017.06.23},
	title = {{Infrared Singularities and Massive Fields}},
	volume = {D11},
	year = {1975},
	Bdsk-Url-1 = {http://dx.doi.org/10.1103/PhysRevD.11.2856}}

@article{Symanzik:1973vg,
	author = {Symanzik, K.},
	doi = {10.1007/BF01646540},
	journal = {Commun. Math. Phys.},
	owner = {trmorris},
	pages = {7-36},
	reportnumber = {DESY-73-06},
	slaccitation = {%%CITATION = CMPHA,34,7;%%},
	timestamp = {2017.06.23},
	title = {{Infrared singularities and small distance behavior analysis}},
	volume = {34},
	year = {1973},
	Bdsk-Url-1 = {http://dx.doi.org/10.1007/BF01646540}}

@article{Bergere:1975tr,
	author = {Bergere, M. C. and Lam, Yuk-Ming P.},
	doi = {10.1103/PhysRevD.13.3247},
	journal = {Phys. Rev.},
	owner = {trmorris},
	pages = {3247-3255},
	reportnumber = {SACLAY-DPh-T/75/65},
	slaccitation = {%%CITATION = PHRVA,D13,3247;%%},
	timestamp = {2017.07.17},
	title = {{Equivalence Theorem and Faddeev-Popov Ghosts}},
	volume = {D13},
	year = {1976},
	Bdsk-Url-1 = {http://dx.doi.org/10.1103/PhysRevD.13.3247}}

@book{Itzykson:1980rh,
	address = {New York},
	author = {Itzykson, C. and Zuber, J. B.},
	isbn = {9780486445687, 0486445682},
	journal = {New York, Usa: Mcgraw-hill (1980) 705 P.(International Series In Pure and Applied Physics)},
	owner = {trmorris},
	publisher = {McGraw-Hill},
	series = {International Series In Pure and Applied Physics},
	slaccitation = {%%CITATION = INSPIRE-159194;%%},
	timestamp = {2017.07.17},
	title = {{Quantum Field Theory}},
	url = {http://dx.doi.org/10.1063/1.2916419},
	year = {1980},
	Bdsk-Url-1 = {http://dx.doi.org/10.1063/1.2916419}}

@article{Weinberg:1964ew,
	author = {Weinberg, Steven},
	doi = {10.1103/PhysRev.135.B1049},
	journal = {Phys. Rev.},
	owner = {trmorris},
	pages = {B1049-B1056},
	slaccitation = {%%CITATION = PHRVA,135,B1049;%%},
	timestamp = {2017.10.11},
	title = {{Photons and Gravitons in s Matrix Theory: Derivation of Charge Conservation and Equality of Gravitational and Inertial Mass}},
	volume = {135},
	year = {1964},
	Bdsk-Url-1 = {http://dx.doi.org/10.1103/PhysRev.135.B1049}}

@article{Weinberg:1965rz,
	author = {Weinberg, Steven},
	doi = {10.1103/PhysRev.138.B988},
	journal = {Phys. Rev.},
	owner = {trmorris},
	pages = {B988-B1002},
	slaccitation = {%%CITATION = PHRVA,138,B988;%%},
	timestamp = {2017.10.11},
	title = {{Photons and gravitons in perturbation theory: Derivation of Maxwell's and Einstein's equations}},
	volume = {138},
	year = {1965},
	Bdsk-Url-1 = {http://dx.doi.org/10.1103/PhysRev.138.B988}}

@article{Ogievetsky:1965,
	author = {Ogievetsky, V. I. and Polubarinov, I. V.},
	journal = {Annals Phys.},
	owner = {trmorris},
	pages = {167},
	timestamp = {2017.10.11},
	title = {Interacting field of spin 2 and the Einstein equations},
	volume = {35},
	year = {1965}}

@article{Boulware:1974sr,
	author = {Boulware, David G. and Deser, Stanley},
	doi = {10.1016/0003-4916(75)90302-4},
	journal = {Annals Phys.},
	owner = {trmorris},
	pages = {193},
	reportnumber = {RLO-1388-675},
	slaccitation = {%%CITATION = APNYA,89,193;%%},
	timestamp = {2017.10.11},
	title = {{Classical General Relativity Derived from Quantum Gravity}},
	volume = {89},
	year = {1975},
	Bdsk-Url-1 = {http://dx.doi.org/10.1016/0003-4916(75)90302-4}}

@article{Deser:1969wk,
	archiveprefix = {arXiv},
	author = {Deser, Stanley},
	doi = {10.1007/BF00759198},
	eprint = {gr-qc/0411023},
	journal = {Gen. Rel. Grav.},
	owner = {trmorris},
	pages = {9-18},
	primaryclass = {gr-qc},
	slaccitation = {%%CITATION = GR-QC/0411023;%%},
	timestamp = {2017.10.11},
	title = {{Selfinteraction and gauge invariance}},
	volume = {1},
	year = {1970},
	Bdsk-Url-1 = {http://dx.doi.org/10.1007/BF00759198}}

@article{Gupta:1954zz,
	author = {Gupta, Suraj N.},
	doi = {10.1103/PhysRev.96.1683},
	journal = {Phys. Rev.},
	owner = {trmorris},
	pages = {1683-1685},
	slaccitation = {%%CITATION = PHRVA,96,1683;%%},
	timestamp = {2017.10.11},
	title = {{Gravitation and Electromagnetism}},
	volume = {96},
	year = {1954},
	Bdsk-Url-1 = {http://dx.doi.org/10.1103/PhysRev.96.1683}}

@article{Kraichnan:1955zz,
	author = {Kraichnan, Robert H.},
	doi = {10.1103/PhysRev.98.1118},
	journal = {Phys. Rev.},
	owner = {trmorris},
	pages = {1118-1122},
	slaccitation = {%%CITATION = PHRVA,98,1118;%%},
	timestamp = {2017.10.11},
	title = {{Special-Relativistic Derivation of Generally Covariant Gravitation Theory}},
	volume = {98},
	year = {1955},
	Bdsk-Url-1 = {http://dx.doi.org/10.1103/PhysRev.98.1118}}

@book{Feynman:1996kb,
	author = {Feynman, R. P.},
	editor = {Morinigo, F. B. and Wagner, W. G. and Hatfield, B.},
	journal = {Reading, USA: Addison-Wesley (1995) 232 p. (The advanced book program)},
	owner = {trmorris},
	slaccitation = {%%CITATION = INSPIRE-427379;%%},
	timestamp = {2017.10.11},
	title = {{Feynman lectures on gravitation}},
	year = {1996}}

@article{Wyss:1965,
	author = {Wyss, W},
	date-modified = {2018-03-02 18:50:52 +0000},
	journal = {Helvetica Physica Acta},
	owner = {trmorris},
	pages = {469},
	timestamp = {2017.10.11},
	title = {Zur {Unizit\"at} der Gravitationstheorie},
	volume = {38},
	year = {1965}}

@article{Boulanger:2000rq,
	archiveprefix = {arXiv},
	author = {Boulanger, Nicolas and Damour, Thibault and Gualtieri, Leonardo and Henneaux, Marc},
	doi = {10.1016/S0550-3213(00)00718-5},
	eprint = {hep-th/0007220},
	journal = {Nucl. Phys.},
	owner = {trmorris},
	pages = {127-171},
	primaryclass = {hep-th},
	reportnumber = {ULB-TH-00-14},
	slaccitation = {%%CITATION = HEP-TH/0007220;%%},
	timestamp = {2017.10.11},
	title = {{Inconsistency of interacting, multigraviton theories}},
	volume = {B597},
	year = {2001},
	Bdsk-Url-1 = {http://dx.doi.org/10.1016/S0550-3213(00)00718-5}}

@article{Fang:1978rc,
	author = {Fang, J. and Fronsdal, C.},
	doi = {10.1063/1.524007},
	journal = {J. Math. Phys.},
	owner = {trmorris},
	pages = {2264-2271},
	reportnumber = {UCLA/78/TEP/6},
	slaccitation = {%%CITATION = JMAPA,20,2264;%%},
	timestamp = {2017.10.11},
	title = {{Deformation of Gauge Groups. Gravitation}},
	volume = {20},
	year = {1979},
	Bdsk-Url-1 = {http://dx.doi.org/10.1063/1.524007}}

@article{Wald:1986bj,
	author = {Wald, Robert M.},
	doi = {10.1103/PhysRevD.33.3613},
	journal = {Phys. Rev.},
	owner = {trmorris},
	pages = {3613},
	reportnumber = {EFI-85-88-CHICAGO},
	slaccitation = {%%CITATION = PHRVA,D33,3613;%%},
	timestamp = {2017.10.11},
	title = {{Spin-2 Fields and General Covariance}},
	volume = {D33},
	year = {1986},
	Bdsk-Url-1 = {http://dx.doi.org/10.1103/PhysRevD.33.3613}}

@article{Bautista:2017enk,
	archiveprefix = {arXiv},
	eprint = {1711.00135},
	owner = {trmorris},
	primaryclass = {hep-th},
	slaccitation = {%%CITATION = ARXIV:1711.00135;%%},
	timestamp = {2017.12.12},
	title = {{Nonlocal Quantum Effective Actions in Weyl-Flat Spacetimes}},
	year = {2017}}

@article{Nieto:2017ddk,
	archiveprefix = {arXiv},
	author = {Nieto, Carlos M. and Percacci, Roberto and Skrinjar, Vedran},
	doi = {10.1103/PhysRevD.96.106019},
	eprint = {1708.09760},
	journal = {Phys. Rev.},
	number = {10},
	owner = {trmorris},
	pages = {106019},
	primaryclass = {gr-qc},
	slaccitation = {%%CITATION = ARXIV:1708.09760;%%},
	timestamp = {2017.12.12},
	title = {{Split Weyl transformations in quantum gravity}},
	volume = {D96},
	year = {2017},
	Bdsk-Url-1 = {https://dx.doi.org/10.1103/PhysRevD.96.106019}}

@article{Codello:2015ana,
	archiveprefix = {arXiv},
	author = {Codello, Alessandro and D'Odorico, Giulio and Pagani, Carlo},
	doi = {10.1103/PhysRevD.91.125016},
	eprint = {1502.02439},
	journal = {Phys. Rev.},
	number = {12},
	owner = {trmorris},
	pages = {125016},
	primaryclass = {hep-th},
	reportnumber = {CP3-ORIGINS-2015-3-DNRF90-AND-DIAS-2015-3},
	slaccitation = {%%CITATION = ARXIV:1502.02439;%%},
	timestamp = {2017.12.12},
	title = {{Functional and Local Renormalization Groups}},
	volume = {D91},
	year = {2015},
	Bdsk-Url-1 = {https://dx.doi.org/10.1103/PhysRevD.91.125016}}

@article{Hayakawa:2008an,
	archiveprefix = {arXiv},
	author = {Hayakawa, Masashi and Uno, Shunpei},
	doi = {10.1143/PTP.120.413},
	eprint = {0804.2044},
	journal = {Prog. Theor. Phys.},
	owner = {trmorris},
	pages = {413-441},
	primaryclass = {hep-ph},
	slaccitation = {%%CITATION = ARXIV:0804.2044;%%},
	timestamp = {2018.02.19},
	title = {{QED in finite volume and finite size scaling effect on electromagnetic properties of hadrons}},
	volume = {120},
	year = {2008},
	Bdsk-Url-1 = {https://dx.doi.org/10.1143/PTP.120.413}}

@book{MR1117903,
	author = {Berndt, Bruce C.},
	doi = {10.1007/978-1-4612-0965-2},
	isbn = {0-387-97503-9},
	keywords = {01A75 (11F11 33E05)},
	mrnumber = {1117903},
	owner = {trmorris},
	pages = {xiv+510},
	publisher = {Springer-Verlag, New York},
	timestamp = {2018.02.21},
	title = {Ramanujan's notebooks. {P}art {III}},
	year = {1997},
	Bdsk-Url-1 = {https://dx.doi.org/10.1007/978-1-4612-0965-2}}

@book{MR0344216,
	author = {Serre, J.-P.},
	keywords = {12-02 (10CXX 10DXX)},
	mrnumber = {0344216},
	note = {Translated from the French, Graduate Texts in Mathematics, No. 7},
	owner = {trmorris},
	pages = {viii+115},
	publisher = {Springer-Verlag, New York-Heidelberg},
	timestamp = {2018.02.21},
	title = {A course in arithmetic},
	year = {1973}}

@article{Shapiro:1972ph,
	author = {Shapiro, Joel A.},
	doi = {10.1103/PhysRevD.5.1945},
	journal = {Phys. Rev.},
	owner = {trmorris},
	pages = {1945-1948},
	slaccitation = {%%CITATION = PHRVA,D5,1945;%%},
	timestamp = {2018.02.23},
	title = {{Loop graph in the dual tube model}},
	volume = {D5},
	year = {1972},
	Bdsk-Url-1 = {https://dx.doi.org/10.1103/PhysRevD.5.1945}}

@article{Polchinski:1985zf,
	author = {Polchinski, Joseph},
	doi = {10.1007/BF01210791},
	journal = {Commun. Math. Phys.},
	owner = {trmorris},
	pages = {37},
	reportnumber = {UTTG-13-85},
	slaccitation = {%%CITATION = CMPHA,104,37;%%},
	timestamp = {2018.02.23},
	title = {{Evaluation of the One Loop String Path Integral}},
	volume = {104},
	year = {1986},
	Bdsk-Url-1 = {https://dx.doi.org/10.1007/BF01210791}}

@article{Igarashi:2009tj,
	archiveprefix = {arXiv},
	author = {Igarashi, Yuji and Itoh, Katsumi and Sonoda, Hidenori},
	doi = {10.1143/PTPS.181.1},
	eprint = {0909.0327},
	journal = {Prog. Theor. Phys. Suppl.},
	owner = {trmorris},
	pages = {1-166},
	primaryclass = {hep-th},
	reportnumber = {KOBE-TH-0907},
	slaccitation = {%%CITATION = ARXIV:0909.0327;%%},
	timestamp = {2018.02.26},
	title = {{Realization of Symmetry in the ERG Approach to Quantum Field Theory}},
	volume = {181},
	year = {2009},
	Bdsk-Url-1 = {https://dx.doi.org/10.1143/PTPS.181.1}}

@article{Becchi:1996an,
	archiveprefix = {arXiv},
	author = {Becchi, C.},
	eprint = {hep-th/9607188},
	journal = {Parma summer school in theoretical physics},
	owner = {trmorris},
	primaryclass = {hep-th},
	reportnumber = {GEF-TH-96-11},
	slaccitation = {%%CITATION = HEP-TH/9607188;%%},
	timestamp = {2018.02.26},
	title = {{On the construction of renormalized gauge theories using renormalization group techniques}},
	year = {1996}}

@article{Gomis:1994he,
	archiveprefix = {arXiv},
	author = {Gomis, Joaquim and Paris, Jordi and Samuel, Stuart},
	doi = {10.1016/0370-1573(94)00112-G},
	eprint = {hep-th/9412228},
	journal = {Phys. Rept.},
	owner = {trmorris},
	pages = {1-145},
	primaryclass = {hep-th},
	reportnumber = {CCNY-HEP-94-03, KUL-TF-94-12, UB-ECM-PF-94-15, UTTG-11-94},
	slaccitation = {%%CITATION = HEP-TH/9412228;%%},
	timestamp = {2018.02.26},
	title = {{Antibracket, antifields and gauge theory quantization}},
	volume = {259},
	year = {1995},
	Bdsk-Url-1 = {https://dx.doi.org/10.1016/0370-1573(94)00112-G}}

@article{Troost:1989cu,
	author = {Troost, W. and van Nieuwenhuizen, P. and Van Proeyen, Antoine},
	doi = {10.1016/0550-3213(90)90137-3},
	journal = {Nucl. Phys.},
	owner = {trmorris},
	pages = {727-770},
	reportnumber = {CERN-TH-5512-89, KUL-TF-89-25},
	slaccitation = {%%CITATION = NUPHA,B333,727;%%},
	timestamp = {2018.02.26},
	title = {{Anomalies and the Batalin-Vilkovisky Lagrangian Formalism}},
	volume = {B333},
	year = {1990},
	Bdsk-Url-1 = {https://dx.doi.org/10.1016/0550-3213(90)90137-3}}

@article{Batalin:1984ss,
	author = {Batalin, I. A. and Vilkovisky, G. A.},
	doi = {10.1016/0550-3213(84)90227-X},
	journal = {Nucl. Phys.},
	owner = {trmorris},
	pages = {106-124},
	slaccitation = {%%CITATION = NUPHA,B234,106;%%},
	timestamp = {2018.02.26},
	title = {{Closure of the Gauge Algebra, Generalized Lie Equations and Feynman Rules}},
	volume = {B234},
	year = {1984},
	Bdsk-Url-1 = {https://dx.doi.org/10.1016/0550-3213(84)90227-X}}

@article{Barnich:1994db,
	archiveprefix = {arXiv},
	author = {Barnich, Glenn and Brandt, Friedemann and Henneaux, Marc},
	doi = {10.1007/BF02099464},
	eprint = {hep-th/9405109},
	journal = {Commun. Math. Phys.},
	owner = {trmorris},
	pages = {57-92},
	primaryclass = {hep-th},
	reportnumber = {ULB-TH-94-06, NIKHEF-H-94-13},
	slaccitation = {%%CITATION = HEP-TH/9405109;%%},
	timestamp = {2018.02.26},
	title = {{Local BRST cohomology in the antifield formalism. 1. General theorems}},
	volume = {174},
	year = {1995},
	Bdsk-Url-1 = {https://dx.doi.org/10.1007/BF02099464}}

@article{Igarashi:2001mf,
	archiveprefix = {arXiv},
	author = {Igarashi, Yuji and Itoh, Katsumi and So, Hiroto},
	doi = {10.1143/PTP.106.149},
	eprint = {hep-th/0101101},
	journal = {Prog. Theor. Phys.},
	owner = {trmorris},
	pages = {149-166},
	primaryclass = {hep-th},
	reportnumber = {NIIG-DP-00-5},
	slaccitation = {%%CITATION = HEP-TH/0101101;%%},
	timestamp = {2018.02.26},
	title = {{BRS symmetry, the quantum master equation, and the Wilsonian renormalization group}},
	volume = {106},
	year = {2001},
	Bdsk-Url-1 = {https://dx.doi.org/10.1143/PTP.106.149}}

@article{Fisch:1989rp,
	author = {Fisch, Jean M. L. and Henneaux, Marc},
	doi = {10.1007/BF02096877},
	journal = {Commun. Math. Phys.},
	owner = {trmorris},
	pages = {627},
	reportnumber = {ULB-TH2-89-02},
	slaccitation = {%%CITATION = CMPHA,128,627;%%},
	timestamp = {2018.02.26},
	title = {{Homological Perturbation Theory and the Algebraic Structure of the Antifield - Antibracket Formalism for Gauge Theories}},
	volume = {128},
	year = {1990},
	Bdsk-Url-1 = {https://dx.doi.org/10.1007/BF02096877}}

@article{Batalin:1981jr,
	author = {Batalin, I. A. and Vilkovisky, G. A.},
	booktitle = {{Second Seminar on Quantum Gravity Moscow, USSR, October 13-15, 1981}},
	doi = {10.1016/0370-2693(81)90205-7},
	journal = {Phys. Lett.},
	note = {[,463(1981)]},
	owner = {trmorris},
	pages = {27-31},
	slaccitation = {%%CITATION = PHLTA,102B,27;%%},
	timestamp = {2018.02.26},
	title = {{Gauge Algebra and Quantization}},
	volume = {102B},
	year = {1981},
	Bdsk-Url-1 = {https://dx.doi.org/10.1016/0370-2693(81)90205-7}}

@article{Batalin:1984jr,
	author = {Batalin, I. A. and Vilkovisky, G. A.},
	doi = {10.1103/PhysRevD.28.2567, 10.1103/PhysRevD.30.508},
	journal = {Phys. Rev.},
	note = {[Erratum: Phys. Rev.D30,508(1984)]},
	owner = {trmorris},
	pages = {2567-2582},
	slaccitation = {%%CITATION = PHRVA,D28,2567;%%},
	timestamp = {2018.02.26},
	title = {{Quantization of Gauge Theories with Linearly Dependent Generators}},
	volume = {D28},
	year = {1983},
	Bdsk-Url-1 = {https://dx.doi.org/10.1103/PhysRevD.28.2567,%2010.1103/PhysRevD.30.508}}

@article{Henneaux:1990rx,
	author = {Henneaux, Marc},
	doi = {10.1007/BF02099287},
	journal = {Commun. Math. Phys.},
	owner = {trmorris},
	pages = {1-14},
	reportnumber = {ULB-TH2-90-02, CECS-PHYS-8-90},
	slaccitation = {%%CITATION = CMPHA,140,1;%%},
	timestamp = {2018.02.26},
	title = {{Space-time Locality of the {BRST} Formalism}},
	volume = {140},
	year = {1991},
	Bdsk-Url-1 = {https://dx.doi.org/10.1007/BF02099287}}

@article{Barnich:1993vg,
	archiveprefix = {arXiv},
	author = {Barnich, Glenn and Henneaux, Marc},
	doi = {10.1016/0370-2693(93)90544-R},
	eprint = {hep-th/9304057},
	journal = {Phys. Lett.},
	owner = {trmorris},
	pages = {123-129},
	primaryclass = {hep-th},
	reportnumber = {ULB-PMIF-93-01},
	slaccitation = {%%CITATION = HEP-TH/9304057;%%},
	timestamp = {2018.02.26},
	title = {{Consistent couplings between fields with a gauge freedom and deformations of the master equation}},
	volume = {B311},
	year = {1993},
	Bdsk-Url-1 = {https://dx.doi.org/10.1016/0370-2693(93)90544-R}}

@article{Henneaux:1997bm,
	archiveprefix = {arXiv},
	author = {Henneaux, Marc},
	booktitle = {{Proceedings, Conference on Secondary Calculus and Cohomological Physics: Moscow, Russia, Aug 24-31, 1997}},
	doi = {10.1090/conm/219/03070},
	eprint = {hep-th/9712226},
	journal = {Contemp. Math.},
	owner = {trmorris},
	pages = {93-110},
	primaryclass = {hep-th},
	slaccitation = {%%CITATION = HEP-TH/9712226;%%},
	timestamp = {2018.02.26},
	title = {{Consistent interactions between gauge fields: The Cohomological approach}},
	volume = {219},
	year = {1998},
	Bdsk-Url-1 = {https://dx.doi.org/10.1090/conm/219/03070}}

@article{Gomez:2015bsa,
	archiveprefix = {arXiv},
	author = {Lucena G{\'o}mez, Gustavo},
	eprint = {1508.07226},
	owner = {trmorris},
	primaryclass = {hep-th},
	slaccitation = {%%CITATION = ARXIV:1508.07226;%%},
	timestamp = {2018.02.26},
	title = {{The Elegance of Cohomological Methods}},
	year = {2015}}

@article{Bakeyev:1996is,
	archiveprefix = {arXiv},
	author = {Bakeyev, Timur D. and Slavnov, A. A.},
	doi = {10.1142/S0217732396001533},
	eprint = {hep-th/9601092},
	journal = {Mod. Phys. Lett.},
	owner = {trmorris},
	pages = {1539-1554},
	primaryclass = {hep-th},
	reportnumber = {SMI-02-96},
	slaccitation = {%%CITATION = HEP-TH/9601092;%%},
	timestamp = {2018.02.27},
	title = {{Higher covariant derivative regularization revisited}},
	volume = {A11},
	year = {1996},
	Bdsk-Url-1 = {https://dx.doi.org/10.1142/S0217732396001533}}

@article{Asorey:1995tq,
	archiveprefix = {arXiv},
	author = {Asorey, Manuel and Falceto, Fernando},
	doi = {10.1103/PhysRevD.54.5290},
	eprint = {hep-th/9502025},
	journal = {Phys. Rev.},
	owner = {trmorris},
	pages = {5290-5301},
	primaryclass = {hep-th},
	reportnumber = {DFTUZ-95-3},
	slaccitation = {%%CITATION = HEP-TH/9502025;%%},
	timestamp = {2018.02.27},
	title = {{On the consistency of the regularization of gauge theories by high covariant derivatives}},
	volume = {D54},
	year = {1996},
	Bdsk-Url-1 = {https://dx.doi.org/10.1103/PhysRevD.54.5290}}

@article{Warr:1986we,
	author = {Warr, Brian J.},
	doi = {10.1016/0003-4916(88)90245-X},
	journal = {Annals Phys.},
	owner = {trmorris},
	pages = {1},
	reportnumber = {CALT-68-1334B},
	slaccitation = {%%CITATION = APNYA,183,1;%%},
	timestamp = {2018.02.27},
	title = {{Renormalization of Gauge Theories Using Effective Lagrangians. 1.}},
	volume = {183},
	year = {1988},
	Bdsk-Url-1 = {https://dx.doi.org/10.1016/0003-4916(88)90245-X}}

@article{Higashi:2007ax,
	archiveprefix = {arXiv},
	author = {Higashi, Takeshi and Itou, Etsuko and Kugo, Taichiro},
	doi = {10.1143/PTP.118.1115},
	eprint = {0709.1522},
	journal = {Prog. Theor. Phys.},
	owner = {trmorris},
	pages = {1115-1125},
	primaryclass = {hep-th},
	reportnumber = {OU-HET584, YITP-07-55},
	slaccitation = {%%CITATION = ARXIV:0709.1522;%%},
	timestamp = {2018.02.27},
	title = {{The BV Master Equation for the Wilson Action in general Yang-Mills Gauge Theory}},
	volume = {118},
	year = {2007},
	Bdsk-Url-1 = {https://dx.doi.org/10.1143/PTP.118.1115}}

@article{Igarashi:2007fw,
	archiveprefix = {arXiv},
	author = {Igarashi, Yuji and Itoh, Katsumi and Sonoda, Hidenori},
	doi = {10.1143/PTP.118.121},
	eprint = {0704.2349},
	journal = {Prog. Theor. Phys.},
	owner = {trmorris},
	pages = {121-134},
	primaryclass = {hep-th},
	slaccitation = {%%CITATION = ARXIV:0704.2349;%%},
	timestamp = {2018.02.27},
	title = {{Quantum master equation for QED in exact renormalization group}},
	volume = {118},
	year = {2007},
	Bdsk-Url-1 = {https://dx.doi.org/10.1143/PTP.118.121}}

@article{ZinnJustin:2002ru,
	author = {Zinn-Justin, Jean},
	journal = {Int. Ser. Monogr. Phys.},
	owner = {trmorris},
	pages = {1-1054},
	slaccitation = {%%CITATION = IMPHA,113,1;%%},
	timestamp = {2018.02.27},
	title = {{Quantum field theory and critical phenomena}},
	volume = {113},
	year = {2002}}

@article{Siegel:1989nh,
	author = {Siegel, W.},
	doi = {10.1142/S0217751X89001618},
	journal = {Int. J. Mod. Phys.},
	owner = {trmorris},
	pages = {3951},
	reportnumber = {ITP-SB-89-14},
	slaccitation = {%%CITATION = IMPAE,A4,3951;%%},
	timestamp = {2018.02.28},
	title = {{Batalin-vilkovisky From Hamiltonian {BRST}}},
	volume = {A4},
	year = {1989},
	Bdsk-Url-1 = {https://dx.doi.org/10.1142/S0217751X89001618}}

@article{Bergshoeff:1991hb,
	author = {Bergshoeff, E. and Kallosh, R. and Van Proeyen, Antoine},
	doi = {10.1088/0264-9381/9/2/003},
	journal = {Class. Quant. Grav.},
	owner = {trmorris},
	pages = {321-360},
	reportnumber = {KUL-TF-91-05, CERN-TH-6020-91, SU-ITP-888},
	slaccitation = {%%CITATION = CQGRD,9,321;%%},
	timestamp = {2018.02.28},
	title = {{Superparticle actions and gauge fixings}},
	volume = {9},
	year = {1992},
	Bdsk-Url-1 = {https://dx.doi.org/10.1088/0264-9381/9/2/003}}

@article{Troost:1994xw,
	archiveprefix = {arXiv},
	author = {Troost, Walter and Van Proeyen, Antoine},
	booktitle = {{Strings and symmetries. Proceedings, 1st Gursey Memorial Conference, Istanbul, Turkey, June 6-10, 1994}},
	doi = {10.1007/3-540-59163-X_268},
	eprint = {hep-th/9410162},
	note = {[Lect. Notes Phys.447,183(1995)]},
	owner = {trmorris},
	primaryclass = {hep-th},
	reportnumber = {KUL-TF-94-27},
	slaccitation = {%%CITATION = HEP-TH/9410162;%%},
	timestamp = {2018.02.28},
	title = {{Regularization, the BV method, and the antibracket cohomology}},
	year = {1994},
	Bdsk-Url-1 = {https://dx.doi.org/10.1007/3-540-59163-X_268}}

@article{Siegel:1989ip,
	author = {Siegel, W.},
	doi = {10.1142/S0217751X89001485},
	journal = {Int. J. Mod. Phys.},
	owner = {trmorris},
	pages = {3705},
	reportnumber = {ITP-SB-89-09},
	slaccitation = {%%CITATION = IMPAE,A4,3705;%%},
	timestamp = {2018.02.28},
	title = {{Relation Between Batalin-vilkovisky and First Quantized Style {BRST}}},
	volume = {A4},
	year = {1989},
	Bdsk-Url-1 = {https://dx.doi.org/10.1142/S0217751X89001485}}

@inproceedings{VanProeyen:1991mp,
	archiveprefix = {arXiv},
	author = {Van Proeyen, Antoine},
	booktitle = {{Strings and Symmetries 1991 Stony Brook, New York, May 20-25, 1991}},
	eprint = {hep-th/9109036},
	owner = {trmorris},
	pages = {0388-406},
	primaryclass = {hep-th},
	reportnumber = {KUL-TF-91-35},
	slaccitation = {%%CITATION = HEP-TH/9109036;%%},
	timestamp = {2018.02.28},
	title = {{Batalin-Vilkovisky Lagrangian quantization}},
	year = {1991}}

@article{Tyutin:1975qk,
	archiveprefix = {arXiv},
	author = {Tyutin, I. V.},
	eprint = {0812.0580},
	owner = {trmorris},
	primaryclass = {hep-th},
	reportnumber = {LEBEDEV-75-39},
	slaccitation = {%%CITATION = ARXIV:0812.0580;%%},
	timestamp = {2018.02.28},
	title = {{Gauge Invariance in Field Theory and Statistical Physics in Operator Formalism}},
	year = {1975}}

@inproceedings{koszul1950type,
	author = {Koszul, Jean-Louis},
	booktitle = {Colloque de Topologie, Bruxelles -or-},
	owner = {trmorris},
	pages = {5 (73--81)},
	series = {Bull. Soc. Math. France},
	timestamp = {2018.03.05},
	title = {Sur un type {d'algebr\'es differ\'entielles} en rapport avec la transgression},
	volume = {78},
	year = {1950}}

@article{borel1953cohomologie,
	author = {Borel, Armand},
	journal = {Ann. Math.},
	owner = {trmorris},
	pages = {115--207},
	publisher = {JSTOR},
	timestamp = {2018.03.05},
	title = {Sur la cohomologie des espaces fibr{\'e}s principaux et des espaces homogenes de groupes de Lie compacts},
	volume = {57},
	year = {1953}}

@article{tate1957homology,
	author = {Tate, John},
	journal = {Illinois J. Math.},
	owner = {trmorris},
	pages = {14--27},
	publisher = {University of Illinois at Urbana-Champaign, Department of Mathematics},
	timestamp = {2018.03.05},
	title = {Homology of Noetherian rings and local rings},
	volume = {1},
	year = {1957}}

@article{Igarashi:1999rm,
	archiveprefix = {arXiv},
	author = {Igarashi, Yuji and Itoh, Katsumi and So, Hiroto},
	doi = {10.1016/S0370-2693(00)00305-1},
	eprint = {hep-th/9912262},
	journal = {Phys. Lett.},
	owner = {trmorris},
	pages = {336-342},
	primaryclass = {hep-th},
	reportnumber = {NIIG-DP-99-2},
	slaccitation = {%%CITATION = HEP-TH/9912262;%%},
	timestamp = {2018.03.06},
	title = {{Exact symmetries realized on the renormalization group flow}},
	volume = {B479},
	year = {2000},
	Bdsk-Url-1 = {https://dx.doi.org/10.1016/S0370-2693(00)00305-1}}

@article{Igarashi:2000vf,
	archiveprefix = {arXiv},
	author = {Igarashi, Yuji and Itoh, Katsumi and So, Hiroto},
	doi = {10.1143/PTP.104.1053},
	eprint = {hep-th/0006180},
	journal = {Prog. Theor. Phys.},
	owner = {trmorris},
	pages = {1053-1066},
	primaryclass = {hep-th},
	reportnumber = {NIIG-DP-00-1},
	slaccitation = {%%CITATION = HEP-TH/0006180;%%},
	timestamp = {2018.03.06},
	title = {{Exact BRS symmetry realized on the renormalization group flow}},
	volume = {104},
	year = {2000},
	Bdsk-Url-1 = {https://dx.doi.org/10.1143/PTP.104.1053}}

@inproceedings{Sonoda:2007av,
	archiveprefix = {arXiv},
	author = {Sonoda, Hidenori},
	eprint = {0710.1662},
	owner = {trmorris},
	primaryclass = {hep-th},
	reportnumber = {KOBE-TH-07-10},
	slaccitation = {%%CITATION = ARXIV:0710.1662;%%},
	timestamp = {2018.03.07},
	title = {{The Exact Renormalization Group: Renormalization theory revisited}},
	url = {http://inspirehep.net/record/763596/files/arXiv:0710.1662.pdf},
	year = {2007},
	Bdsk-Url-1 = {http://inspirehep.net/record/763596/files/arXiv:0710.1662.pdf}}

@article{Ellwanger:1994iz,
	archiveprefix = {arXiv},
	author = {Ellwanger, Ulrich},
	doi = {10.1016/0370-2693(94)90365-4},
	eprint = {hep-th/9402077},
	journal = {Phys. Lett.},
	owner = {trmorris},
	pages = {364-370},
	primaryclass = {hep-th},
	reportnumber = {HD-THEP-94-2},
	slaccitation = {%%CITATION = HEP-TH/9402077;%%},
	timestamp = {2018.03.07},
	title = {{Flow equations and BRS invariance for Yang-Mills theories}},
	volume = {B335},
	year = {1994},
	Bdsk-Url-1 = {https://dx.doi.org/10.1016/0370-2693(94)90365-4}}

@article{Bonini:1994kp,
	archiveprefix = {arXiv},
	author = {Bonini, M. and D'Attanasio, M. and Marchesini, G.},
	doi = {10.1016/0550-3213(94)00569-Z},
	eprint = {hep-th/9410138},
	journal = {Nucl. Phys.},
	owner = {trmorris},
	pages = {163-186},
	primaryclass = {hep-th},
	reportnumber = {UPRF-94-412},
	slaccitation = {%%CITATION = HEP-TH/9410138;%%},
	timestamp = {2018.03.07},
	title = {{BRS symmetry for Yang-Mills theory with exact renormalization group}},
	volume = {B437},
	year = {1995},
	Bdsk-Url-1 = {https://dx.doi.org/10.1016/0550-3213(94)00569-Z}}

@article{Efremov:2017sqi,
	archiveprefix = {arXiv},
	author = {Efremov, Alexander N. and Guida, Riccardo and Kopper, Christoph},
	doi = {10.1063/1.5000041},
	eprint = {1704.06799},
	journal = {J. Math. Phys.},
	number = {9},
	owner = {trmorris},
	pages = {093503},
	primaryclass = {math-ph},
	reportnumber = {RR-018-04-2017, T17-055},
	slaccitation = {%%CITATION = ARXIV:1704.06799;%%},
	timestamp = {2018.03.07},
	title = {{Renormalization of SU(2) Yang-Mills theory with flow equations}},
	volume = {58},
	year = {2017},
	Bdsk-Url-1 = {https://dx.doi.org/10.1063/1.5000041}}

@article{Frob:2015uqy,
	archiveprefix = {arXiv},
	author = {{Fr\"ob}, Markus B. and Holland, Jan and Hollands, Stefan},
	doi = {10.1063/1.4967747},
	eprint = {1511.09425},
	journal = {J. Math. Phys.},
	number = {12},
	owner = {trmorris},
	pages = {122301},
	primaryclass = {math-ph},
	slaccitation = {%%CITATION = ARXIV:1511.09425;%%},
	timestamp = {2018.03.07},
	title = {{All-order bounds for correlation functions of gauge-invariant operators in Yang-Mills theory}},
	volume = {57},
	year = {2016},
	Bdsk-Url-1 = {https://dx.doi.org/10.1063/1.4967747}}

@article{Alvarez:2006uu,
	archiveprefix = {arXiv},
	author = {Alvarez, E. and Blas, D. and Garriga, J. and Verdaguer, E.},
	doi = {10.1016/j.nuclphysb.2006.08.003},
	eprint = {hep-th/0606019},
	journal = {Nucl. Phys.},
	owner = {trmorris},
	pages = {148-170},
	primaryclass = {hep-th},
	reportnumber = {IFT-UAM-CSIC-05-43},
	slaccitation = {%%CITATION = HEP-TH/0606019;%%},
	timestamp = {2018.05.23},
	title = {{Transverse Fierz-Pauli symmetry}},
	volume = {B756},
	year = {2006},
	Bdsk-Url-1 = {https://doi.org/10.1016/j.nuclphysb.2006.08.003}}

@article{Buchmuller:1988wx,
	author = {Buchmuller, W. and Dragon, N.},
	doi = {10.1016/0370-2693(88)90577-1},
	journal = {Phys. Lett.},
	owner = {trmorris},
	pages = {292-294},
	reportnumber = {DESY-88-029, ITP-UH-2/88},
	slaccitation = {%%CITATION = PHLTA,B207,292;%%},
	timestamp = {2018.05.23},
	title = {{Einstein Gravity From Restricted Coordinate Invariance}},
	volume = {B207},
	year = {1988},
	Bdsk-Url-1 = {https://doi.org/10.1016/0370-2693(88)90577-1}}

@inproceedings{Weinberg:1976xy,
	author = {Weinberg, Steven},
	booktitle = {{14th International School of Subnuclear Physics: Understanding the Fundamental Constitutents of Matter Erice, Italy, July 23-August 8, 1976}},
	doi = {10.1007/978-1-4684-0931-4_1},
	owner = {trmorris},
	pages = {1},
	reportnumber = {HUTP-76-160},
	slaccitation = {%%CITATION = HUTP-76-160;%%},
	timestamp = {2018.12.05},
	title = {{Critical Phenomena for Field Theorists}},
	url = {https://www.quantamagazine.org/why-an-old-theory-of-everything-is-gaining-new-life-20180108},
	year = {1976},
	Bdsk-Url-1 = {https://www.quantamagazine.org/why-an-old-theory-of-everything-is-gaining-new-life-20180108},
	Bdsk-Url-2 = {https://doi.org/10.1007/978-1-4684-0931-4_1}}

@phdthesis{Morgan1991,
	author = {Morgan, D.},
	date-modified = {2020-06-18 19:23:39 +0100},
	owner = {trmorris},
	school = {University of Texas, Austin},
	timestamp = {2018.12.05},
	title = {Quartet: Baryogenesis, Bubbles of False Vacuum, Quantum Black Holes, and the Renormalization Group},
	year = {1991}}

@article{Falls:2017nnu,
	archiveprefix = {arXiv},
	author = {Falls, Kevin and Morris, Tim R.},
	doi = {10.1103/PhysRevD.97.065013},
	eprint = {1712.05011},
	journal = {Phys. Rev.},
	number = {6},
	owner = {trmorris},
	pages = {065013},
	primaryclass = {hep-th},
	slaccitation = {%%CITATION = ARXIV:1712.05011;%%},
	timestamp = {2018.12.05},
	title = {{Conformal anomaly from gauge fields without gauge fixing}},
	volume = {D97},
	year = {2018},
	Bdsk-Url-1 = {https://doi.org/10.1103/PhysRevD.97.065013}}

@article{Rosten:2018cyr,
	archiveprefix = {arXiv},
	author = {Rosten, Oliver J.},
	eprint = {1807.02181},
	owner = {trmorris},
	primaryclass = {hep-th},
	slaccitation = {%%CITATION = ARXIV:1807.02181;%%},
	timestamp = {2018.12.05},
	title = {{The Conformal Anomaly and a new Exact RG}},
	year = {2018}}

@article{Rosten:2010pc,
	archiveprefix = {arXiv},
	author = {Rosten, Oliver J.},
	doi = {10.1088/1751-8113/44/19/195401},
	eprint = {1010.1530},
	journal = {J. Phys.},
	owner = {trmorris},
	pages = {195401},
	primaryclass = {hep-th},
	slaccitation = {%%CITATION = ARXIV:1010.1530;%%},
	timestamp = {2018.12.05},
	title = {{Equivalent Fixed-Points in the Effective Average Action Formalism}},
	volume = {A44},
	year = {2011},
	Bdsk-Url-1 = {https://doi.org/10.1088/1751-8113/44/19/195401}}

@article{Wetterich:2016ewc,
	archiveprefix = {arXiv},
	author = {Wetterich, C.},
	doi = {10.1016/j.nuclphysb.2018.04.020},
	eprint = {1607.02989},
	journal = {Nucl. Phys.},
	owner = {trmorris},
	pages = {262-282},
	primaryclass = {hep-th},
	slaccitation = {%%CITATION = ARXIV:1607.02989;%%},
	timestamp = {2018.12.08},
	title = {{Gauge invariant flow equation}},
	volume = {B931},
	year = {2018},
	Bdsk-Url-1 = {https://doi.org/10.1016/j.nuclphysb.2018.04.020}}

@article{Wetterich:2017aoy,
	archiveprefix = {arXiv},
	author = {Wetterich, C.},
	doi = {10.1016/j.nuclphysb.2018.07.002},
	eprint = {1710.02494},
	journal = {Nucl. Phys.},
	owner = {trmorris},
	pages = {265-316},
	primaryclass = {hep-th},
	slaccitation = {%%CITATION = ARXIV:1710.02494;%%},
	timestamp = {2018.12.08},
	title = {{Gauge-invariant fields and flow equations for Yang-Mills theories}},
	volume = {B934},
	year = {2018},
	Bdsk-Url-1 = {https://doi.org/10.1016/j.nuclphysb.2018.07.002}}

@article{Sonoda:2007dj,
	archiveprefix = {arXiv},
	author = {Sonoda, Hidenori},
	doi = {10.1088/1751-8113/40/31/034},
	eprint = {hep-th/0703167},
	journal = {J. Phys.},
	owner = {trmorris},
	pages = {9675-9690},
	primaryclass = {HEP-TH},
	reportnumber = {KOBE-TH-06-06},
	slaccitation = {%%CITATION = HEP-TH/0703167;%%},
	timestamp = {2018.12.21},
	title = {{On the construction of QED using ERG}},
	volume = {A40},
	year = {2007},
	Bdsk-Url-1 = {https://doi.org/10.1088/1751-8113/40/31/034}}

@article{Morris:2018upm,
	archiveprefix = {arXiv},
	author = {Morris, Tim R.},
	doi = {10.1142/S021827181847003X},
	eprint = {1804.03834},
	journal = {Int. J. Mod. Phys.},
	number = {14},
	owner = {trmorris},
	pages = {1847003},
	primaryclass = {hep-th},
	slaccitation = {%%CITATION = ARXIV:1804.03834;%%},
	timestamp = {2019.02.15},
	title = {{Perturbatively renormalizable quantum gravity}},
	volume = {D27},
	year = {2018},
	Bdsk-Url-1 = {https://doi.org/10.1142/S021827181847003X}}

@article{Morris:toappear,
	author = {Morris, Tim R.},
	owner = {trmorris},
	timestamp = {2019.02.15},
	title = {to appear}}

@article{Bonini:1993sj,
	archiveprefix = {arXiv},
	author = {Bonini, M. and D'Attanasio, M. and Marchesini, G.},
	doi = {10.1016/0550-3213(94)90335-2},
	eprint = {hep-th/9312114},
	journal = {Nucl. Phys.},
	owner = {trmorris},
	pages = {429-455},
	primaryclass = {hep-th},
	reportnumber = {UPRF-93-388},
	slaccitation = {%%CITATION = HEP-TH/9312114;%%},
	timestamp = {2019.04.04},
	title = {{Renormalization group flow for SU(2) Yang-Mills theory and gauge invariance}},
	volume = {B421},
	year = {1994},
	Bdsk-Url-1 = {https://doi.org/10.1016/0550-3213(94)90335-2}}

@article{Ellwanger:1995qf,
	archiveprefix = {arXiv},
	author = {Ellwanger, Ulrich and Hirsch, Manfred and Weber, Axel},
	doi = {10.1007/s002880050073},
	eprint = {hep-th/9506019},
	journal = {Z. Phys.},
	owner = {trmorris},
	pages = {687-698},
	primaryclass = {hep-th},
	reportnumber = {LPTHE-ORSAY-95-39},
	slaccitation = {%%CITATION = HEP-TH/9506019;%%},
	timestamp = {2019.04.04},
	title = {{Flow equations for the relevant part of the pure Yang-Mills action}},
	volume = {C69},
	year = {1996},
	Bdsk-Url-1 = {https://doi.org/10.1007/s002880050073}}

@article{Freire:2000bq,
	archiveprefix = {arXiv},
	author = {Freire, Filipe and Litim, Daniel F. and Pawlowski, Jan M.},
	doi = {10.1016/S0370-2693(00)01231-4},
	eprint = {hep-th/0009110},
	journal = {Phys. Lett.},
	owner = {trmorris},
	pages = {256-262},
	primaryclass = {hep-th},
	reportnumber = {DIAS-STP-00-17, HD-THEP-00-45},
	slaccitation = {%%CITATION = HEP-TH/0009110;%%},
	timestamp = {2019.04.04},
	title = {{Gauge invariance and background field formalism in the exact renormalization group}},
	volume = {B495},
	year = {2000},
	Bdsk-Url-1 = {https://doi.org/10.1016/S0370-2693(00)01231-4}}

@article{Corell:2018yil,
	archiveprefix = {arXiv},
	author = {Corell, Lukas and Cyrol, Anton K. and Mitter, Mario and Pawlowski, Jan M. and Strodthoff, Nils},
	doi = {10.21468/SciPostPhys.5.6.066},
	eprint = {1803.10092},
	journal = {SciPost Phys.},
	owner = {trmorris},
	pages = {066},
	primaryclass = {hep-ph},
	slaccitation = {%%CITATION = ARXIV:1803.10092;%%},
	timestamp = {2019.04.04},
	title = {{Correlation functions of three-dimensional Yang-Mills theory from the FRG}},
	volume = {5},
	year = {2018},
	Bdsk-Url-1 = {https://doi.org/10.21468/SciPostPhys.5.6.066}}

@article{Cyrol:2017ewj,
	archiveprefix = {arXiv},
	author = {Cyrol, Anton K. and Mitter, Mario and Pawlowski, Jan M. and Strodthoff, Nils},
	doi = {10.1103/PhysRevD.97.054006},
	eprint = {1706.06326},
	journal = {Phys. Rev.},
	number = {5},
	owner = {trmorris},
	pages = {054006},
	primaryclass = {hep-ph},
	slaccitation = {%%CITATION = ARXIV:1706.06326;%%},
	timestamp = {2019.04.04},
	title = {{Nonperturbative quark, gluon, and meson correlators of unquenched QCD}},
	volume = {D97},
	year = {2018},
	Bdsk-Url-1 = {https://doi.org/10.1103/PhysRevD.97.054006}}

@article{Cyrol:2016tym,
	archiveprefix = {arXiv},
	author = {Cyrol, Anton K. and Fister, Leonard and Mitter, Mario and Pawlowski, Jan M. and Strodthoff, Nils},
	doi = {10.1103/PhysRevD.94.054005},
	eprint = {1605.01856},
	journal = {Phys. Rev.},
	number = {5},
	owner = {trmorris},
	pages = {054005},
	primaryclass = {hep-ph},
	slaccitation = {%%CITATION = ARXIV:1605.01856;%%},
	timestamp = {2019.04.04},
	title = {{Landau gauge Yang-Mills correlation functions}},
	volume = {D94},
	year = {2016},
	Bdsk-Url-1 = {https://doi.org/10.1103/PhysRevD.94.054005}}

@article{Mitter:2014wpa,
	archiveprefix = {arXiv},
	author = {Mitter, Mario and Pawlowski, Jan M. and Strodthoff, Nils},
	doi = {10.1103/PhysRevD.91.054035},
	eprint = {1411.7978},
	journal = {Phys. Rev.},
	owner = {trmorris},
	pages = {054035},
	primaryclass = {hep-ph},
	slaccitation = {%%CITATION = ARXIV:1411.7978;%%},
	timestamp = {2019.04.04},
	title = {{Chiral symmetry breaking in continuum QCD}},
	volume = {D91},
	year = {2015},
	Bdsk-Url-1 = {https://doi.org/10.1103/PhysRevD.91.054035}}

@article{Fischer:2008uz,
	archiveprefix = {arXiv},
	author = {Fischer, Christian S. and Maas, Axel and Pawlowski, Jan M.},
	doi = {10.1016/j.aop.2009.07.009},
	eprint = {0810.1987},
	journal = {Annals Phys.},
	owner = {trmorris},
	pages = {2408-2437},
	primaryclass = {hep-ph},
	slaccitation = {%%CITATION = ARXIV:0810.1987;%%},
	timestamp = {2019.04.04},
	title = {{On the infrared behavior of Landau gauge Yang-Mills theory}},
	volume = {324},
	year = {2009},
	Bdsk-Url-1 = {https://doi.org/10.1016/j.aop.2009.07.009}}

@article{Bonini:1996bk,
	archiveprefix = {arXiv},
	author = {Bonini, M. and Marchesini, G. and Simionato, M.},
	doi = {10.1016/S0550-3213(96)00571-8},
	eprint = {hep-th/9604114},
	journal = {Nucl. Phys.},
	owner = {trmorris},
	pages = {475-494},
	primaryclass = {hep-th},
	reportnumber = {UPRF-96-464, IFUM-525-FT},
	slaccitation = {%%CITATION = HEP-TH/9604114;%%},
	timestamp = {2019.04.16},
	title = {{Beta function and infrared renormalons in the exact Wilson renormalization group in Yang-Mills theory}},
	volume = {B483},
	year = {1997},
	Bdsk-Url-1 = {https://doi.org/10.1016/S0550-3213(96)00571-8}}

@article{Bevan:2014iga,
	archiveprefix = {arXiv},
	author = {Bevan, A. J. and others},
	collaboration = {BaBar, Belle},
	doi = {10.1140/epjc/s10052-014-3026-9},
	eprint = {1406.6311},
	journal = {Eur. Phys. J.},
	owner = {trmorris},
	pages = {3026},
	primaryclass = {hep-ex},
	reportnumber = {SLAC-PUB-15968, KEK-PREPRINT-2014-3, FERMILAB-PUB-14-262-T},
	slaccitation = {%%CITATION = ARXIV:1406.6311;%%},
	timestamp = {2019.05.01},
	title = {{The Physics of the B Factories}},
	volume = {C74},
	year = {2014},
	Bdsk-Url-1 = {https://doi.org/10.1140/epjc/s10052-014-3026-9}}

@article{Shomer:2007vq,
	archiveprefix = {arXiv},
	author = {Shomer, Assaf},
	eprint = {0709.3555},
	owner = {trmorris},
	primaryclass = {hep-th},
	slaccitation = {%%CITATION = ARXIV:0709.3555;%%},
	timestamp = {2019.05.01},
	title = {{A Pedagogical explanation for the non-renormalizability of gravity}},
	year = {2007}}

@article{Wilson:1974sk,
	author = {Wilson, Kenneth G.},
	doi = {10.1103/PhysRevD.10.2445},
	journal = {Phys. Rev.},
	note = {[,319(1974)]},
	owner = {trmorris},
	pages = {2445-2459},
	reportnumber = {CLNS-262},
	slaccitation = {%%CITATION = PHRVA,D10,2445;%%},
	timestamp = {2019.05.16},
	title = {{Confinement of Quarks}},
	volume = {D10},
	year = {1974},
	Bdsk-Url-1 = {https://doi.org/10.1103/PhysRevD.10.2445}}

@article{Aharony:1998tt,
	__markedentry = {[trmorris:]},
	archiveprefix = {arXiv},
	author = {Aharony, Ofer and Banks, Tom},
	doi = {10.1088/1126-6708/1999/03/016},
	eprint = {hep-th/9812237},
	journal = {JHEP},
	owner = {trmorris},
	pages = {016},
	primaryclass = {hep-th},
	reportnumber = {RU-98-52},
	slaccitation = {%%CITATION = HEP-TH/9812237;%%},
	timestamp = {2019.05.16},
	title = {{Note on the quantum mechanics of M theory}},
	volume = {03},
	year = {1999},
	Bdsk-Url-1 = {https://doi.org/10.1088/1126-6708/1999/03/016}}

@article{Matt1,
	author = {Kellett, Matthew P and Morris, Tim R.},
	date-added = {2017-10-04 10:43:33 +0000},
	date-modified = {2017-10-04 10:53:47 +0000},
	owner = {trmorris},
	timestamp = {2019.06.24},
	title = {in preparation},
	year = {2017}}

@article{morr4,
	author = {Tim R. Morris},
	owner = {trmorris},
	timestamp = {2019.12.02},
	title = {To appear}}

@article{Morris:2018zgy,
	__markedentry = {[trmorris:6]},
	archiveprefix = {arXiv},
	author = {Morris, Tim R. and Percacci, Roberto},
	doi = {10.1103/PhysRevD.99.105007},
	eprint = {1810.09824},
	journal = {Phys. Rev.},
	number = {10},
	owner = {trmorris},
	pages = {105007},
	primaryclass = {hep-th},
	slaccitation = {%%CITATION = ARXIV:1810.09824;%%},
	timestamp = {2019.12.09},
	title = {{Trace anomaly and infrared cutoffs}},
	volume = {D99},
	year = {2019},
	Bdsk-Url-1 = {https://doi.org/10.1103/PhysRevD.99.105007}}

@article{Falls:2017lst,
    author = "Falls, Kevin and King, Callum R. and Litim, Daniel F. and Nikolakopoulos, Kostas and Rahmede, Christoph",
    title = "{Asymptotic safety of quantum gravity beyond Ricci scalars}",
    eprint = "1801.00162",
    archivePrefix = "arXiv",
    primaryClass = "hep-th",
    doi = "10.1103/PhysRevD.97.086006",
    journal = "Phys. Rev. D",
    volume = "97",
    number = "8",
    pages = "086006",
    year = "2018"
}
@article{Christiansen:2017bsy,
    author = "Christiansen, Nicolai and Falls, Kevin and Pawlowski, Jan M. and Reichert, Manuel",
    title = "{Curvature dependence of quantum gravity}",
    eprint = "1711.09259",
    archivePrefix = "arXiv",
    primaryClass = "hep-th",
    doi = "10.1103/PhysRevD.97.046007",
    journal = "Phys. Rev. D",
    volume = "97",
    number = "4",
    pages = "046007",
    year = "2018"
}
@article{Alkofer:2018baq,
    author = "Alkofer, Nat\'alia",
    title = "{Asymptotically safe $f(R)$-gravity coupled to matter II: Global solutions}",
    eprint = "1809.06162",
    archivePrefix = "arXiv",
    primaryClass = "hep-th",
    doi = "10.1016/j.physletb.2018.12.061",
    journal = "Phys. Lett. B",
    volume = "789",
    pages = "480--487",
    year = "2019"
}
@article{Burger:2019upn,
    author = {B\"urger, Benjamin and Pawlowski, Jan M. and Reichert, Manuel and Schaefer, Bernd-Jochen},
    title = "{Curvature dependence of quantum gravity with scalars}",
    eprint = "1912.01624",
    archivePrefix = "arXiv",
    primaryClass = "hep-th",
    month = "12",
    year = "2019"
}
@article{DeBrito:2018hur,
    author = "De Brito, Gustavo P. and Ohta, Nobuyoshi and Pereira, Antonio D. and Tomaz, Anderson A. and Yamada, Masatoshi",
    title = "{Asymptotic safety and field parametrization dependence in the $f(R)$ truncation}",
    eprint = "1805.09656",
    archivePrefix = "arXiv",
    primaryClass = "hep-th",
    doi = "10.1103/PhysRevD.98.026027",
    journal = "Phys. Rev. D",
    volume = "98",
    number = "2",
    pages = "026027",
    year = "2018"
}
@book{Hawking:1979ig,
    author = "Hawking, S. W. and Israel, W.",
    title = "{General Relativity}: {An Einstein Centenary Survey}",
    isbn = "978-0-521-29928-2",
    publisher = "Univ. Pr.",
    address = "Cambridge, UK",
    year = "1979"
}
@article{Falls:2018ylp,
    author = {Falls, Kevin G. and Litim, Daniel F. and Schr\"oder, Jan},
    title = "{Aspects of asymptotic safety for quantum gravity}",
    eprint = "1810.08550",
    archivePrefix = "arXiv",
    primaryClass = "gr-qc",
    doi = "10.1103/PhysRevD.99.126015",
    journal = "Phys. Rev. D",
    volume = "99",
    number = "12",
    pages = "126015",
    year = "2019"
}
@article{Kluth:2019vkg,
    author = "Kluth, Yannick and Litim, Daniel F.",
    title = "{Heat kernel coefficients on the sphere in any dimension}",
    eprint = "1910.00543",
    archivePrefix = "arXiv",
    primaryClass = "hep-th",
    doi = "10.1140/epjc/s10052-020-7784-2",
    journal = "Eur. Phys. J. C",
    volume = "80",
    number = "3",
    pages = "269",
    year = "2020"
}
@article{Mitchell:2021qjr,
    author = "Mitchell, Alex and Morris, Tim R. and Stulga, Dalius",
    title = "{Provable properties of asymptotic safety in f(R) approximation}",
    eprint = "2111.05067",
    archivePrefix = "arXiv",
    primaryClass = "hep-th",
    doi = "10.1007/JHEP01(2022)041",
    journal = "JHEP",
    volume = "01",
    pages = "041",
    year = "2022"
}
@article{Bonanno:2020bil,
    author = "Bonanno, Alfio and Eichhorn, Astrid and Gies, Holger and Pawlowski, Jan M. and Percacci, Roberto and Reuter, Martin and Saueressig, Frank and Vacca, Gian Paolo",
    title = "{Critical reflections on asymptotically safe gravity}",
    eprint = "2004.06810",
    archivePrefix = "arXiv",
    primaryClass = "gr-qc",
    doi = "10.3389/fphy.2020.00269",
    journal = "Front. in Phys.",
    volume = "8",
    pages = "269",
    year = "2020"
}
@article{Souma:1999at,
    author = "Souma, Wataru",
    title = "{Nontrivial ultraviolet fixed point in quantum gravity}",
    eprint = "hep-th/9907027",
    archivePrefix = "arXiv",
    reportNumber = "KUCP-0128",
    doi = "10.1143/PTP.102.181",
    journal = "Prog. Theor. Phys.",
    volume = "102",
    pages = "181--195",
    year = "1999"
}
@article{Pawlowski:2020qer,
    author = "Pawlowski, Jan M. and Reichert, Manuel",
    title = "{Quantum Gravity: A Fluctuating Point of View}",
    eprint = "2007.10353",
    archivePrefix = "arXiv",
    primaryClass = "hep-th",
    doi = "10.3389/fphy.2020.551848",
    journal = "Front. in Phys.",
    volume = "8",
    pages = "551848",
    year = "2021"
}
@article{Mandric:2022dte,
    author = "Mandric, Vlad-Mihai and Morris, Tim R.",
    title = "{Properties of a proposed background independent exact renormalization group}",
    eprint = "2210.00492",
    archivePrefix = "arXiv",
    primaryClass = "hep-th",
    month = "10",
    year = "2022"
}
@article{Kluth:2020bdv,
    author = "Kluth, Yannick and Litim, Daniel F.",
    title = "{Fixed Points of Quantum Gravity and the Dimensionality of the UV Critical Surface}",
    eprint = "2008.09181",
    archivePrefix = "arXiv",
    primaryClass = "hep-th",
    month = "8",
    year = "2020"
}

@article{Becker:2021pwo,
    author = "Becker, Maximilian and Reuter, Martin",
    title = "{Background independent field quantization with sequences of gravity-coupled approximants. II. Metric fluctuations}",
    eprint = "2109.09496",
    archivePrefix = "arXiv",
    primaryClass = "hep-th",
    doi = "10.1103/PhysRevD.104.125008",
    journal = "Phys. Rev. D",
    volume = "104",
    number = "12",
    pages = "125008",
    year = "2021"
}

%\printbibliography

%\newpage
%\printbibliography
%\printindex
\end{document}